\documentclass[12pt, a4paper, twoside, fleqn]{report}
\usepackage{amsfonts}
\usepackage{epsf, amsmath, commands}
\usepackage{fancyheadings}
\usepackage{exscale}
\usepackage{times}
\usepackage[backref]{hyperref}

\font\trm=cmr5 at 8 pt 

\def \ed{\end{document}}
\def \half{\frac{1}{2}}         
\renewcommand{\(}{\left(}
\renewcommand{\)}{\right)}

\newcommand{\pa}{\partial}
\newcommand{\non}{\nonumber}

\newcommand{\eps}{\epsilon}
\newcommand{\lam}{\lambda}

\newcommand{\df}[2]{ \frac{\partial {#1}}{\partial {#2}} }
\newcommand{\dff}[2]{ \frac{\partial^2 {#1}}{\partial {#2}^2} }

\newcommand{\I}{\mbox{\trm i}}

\newcommand{\wh}{\widehat}
\def \d {\partial}

\def \rw{{\cal\char'131}}
\def \mz{\widehat{\cal\char'131}}
\def \= {\;&=\;}

\let\abs=\envert

\begin{document}
\pagestyle{empty}
\topmargin2cm
\begin{center}
{\Huge\sf The Numerical Evolution\\of\\Neutron Star Oscillations\\[2cm]}
{\LARGE\sc Dissertation}\\[1cm]
{\Large zur Erlangung des Grades eines\\
Doktors der Naturwissenschaften\\[1cm]
der Fakult\"at f\"ur Physik\\
der Eberhard-Karls-Universit\"at zu T\"ubingen\\[1.5cm]
vorgelegt von\\[5mm]
{\LARGE \bf Johannes O. Ruoff}\\[5mm]aus Sindelfingen\\[2cm]
\bf 2000}
\end{center}

\newpage
\thispagestyle{empty}
\strut
\vfill
{\Large
\begin{tabular}{ll}
Tag der m\"undlichen Pr\"ufung:~~~& 28.~2.~2000   \\[1.0ex]
Dekan:& Prof.~Dr.~Klaus Werner \\[1.0ex]
1. Berichterstatter:& Prof.~Dr.~Hanns Ruder \\[1.0ex]
2. Berichterstatter:& Prof.~Dr.~Pablo Laguna \\[1.0ex]
\end{tabular}
}
\thispagestyle{empty}
\vspace*{10cm}
\newpage
\topmargin1.5cm
\renewcommand{\textheight}{22cm}
\topmargin-1cm
\newpage
\begin{center}
{\Huge\bf Abstract\\[1cm]}
\end{center}
{\small
The present work investigates the numerical evolution of linearized
oscillations of non-rotating, spherically symmetric neutron stars within
the framework of general relativity.

We derive the appropriate equations using the (3+1)-formalism. We
first focus on the evolution of radial oscillations, which do not emit
gravitational waves. The associated system of equations being quite
simple, we demonstrate how to handle a numerical instability that also
occurs in the non-radial case, when the stellar model is constructed
based on a realistic equation of state. We devise a coordinate
transformation that not only removes this instability but also provides
much more accurate results. For comparison reasons, we compute the
eigenfrequencies of the radial modes with an eigenvalue code and
thereby confirm the results of V\"ath \& Chanmugam, which differ from
previous calculations that were performed by Glass \& Lindblom.

The main part deals with the evolution of non-radial oscillations
($l\ge2$) of neutron stars. Here, we compare different formulations of
the equations and discuss how they have to be numerically dealt with
in order to avoid instabilities at the origin. We present results for
various polytropic stellar models and different initial data. They
show that the quasi-normal modes of the star, such as the $f$-, the
$p$-, and the $w$-modes, can, indeed, be excited by suitable initial
data. However, the excitation strength of the $w$-modes strongly
depends on the chosen initial data. For some initial data the
occurrence of the $w$-modes can be totally suppressed.

For ultra-compact models we find the interesting feature that the
first ring-down phase of the wave signal cannot be associated with any
of the known quasi-normal modes that belong to the star itself; the
frequency and damping time rather correspond to the first quasi-normal
mode of an equal mass black hole.

When switching to realistic equations of state, we find that we face
the same numerical problems as in the radial case. Here, too, we can
get rid of them by means of the same coordinate transformation inside
the star, but things are more complicated because the fluid equation
is coupled to the metric equations, which propagate the gravitational
waves. For those equations the transformation is not defined,
therefore we have to interpolate between the different grids. For
polytropic equations of state this can give rise to a quite strong
violation of the Hamiltonian constraint, which can spoil the resulting
wave signal. However, for realistic equations of state, and it is only
for those that the transformation is necessary, this does not
happen and the equations can be integrated in a stable way and also
provide quite accurate results.

In the last part of this thesis we consider a physical mechanism for
exciting oscillations of neutron stars. We use the time dependent
gravitational field of a small point mass $\mu$ that orbits the
neutron star to induce stellar oscillations. Hereby, we assume $\mu$
to be much smaller than the mass $M$ of the neutron star. In this
so-called particle limit, the gravitational field of the moving
particle is considered to be a perturbation of the background field of
the neutron star. With this particle we have a physical means which
removes the arbitrariness in choosing the initial data.

However, even with the presence of the particle, there is still too
much freedom in constructing the initial data, which is due to the
fact that in addition to the field of the particle we always can
superpose any arbitrary amount of gravitational waves. Our task is to
minimize this additional gravitational-wave content and find initial
data which correspond to the pure gravitational field of the particle.
By looking at the flat space case we can construct analytic initial
data that satisfy the above requirements. Those then will serve as a
good approximation for the ``real'' initial data in the curved
spacetime of the neutron star. By sampling various orbital parameters
of the particle we show that in general the particle is not able to
excite any $w$-modes. It is only for speeds very close to the speed of
light that the $w$-mode is a significant part of the wave signal.
This result indicates that it is rather improbable that any physical
mechanism which can be simulated by an orbiting particle can excite
the $w$-modes of neutron stars in a significant manner.\\}

\vspace*{10cm}
\newpage
\pagestyle{fancy}
\renewcommand{\chaptermark}[1]{\markboth{\chaptername\ \thechapter.\
    #1}{}}
\renewcommand{\sectionmark}[1]{\markright{\thesection.\ #1}{}}
\setlength{\headrulewidth}{1pt}
\lhead[\rm\thepage]{\sl\rightmark}
\rhead[\sl\leftmark]{\rm\thepage}
\cfoot{}

\setcounter{page}{1}
\addtocontents{toc}{\protect\markright{Contents}}
\tableofcontents
\cleardoublepage
\newpage
\chapter{Introduction}

The investigation of stellar oscillations has a quite long history,
and is a whole astrophysical branch on its own. The evaluation of the
oscillation modes of stellar objects has helped to reveal a lot of
information about their interior structure. With the use of
asteroseismological methods we now have a quite detailed understanding
of the physics of a whole range of stellar objects, be it our own sun
or a white dwarf somewhere in our galaxy.

Most stellar objects can be adequately described by Newtonian theory,
however, for very compact objects such as neutron stars, the effects
of general relativity cannot be longer neglected. In fact, they are
very important, for now, stellar oscillations will be associated with
the emission of gravitational waves, which can carry important
information about the physics inside those compact objects. These
waves, when detected, will open a totally new observational window for
astrophysicists. Gravitational waves are not affected by events at the
surface of the star, which is what happens to electromagnetic
radiation and they also remain practically unaffected by any kind of
matter while travelling through space. Thus, they carry ``clean''
information of the physical properties of the neutron stars.

It is well known that lowest gravitational-wave multipole is the
quadrupole radiation, hence the radial and dipole oscillations do not
emit gravitational waves. From the gravitational-wave astronomical
point the latter are therefore quite uninteresting, however, they
could make themselves visible through tiny undulations in the
electromagnetic radiation signal of the neutron stars.

The study of the non-radial oscillations modes of non-rotating neutron
stars within the framework of general relativity was initiated in 1967
by a series of papers by Thorne and several coauthors
\cite{TC67,Th68,PT69,Th69a,Th69b}. In the following decades a lot of
authors made endeavors to study the relativistic oscillation
spectrum of neutron stars, which turned out to be particularly
rich. 

From the general Newtonian theory of non-radial linear oscillations it
follows that the oscillations of a spherical stellar object can be
divided into two classes according to their transformation behavior
under space reflection. One class consists of even parity or polar
modes; the odd parity or axial modes belong to the other class.

If the neutron star is modeled with a perfect fluid then from
Newtonian theory it follows that any non-radial oscillations such as
the $f$undamental mode \cite{Det75a}, the $p$ressure modes, and the
$g$ravitational modes \cite{MvHS83}, must belong to the polar
class. This means that for Newtonian perfect fluid stellar models
there are no axial modes at all. It is only with the inclusion of a
solid crust that there can exist axial $t$orsional modes.

In the first 20 years after the fundamental paper of Thorne, it was
common belief that the oscillation spectra of relativistic neutron
stars would not differ much from Newtonian stars. The only new effect
of general relativity was thought to be the radiation-damping of the
oscillation modes due to the emission of gravitational waves.
Therefore the main focus was to compute the frequencies and damping
times of those polar modes which also exist in the Newtonian theory,
but now within the framework of general relativity. The only axial
modes under consideration were the $t$-modes of neutron stars with a
solid crust \cite{ST83,MHvHB85,Finn90}.

It was only in 1988 that another family of polar modes was found by
Kojima \cite{Koj88}, whose existence has already been anticipated two
years earlier in a toy model by Kokkotas \& Schutz \cite{KS86}. Those
new modes, which were termed $w$ave-modes by Kokkotas \& Schutz
\cite{KS92}, are of purely relativistic origin and do not have a
Newtonian counterpart. They are predominantly metric oscillations and
couple only weakly to the motion of the matter.

Around the same time Chandrasekhar \& Ferrari \cite{CF91b} showed that
for ultra-relativistic stellar models, there exists also a family of
axial perturbations, which do not couple to the matter at all. This is
because the spacetime curvature inside the star can be so
strong that it can trap impinging gravitational waves. Those
``trapped'' modes are quite long-lived since they correspond to
quasi-bound states inside the gravitational potential of the neutron
star, which only slowly leak out.

A little later, it was found by Leins et al.~\cite{LNS93} that the
polar $w$-modes are split into two branches, one with only a few
strongly damped modes (called $w_{II}$- or interface modes), and the
other one with an infinite number of modes with increasing frequency
and damping times.

In 1994 Kokkotas \cite{Kok94} showed that for less relativistic stars,
which do not possess any trapped modes, there also exists a family of
axial modes, which are quite similar to the polar $w$-modes. Finally,
in 1996 Andersson et al.~\cite{And96,AKK96} established that all three
kinds of gravitational-wave modes (the $w$-modes, the $w_{II}$-modes
and the trapped modes) exist for both the polar and the axial cases,
the trapped modes, however, only for ultra-relativistic star.

This rich set of various pulsation modes having been found, it then
was natural to ask whether they would indeed be excited in a real
pulsation scenario. Allen et al.~\cite{AAKS98} took a first step in
answering this question by numerically integrating the evolution
equations for several different sets of initial data. By looking at
the emitted gravitational radiation, they indeed found that both the
fluid $f$- and $p$-modes and the gravitational $w$-modes can be
excited. However, the excitation strength depends strongly on the
choice of initial data, which, of course, were constructed ad hoc and
had no astrophysical meaning whatsoever.

Andersson \& Kokkotas \cite{AK98b} have shown that, once the
parameters of the $f$- and the first $w$-mode are known, it is in
principle possible to obtain the mass and the radius of a neutron star
quite accurately. Those quantities could then be used to determine the
equation of state for the neutron star matter. Because of the still
quite poor understanding of the physics at the subnuclear density
level, there exists a plethora of different equations of state in the
literature, all based on different physical models of the nuclear
interaction. But once the mass and radius of a neutron star are known
with a certain accuracy, it is possible to rule out those equations of
state that predict a different radius for a given mass, or vice
versa. And this in turn would hint at the physics that governs the
behavior of the neutron star matter in the center of the star.

It is therefore of particular interest whether or not the $w$-modes
will be excited in a real astrophysical scenario such as the collapse
of a progenitor star to a neutron star. Whereas the $f$- and some
fluid $p$-modes will always be present in an wave signal, it is not
clear if this is also the case for the $w$-modes. The numerical
experiments show that it is possible to construct initial data which
do not excite any $w$-modes at all.

To satisfactorily answer the question what a real signal would look
like, one would have to follow the appropriate scenario (e.g.~the
collapse of a burned out star core) in full detail from the point on
where relativistic effects become important. Unfortunately, this is,
for the time being, still impossible. This is mainly because one would
have to use the full set of the nonlinear Einstein equations together
with the whole machinery of nuclear physics to simulate a core
collapse. However, when the newborn neutron star has almost settled,
the study of stellar pulsations can be done within the linearized
theory, which makes it tractable. But even there one usually restricts
oneself to very simple neutron star models. In our case, we will
neglect the spin of the star, which makes the equations much simpler,
but also prevents us from studying the effects of rotation, which
might be very important.

There are no unstable non-radial modes for non-rotating stars
as long as the Schwarzschild discriminant
\begin{align*}
        S(r) \= \frac{dp}{dr} - \(\df{p}{\eps}\)_s\frac{d\eps}{dr}
\end{align*}
is positive \cite{DI73,Lind95}. For negative $S$ the oscillations become
unstable with respect to convection. In this thesis we always use
barotropic equations of state, for which it is $S \equiv 0$, therefore
all non-radial modes will be stable.

However, things can change dramatically when rotation is included.
There will be always some modes which will undergo the CFS-instability
\cite{Chandra70a,Chandra70b,FS78}, and recently a whole set of 
unstable modes ($r$-modes) has been discovered. They might be
responsible for slowing down rapidly rotating neutron stars
\cite{Andersson98}. For a recent review of the stability properties of
rotating neutron stars, see \cite{Sterg98}. Other recent reviews on
oscillations of neutron stars and black holes can be found in
\cite{KS99,Nollert99}.

The thesis is organized as follows:

In chapter 2 we will derive the general form of the perturbation
equations using the (3+1)-split.

In chapter 3 we will specialize on the radial case, where we
investigate both the evolution and the eigenvalue problem. We discuss
the occurrence of a numerical instability, which occurs both in the
radial and non-radial cases when the neutron star is modeled with a
realistic equation of state. We also present a remedy which not only
removes the instability but also automatically increases the accuracy
of the evolution as compared to the old set of equations.

In chapter 4 we focus on the non-radial oscillations. We present
various forms of the equations and discuss which of them are the most
suitable set for the numerical evolution. We show how to treat the
boundary conditions in order to obtain stable evolutions for any value
of $l$. Here, too, we demonstrate how to get rid of the same numerical
instability that occurs with the use of realistic equations of state.

In chapter 5 we examine the excitation of neutron star oscillations by
the scattering of a small point mass $\mu$ in the particle limit, where
we assume $\mu$ to be much smaller than the mass $M$ of the neutron
star. The gravitational field of the particle, which follows a
geodesic path in the background metric of the neutron star, can then
be treated as a perturbation of the background. We sample a variety of
orbital parameters to see whether or not the particle is able to
excite the $w$-modes of the neutron star.\\[2cm]

\noindent {\Large \bf Notations and conventions\\}

\begin{itemize}
\item We use Einstein's sum convention. Greek indices run from 0 to 3, Latin
indices from 1 to 3. The time component is the 0-component.
\item The signature of the metric is $(-,+,+,+)$, i.e.~time-like 4-vectors
have negative norm.
\item We use geometric units, i.e.~we set $c = G = 1$.
\item Covariant derivatives with respect to $x^\mu$ are denoted by
$D_\mu$ and partial derivatives $\df{}{x^\mu}$ are sometimes
abbreviated by $\d_\mu$.
\item Derivatives with respect to the radial coordinate $r$ are often
denoted by a prime, and time derivatives by an over-dot.
\item The perturbation variables should be properly denoted by $Q_{lm}(t,r)$,
however, we will often omit the indices $l$ and $m$.
\end{itemize}


\chapter{Linearizing Einstein's equations}
\label{chap2}

Within perturbation theory we first construct a stationary
non-pulsating and non-rotating neutron star model by solving Einstein's
equations, which in this case reduce to a set of coupled ordinary
differential equations.

Having found this (numerically) exact background solution $g_{\mu\nu}$,
we describe the pulsations by ``small'' deviations from the original
metric. Thus, the metric of the perturbed spacetime will be written as
$\bar{g}_{\mu\nu} = g_{\mu\nu} + h_{\mu\nu}$, where $h_{\mu\nu}$ is
considered to be a small perturbation of the background metric $g_{\mu\nu}$. 

To find the relevant equations for $h_{\mu\nu}$, we plug the perturbed
metric $\bar{g}_{\mu\nu}$ into Einstein's equations and neglect all
terms that are quadratic or of a higher power in the $h_{\mu\nu}$. We
will thus obtain a {\it linear} system of partial differential
equations for the $h_{\mu\nu}$, which is much easier to solve than the
full nonlinear set. Still, this set of equations depends on all four
coordinates, which, even with the presently available computer power,
would be too time-consuming to be solved in its fullfledged form.

However, because of the spherical symmetry of the background metric we
can get rid of the angular dependence by expanding the perturbation
equations into spherical tensor harmonics. It is thus possible to
reduce the equations into a (1+1)-dimensional evolution system, which
can be numerically solved on present day PCs.

\section{The unperturbed stellar model}

The simplest model for a neutron star is a non-rotating zero
temperature perfect fluid sphere, whose static spherically symmetric
geometry is given by a line element of the form
\begin{align}\label{ds}
        ds^2 \= -e^{2\nu} dt^2 + e^{2\lam} dr^2 + r^2(d\theta^2
               + \sin^2\theta d\phi^2)\;,
\end{align}
where the two functions $\nu$ and $\lam$ only depend on the radial
coordinate $r$ and have to be determined by the field equations. 
The appropriate energy-momentum tensor is given by
\begin{align}
        T_{\mu\nu} \= \(\eps + p\)u_\mu u_\nu + pg_{\mu\nu}\;,
\end{align}
where $p$ is the pressure, $\eps$ the energy density, and $u_\mu$ the
covariant 4-velocity of the fluid. In the rest frame of the fluid,
which is static, the only non-vanishing component is the time
component $u_0 = -e^{\nu}$, and Einstein's equations $G_{\mu\nu} =
8\pi T_{\mu\nu}$ and the conservation equations $D_\nu T^{\mu\nu} = 0$
yield the following three structure equations for the four unknown
$\lam,\,\nu,\, p$, and $\eps$:
\begin{subequations}
\label{toveqns}
\begin{align}
        \label{tov1}\lam' \= \frac{1 - e^{2\lam}}{2r}
        + 4\pi r e^{2\lam}\epsilon \\
        \label{tov2}\nu' \= \frac{e^{2\lam} - 1}{2r}
        + 4\pi r e^{2\lam}p\\
        \label{tov3}p' \= - \nu'(p + \epsilon)\;.
\end{align}
\end{subequations}
To fully determine this system of equations, an equation of state
\begin{align}
        p \= p(s,n)\\
        \eps \= \eps(s,n)
\end{align}
must be supplemented. We will refer to the equations \eqref{toveqns}
as Tolman-Oppenheimer-Volkov or TOV equations, even if the original
TOV equations are written in a slightly different form.

Throughout this work we always assume the neutron star to have zero
temperature, which is quite reasonable since the pressure inside the
neutron star is mainly maintained by a Fermi gas of degenerate
neutrons. Hence, the specific entropy $s$ can be set to zero, too, and
we can eliminate the baryon density $n$ and obtain a barotropic
equation of state, where the pressure is a function of the energy
density alone:
\begin{align}\label{eos}
p \= p(\eps)\;.
\end{align}
The simplest and therefore quite often used form is
given by a so-called polytropic equation of state
\begin{align}
p \= \kappa\eps^\Gamma\;,
\end{align}
where $\kappa$ and $\Gamma$ are the polytropic constant and polytropic
index, respectively. More realistic equations of state have to include
the microphysics that dictates the interplay between $p$ and $\eps$ on
the nuclear and subnuclear levels for the neutron star matter.
However, the physics under the extreme conditions of the high
pressures that prevail in the center of neutron stars is not yet fully
understood. In the literature there are therefore quite a few
different realistic equations of state, which were calculated based on
various different microscopic models of (sub)nuclear interactions.

It is clear that the zero temperature assumption neglects all kinds of
thermal and viscous effects which can affect the stellar
oscillations. On the one hand it will suppress the existence of the
whole family of $g$-modes, and on the other hand we ignore the effects
of viscosity and internal friction, which would normally damp out any
oscillation. However, the damping times \cite{CL87} are such that
their neglect will have no effect on the numerical evolutions we
are investigating.

If we introduce an additional function $m$, which is related to
$\lam$ by
\begin{align}
        e^{-2\lam} \;&\equiv\; 1 - \frac{2m}{r}\;,
\end{align}
we obtain from \eqref{tov1}
\begin{align}
        \label{tov4}m' \= 4\pi r^2\epsilon\;.
\end{align}
Integration immediately leads to 
\begin{align}
        m(r) \= \int^r_0 4\pi {\tilde r}^2 \eps\,d\tilde r\;,
\end{align}
which makes it clear that $m(r)$ represents the total gravitational
mass enclosed inside the radius $r$.

To obtain the stellar model, we have to integrate the TOV equations
\eqref{toveqns} together with \eqref{eos} from the center up to the
point where the pressure $p$ vanishes. This then defines the surface
$R$ of the star.

From Birkhoff theorem it follows that the exterior vacuum region of
the star is described by the Schwarzschild metric with the mass
parameter $M \equiv m(R)$.

\section{The perturbation equations}

A somewhat more elaborate derivation of the relevant equations can be
found in my Diploma thesis \cite{Ruoff96}. Here, we will only briefly
discuss the necessary steps and not go into very great details. Our
starting point is the ADM-formulation \cite{ADM} or
(3+1)-decomposition of the Einstein equations. In this formulation the
10 field equations are first split into 6 dynamical equations and 4
constraint equations. The dynamical equations, which have second order
time derivatives, are then cast into a set of 12 evolution equations,
which are first order in time. Those are the 6 evolution equations for
the 3-metric $\gamma_{ij}$ of a space-like 3-dimensional hypersurface
$\Sigma$ and another 6 equations for the time development of its
extrinsic curvature $K_{ij}$:
\begin{align}
        \label{dtgij}
        \d_t \gamma_{ij} \= - 2\alpha K_{ij}
        + \beta^k\d_k\gamma_{ij}
        + \gamma_{ki}\d_j\beta^k + \gamma_{kj}\d_i\beta^k\\
\begin{split}\label{dtKij}
        \d_t K_{ij} \= \alpha \left[R_{ij} + KK_{ij} -
        2K_{ik}K^k_{\phantom{i}j} - 8\pi
        \(T_{ij}-\frac{1}{2}T\gamma_{ij}\) \right] \\
        &\quad{} -D_iD_j\alpha + \beta^k\d_kK_{ij}
        + K_{ik}\d_j\beta^k + K_{jk}\d_i\beta^k\;.
\end{split}
\end{align}
Herein, $\alpha$ denotes the lapse function and $\beta^i$ is the shift
vector. The remaining 4 constraint equations, which have to be
satisfied by any physically acceptable initial data, are given by
\begin{align}
        R - K_{ij}K^{ij} + K^2 \= 16\pi \rho \label{hci}\\
        D_iK - D_jK^j_{\phantom{j}i} \= 8\pi j_i \label{mci}\;.
\end{align}
Here, $\rho$ and $j_i$ are the energy density and momentum density
measured by a momentarily stationary observer, whose 4-velocity coincides
with the time-like normal vector $n^\mu$ of the space-like hypersurface:
\begin{align}
        \rho \= T_{\mu\nu}n^\mu n^\nu\\
        j_i  \= T_{i\nu}n^\nu\;.
\end{align}
The scalar equation \eqref{hci} is usually called the {\it Hamiltonian
constraint} and the vector equation \eqref{mci} is called {\it
momentum constraint}. Once satisfied on the initial hypersurface, they
will be automatically preserved throughout the evolution by virtue of
the Bianchi-identities.

As we have already mentioned above, the oscillations of the neutron
star will be described within the framework of perturbation theory,
i.e.~we will treat them as small perturbations around the fixed
background which is given by the unperturbed stellar model. Hence, we
write the perturbed metric $\bar g_{\mu\nu}$ as a sum of the static
background $g_{\mu\nu}$ and the time-dependent perturbations
$h_{\mu\nu}$:
\begin{align}\label{metric}
\bar g_{\mu\nu} \= g_{\mu\nu} + h_{\mu\nu}\,.
\end{align}
Since we are using the (3+1)-decomposition, we have to write down the
perturbed metric \eqref{metric} in terms of lapse function, shift
vector, and 3-metric. Furthermore, we have to deal with the extrinsic
curvature as a dynamical variable. Their respective perturbations will
be denoted by $\alpha$, $\beta^i$, $h_{ij}$, and $k_{ij}$. For the
background metric \eqref{ds}, shift and extrinsic curvature vanish and
the unperturbed lapse function $A$ is given by
\begin{align}
        A \= \sqrt{-g_{00}} \;=\; e^\nu\,.
\end{align}
In addition to the metric perturbations, we have to describe the
perturbations of the energy-momentum tensor $T_{\mu\nu}$. For a perfect
fluid, the only quantities that can be perturbed are energy density
$\eps$, pressure $p$, and 4-velocity $u_\mu$, whose perturbations will
be denoted by $\delta\eps$, $\delta p$, and $\delta u_\mu$,
respectively. For a barotropic equation of state, however,
$\delta\eps$ and $\delta p$ are not independent but are rather related
through
\begin{align}
        \delta p \= \frac{dp}{d\eps}\,\delta\eps\;,
\end{align}
where $\frac{dp}{d\eps} =: C_s^2$ is the square of the sound speed $C_s$
inside the fluid.

Due to the simple form of \eqref{ds}, the evolution
equations for the 3-metric and the extrinsic curvature perturbations
do not become as messy as they usually do in perturbation theory:
\begin{align}\label{hij}
        \d_t h_{ij} \= \beta^k\d_k\gamma_{ij}
        + \gamma_{ki}\d_j\beta^k + \gamma_{kj}\d_i\beta^k
        - 2e^\nu k_{ij}\\
\begin{split}
        \label{kij}
        \d _t k_{ij} \= - \d_i \d_j \alpha
        + \Gamma^k_{\phantom{i}ij}\d _k\alpha
        + \delta \Gamma^k_{\phantom{i}ij}\d _k e^\nu
        + \alpha\left[R_{ij} + 4\pi(p - \eps)\gamma_{ij}\right]\\
        &\quad{} + e^\nu\left[\delta R_{ij} + 4\pi\((p - \eps)h_{ij} +
        \delta\eps(C_s^2 - 1)\gamma_{ij}\) \right].
\end{split}
\end{align}
Herein, $\gamma_{ij}$ denotes the spatial part of \eqref{ds}, its inverse is
$\gamma^{ij}$, and the perturbed Christoffel symbols are defined as
\begin{align}
        \delta \Gamma^k_{\phantom{i}ij} \= \frac{1}{2}\gamma^{km}
        \(D_ih_{mj} + D_jh_{mi} - D_mh_{ij}\)\non\\
        \= \half\gamma^{km}\(\d_ih_{mj} + \d_jh_{mi} - \d_mh_{ij}
        - 2\Gamma^{l}_{\phantom{i}ij} h_{lm}\)\;.
\end{align}
The perturbed Ricci tensor is given by
\begin{align}
        \delta R_{ij} \= 
        D_k\delta\Gamma^k_{\phantom{i}ij}
        - D_j\delta\Gamma^k_{\phantom{i}ik}\non\\
        \= \d_k\delta\Gamma^k_{\phantom{i}ij}
        - \d_j\delta\Gamma^k_{\phantom{i}ik}
        + \Gamma^l_{\phantom{i}ij}\delta\Gamma^k_{\phantom{i}lk}
        + \Gamma^k_{\phantom{i}lk}\delta\Gamma^l_{\phantom{i}ij}
        - \Gamma^l_{\phantom{i}ik}\delta\Gamma^k_{\phantom{i}lj}
        - \Gamma^k_{\phantom{i}lj}\delta\Gamma^l_{\phantom{i}ik}\,.
\end{align}
To first order in the perturbations, the constraints \eqref{hci} and
\eqref{mci} read
\begin{align}\label{hhc}
        \gamma^{ij}\delta R_{ij} - h^{ij}R_{ij} \= 16 \pi \delta \eps\\
        \gamma^{jk}\(\d _i k_{jk} - \d _j k_{ik}
        - \Gamma^l_{\phantom{i}ik} k_{jl} + \Gamma^l_{\phantom{i}jk} k_{il}\)
        \= -8\pi(p + \eps)\delta u_i\,.\label{mmc}
\end{align}
It is interesting to note that it is possible to eliminate
$\delta\eps$ in \eqref{kij} by virtue of the Hamiltonian constraint
\eqref{hhc} and to obtain thus a consistent system of evolution
equations for the metric and extrinsic curvature alone. The
constraints then can serve to compute the matter perturbations
$\delta\eps$ and $\delta u_i$. In this case we would not have to use
the equations of motion for the matter perturbations, which follow
from the conservation of energy-momentum $D_\nu T^{\mu\nu}$.

Our next step consists of expanding the equations in spherical tensor
harmonics. Any symmetric tensor $A_{\mu\nu}$ can be expressed in terms
of a set of ten spherical tensor harmonics
$\lbrace[\rw^A_{lm}]_{\mu\nu} (\theta,\phi)\rbrace_{A=1,...,10}$,
namely
\begin{align}\label{exp}
        A_{\mu \nu}(t,r,\theta,\phi) \= \sum_{l = 0}^{\infty}
        \sum_{m = -l}^l \sum_{A = 1}^{10}
        a_{A}^{lm}(t,r)\; [\rw^A_{lm}]_{\mu\nu}(\theta,\phi)\;.
\end{align}
Due to the spherical symmetry of the background we thereby can
eliminate the angular dependence and obtain equations for the
coefficients $a_A^{lm}(t,r)$. The field equations thus will be reduced
to partial differential equations in $t$ and $r$. 

There are various ways to define those tensor harmonics $\rw^A_{lm}$,
and an exhaustive overview can be found in \cite{Th80}. The set first
given by Regge and Wheeler \cite{RW57} is widely used throughout the
literature and we will follow this tradition, even if this set has the
disadvantage of not being orthonormal. However, this is of no account
in the following proceeding. It is only with the inclusion of an
orbiting particle that this fact causes some minor inconveniences.

The Regge-Wheeler harmonics can be divided into two subsets that
behave in different ways under parity transformation. Under space
reflection the {\it polar} or {\it even parity} harmonics change sign
according to $(-1)^l$, whereas the {\it axial} or {\it odd parity}
harmonics transform like $(-1)^{l+1}$.

We now expand the metric as follows:
\begin{align}\label{exp_h}
\begin{split}
        h_{\mu\nu} \= -2e^\nu \widehat{S}_1^{lm}[\rw^1_{lm}]_{\mu\nu}
        + \widehat{S}_2^{lm}[\rw^2_{lm}]_{\mu\nu}
        + \widehat{S}_3^{lm}[\rw^5_{lm}]_{\mu\nu}\\
        &\quad{} + \widehat{V}_1^{lm}[\rw^3_{lm}]_{\mu\nu}
        + \widehat{V}_2^{lm}[\rw^4_{lm}]_{\mu\nu}
        + \widehat{V}_3^{lm}[\rw^6_{lm}]_{\mu\nu} 
        + \widehat{V}_4^{lm}[\rw^7_{lm}]_{\mu\nu}\\
        &\quad{} + \widehat{T}_1^{lm}[\rw^8_{lm}]_{\mu\nu} 
        + \widehat{T}_2^{lm}[\rw^9_{lm}]_{\mu\nu}
        + \widehat{T}_3^{lm}[\rw^{10}_{lm}]_{\mu\nu}\;.
\end{split}
\end{align}
The notation has been chosen such that the coefficients
$\widehat{S}_i$ represent the scalar parts of $h_{\mu\nu}$, namely
$\alpha, \beta_r$, and $h_{rr}$, whereas the $\widehat{V}_i$ stand for
the vector components $\beta_\theta, \beta_\phi, h_{r\theta}$ and
$h_{r\phi}$. Lastly, the $\widehat{T}_i$ represent the tensorial
components $h_{\theta\theta}, h_{\theta\phi}$ and $h_{\phi\phi}$. Note
that this expansion includes both polar and axial harmonics.
Similarly, the extrinsic curvature tensor $k_{ij}$ will be expanded as
\begin{align}\label{exp_k}
\begin{split}
        k_{ij} \= \widehat{K}_1^{lm}[\rw^5_{lm}]_{ij}
        + \widehat{K}_2^{lm}[\rw^6_{lm}]_{ij}
        + \widehat{K}_3^{lm}[\rw^7_{lm}]_{ij}\\
        &\quad{} + \widehat{K}_4^{lm}[\rw^8_{lm}]_{ij}
        + \widehat{K}_5^{lm}[\rw^9_{lm}]_{ij}
        + \widehat{K}_6^{lm}[\rw^{10}_{lm}]_{ij}\;.
\end{split}
\end{align}
Last not least, we need the matter variables
\begin{align}
        \delta\eps \= \hat{\rho}^{lm}Y_{lm}\label{delta_eps}\\
        u_i \= \hat{u}_1^{lm}[\rw^5_{lm}]_{i1}
        + \hat{u}_2^{lm}[\rw^6_{lm}]_{i1}
        + \hat{u}_3^{lm}[\rw^7_{lm}]_{i1}\label{u_i}\;.
\end{align}
Here, $Y_{lm}$ in the expansion \eqref{delta_eps} for $\delta\eps$ is
just the ordinary spherical harmonic. Using this expansion, we will
obtain 12 evolution equations for the coefficients of the 3-metric
$h_{ij}$ and the extrinsic curvature $k_{ij}$. However, it is clear
that this set cannot be used for the evolution, for we have not
specified any gauge yet. Picking a specific gauge means prescribing
the coefficients of lapse $\alpha$ and shift $\beta_i$ for which we do
not have any evolution equations. Of course, there are various ways to
choose lapse and shift, and for the numerical evolution of the full
nonlinear Einstein equations it is crucial to make a good choice
because inadequate lapse or shift may cause the code to crash at very
early times due to the blow up of some variables.

However, we are dealing with the linearized field equations and we do
not have to worry too much about some of the nasty things that might
happen in the nonlinear case. We will choose lapse and shift in a
way as to simplify the equations as much as possible.

We could now proceed and write down the general form of the
perturbation equations, but we will not do so because on the one hand
the equations are quite lengthy and on the other hand they are not
very useful as such. Besides, they can be found in Appendix C of my
diploma thesis \cite{Ruoff96}. However, it seems to be appropriate to
make some comments on the equations.

First of all, the equations are independent of the spherical harmonic
index $m$ because of the spherical symmetry of the background.
Secondly, they decouple with respect to their behavior under parity
transformation. Hence we obtain two sets of equations, one describing
polar perturbations, the other axial ones. The latter cannot generate
any density or pressure changes in the neutron star for those
quantities are scalars and therefore have even parity. The equations
can be further subdivided according to their value of $l$. The {\em
radial} and {\em dipole} modes are characterized by $l=0$ and $l=1$,
respectively. Those oscillations do not give rise to gravitational
radiation, hence the equations are only meaningful in the interior of
the star itself. Of course, they are also valid in the exterior, but
we can make them identically vanish by means of an appropriate gauge
transformation. Furthermore, for $l=0$ we do not have any angular
dependence at all, hence all the metric and extrinsic curvature
coefficients that are proportional to a derivative of $Y_{lm}$ vanish
identically. We will consider this case in the next chapter. The
dipole modes, which have been studied in \cite{CT70,Det75b,LS89}, will
not be treated in this thesis.

For $l\ge2$ it is not possible any more to make the exterior equations
vanish, hence those equations describe real gravitational waves, which
propagate through the spacetime and carry information about the
oscillations of the neutron star. This case is the main point of
investigation of this thesis and is presented in chapter 4.

\chapter{Radial oscillations of neutron stars}

As they are the simplest oscillation modes of neutron stars, radial
modes have been the first under investigation \cite{MT66}. More
important, they can give information about the stability of the
stellar model under consideration \cite{Chandra64a,Chandra64b}. Since
they do not couple to gravitational waves, the appropriate equations
are quite simple and it is relatively easy to numerically solve the
eigenvalue problem that leads to the discrete set of oscillation
frequencies of a neutron star. In the absence of any dissipative
processes the oscillation spectrum of a stable stellar model forms a
complete set; it is therefore possible to describe any arbitrary
periodic radial motion of a neutron star as a superposition of its
various eigenmodes. It hence seems quite superfluous to explicitly
solve the time dependent equations, for we do not expect to gain new
physical insight. This might be true indeed. There are, however, quite
unexpected numerical problems that are associated with the evolution
of the radial oscillations of realistic neutron star models.

Those problems are quite generic, for they also occur for the non-radial
oscillations and probably also in the case of rotating neutron stars,
and once we have them under control in the radial case it should be
straightforward to confer the appropriate numerical treatment to other
cases.

The above mentioned numerical problems are instabilities that occur
when the neutron star model is constructed with a realistic equation
of state. However, they do not appear, if one uses polytropic
equations of state, which is often done for the sake of simplicity. As
it turned out, this instability is not a general instability of the
numerical scheme, for it is dependent on the resolution and can be
made to disappear if the resolution exceeds a certain
threshold. However, this threshold strongly depends on the numerical
scheme that is used to evolve the equations, and can be so high that
it prevents any evolutions within a reasonable time limit. For other
discretizations, the threshold can be relatively low and does not
represent a real obstacle to obtaining numerical results. However,
this is mainly because the radial case is a (1+1)-d problem and one
only has to consider the stellar interior, since the exterior
spacetime remains totally unaffected. For non-radial oscillations with
$l \ge 2$ this is not true any more, for here the oscillations will
generate gravitational waves, which propagate towards infinity. This
means that we have to include the exterior domain in our numerical
evolution as well, which will result in a much bigger computational
expenditure. Here, the required minimum resolution can extend the
computation time of a single run to quite large values. Still, even
the non-radial case is a (1+1)-d problem, and if we include rotation
we will face a (2+1)-d problem, which can become totally intractable
if a high resolution is required to obtain a stable evolution.

As we shall see, the cause of the instability is confined to a small
region close to the surface of the star, and it is only here that the
high resolution is needed in order to obtain stability. For the more
time-consuming cases, we then could only locally refine the grid in
this region instead of using high resolution for the whole domain.
However, for each stellar model, we would have to localize the
troublesome region and find the required resolution, which is more
or less a matter of trial and error.

Therefore, we seek for a better way to solve this problem and we find
it in reformulating the equations in such a way that the instability
totally disappears, and the equations can be evolved for any stellar
model and for any desired resolution in a stable way.

\section{Derivation of the evolution equations}

The first one to write down the equations was Chandrasekhar
\cite{Chandra64a,Chandra64b}, later on various authors rewrote them in
many different ways \cite{BTM66,Chan77,GL83,GHZ98}. Since the
(3+1)-formalism is particularly suitable for the numerical evolution we
will rederive the equations using the framework of the
(3+1)-decomposition.

From the last chapter we know that radial perturbations are described
by $l=0$. Since the appropriate harmonic is just a number $Y_{00} =
1/\sqrt{4\pi}$ there is no angular dependence at all and any
derivative of $Y_{00}$ with respect to $\theta$ or $\phi$ vanishes. In
this case we can absorb $Y_{00}$ in the perturbation variables, which
are then functions of $t$ and $r$ only, and the expansion of the
perturbations reads
\begin{subequations}
\begin{align}
        \alpha \= e^{\nu}S_1(t,r)\\
        \beta_r \= re^{2\lam}S_2(t,r)\\
        h_{ij} \= \(\begin{array}{ccc}
        re^{2\lam}S_3(t,r) & 0 & 0\\
        0 & r^2T(t,r) & 0\\
        0 & 0 & r^2\sin^2\theta\,T(t,r)\end{array}\)\;.
\end{align}
\end{subequations}
Similarly, we have for the extrinsic curvature 
\begin{align}
        k_{ij} \= -e^{-\nu}\(\begin{array}{ccc}
        e^{2\lam}K_1(t,r) & 0 & 0\\
        0 & \half r^2K_2(t,r) & 0\\
        0 & 0 & \half r^2\sin^2\theta K_2(t,r)\end{array}\)\;.
\end{align}
This particular decomposition has been chosen in order to obtain a
more convenient set of equations. The matter perturbations are
characterized by the perturbation of the energy density $\delta\eps$
and the (covariant) radial component of the 4-velocity $u_r$, which
are expanded as
\begin{align}
        \delta\eps \= \rho(t,r)\\
        \delta u_r \= -e^\nu u(t,r)\;.\label{ur}
\end{align}
We then obtain the following four evolution equations for the metric
perturbations and the perturbations of the extrinsic curvature:
\begin{subequations}
\begin{align}
        \df{S_3}{t} \= 2\(\df{S_2}{r} + \(\lam' + \frac{1}{r}\)S_2
        + \frac{K_1}{r}\)\\
        \df{T}{t} \= 2S_2 + K_2\\
\begin{split}
        \df{K_1}{t} \= e^{2\nu - 2\lam}\bigg[\,\dff{T}{r} + \dff{S_1}{r}
        + \(\frac{2}{r} - \lam'\)\df{T}{r}
        + \(2\nu' - \lam'\)\df{S_1}{r}\\
        &{}\qquad\qquad - \(\half r\nu' + 1\)\df{S_3}{r}
        + \(\lam' - \frac{3}{2}\nu'
        + \frac{e^{2\lam} - 2}{r}\)S_3\,\bigg]\\
        &{}\quad + 4\pi e^{2\nu}\(1 - C_s^2\)\rho
\end{split}\\
\begin{split}
        \df{K_2}{t} \= e^{2\nu - 2\lam}\bigg[\,\dff{T}{r}
        + \(\nu' - \lam' + \frac{4}{r}\)\df{T}{r}
        + \frac{2}{r}\df{S_1}{r} - \df{S_3}{r}\\
        &\qquad\qquad{} + 2\frac{e^{2\lam}}{r^2}T
        + \(2\lam' - 2\nu' - \frac{3}{r}\)S_3\,\bigg]\\
        &\quad{} + 8\pi e^{2\nu}\(1 - C_s^2\)\rho\;.
\end{split}
\end{align}
\end{subequations}
In addition we have the Hamiltonian constraint
\begin{align}
        8\pi\,e^{2\lam}\rho \= -\dff{T}{r}
        + \(\lam' - \frac{3}{r}\)\df{T}{r} - \frac{e^{2\lam}}{r^2}T
        + \df{S_3}{r} + 2\(\frac{1}{r} - \lam'\)S_3
\end{align}
and the momentum constraint
\begin{align}\label{MC_rad}
        8\pi\(p + \eps\)e^{2\nu}u \= -\df{K_2}{r} + \(\nu' - \frac{1}{r}\)K_2
        + \frac{2}{r}K_1\;.
\end{align}
Note, that in the above equations we have not yet specified any gauge.
To perform numerical evolutions, we have to fix the gauge, that is we
need some additional prescriptions for lapse $S_1$ and shift $S_2$.
Let us choose vanishing shift
\begin{align}
        S_2 \= 0\;,
\end{align}
but, for the moment, let us keep the lapse $S_1$ still unspecified.
Instead, let us pick $T = 0$ at the initial slice at $t = 0$. We then
obtain a much simpler set of evolution equations:
\begin{subequations}
\begin{align}\label{S3_rad}
        \df{S_3}{t} \= \frac{2}{r}K_1\\
        \df{T}{t} \= K_2\\
\begin{split}\label{K1_rad}
        \df{K_1}{t} \= e^{2\nu - 2\lam}\bigg[\,\dff{S_1}{r}
        + \(2\nu' - \lam'\)\df{S_1}{r}
        - \(\half r\nu' + 1\)\df{S_3}{r}\\
        &{}\qquad\qquad + \(\lam' - \frac{3}{2}\nu'
        + \frac{e^{2\lam} - 2}{r}\)S_3\,\bigg]
        + 4\pi e^{2\nu}\(1 - C_s^2\)\rho
\end{split}
\end{align}
\begin{align}
\begin{split}\label{K2_rad}
        \df{K_2}{t} \= e^{2\nu - 2\lam}\bigg[\,
        \frac{2}{r}\df{S_1}{r} - \df{S_3}{r}
        +  \(2\lam' - 2\nu' - \frac{3}{r}\)S_3\,\bigg]
        + 8\pi e^{2\nu}\(1 - C_s^2\)\rho\;,
\end{split}
\end{align}
\end{subequations}
and the Hamiltonian constraint reduces to
\begin{align}\label{HC_rad}
        8\pi e^{2\lam}\rho \= \df{S_3}{r} + 2\(\frac{1}{r} - \lam'\)S_3\;.
\end{align}
Note that we still have an equation for $T$. Our goal is to use the
remaining gauge freedom in such a way that we can get rid of this
equation.  This means that we somehow have to make the extrinsic
curvature variable $K_2$ vanish. To do so, we first solve the
Hamiltonian constraint \eqref{HC_rad} for $S_3'$ and plug it in into
\eqref{K2_rad}, which then reads:
\begin{align}\label{dK2_rad}
        \df{K_2}{t} \= e^{2\nu - 2\lam}
        \(\frac{2}{r}\df{S_1}{r} - \(2\nu' + \frac{1}{r}\)S_3\)
        - 8\pi e^{2\nu}C_s^2\rho\;.
\end{align}
We finally fix the gauge completely by choosing the lapse $S_1$ such
that we make the righthand side of \eqref{dK2_rad} vanish. This can be
accomplished by requiring that $S_1$ should satisfy the following
ordinary differential equation:
\begin{align}\label{S1_rad}
        \df{S_1}{r} \= \(r\nu' + \frac{1}{2}\)S_3
        + 4\pi re^{2\lam}C_s^2\,\rho\;.
\end{align}
This condition can also be used to replace $S_1'$ and $S_1''$ in
\eqref{K1_rad}. Moreover, if we use the Hamiltonian constraint
\eqref{HC_rad} to replace the fluid variable $\rho$, we end up
with just two equations for the metric variable $S_3$ and the
extrinsic curvature variable $K_1$. We can even go one step further
and combine them to yield a single wave equation for $S_3$. However,
we do not write down this equation because we will present a somewhat
simpler system below. 

Before doing so, we would like to prove that radial gravitational
waves do not exist. The particular feature of this wave equation for
$S_3$ is that its propagation speed is not the speed of light alone,
but it is multiplied with the square of the sound speed $C^2_s$. Hence,
outside the star this equation loses its wavelike character because
$C_s = 0$. In fact, in the exterior region, where in addition to $C_s =
0$ we have $\lam' = -\nu'$, the equation for $S_3$ reduces to
\begin{align}
        \dff{S_3}{t} \= \nu'e^{2\nu-2\lam}\(\df{S_3}{r}
        + 2\(\nu' + \frac{1}{r}\)S_3\)\;,
\end{align}
which by virtue of the Hamiltonian constraint \eqref{HC_rad} with
$\rho = 0$ becomes 
\begin{align}
        \dff{S_3}{t} \= 0\;.
\end{align}
This shows us that radial oscillations of neutron stars cannot give
rise to any gravitational waves.

By the way, we could have inferred this much more easily by looking at
the momentum constraint \eqref{MC_rad}, which without $K_2$ just reads
\begin{align}\label{MC2_rad}
        K_1 \= 4\pi re^{2\nu}\(p + \eps\)u
        \;=\; e^{2\nu-2\lam}\(\lam' + \nu'\)u\;.
\end{align}
Since $u$ vanishes outside the star, so must $K_1$ and therefore
$S_3$.

To obtain a simpler set of equations than the above mentioned wave
equation, we look at the so far neglected equations which follow from
the conservation law $D_\nu T^{\mu\nu} = 0$. Here,
instead of $\rho$ we rather use
\begin{align}\label{Hdef}
        H(t,r) \= \frac{C^2_s}{p + \eps}\,\rho(t,r)
\end{align}
because the resulting equations are somewhat nicer. We obtain the two
matter equations
\begin{subequations}
\begin{align}
        \df{H}{t} \= e^{2\nu - 2\lam}C^2_s\,\df{u}{r}
        + e^{2\nu - 2\lam}\(C^2_s\(2\nu' - \lam' + \frac{2}{r}\) - \nu'\)u
        - C^2_s\,K_1\label{Ht_rad}\\
        \label{ut_rad}\df{u}{t}\= \df{H}{r} + \df{S_1}{r}\;.
\end{align}
\end{subequations}
Here, we have already made use of $T = S_2 = K_2 = 0$. We now use the
momentum constraint \eqref{MC2_rad} to eliminate $K_1$ from equation
\eqref{Ht_rad}, and our gauge condition \eqref{S1_rad} together with
\eqref{Hdef} will serve to replace $S_1$ in \eqref{ut_rad}. The
resulting equations still contain the metric variable $S_3$, and in
order to obtain a closed system of equations, we need the evolution
equation for $S_3$, which is given by \eqref{S3_rad}, but with $K_1$
replaced by \eqref{MC2_rad}. We finally end up with the following
quite simple set of equations:
\begin{subequations}
\label{evol_rad}
\begin{align}
        \df{H}{t} \= e^{2\nu - 2\lam}\left[C^2_s\,\df{u}{r}
        + \(C^2_s\(\nu' - 2\lam' + \frac{2}{r}\) - \nu'\)u\right]
        \label{dtH_rad}\\
        \df{u}{t} \= \df{H}{r} + (\nu' + \lam')H
        + \(r\nu' + \frac{1}{2}\)S_3\label{dtu_rad}\\
        \df{S_3}{t} \= 8\pi(p + \eps)e^{2\nu}u\;.\label{dtS3_rad}
\end{align}
\end{subequations}
Additionally, $H$ and $S_3$ have to satisfy the Hamiltonian constraint
\eqref{HC_rad}
\begin{align}
        8\pi e^{2\lam}\frac{p + \eps}{C^2_s}H \= 
        \df{S_3}{r} + 2\(\frac{1}{r} - \lam'\)S_3\;.
\end{align}
In order to obtain physical solutions of the equations, we have to
impose boundary conditions at the origin and at the stellar
surface. At the origin we have to require the perturbation variables
to be regular. By inspection we find that
\begin{align}
        H(t,r) \= H^0(t) + {\cal O}(r^2)\\
        u(t,r) \= u^0(t)\,r + {\cal O}(r^3)\\
        S_3(t,r) \= S_3^0(t)\,r + {\cal O}(r^3)\;.
\end{align}
The boundary condition at the surface is given by the requirement that
the Lagrangian pressure perturbation $\Delta p$ has to vanish. This
condition requires the concept of the radial displacement function
$\xi(t,r)$, which describes the displacement of a fluid element
from its equilibrium position as a function of time and position and
will be discussed in more detail in chapter 3. For the radial case the
relation between $\xi$ and $\delta u_r$ is given by equation (26.6) of MTW
\cite{MTW}:
\begin{align}\label{xiu}
        \df{\xi}{t} \= e^{\nu-2\lam}\,\delta u_r\;.
\end{align}
The authors then introduce the renormalized displacement function
\begin{align}\label{zeta}
        \zeta \;&:=\; r^2e^{-\nu}\xi\;,
\end{align}
for which the Lagrangian pressure perturbation $\Delta p$ can be written
as
\begin{align}\label{Delta_p}
        r^2\Delta p \= - \Gamma_1 p e^{-\nu}\df{\zeta}{r}\;,
\end{align}
where $\Gamma_1p \equiv \(p + \eps\)C^2_s$ for barotropic equations
of state. The condition of vanishing Lagrangian pressure perturbation
at the stellar surface $r = R$ then translates to
\begin{align}\label{bc_zeta}
        \df{\zeta}{r}(R)=0\;,
\end{align}
in the case that $\Gamma_1p$ does not vanish at the surface.
For vanishing $\Gamma_1p$ this is not necessarily true anymore,
instead, we just have to require the boundedness of $\zeta'$ and
$\zeta$ itself. For a more detailed discussion of the boundary
condition at the surface, see \cite{BTM66}.

As it is only for the special cases of polytropic equations of
state that $\Gamma_1p = 0$ at the surface, we will always use
$\zeta'(R)=0$ as our boundary condition, regardless of the actual
equation of state. To use this boundary condition for our system of
evolution equations \eqref{evol_rad}, we have to translate it into a
condition for $u$, which can be easily obtained by making use of
\eqref{xiu} and \eqref{ur}
\begin{align}
        0 \= \(\df{\zeta}{t}\)'_{r=R} \;=\; \(r^2e^{-\nu}\df{\xi}{t}\)'_{r=R}
        \;=\; \(r^2e^{-2\lam}\delta u_r\)'_{r=R}
        \;=\; -\(r^2e^{\nu-2\lam}u\)'_{r=R}\;.
\end{align}
Explicitly we have
\begin{align}
        u'(R) \= \(2\lam'(R) - \nu'(R) - \frac{2}{R}\)u(R)\;,
\end{align}
which can be used in the numerical code to update the value of $u$ at
the surface. This is the only relevant boundary condition because the
values of the remaining quantities $H$ and $S_3$ directly follow
from the evolution equations. From \eqref{dtS3_rad} we deduce that
$S_3(R) = 0$, and because of $C_s(R) = 0$ equation \eqref{dtH_rad}
reduces to an ordinary differential equation for $H$ at the stellar
surface.

Finally, we should mention that our system is equivalent to equation
(26.19) of MTW, which is a single wave equation for the renormalized
displacement function $\zeta$. It can be written in a very compact
form with the righthand side having the form of a self adjoint
differential operator:
\begin{subequations}
\label{MTWeq}
\begin{align}
        W\,\dff{\zeta}{t} \= \df{}{r}\(P\,\df{\zeta}{r}\) + Q\,\zeta
\end{align}
with
\begin{align}
        r^2 W \= \(p + \eps\)e^{3\lam + \nu}\phantom{\bigg(}\\
        r^2 P \= \(p + \eps\)C^2_s\,e^{\lam + 3\nu}\\
        r^2 Q \= e^{\lam + 3\nu}\(p + \eps\)\((\nu')^2 + 4\frac{\nu'}{r}
        - 8\pi e^{2\lam}p\)\;.
\end{align}
\end{subequations}

\section{The eigenvalue problem}

By applying the Fourier transformation to the numerical evolution we
should be able to find the frequencies of the eigenmodes. However, it
is certainly reasonable to calculate the eigenfrequencies directly
from equation \eqref{MTWeq}, too, by making the harmonic time ansatz
\begin{align}
        \zeta(t,r) \= e^{\I\omega t}\chi(r)\;.
\end{align}
This then gives us a linear ordinary differential equation for
$\chi(r)$ with $\omega^2$, the square of the oscillation
frequency, as a free parameter:
\begin{align}\label{MTWev}
        0 \= \frac{d}{dr}\(P\frac{d\chi}{dr}\)
        + \(Q + \omega^2W\)\chi\;.
\end{align}
Together with boundary condition \eqref{bc_zeta} this defines a
Sturm-Liouville eigenvalue problem, which has solutions only for a
countable set of real eigenvalues $\omega^2$. For $\omega^2$ positive,
$\omega$ itself is real and thus, the solution is purely oscillatory.
However, for negative $\omega^2$ we have an imaginary frequency
$\omega$, which corresponds to an exponentially growing or damped
solution. Since the general solution is always a superposition of both
the growing and the damped modes, this means that the occurrence of a
negative value of $\omega^2$ corresponds to an instability with
respect to radial oscillations of the stellar model under
consideration. For neutron stars this will, indeed, happen for central
densities $\eps_0$ larger than the critical central density
$\eps_{crit}$ at which the stellar mass $M$ as a function of $\eps_0$
has its maximum. In this case the star will ultimately collapse to a
black hole. For $\eps_0 = \eps_{crit}$ there must be a neutral mode
with the corresponding eigenvalue $\omega^2 = 0$ \cite{ZN}.

To solve the eigenvalue equation \eqref{MTWev} numerically, we will
write it as a system of two first order equations in $\chi$ and $\eta
:= P\chi'$:
\begin{subequations}
\label{ev_eq}
\begin{align}
        \label{xi2}\frac{d\chi}{dr} \= \frac{\eta}{P}\\
        \frac{d\eta}{dr} \= -\(\omega^2W + Q\)\chi\;.
\end{align}
\end{subequations}
By inspection we find that close to the origin we have $\chi(r) =
\chi_0\,r^3 + {\cal O}(r^5)$ and $\eta(r) = \eta_0 + {\cal
O}(r^2)$. From \eqref{xi2} it then follows that the leading order
coefficients are related by $3\chi_0 = \eta_0/P_0$, where $P_0 = \(p(0)
+ \eps(0)\)C^2_s(0)\,e^{\lam(0) + 3\nu(0)}$. Choosing $\eta_0 = 1$ we
obtain $\chi_0 = 1/(3P_0)$, which gives us the initial values for the
integration.

To find the eigenvalues, we will choose some arbitrary $\omega$ and
integrate the equations from the origin $r=0$ outwards to the stellar
surface at $r=R$, where we have to check whether the boundary condition
$\chi'(R) = 0$ is satisfied. If so, the chosen $\omega$ corresponds
to the desired eigenfrequency.

Numerically, the boundary condition $\chi'(R) = 0$ will never be
fulfilled exactly. However, if we consider $\chi'(R ,\omega)$ as a
function of $\omega$ we can find that the zeroes of $\chi'(R ,\omega)$
are of first order, which means that they are associated with a sign
change of $\chi'(R ,\omega)$. Numerically it is quite easy to look for
a sign change in a given interval, and we can quickly locate the exact
value of $\omega$ up to the desired precision, for example by the
method of bisection. Of course, there exist other methods of computing
the eigenvalues and some of them are compiled in \cite{BTM66}.

To check our numerical eigenvalue code, we compare our modes with
results available in the literature. An extended survey of the first
two radial modes for a wide range of different equations of state was
given by Glass \& Lindblom \cite{GL83}. However, it seems that
an error sneaked into their computer code, for their equations are
correct, but the results are erroneous. Instead, we agree with results
obtained by V\"ath \& Chanmugam \cite{VC92}, who also pointed out that
the results of Glass \& Lindblom are flawed. The strongest argument in
favor of their (and therefore our) results being correct is that they
indeed obtain the neutral modes right at the maximum of the mass
function $M(\eps_0)$.

\begin{figure}[t]
\leavevmode
\epsfxsize=\textwidth
\epsfbox{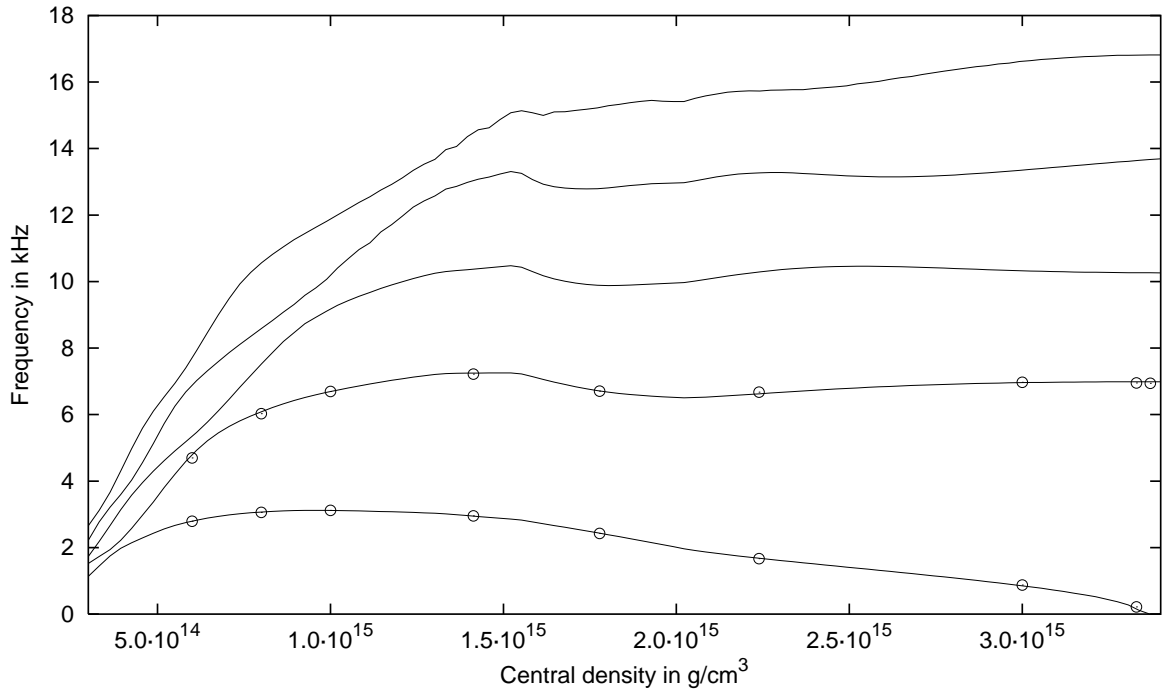}
\caption{\label{eosD}The first 5 radial pulsation modes as a function of
the central energy density for the EOS of model V of Bethe \& Johnson.
The results of V\"ath \& Chanmugam are represented by circles.}
\end{figure}

In Fig.~\ref{eosD} we show the frequencies of the first 5 radial
oscillation modes as a function of the central density $\eps_0$. In
addition, we include the values obtained by V\"ath \& Chanmugam, which
perfectly agree with our results. The stellar models were obtained
using a realistic equation of state described by the model V of Bethe
\& Johnson \cite{BJ74} (EOS D in the list compiled by Arnett \& Bowers
\cite{AB77}).

\section{Numerical results for polytropes}

For the first evolution runs we use a polytropic equation of state
with $\Gamma = 2$ and $\kappa = 100\,$km$^2$. We discretize the
system \eqref{evol_rad} with a two-step Lax-Wendroff scheme (see
e.g. \cite{NumRec}), where we first perform a half time step to compute
intermediate values and then perform a full time step to obtain the
values at the next time level.

We show the evolution of a narrow Gaussian profile in $S_3$ for three
different stellar models. Model 1 has a central energy density of
$\eps_0 = 3\!\cdot\!10^{15}\,$g/cm$^3$, which corresponds to a mass of $M
= 1.27\, M_\odot$ and radius of $R = 8.86\,$km. Model 2 with $\eps_0
= 5.65\!\cdot\!10^{15}\,$g/cm$^3$ is right below the stability limit and
model 3 with $\eps_0 = 5.67\!\cdot\!10^{15}\,$g/cm$^3$ is above it. For
models 1 and 2 we expect periodic time evolutions with the signal
being a superposition of the various eigenmodes. Model 3, which is
unstable with respect to radial collapse, should show an exponential
growing mode.

\begin{figure}[t]
\begin{minipage}{\textwidth}
\leavevmode
\epsfxsize=\textwidth
\epsfbox{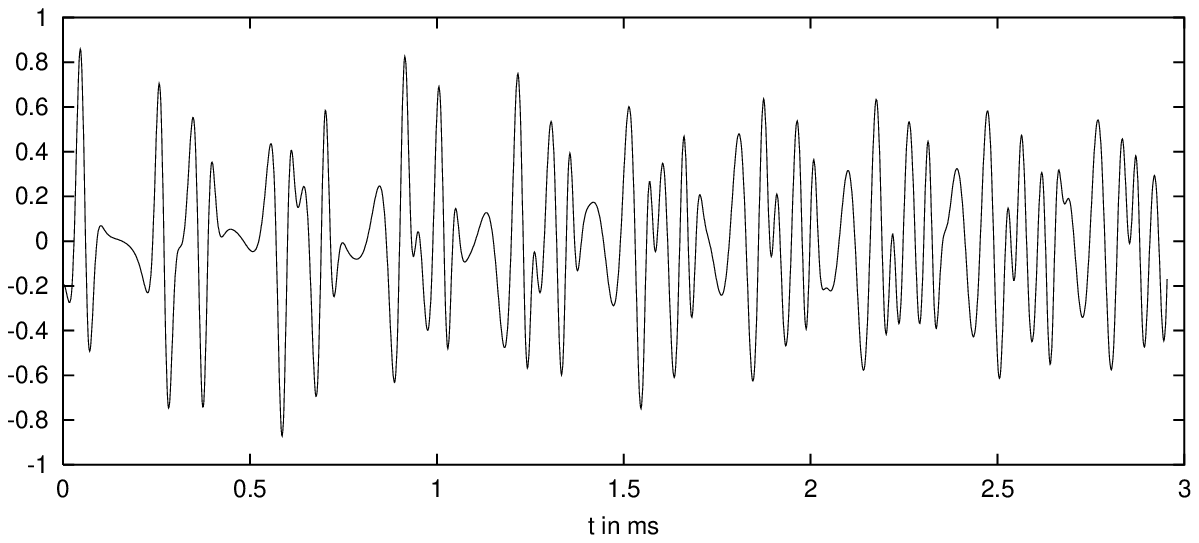}
\caption{\label{3e15}Evolution of $u$ for a polytropic stellar model
with central density $\eps_0 = 3\!\cdot\!10^{15}\,$g/cm$^3$.}
\vspace*{5mm}
\end{minipage}
\begin{minipage}{\textwidth}
\leavevmode
\epsfxsize=\textwidth
\epsfbox{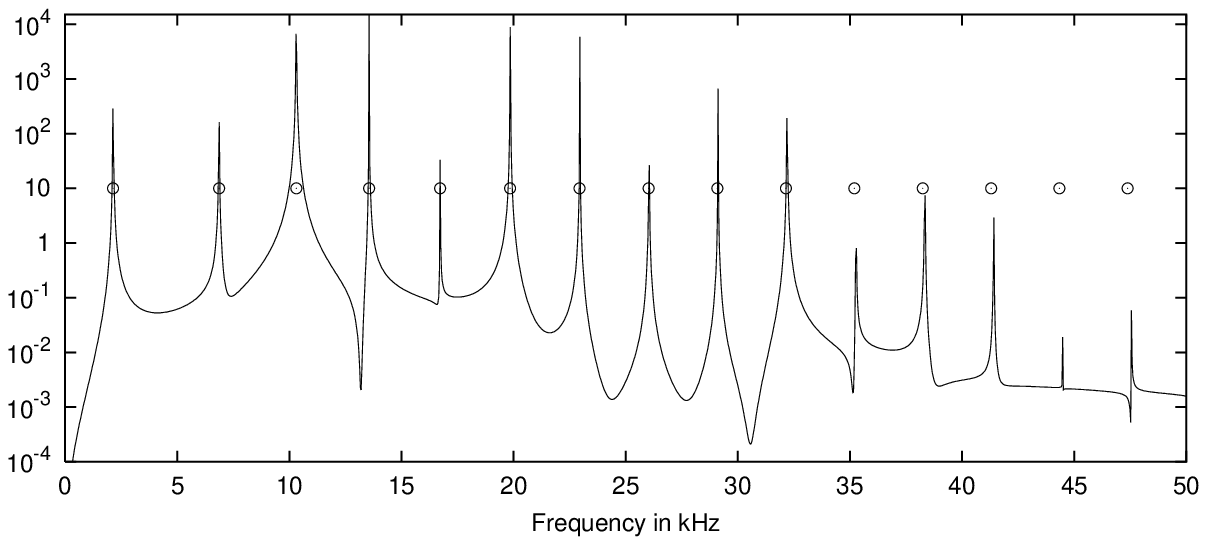}
\caption{\label{3e15.fft}The power spectrum of the above wave signal
shows that at least 15 eigenmodes are present. The circles represent
the eigenfrequencies obtained by solving \eqref{ev_eq}.}
\end{minipage}
\end{figure}

Our expectations are fully met by the numerical evolution of equations
\eqref{evol_rad}. In Fig.~\ref{3e15} we show the time evolution for
model 1 with its Fourier transformation in Figure \ref{3e15.fft}. The
spectrum shows that the chosen initial data excite many of the
eigenfrequencies of the stellar model; at least 15 modes are clearly
present. In the spectrum we also include the eigenfrequencies computed
by directly solving the eigenvalue problem \eqref{ev_eq}, which agree
perfectly.  It is only for the higher frequency modes that the peaks
in the spectrum are systematically located at higher frequencies
than the actual eigenfrequencies, which is due to insufficient
resolution of the evolution. By increasing the number of grid points
the peaks converge to the right frequencies.

\begin{figure}[t]
\begin{minipage}{\textwidth}
\leavevmode
\epsfxsize=\textwidth
\epsfbox{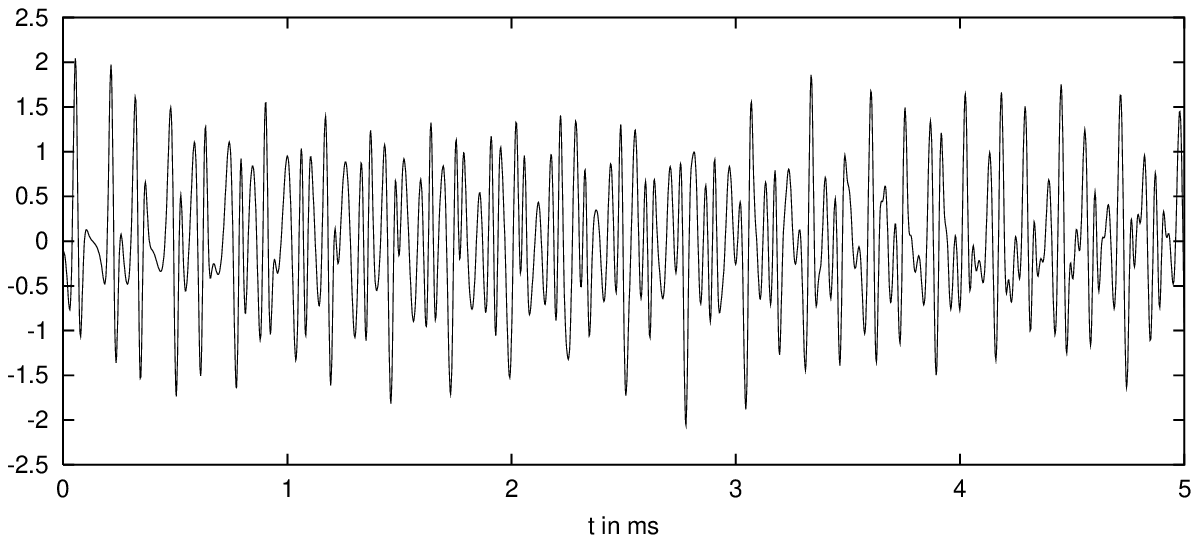}
\caption{\label{5.65e15}Section of the evolution of $u$ for a polytropic
stellar model with central density $\eps_0 = 5.65\!\cdot\!10^{15}\,$g/cm$^3$,
which is right below the stability limit.}
\vspace*{5mm}
\end{minipage}
\begin{minipage}{\textwidth}
\leavevmode
\epsfxsize=\textwidth
\epsfbox{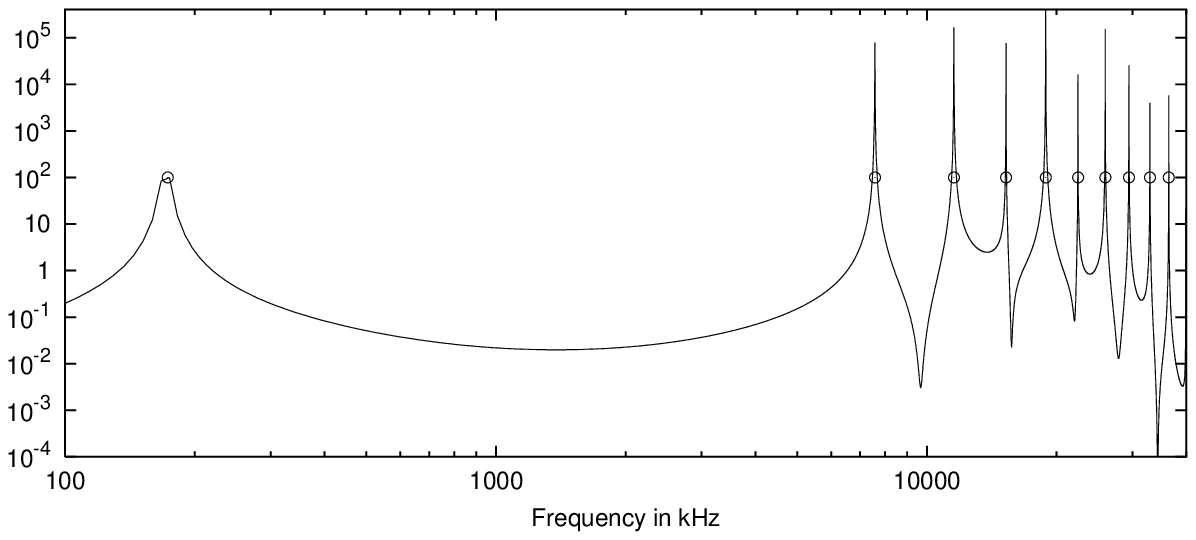}
\caption{\label{5.65e15.fft}Fourier transformation of the above wave signal.
Here, the lowest mode, which lies at $\nu_1 = 173\,$Hz is clearly visible.
The second mode lies at $\nu_2 = 7580\,$Hz.}
\end{minipage}
\end{figure}

As the central density increases, the star approaches its stability
limit. At the same time the frequency of the lowest mode starts to
migrate towards zero. The stability limit itself is characterized by
the presence of an eigenmode with zero frequency. As was already
stated above, at this point the total mass $M$ as function of the
central density $\eps_0$ exhibits a local maximum. In Figure
\ref{5.65e15} we show the time evolution of $H$ for model 2. The
evolution does not really look different from the one for model 1, but
in the signal there should be a very low frequent oscillation, which
corresponds to the lowest eigenmodes. The Fourier transformation in
Fig.~\ref{5.65e15.fft} confirms the presence of a very low frequency
mode, which for this model has slipped down to a frequency of $\nu_1 =
173\,$Hz. The second mode resides at the much higher frequency of
$\nu_2 = 7580\,$Hz. We should mention that in order to obtain the
spectrum in Fig.~\ref{5.65e15.fft}, where we have a resolution of
about 10 Hz, we have to evolve up to $t = 100\,$ms. With a time step
size that is somewhat smaller than $1\!\cdot\!10^{-4}$ms we then need
more than one million integration steps!

\begin{figure}[t]
\leavevmode
\epsfxsize=\textwidth
\epsfbox{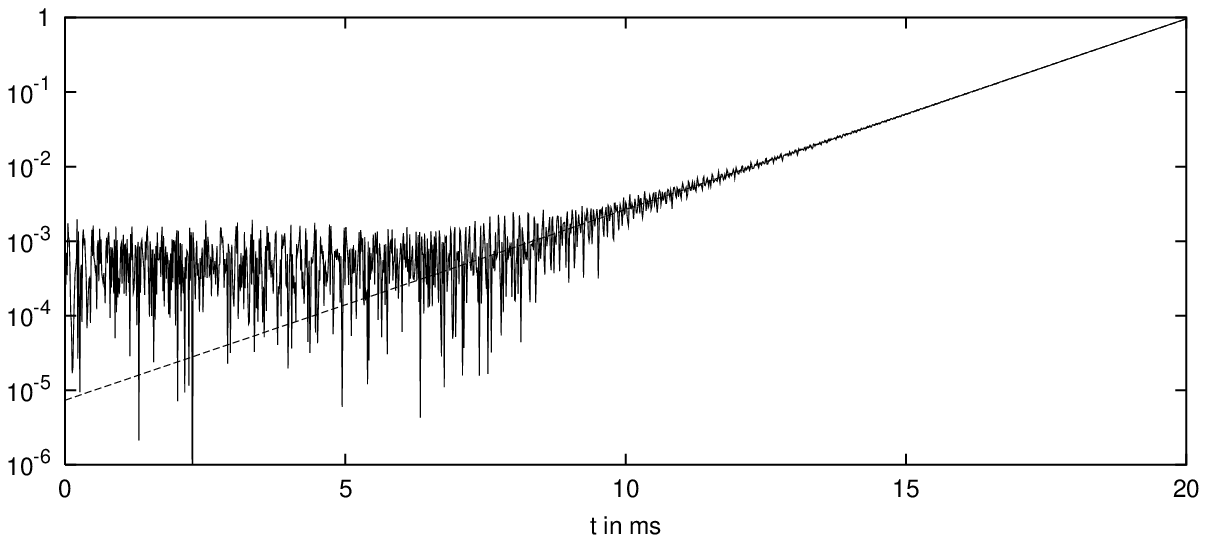}
\caption{\label{5.67e15}Evolution of $u$ the unstable polytropic stellar
model with central density $\eps_0 = 5.67\!\cdot\!10^{15}\,$g/cm$^3$.}
\end{figure}

Model 3 is unstable, which is nicely confirmed by the evolution. With
the eigenvalue code we find an imaginary frequency with Im$(\omega) =
606\,$Hz, which corresponds to an $e$-folding time of $\tau =
1.65\,$ms. In the logarithmic plot of Fig.~\ref{5.67e15} the
exponential growth shows up as a linear increase in the amplitude.
From a fit of an exponential function to the numerical data, we find
$\tau = 1.69\,$ms.

\section{Getting into trouble: Using realistic equations of state}
\label{troubles}

So far we have used polytropic equations of state, which are quite
decent approximations to realistic equations of state as far as
general features of neutron stars like mass and radius are
concerned. However, it is in particular the oscillations of neutron
stars that are very sensitive to local changes in the equation of
state, which are due to the different behavior of the neutron star
matter under varying pressure. It is therefore much more interesting
to use realistic equations of state that take into account the
underlying microphysics which determines the state of the matter as a
function of pressure and temperature. For comprehensive overviews
on realistic equations of state, see \cite{ST,RMHR,Glenn}.

As was already mentioned in the first chapter, for the sake of
simplicity we will resort to zero temperature equations of state
only. Of course, if we were interested in damping times, which are due
to internal friction and other viscous effects that result from the
finite temperature inside the star, we would have to abandon the zero
temperature approximation.

Realistic equations of state cannot be given in analytic terms over the
whole pressure range inside the neutron star, hence they usually exist
in tabulated form only. To solve the TOV equations in this case, one
has to interpolate between the given values in order to obtain the
stellar model with continuous functions of radius $r$.

The thus obtained stellar models are not as smooth as those with
polytropic equations of state, for it is clear that the energy density
and other matter functions will have bumps and edges as the matter
undergoes phase transitions for increasing pressure, which can change
its stiffness quite abruptly. This is particularly the case close to
the stellar surface, where the pressure is zero and increases
extremely as one moves into the stellar interior. It is clear that it
is here that the matter undergoes a couple of phase transitions.
First the crystal lattice gets destroyed and changes into a plasma of
nuclei and free electrons. As the pressure further increases, the
electrons get captured by the protons of the nuclei, which yields more
and more neutron rich nuclei. Eventually the nuclei start to dissolve
and neutrons begin to ``drip'' out of the nuclei. At this neutron drip
point, which occurs at $7.8\!\cdot\!10^{29}\,$dyn/cm$^2$, or at a
density of about $4.3\!\cdot\!10^{11}\,$g/cm$^3$, the equation of
state has its most drastic change. For increasing pressure the matter
then stays in the form of a degenerate Fermi gas until it reaches the
point beyond nuclear density, where, again, phase transitions may take
place.

This is the point where the underlying physics is the least known, and
one has to rely on physical models with simplifying assumptions in
order to theoretically compute the equation of state. Of course,
different nuclear models lead to different equations of state and it
is therefore at the nuclear density level and beyond that the
equations of state that are given in the literature differ the most
strongly from each other. For a more detailed description of the
physics at high densities, we refer the reader to \cite{RMHR}.

In the following we will make use of an equation of state called MPA
\cite{Wu91}, which yields a maximal mass model of $1.56 M_\odot$. We can
obtain a typical stellar model by taking a central density of $\eps_0
= 4\!\cdot\!10^{15}\,$g/cm$^3$, which yields a radius and a mass of $R =
8.18\,$km and $M = 1.55M_\odot$, respectively.

If we try to repeat the evolution of an initial perturbation of the
radial velocity field $u$ for the above stellar model using the
Lax-Wendroff scheme for the discretization of the relevant equations
\eqref{evol_rad}, we will witness a stellar explosion. That is, 
after a few oscillations we will suddenly find an exponentially
growing mode that immediately swamps the whole evolution. This cannot
be a physical instability, for we have taken a stable background model,
and by playing around with different resolutions we quickly convince
ourselves that it has to be a numerical instability, for neither the
$e$-folding time nor the frequency seem to converge to some limiting
values.

Apparently, it has to be a problem of the numerical discretization
scheme we use, we can therefore try to switch to a different one. For
instance, we could try to discretize the system \eqref{evol_rad} on
a staggered mesh. If we write the equations schematically as
\begin{align}
        \dot Q \= aP' + bP\\
        \dot P \= cQ' + dQ\;,
\end{align}
the discretized form reads
\begin{align}
        Q^{n+1}_i \= Q^n_i
        + a_i\frac{\Delta t}{\Delta r}\(P^n_{i + 1/2} - P^n_{i - 1/2}\)
        + b_i\frac{\Delta t}{2}\(P^n_{i + 1/2} + P^n_{i - 1/2}\)\\
        P^{n+1}_{i + 1/2} \= P^n_{i + 1/2}
        + c_{i + 1/2}\frac{\Delta t}{\Delta r}\(Q^{n+1}_{i+1} - Q^{n+1}_{i}\)
        + d_{i + 1/2}\frac{\Delta t}{2}\(Q^{n+1}_{i+1} + Q^{n+1}_{i}\)\;.
\end{align}
We can see that $Q$ lives on integer grid points, whereas $P$ lives on
half integer ones, that is the $P$-grid is shifted by half a grid
point both in $r$- and $t$-direction with respect to the $Q$-grid. In
\eqref{evol_rad} we put $H$ on the regular grid and $u$ and $S_3$ on
the shifted grid.

\begin{figure}[t]
\leavevmode
\epsfxsize=\textwidth
\epsfbox{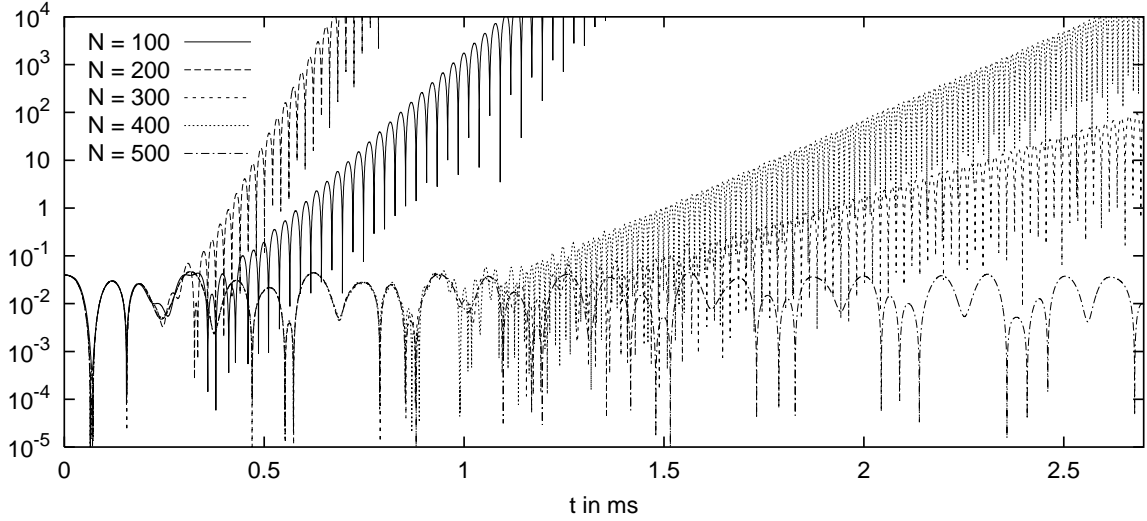}
\caption{\label{MPA_sg}Evolution using the first order system
\eqref{evol_rad} discretized on a staggered grid for resolutions ranging
from N = 100 up to 500 grid points. The key is the same as in the figure
below.}
\end{figure}

In Fig.~\ref{MPA_sg} we show the evolution for various resolutions.
Again, we find a growing mode for low resolutions, but as the number
of grid points $N$ is increased, the occurrence of the instability gets
more and more delayed, and the slope of the exponential growth
decreases, too. And for $N = 500$ it suddenly disappears.
Interestingly, it is even possible to pin the point where the
instability vanishes down to $N = 469$.  For $N \le 469$ we still have
exponential growth, but if we add just another grid point we can evolve
the equations arbitrarily long.

It seems that also for the Lax-Wendroff discretization the instability
can be made to vanish by increasing the resolution, but the required 
resolution is extremely high. Too high to allow for numerical runs 
within a reasonable time scale.

\begin{figure}[t]
\leavevmode
\epsfxsize=\textwidth
\epsfbox{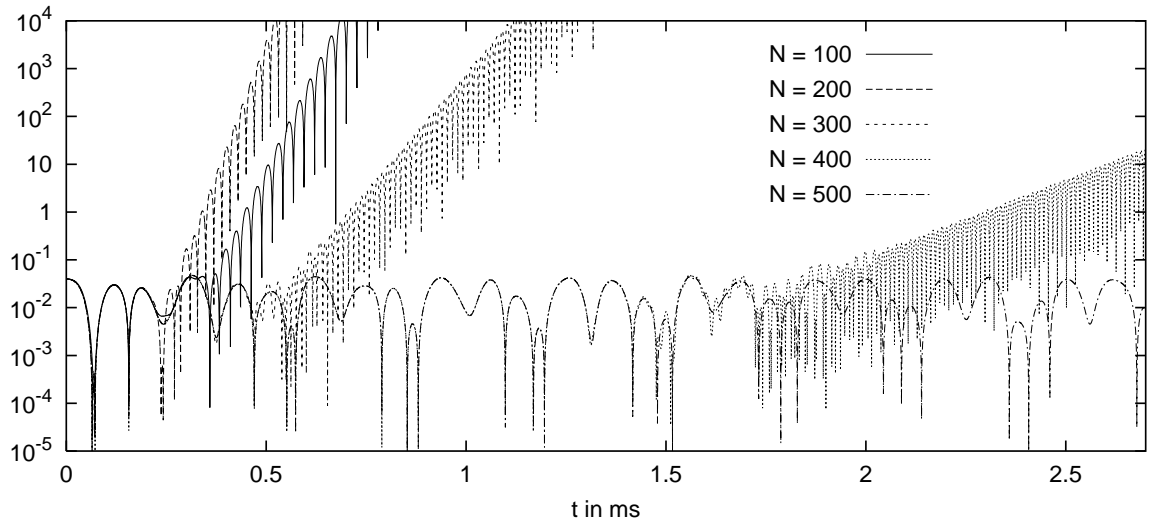}
\caption{\label{MPA_wave}Evolution using the wave equation \eqref{wave_w}
discretized with central differences for resolutions ranging from N =
100 up to 500 grid points.}
\end{figure}

We now try a third possible discretization, where we transform
\eqref{evol_rad} into a single second order wave equation. This can
only be performed for either $u$ or $S_3$, but not for $H$, since in
this case we cannot totally eliminate the remaining variable
$S_3$. Here we choose $u$, but since the resulting wave equation is
somewhat lengthy, we define another variable $w$ through
\begin{align}
        w(t,r) \= re^{\nu-2\mu}u(t,r)\;,
\end{align}
for which we have a quite transparent wave equation
\begin{align}
\begin{split}\label{wave_w}
        \dff{w}{t} \= e^{2\nu-2\mu}\bigg[\,
        C_s^2\dff{w}{r} + \(C_s^2\(2\nu' + \mu'\)
        + (C_s^2)' - \nu'\)\df{w}{r}\\
        &{} \quad
        + \(C_s^2\(2\frac{\nu'}{r} + \frac{\mu'}{r} - \frac{2}{r^2}\)
        + \frac{(C_s^2)'}{r} + \frac{e^{2\mu} - 1}{r^2}
        + \nu'\(\nu' + \frac{1}{r}\)\)w\,\bigg]\;.
\end{split}
\end{align}
At the stellar surface, we have the following boundary condition
\begin{align}\label{bcw}
        (rw)'(R) \= 0\;.
\end{align}
We discretize \eqref{wave_w} with central differences, which again
will be demonstrated for the schematic equation
\begin{align}
        \ddot{Q} \= aQ'' + bQ' + cQ\;.
\end{align}
A second order discretization scheme is the following
\begin{align}
\begin{split}
        Q^{n+1}_i \= 2Q^n_i + Q^{n-1}_i
        + (\Delta t)^2\(        
        \frac{a_i}{(\Delta r)^2}\(Q^n_{i+1} - 2Q^n_i + Q^n_{i-1}\)\right.\\
        &\quad \left.+ \frac{b_i}{2\Delta r}\(Q^n_{i+1} - Q^n_{i-1}\)
        + c_i Q^n_i\)\;.
\end{split}
\end{align}
In Fig.~\ref{MPA_wave} we show the evolution for different
resolutions. The plots are quite similar to Fig.~\ref{MPA_sg}. Here,
too, the instability goes away for resolutions of 500 or more
grid points. This similarity of the behavior between those two
discretizations is understandable, since the discretization of
$\eqref{evol_rad}$ on the staggered grid is equivalent to the above
discretization of the wave equation.

The question now is, why is there an instability at all? Why is it
that for polytropic equations of state there is no problem at all,
whereas by using a realistic equation of state the numerical evolution
can blow up? Which part of the equations is responsible for this
peculiar behavior? This is all part of the investigation in the next
section.

\subsection{The physical problem}

\begin{figure}[t]
\begin{minipage}{\textwidth}
\leavevmode
\epsfxsize=\textwidth
\epsfbox{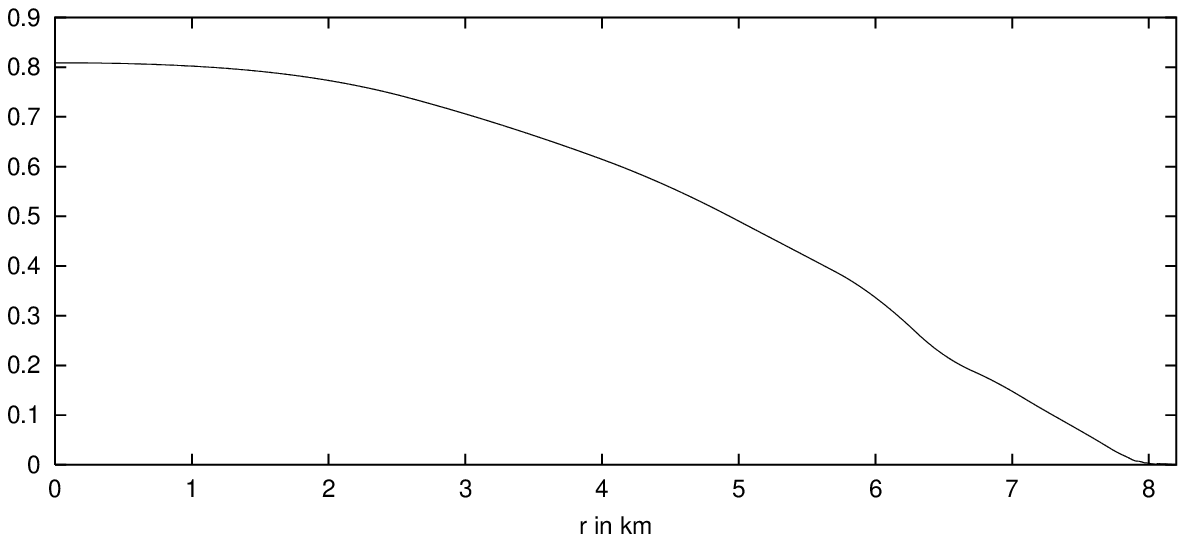}
\end{minipage}
\begin{minipage}{\textwidth}
\leavevmode
\epsfxsize=\textwidth
\epsfbox{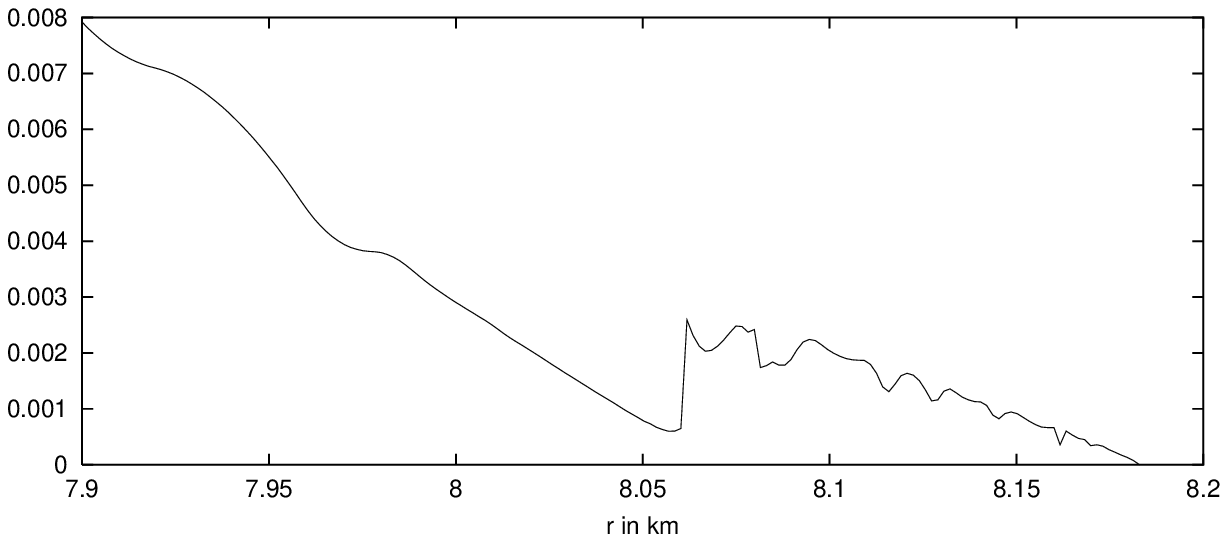}
\caption{\label{dpde_MPA}Square of the sound speed $C_s^2$ inside the neutron
star model using eos MPA (above) and a section near the surface (below).}
\end{minipage}
\end{figure}

It becomes clear very soon that the culprit has to be the profile of
the sound speed inside the star. For if we compute the stellar models
with a realistic equation of state, but then replace the profile of
the sound speed with a profile that results from a polytropic equation
of state, the instability does not occur any more.
 
For polytropic equations of state, the sound speed is a smooth,
monotonically decreasing function of the radius $r$ that is zero only
at the surface of the star, whereas for realistic equations of state
there are regions in the outer layers of the star, where the sound
speed can have sharp drops. In particular, close to the neutron drip
point, the equation of state becomes very soft, which results in a
drastic decrease of the sound speed. This sharp drop then can lead to
the observed numerical instabilities due to an interplay between the
boundary condition at the surface of the star and the small value of
the sound speed in this particular region.

In Fig.~\ref{dpde_MPA} we show the square of the sound speed $C_s^2 =
\frac{dp}{d\eps}$ for a stellar model using eos MPA with central
density of $\eps_0 = 4\!\cdot\!10^{15}$ g/cm$^3$. It can be clearly seen
that at $r \approx 8.06\,$km there is a local minimum of the sound
speed, where it drops down to $\frac{dp}{d\eps} \approx 0.0005$. For
larger $r$ we can see a series of much smaller dips, which is an
artefact of the numerical spline interpolation between the tabulated
points. But the dip at $r \approx 8.06\,$km is physical and is present
for any realistic equation of state.

\subsection{A toy model}
The wave equations that describe the non-radial metric oscillations,
e.g.~the Regge-Wheeler equation \eqref{RWex} and the Zerilli equation 
\eqref{Zeqn} in chapter 4 are of the form
\begin{align}\label{Phitilde}
        \dff{\tilde\Phi}{t} \= c(r)\df{}{r}\(c(r)\df{}{r}\tilde\Phi\)
        + \tilde V\tilde\Phi
\end{align}
or
\begin{align}
        \dff{\tilde\Phi}{t} \= \dff{}{r_*}\tilde\Phi
        + \tilde V\tilde\Phi\;,
\end{align}
where 
\begin{align}
        \df{}{r_*}\= c(r)\df{}{r}\;.
\end{align}
Here $c(r)$ is the local propagation speed, in this case the speed of
light, and $V$ represents a potential. Besides, the equation can
contain additional terms from the coupling to other quantities, but
they are not relevant for our discussion. By choosing another
variable $\Phi = \sqrt{c(r)}\tilde\Phi$, we can further change
\eqref{Phitilde} to
\begin{align}\label{Phi}
        \dff{\Phi}{t} \= c^2(r)\dff{\Phi}{r} + V\Phi\;,
\end{align}
which is the standard form of a wave equation with a variable
propagation speed $c(r)$ and a potential term $V$. However, this is
not possible for the fluid wave equations. If we let $\Psi$ represent
any of the fluid variables $u$, $w$ or $H$, which have to be multiplied
by appropriate factors $e^\nu$ or $e^\lam$, we cannot find a wave
equation of the form \eqref{Phi}, with $c^2(r)$ now replaced by the
sound speed $C_s^2(r)$. Instead, we always will have an additional term,
which is proportional to the first derivative $\Psi'$ and which we
cannot make disappear (we suppress the overall factor $e^{2\nu -
2\lam}$)
\begin{align}\label{Psi}
        \dff{\Psi}{t} \= C^2_s(r)\dff{\Psi}{r} - \nu'\df{}{r}\Psi + V\Psi\;.
\end{align}
At first glance this does not seem to be bothersome at all, but it is
in the case of the sound speed becoming almost zero that the problems
start to arise because then the term proportional to $\Psi'$ starts to
dominate.

For the following discussion we will not use \eqref{Psi} but rather
the fluid equation \eqref{waveH} that governs the non-radial
oscillations, which are investigated in chapter 4. From numerical
experiments we find that the problematic terms in \eqref{waveH} are
the first two ones:
\begin{align}\label{Hprob}
\dff{H}{t} \= e^{2\nu - 2\lam}\bigg[\,C^2_s\dff{H}{r} 
        + \(C^2_s \(2\nu' - \lam'\) - \nu'\)\df{H}{r}\,\bigg]\;.
\end{align}
The wave equation \eqref{wave_w} that we use for the radial
oscillations is somewhat different, but if we were to transform
\eqref{evol_rad} into a wave equation for $H$ instead of $u$, the first
two terms would be exactly the same as in \eqref{Hprob}. To further
simplify \eqref{Hprob}, we now set
\begin{align*}
        e^{2\nu - 2\lam} \= 1\\
        C_s^2 \= c^2\\
        \lam' \= 0\\
        \nu' \= a\;,
\end{align*}
which gives
\begin{align}\label{Ht}
        \dff{H}{t} \= c^2\dff{H}{r} + a\(2c^2 - 1\)\df{H}{r}\;.
\end{align}
Our goal now is to see if we can get the same kinds of instabilities
with this simplified version.

It turns out that as long as we do not require periodic boundary
conditions, the actual choice of the boundary conditions does not have
a great influence on the stability behavior of \eqref{Ht}. For the
sake of simplicity we therefore choose $H(0) = H(R) = 0$. We could as
well choose the first derivative of $H$ to vanish at either
boundary. Finally, we set $R = 1$ and discretize our toy model
\eqref{Ht} with central differences:
\begin{align}
\begin{split}\label{Hnum}
        H^{n+1}_i \= 2H^n_i - H^{n-1}_i
        + c_i^2\frac{(\Delta t)^2}{(\Delta r)^2}\(H^n_{i+1} - 2H^n_i    
        + H^n_{i-1}\)\\
        &{}\quad + a\(2c_i^2 - 1\)\frac{(\Delta t)^2}{2\Delta r}
        \(H^n_{i+1} - H^n_{i-1}\)\;.
\end{split}
\end{align}
To simulate the dip in the sound speed, we assume a constant sound speed
with a small Gaussian well with depth $h$ and width $w$
located at $r_c$:
\begin{align}
        c^2(r) \= c_0 - he^{-\(\frac{r - r_c}{w}\)^2}\;.
\end{align}

To infer the stability behavior of our numerical scheme, we have to
look at the eigenvalues of the discretized time evolution operator.
Let this operator, which advances data from the time level $n$ to the
next time level $n+1$, be denoted by ${\bf B}$. If we furthermore
collect all values $H^n_i$ on the time level $n$ in the vector $H^n$,
we can formally write
\begin{align}
        H^{n+1} \= {\bf B}H^n\;.
\end{align}
Consequently, the inverse of ${\bf B}$ leads to the previous level
\begin{align}
        H^{n-1} \= {\bf B}^{-1}H^n\;.
\end{align}
Our numerical scheme \eqref{Hnum} can be written in vector form as
\begin{align*}
        H^{n+1} + H^{n-1} \= 2{\bf G}H^n \\
        \Leftrightarrow \quad\({\bf B} + {\bf B}^{-1}\)H^n
        \= 2{\bf G}H^n\;,
\end{align*}
where ${\bf G}$ is the matrix that includes the discretized spatial
derivatives. Suppose now that $H^n$ is an eigenvector of ${\bf B}$
with eigenvalue $b$. Hence the eigenvalue of ${\bf B}^{-1}$ is
$b^{-1}$ and we have
\begin{align}
        \(b + b^{-1}\)H^n \= 2{\bf G}H^n\;,
\end{align}
that is, $H^n$ is also an eigenvector of ${\bf G}$ with eigenvalue
\begin{align}
        g \= \frac{1}{2}\(b + b^{-1}\)\;.
\end{align}
Solving for $b$ yields 
\begin{align}\label{bg}
        b \= g \pm \sqrt{g^2 - 1}\;.
\end{align}
For stability, we have to have $|b| \le 1$. If we denote the two solutions
of \eqref{bg} by
\begin{subequations}
\begin{align}
        b_+ \= g + \sqrt{g^2 - 1}\\
        b_- \= g - \sqrt{g^2 - 1}\;,
\end{align}
\end{subequations}
we find that $b_- = b_+^{-1}$. Hence, if $|b_+| < 1$ it is $|b_-| > 1$
and vice versa, i.e. we cannot have the moduli of both eigenvalues
smaller than one at the same time. Therefore it is only for $|b_+| =
|b_-| = 1$ that we can have stability. This means that $b$ has to lie
on the complex unit circle:
\begin{align}
        b = e^{\I\phi}\;.
\end{align}
It then follows for $g$ that 
\begin{align}
        g \= \frac{1}{2}\(e^{\I\phi} + e^{-\I\phi}\) \;=\; \cos\phi\;.
\end{align}
Thus, $g$ is real and only assumes values in the interval $[-1,1]$.

Hence, if we find any eigenvalue $g$ of ${\bf G}$ that is either
complex or whose modulus is larger than one, we always have one $|b|
> 1$ and therefore the numerical scheme is unstable. Now, ${\bf G}$ is
a tridiagonal matrix with the band given by the triple
\begin{align*}
\begin{split}
        &\(G_{i,i-1}, G_{ii}, G_{i,i+1}\) \;= \\
        &\qquad
        \(\frac{1}{2}c^2f^2 - \frac{a}{4}f^2\Delta x\(2c^2 - 1\),
        \quad 1 - c^2f^2, \quad
        \frac{1}{2}c^2f^2 + \frac{a}{4}f^2\Delta x\(2c^2 - 1\)\)\;,
\end{split}
\end{align*}
where we have defined
\begin{align}
        f := \frac{\Delta t}{\Delta x}\;.
\end{align}
A necessary condition for stability turns out to be
\begin{align}
        cf \le 1\;,
\end{align}
which is the usual CFL criterion.\\

Of course, there are a lot of parameters to play around with. It would
not make sense to try out all the possible combinations. Instead, we
choose some parameters to be fixed. For example, we choose $a = f =
1$. Indeed, we have found that the onset of instability is
independent of $f$ (always assuming $f \le 1$, of course), which means
that smaller time steps with fixed grid spacing cannot cure the
instability. Too large a value of $a$ combined with too low a
resolution also can give rise to instabilities. The choice of $a=1$
ensures that the scheme is stable if $c=1$ throughout the whole
domain.

We now set $c_0 = 1$, and $r_c = 0.5$, that is, the dip is in the
middle of the domain. Interestingly, for a given resolution, the
instabilities may vanish, if we move $r_c$ towards the left boundary,
whereas a stable set of parameters can get unstable, if we move $r_c$
towards the right boundary. That means if an instability occurs at
$r_c = 0.5$, it will necessarily also occur at $r_c > 0.5$. (In the
actual case of a neutron star, the dip sits pretty close to the star's
surface). Having now fixed the parameters $f, a, c_0$ and $r_c$, we
are left with $h, w$ and $N$.

\begin{figure}[t]
\leavevmode
\epsfxsize=\textwidth
\epsfbox{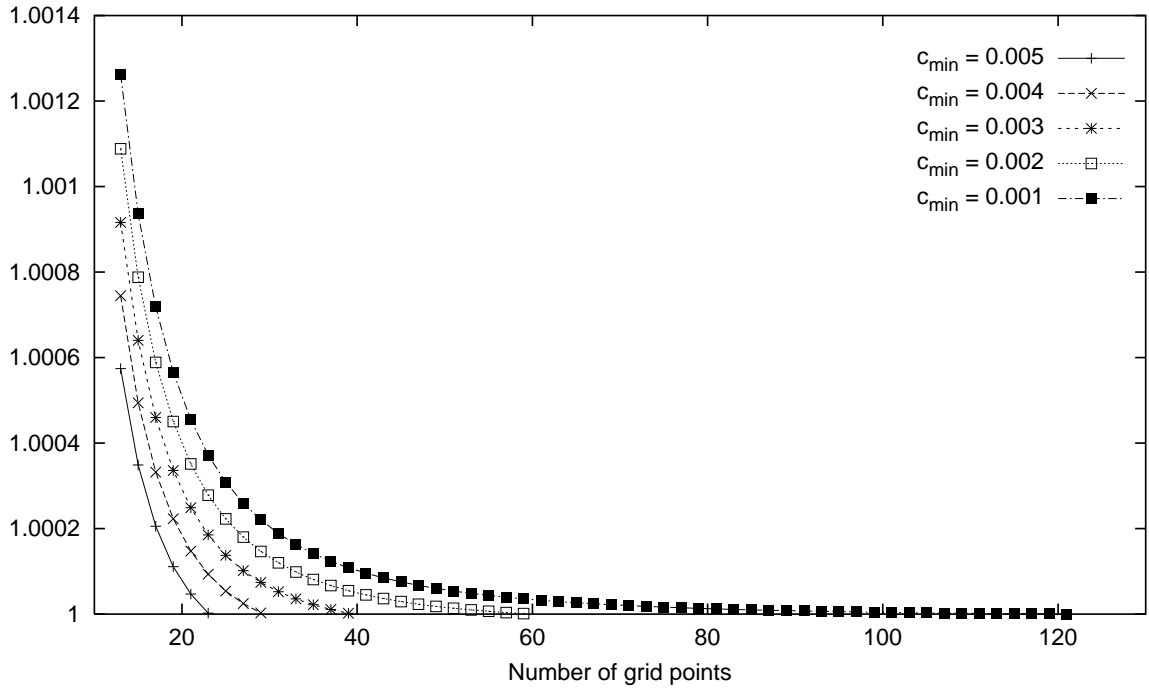}
\caption{\label{ews} Largest eigenvalue as a function of the number of grid
points for different values of $c_{min}$.}
\end{figure}

Our results can be summarized as follows. First, let us choose $w$ to
be so small that the considered grids will not be able to resolve the
Gaussian. Thus, on our grids we have $c = 1$ everywhere except at the
middle grid point, where we have $c_{min} = 1 - h$. (We then have to
take an odd number of grid points, for only in this case there is a
grid point located at $r = r_c = 0.5$). By explicitly computing the
eigenvalues of ${\bf G}$ we find that they are indeed real, but they
are not necessarily smaller than one. It is rather the case that for a
given grid size $N$ there exists a certain critical value
$c_{crit}$, for which at least one eigenvalue will be greater than
one. For $c_{min}$ only slightly larger than $c_{crit}$ all
eigenvalues are $\le 1$, whereas for $c_{min} < c_{crit}$
there is always at least one eigenvalue $> 1$.

The higher the resolution, i.e. the larger $N$ and the smaller $\Delta
r$, the smaller $c_{crit}$. Conversely, for a given $c_{min}$,
there exists a minimum grid size $N_{crit}$, above which all the
eigenvalues are $\le 1$ and below which there will always be at least
one eigenvalue $> 1$. And the smaller $c_{min}$, the larger the
required grid size $N_{crit}$ to obtain a stable evolution. However,
for $c_{min} = 0$ there always seems to be one eigenvalue $> 1$,
regardless of the chosen resolution. Thus, in this case, the scheme is
unconditionally unstable.

In Fig.~\ref{ews} we show the eigenvalues of ${\bf G}$ as a function
of the grid size $N$ for various values of $c_{min}$. It can be
clearly seen that for smaller values of $c_{min}$ it takes a larger
grid size to squeeze the largest eigenvalue below one. For $c_{min} =
0.002$ it is for $N \ge 61$ that all eigenvalues are $\le 1$, whereas
for $c_{min} = 0.001$ we must have $N \ge 121$.

\begin{figure}[t]
\begin{center}
\vspace*{-8cm}
\hspace*{1cm}
\epsfxsize=14cm
\epsfbox{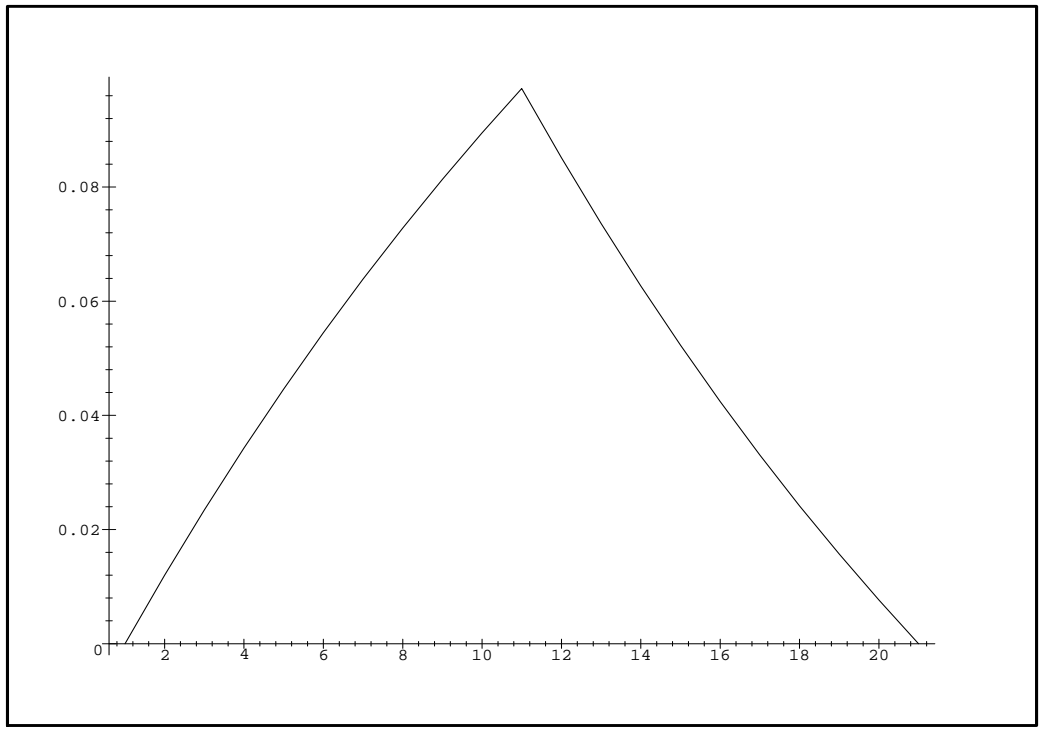}
\vspace*{-1mm}
\caption{\label{ev}Eigenvector corresponding to the eigenvalue larger than one
for grid size $N = 21$ and $c_{min} = 0.004$. Compare with numerical results
in Fig.~\ref{evol59}.}
\end{center}
\vspace*{-3mm}
\end{figure}

In Fig.~\ref{ev} we show the eigenvector corresponding to the largest
eigenvalue for the parameters $c_{min} = 0.004$ and $N = 21$.

\begin{figure}[p]
\begin{minipage}{\textwidth}
\vspace*{-1cm}
\leavevmode
\epsfxsize=\textwidth
\epsfbox{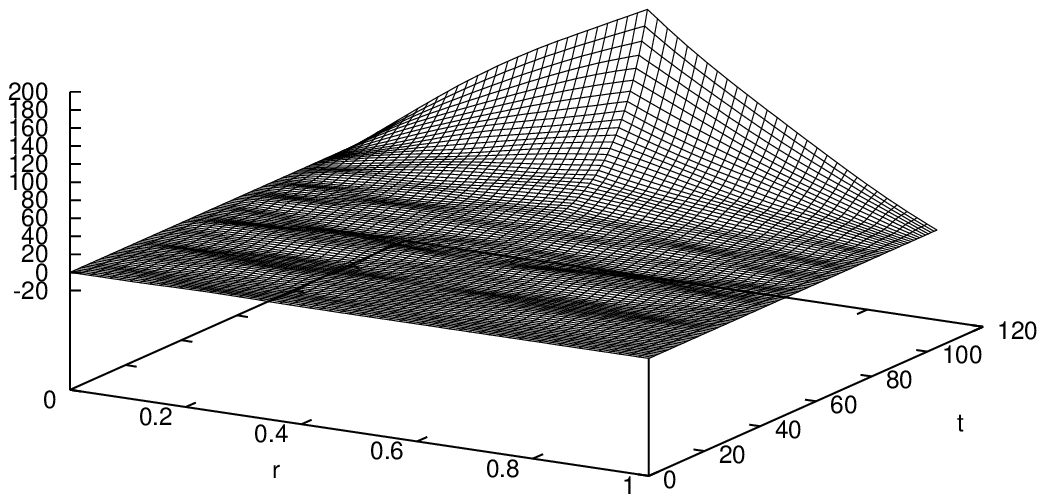}
\caption{\label{evol59}Numerical evolution for $c_{min} = 0.002$ and
$N = 59$. Exponential growth of that eigenmode whose related
eigenvalue is $>1$ is observed. Compare the shape of the
mode with Fig.~\ref{ev}}
\end{minipage}
\begin{minipage}{\textwidth}
\leavevmode
\epsfxsize=\textwidth
\epsfbox{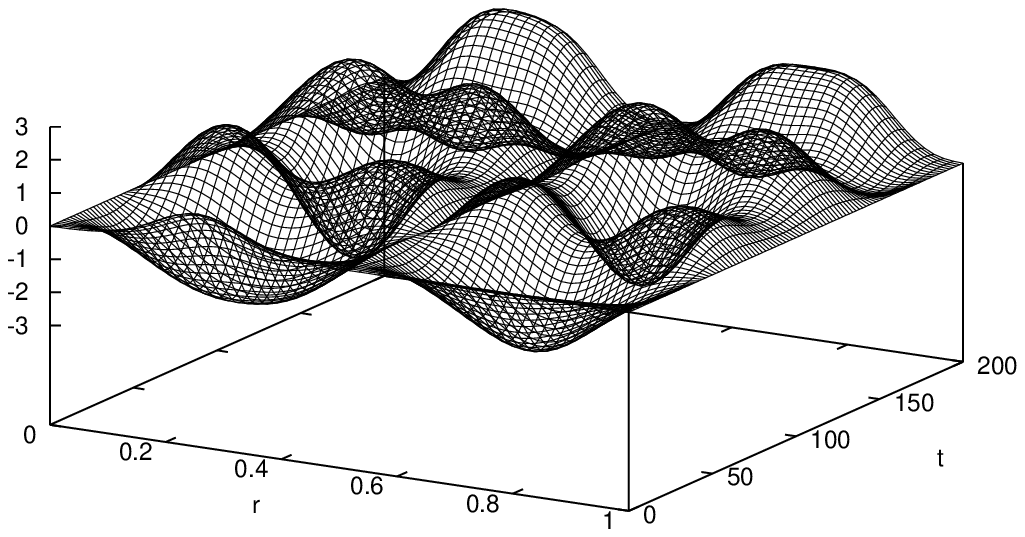}
\caption{\label{evol61}Numerical evolution for $c_{min} = 0.002$
and $N = 61$. All eigenvalues are $< 1$, and the numerical
solution remains bounded.}
\end{minipage}
\end{figure}

Figures~\ref{evol59} and \ref{evol61} show the numerical evolution for
$c_{min} = 0.002$ for grid sizes of $N = 59$ and $N = 61$,
respectively. From Fig.~\ref{ews} we see that for $N = 59$ there
exists one eigenvalue that is $> 1$, whereas for grid sizes of $N =
61$ or greater, all eigenvalues remain $\le 1$. As expected, the
numerical evolution shows a growing mode for $N = 59$, whereas for $N =
61$ the evolution remains bounded. From the shape of the growing
solution we find, indeed, that it corresponds to the eigenvector of
Fig.~\ref{ev}.

The influence of the dip width $w$ is rather counterintuitive. The
larger $w$, the wider and smoother the dip. If the instability should
be due to the discontinuity in $c$, it should vanish for increasing
$w$. The contrary is the case. For a given resolution and a value of
$c_{min}$ just above $c_{crit}$, a widening of the width, i.e. an
increase in $w$ will eventually result in a transition to instability.

We would also like to stress that the instability is not invariably
related to the dip, but is rather due to the interplay of the dip with
the prescribed boundary conditions. For if we were to choose periodic
boundary conditions, all the previously unstable cases suddenly would
become stable! This is due to the fact that with periodic boundary
conditions our domain is not bounded any more, in fact, it becomes
infinite.

This peculiar behavior is apparently present for any numerical
discretization, however, for the Lax-Wendroff scheme the required
resolution in order to suppress the instability can be so high that it
can prevent one from performing evolutions within a reasonable time
frame. It is not clear to us why for the Lax-Wendroff scheme the
instability is much more persistent than for the other two schemes we
were investigating. But this shall be of no concern, for we will
rewrite the equations in such a way that they can be integrated in a
stable way for any resolution.

\subsection{The solution of the problem: Rewrite the fluid equation}

From the previous section it has become clear that the exploding modes
are due to the smallness of the sound speed combined with insufficient
resolution. Increasing the resolution for a given profile of the sound
speed or increasing the minimum value of the sound speed for a given
resolution will eventually result in numerical stability.

However, for the physical equations the profile of the sound speed is
fixed for a given stellar model and it is only by further increasing
the resolution that we can prevent the occurrence of the instability.

Yet, the nature of the instability is such that we only need the high
resolution in the small region close to the surface of the star, where
the sharp drop in the sound speed occurs. If we use a uniform grid, we
have to use the same resolution throughout the whole domain even in
those places where it would not be necessary. And in our case, this
would be 99\% of our domain!

The natural way out one might first think of is to locally refine
the grid and to provide the required resolution only in the region where
it is really needed. This could be accomplished by fixed mesh
refinement since this region is determined by the profile of the sound
speed, which does not change throughout the evolution. However, we then
would have to deal with the transition from the coarse grid to the
fine grid and vice versa, which might be troublesome. Another drawback
is that for different stellar models we would need a different grid
refinement and it would be a matter of trial and error to find the
appropriate refinements for a stable evolution.

Yet, there is a better way out. We can try to find a new radial
coordinate $x$ which is related to the actual radial coordinate $r$ in
such a way that an equidistant grid in $x$ would correspond to a grid
in $r$ that becomes automatically denser in regions where the sound
speed assumes small values. A simple relation between the grid
spacings $\Delta x$ and $\Delta r$ that would have the desired
properties is
\begin{align}\label{DrDx}
        \Delta r(r)= C_s(r)\Delta x\;.
\end{align}
An equidistant discretization with a constant grid spacing $\Delta x$
would correspond to a coarse grid in $r$ for large values of $C_s$,
which becomes finer and finer as the sound speed $C_s$ decreases.

From \eqref {DrDx} we can immediately deduce the form of our new
coordinate $x$ as a function of $r$. By replacing in \eqref {DrDx}
the $\Delta$'s by differentials we obtain
\begin{align}\label{transf}
        \frac{dx}{dr} \= \frac{1}{C_s(r)}
\end{align}
or 
\begin{align}\label{int_transf}
        x(r) \= \int_0^r\frac{dr'}{C_s(r')}\;.
\end{align}
As a consequence, the derivatives transform as
\begin{align}\label{diff_transf}
        \frac{d}{dx} \= C_s\frac{d}{dr}
\end{align}
and
\begin{align}
        \frac{d^2}{dx^2} \= C_s^2\frac{d^2}{dr^2} + C_s(C_s)'\frac{d}{dr}\;.
\end{align}
From the last relation we see that the thus defined coordinate
transformation will transform the wave equation in such a way that the
propagation speed with respect to the $x$-coordinate will be one
throughout the whole stellar interior.

Of course, we have to use \eqref{transf} with some caution, for if
$C_s = 0$ this transformation becomes singular. And this is what
happens at the stellar surface. If, for instance, the profile of the
sound speed is given in the form $C_s = C_0(1 - r/R)$, where we have
$C_s(R) = 0$, we can find an analytic expression for $x$
\begin{align}
        x(r) \= -\frac{R}{C_0}\log\(1 - \frac{r}{R}\)\;,
\end{align}
which tells us that at the surface $r = R$ we obtain $x(R) = \infty$.
In this case the coordinate transformation seems to be quite useless
since numerically we cannot deal with a grid that extends up to
infinity. We would have to truncate it somewhere. But from a numerical
point of view this is not that bad since going to infinity in the
$x$-coordinate would mean to have an infinitely fine resolution in the
$r$-coordinate at the stellar surface. But this is numerically
impossible as well, so truncating the $x$-coordinate at some point
means to define a maximal resolution in $r$ at the surface.

It should be noted that the above transformation is of the same kind
as in the definition of the tortoise coordinate $r_*$, which we will
briefly encounter in the next chapter. Here, too, the definition
\begin{align}
        \frac{dr}{dr_*} \= e^{\nu -\lam}
\end{align}
leads to wave equations with propagation speeds of one throughout the
whole domain (cf. the definitions of the Regge-Wheeler \eqref{RWex} and
Zerilli equation \eqref{Zeqn}). Thus, we may call $x$ a hydrodynamical
tortoise coordinate.

\begin{figure}[t]
\leavevmode
\epsfxsize=\textwidth
\epsfbox{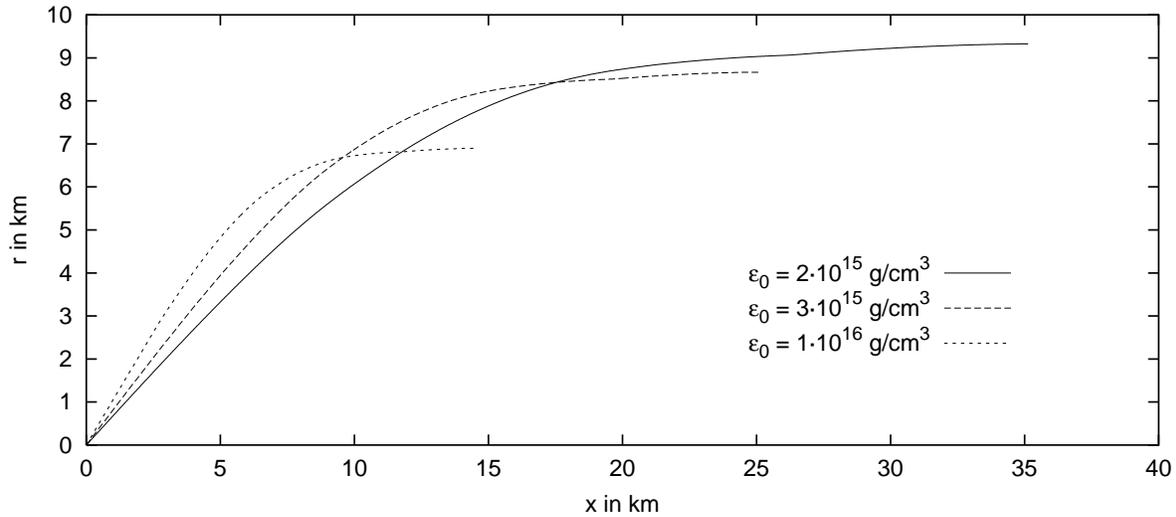}
\caption{\label{r_x}The $r$-coordinate as a function of the $x$-coordinate
for three different stellar models.}
\end{figure}

To obtain the background data on the $x$-grid, we have to solve the
TOV equations \eqref{toveqns} with respect to $x$. Since we also need
$r$ as a function of $x$, we simultaneously solve \eqref{transf},
too. The transformed set of equations then reads:
\begin{subequations}
\label{tovx}
\begin{align}
        \label{tovx1}\frac{d\lam}{dx} \= C_s\(\frac{1 - e^{2\lam}}{2r}
        + 4\pi r e^{2\lam}\epsilon\)\\
        \label{tovx2}\frac{d\nu}{dx} \= C_s\(\frac{e^{2\lam} - 1}{2r}
        + 4\pi r e^{2\lam}p\)\\
        \label{tovx3}\frac{dp}{dx} \= -\frac{d\nu}{dx}(p + \eps)\\
        \label{tovx4}\frac{dr}{dx} \= C_s\;.
\end{align}
\end{subequations}
Of course, we also have to rewrite the wave equation \eqref{wave_w}
in the new coordinate:
\begin{align}
\begin{split}\label{wave_x}
        \dff{w}{t} \= e^{2\nu-2\mu}\bigg[\,
        \dff{w}{x} + \(2\nu_{,x} + \mu_{,x} + \frac{C_{s,x}}{C_s}
        - \frac{\nu_{,x}}{C^2_s}\)\df{w}{x}\\
        &{} \quad
        + \(C_s\(2\frac{\nu_x}{r} + \frac{\mu_x}{r} - C_s\frac{2}{r^2}\)
        + \frac{C_{s,x}}{r} + \frac{e^{2\mu} - 1}{r^2}
        + \frac{\nu_x}{C_s}\(\frac{\nu_x}{C_s} + \frac{1}{r}\)\)w\,\bigg]\;.
\end{split}
\end{align}
Here, the subscript $x$ denotes a derivative with respect to $x$. The
last missing thing is the transformation of the boundary condition
\eqref{bcw}. Unfortunately, we cannot transform \eqref{bcw} in a
straightforward way since at the surface it is $C_s = 0$, and therefore
the transformation of the derivative $d/dr = (C_s)^{-1} d/dx$ is not
defined. However, the inverse transformation \eqref{diff_transf} can
make sense if we note that if $C_s = 0$ then any derivative with
respect to $x$ has to vanish. This is in particular true for $w$
itself. Thus, at the stellar surface $r = R$, where the sound speed
$C_s$ vanishes we can impose the following boundary condition for $w$:
\begin{align}\label{bcx}
        \df{w}{x}|_{x(R)} \= 0\;.
\end{align}
This corresponds to reflection at a loose end on the $x$-grid.
Actually, this boundary condition is more general than the old one
\eqref{bcw} since any finite boundary condition for $w'$ on the $r$-grid
would translate into \eqref{bcx}. Conversely, the above boundary
condition \eqref{bcx} ensures the finiteness of $w'$ and $w$ at the
surface, which is what actually follows from the vanishing of the
Lagrangian pressure perturbation in \eqref{Delta_p}. We do not even
need to know the actual boundary condition for $w'$, the equations
will automatically lead to the correct one.

Figure \ref{r_x} shows the $r$-coordinate as a function of the
$x$-coordinate for three different stellar models. From the curves it
is clear that an equidistant grid spacing in $x$ will result in an
equivalent spacing in $r$ that gets very dense towards the surface of
the star, which means that this part gets highly resolved. And this is
exactly what we need in order to overcome the instability and to
obtain a decent accuracy.

\begin{figure}[p]
\vspace*{-2cm}
\begin{minipage}{\textwidth}
\leavevmode
\epsfxsize=\textwidth
\epsfbox{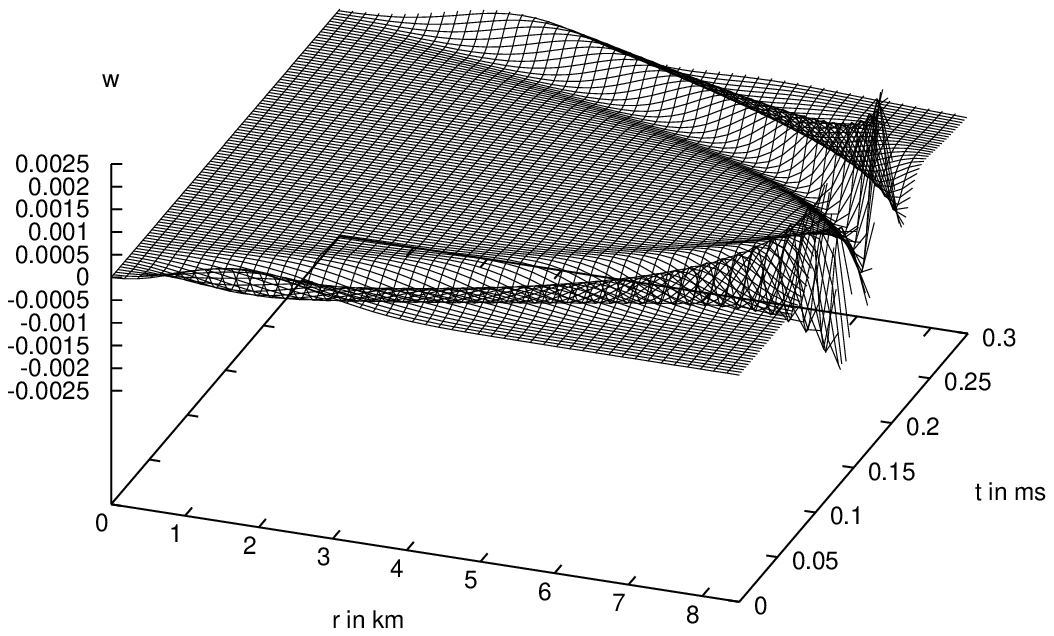}
\vspace*{-2cm}
\end{minipage}
\begin{minipage}{\textwidth}
\leavevmode
\epsfxsize=\textwidth
\epsfbox{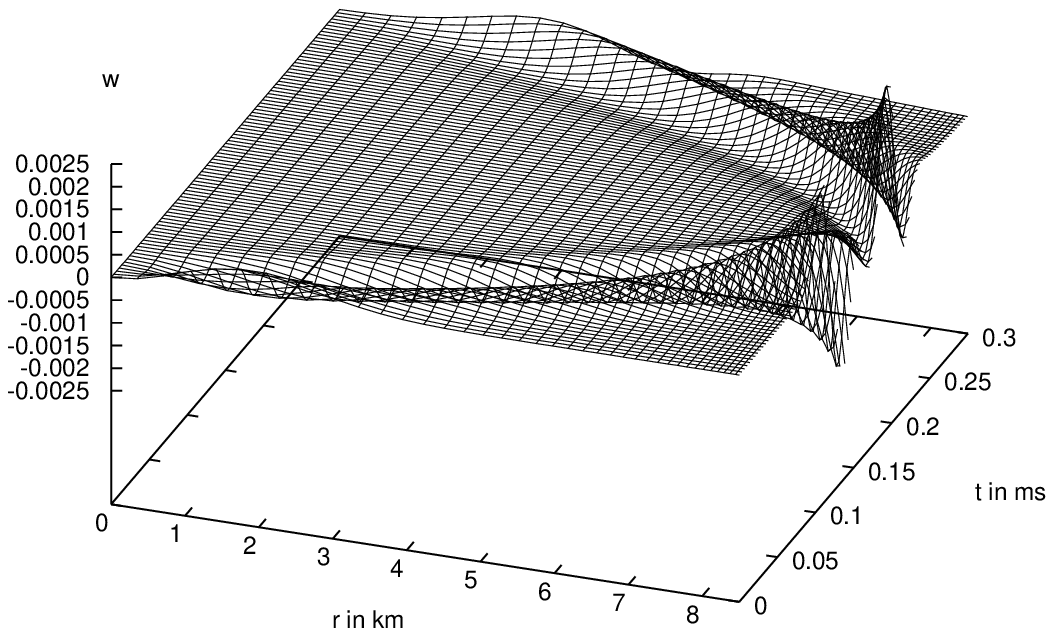}
\caption{\label{w_evol}Evolution of $w$ on the $r$-grid (upper panel)
and on the  $x$-grid (lower panel). In both cases we have excised the part
close to the surface.}
\end{minipage}
\end{figure}

In Fig.~\ref{w_evol} we compare the evolution of $w$ using the wave
equation \eqref{wave_w} on the $r$-grid with the transformed wave
equation \eqref{wave_x} on the $x$-grid. We show the propagation of a
perturbation that starts travelling from the origin towards the stellar
surface and then gets reflected. In both cases we use a resolution of
50 grid points inside the star. For this low resolution a realistic
stellar model would cause the evolution on the $r$-grid to be unstable,
hence here we use a polytropic equation of state. 

In the lower panel of Fig.~\ref{w_evol} we evolve $w$ on the
$x$-grid, but we plot it as a function of $r_i = r(x_i)$. We can
clearly see that the density of grid points increases as one
approaches the stellar surface.

\begin{figure}[t]
\leavevmode
\epsfxsize=\textwidth
\epsfbox{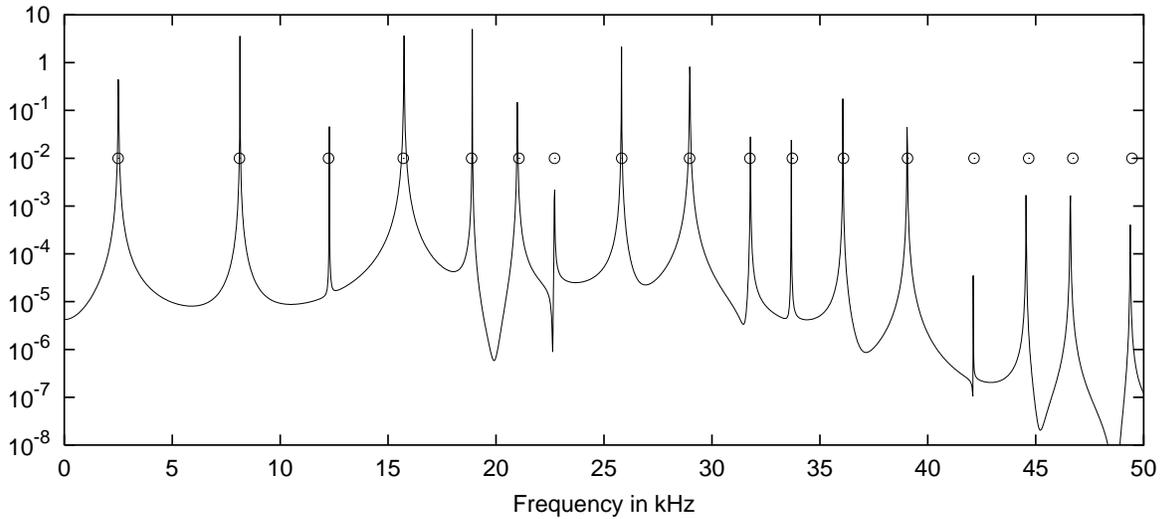}
\caption{\label{MPAfft}Spectrum of an evolution using the transformed
equation \eqref{wave_x} for the MPA eos with $N = 200$. The circles represent
the frequencies of the eigenmodes obtained from \eqref{ev_eq}.}
\end{figure}
\clearpage

\begin{figure}[p]
\vspace*{-2cm}
\begin{minipage}{\textwidth}
\leavevmode
\epsfxsize=\textwidth
\epsfbox{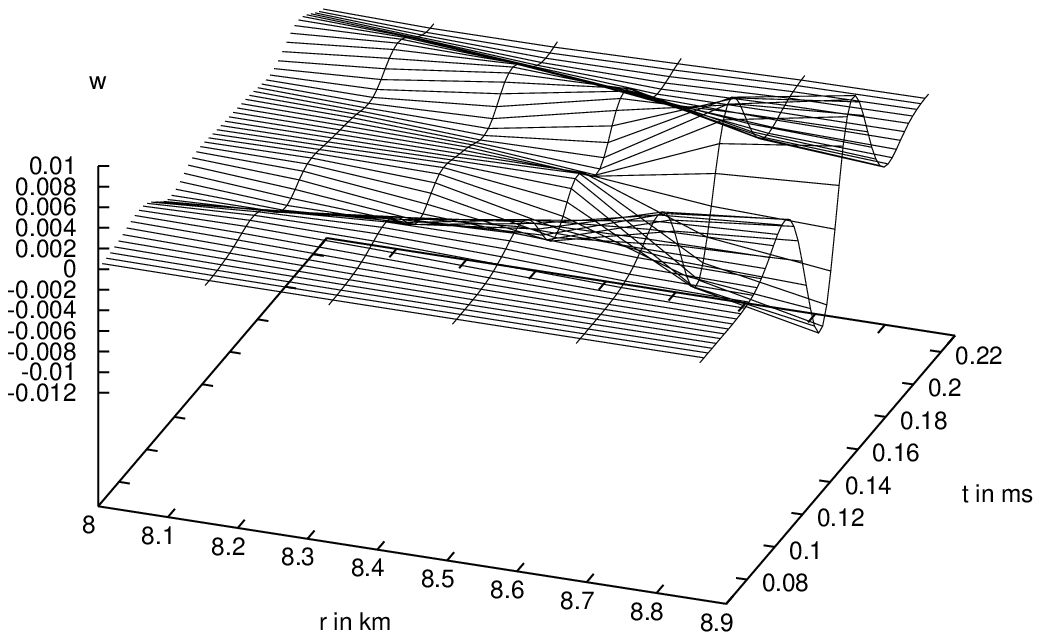}
\vspace*{-2cm}
\end{minipage}
\begin{minipage}{\textwidth}
\leavevmode
\epsfxsize=\textwidth
\epsfbox{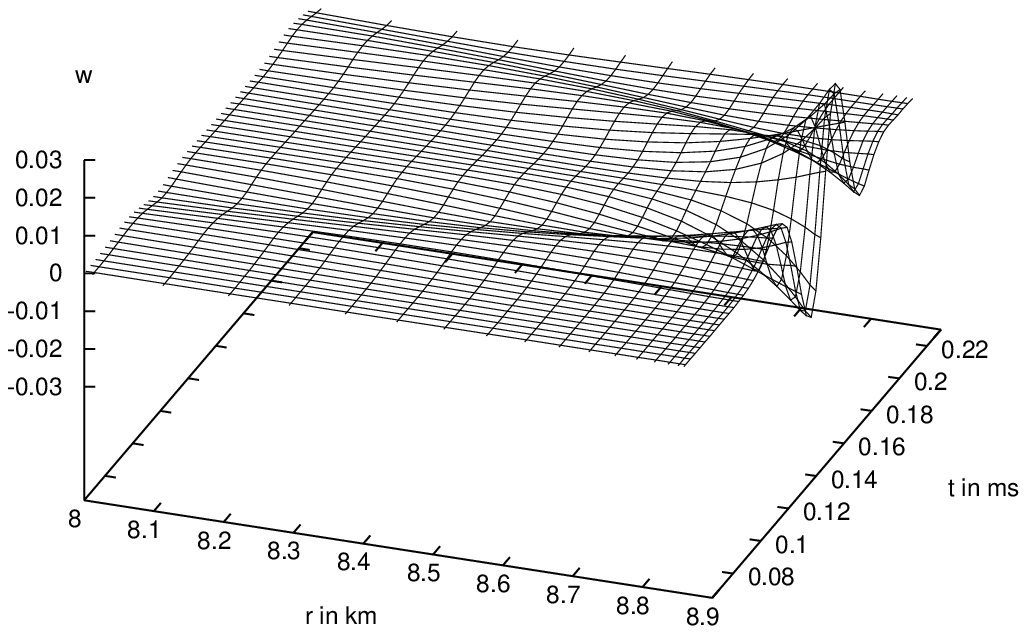}
\vspace*{-5mm}
\caption{\label{w_evol_zoom}The excised part of Fig.~\ref{w_evol}.
The region close to the surface gets much better resolved on the
$x$-grid (lower panel) than on the $r$-grid (upper panel).}
\end{minipage}
\end{figure}

We have cut off the part close to the surface, since here the value of
$w$ drastically increases. In the part shown the two evolutions look
quite alike and it seems that there is no real advantage of one over
the other. It is only at the surface of the star that the advantage
of the better resolution comes to light, which is shown in
Fig.~\ref{w_evol_zoom}. Here the amplitude of $w$ is much higher
because it can be much better resolved.

Finally, we want to demonstrate the effectiveness of our coordinate
transformation by evolving an initial perturbation for the MPA
equation of state. We choose the central energy density to be
$3\!\cdot\!10^{15}\,$g/cm$^3$, which yields a stellar mass of $M = 1.49
M_\odot$. The resolution is N = 200 grid points, which would be
too low for the other cases to yield a stable evolution. Here we do
not have any problems, the evolution is stable for {\it any} chosen
resolution.

In Fig.~\ref{MPAfft} we only show the spectrum that results from the
evolution. We can see many sharp peaks, which perfectly agree with the
corresponding eigenfrequencies computed by solving the eigenvalue
equations \eqref{ev_eq}.

\clearpage

\chapter{Non-radial oscillations of neutron stars}

Starting from the general expansions of the metric \eqref{exp_h} and
extrinsic curvature \eqref{exp_k}, we will choose shift and lapse in
such a way that we can obtain the famous Regge-Wheeler gauge, which
was first introduced by Regge and Wheeler \cite{RW57} in the context
of a stability analysis of black holes. This gauge was also used by
Thorne et al.~\cite{TC67,Th68,PT69,Th69a,Th69b,IT73} in the their
pioneering papers on neutron star oscillations and still is widely
used by many authors.

The main focus at that time was to investigate the different kinds of
oscillation modes that a neutron star possesses. Therefore the time
dependence of the equations was assumed to be given by $e^{\I\omega
t}$, which reduces the equations to a set of ordinary differential
equations. Those had to be solved together with the appropriate
boundary conditions. The first system that was presented was of fifth
order, but it was later discovered that it was erroneous, and instead a
fourth order system \cite{IT73} could fully describe the non-radial
oscillations of a neutron star. Detweiler \& Lindblom \cite{LD83} then
used this set of equations (which was actually slightly modified due
to numerical reasons) to compute the $f$- and some $p$-modes modes for
stellar models with various equations of state.

Chandrasekhar \& Ferrari \cite{CF91a}, however, chose a different
gauge, the so-called diagonal gauge, and were able to describe the
oscillations in terms of pure metric perturbations similar to the
black hole case. This enabled them to treat those oscillations as the
scattering of gravitational waves at the given background metric.

However, the resulting system of equations was of fifth order and
therefore allowed for an additional solution that had to be rejected
because it was divergent at the origin. The diagonal gauge was already
used much earlier by Chandrasekhar to describe the perturbations of
black holes \cite{Chandra}. Here, too, he obtained systems of
equations whose degrees were higher by one than the equations that
were obtained in the Regge-Wheeler gauge. Of course, they also allowed
for spurious solutions, which had to be rejected on physical grounds.

In 1981, Xanthopoulos \cite{Xan81a,Xan81b} then showed that the
spurious solution (which was later called Xanthopoulos solution)
could actually be used to reduce the degree of those systems by
one. However, it was still not clear why in the diagonal gauge one
was always led to equations, which were higher by one degree than in
the Regge-Wheeler gauge.

Shortly after the papers of Chandrasekhar \& Ferrari, Ipser \& Price
\cite{IP91} demonstrated that also in the Regge-Wheeler gauge it is
possible to construct a fourth order system that does not use any
fluid perturbations. Furthermore, they showed \cite{PI91} that the
reason of the diagonal gauge yielding a fifth order system instead of
a fourth order system was that this gauge still allowed for an
additional nontrivial gauge transformation, of which nobody has been
aware before.

The Regge-Wheeler gauge in turn is complete in the sense that it does
not allow for additional gauge transformations. The main advantage of
this gauge is that it halves the number of relevant metric
coefficients in \eqref{exp_h}. Since some of the vanishing metric
components belong to the spatial part of the metric, it is clear that
we have to choose shift and lapse in such a way that during the 
evolution those components remain zero. 

If we make the following ansatz for lapse and shift (we omit the
indices $l$ and $m$):
\begin{subequations}
\begin{align}
        \widehat{S}_1 \= -\half e^{\nu - 2\lam}\widehat{S}_3\\
        \widehat{S}_2 \= 2e^\nu \widehat{K}_2\\
        \widehat{V}_1 \= 0\\
        \widehat{V}_2 \= e^\nu \widehat{K}_6\;,
\end{align}
\end{subequations}
the evolution equations for the spatial metric coefficients
$\widehat{V}_3, \widehat{T}_1$ and $\widehat{T}_3$
reduce to
\begin{align}
        \df{}{t}\widehat{V}_3 \= 0\\
        \label{T1}\df{}{t}\widehat{T}_1 \= -2e^\nu \widehat{K}_4\\
        \df{}{t}\widehat{T}_3 \= 0\;.
\end{align}
In addition the evolution equation for $\widehat{K}_4$ depends
only on $\widehat{T}_1$ and $\widehat{V}_3$. Thus, if we
initially set $\widehat{V}_3 = \widehat{T}_1 =
\widehat{T}_3 = \widehat{K}_4 = 0$, we obtain
$\d\widehat{V}_3/\d t = \d\widehat{T}_1/\d t =
\d\widehat{T}_3/\d t = \d\widehat{K}_4/\d t = 0$, which ensures
the vanishing of all those coefficients for all times.

This leaves us with only three dynamic metric variables namely the
axial variable $\widehat{V}_4$ and the two polar variables
$\widehat{S}_3$ and $\widehat{T}_2$. A glance at the
Hamiltonian constraint reveals that in the exterior region, the two
polar variables $\widehat{S}_3$ and $\widehat{T}_2$ are not
even independent from each other. This means that in the exterior
region, axial and polar perturbations each only have one degree of
freedom. This is in agreement with the general observation that any
gravitational wave can always be described by a superposition of two
independent polarization states.

\section{Axial perturbations}\label{odd}

Due to the spherical symmetry of the background the polar and axial
equations are totally decoupled from each other. As was already
mentioned, the axial perturbations are characterized by just one
metric variable $\widehat{V}_4$. Furthermore, they do not couple to the
fluid of the neutron star since energy density and pressure are scalar
quantities, and therefore their perturbations belong to the polar
class. Still, there could be some motion of the fluid, described by
the axial coefficient $\hat{u}_3$ of the 4-velocity, but the
dynamical equation for $\hat{u}_3$ turns out to be
\begin{align}
        \df{}{t}\hat{u}_3 \= 0\;,
\end{align}
which shows that nonzero $\hat{u}_3$ would describe at most some
stationary fluid motion. 

Nevertheless, the axial perturbations are not uninteresting and have
been studied in much detail \cite{CF91b,Kok94,AKK96}. It has been
shown that each neutron star model possesses a characteristic spectrum
of quasi-normal modes, which depends mainly on the compactness of the
model. It is clear that those modes cannot be excited be means of
fluid perturbations of the neutron star, but they can very well be
induced by impinging gravitational waves or by the gravitational
potential of a large mass moving on a close orbit. In this thesis,
however, we mainly focus on polar perturbations, but for sake of
completeness we will also write down the equations for axial
perturbations.

Yet, we will use a slightly different expansion from the one in
\eqref{exp_h} and \eqref{exp_k}, which somewhat simplifies the resulting
equations. The nonzero axial components of the metric are (summation
over all $l$ and $m$ is implied):
\begin{subequations}
\begin{align}
        \(\beta_\theta,\,\beta_\phi\) \= e^{\nu-\lam}K_6^{lm}
        \(-\sin^{-1}\theta\df{}{\phi}Y_{lm},\,
        \sin\theta\df{}{\theta}Y_{lm}\)\\
        \(h_{r\theta},\,h_{r\phi}\) \= e^{\lam-\nu}V_4^{lm}
        \(-\sin^{-1}\theta\df{}{\phi}Y_{lm},\,
        \sin\theta\df{}{\theta}Y_{lm}\)\;,
\end{align}
\end{subequations}
and the ones of the extrinsic curvature read
\begin{subequations}
\begin{align}
        \(k_{r\theta},\,k_{r\phi}\) \= \half e^{\lam}K_3^{lm}
        \(-\sin^{-1}\theta\df{}{\phi}Y_{lm},\,
        \sin\theta\df{}{\theta}Y_{lm}\)\\
        \(\begin{array}{cc}
        k_{\theta\theta} & k_{\theta\phi}\\
        k_{\phi\theta} & k_{\phi\phi}\end{array}\) \= \half e^{-\lam}K_6^{lm}
        \(\begin{array}{cc}
        -\sin^{-1}\theta\,X_{lm} & \sin\theta\,W_{lm}\\
        \sin\theta\,W_{lm} & \sin\theta\,X_{lm}\end{array}\)\;,
\end{align}
\end{subequations}
with $X_{lm}$ and $W_{lm}$ defined in Equations \eqref{Xdef} and
\eqref{Wdef} of Appendix A. This gives us the following set of
evolution equations (we again omit the indices $l$ and $m$):
\begin{subequations}
\label{axial}
\begin{align}
        \label{V4} \df{V_4}{t} \= e^{2\nu-2\lam}\(\df{K_6}{r}\
         + \(\nu' - \lam' - \frac{2}{r}\)K_6 - e^{2\lam}K_3\)\\
        \df{K_3}{t} \= \frac{l(l+1) - 2}{r^2}\,V_4\\
        \label{K6} \df{K_6}{t} \= \df{V_4}{r}\;,
\end{align}
\end{subequations}
and one constraint equation:
\begin{align}\label{oddMC}
        \df{K_3}{r} + \frac{2}{r}K_3 - \frac{l(l+1) - 2}{r^2}K_6
        \= 16\pi e^{\lam}(p + \eps){\hat u}_3\;.
\end{align}
The three evolution equations can be combined to yield a single wave
equation for the metric perturbation $V_4$. If instead of $V_4$ we use
\begin{align}
        Q \;&:=\; \frac{V_4}{r}\;,
\end{align}
this wave equation reads
\begin{align}\label{RWin}
        \dff{Q}{t} \= \frac{\pa^2Q}{\pa r^2_*}
        + e^{2\nu}\( 4\pi\(p - \eps\) + \frac{6m}{r^3}
        - \frac{l(l+1)}{r^2}\)Q\;,
\end{align}
where $r_*$ is the tortoise coordinate defined by
\begin{align}
        \frac{dr}{dr_*} \= e^{\nu-\lam}\;.
\end{align}
In the exterior, we have $m = M$ the total mass and $\eps = p = 0$ and
\eqref{RWin} reduces to the famous Regge-Wheeler equation
\begin{align}\label{RWex}
        \dff{Q}{t} \= \frac{\pa^2Q}{\pa r^2_*}
        + e^{2\nu}\(\frac{6M}{r^3} - \frac{l(l+1)}{r^2}\)Q\;.
\end{align}
In this case we can give an analytic relation between $r_*$ and $r$:
\begin{align}
        r_* \= r + 2M\log\(\frac{r}{2M} - 1\)\;.
\end{align}

\section{Polar perturbations}\label{polar}

Here, too, we will use slightly different expansion coefficients from
\eqref{exp_h}, which will give us a set of evolution equations that is
well suited for numerical treatment. We decompose the metric as
follows (again, summation over $l$ and $m$ is implied):
\begin{align}
        \alpha \= -\half e^{\nu}\(\frac{T^{lm}}{r} + rS^{lm}\)Y_{lm}\\
        \beta_r \= e^{2\lam}K_2^{lm}Y_{lm}\\
        h_{ij} \= 
        \(\begin{array}{ccc}
        e^{2\lam}\(\frac{T^{lm}}{r} + rS^{lm}\) & 0 & 0\\
        0 & r\,T^{lm} & 0\\
        0 & 0 & r\sin^2 T^{lm}\end{array}\)Y_{lm}\;,\label{p_metric}
\end{align}
and for the extrinsic curvature (symmetric components are denoted by an
asterisk)
\begin{align}
        \!\!\!\!k_{ij} \= 
        -\half e^{-\nu}\(\begin{array}{ccc}
        \frac{e^{2\lam}}{r}K_1^{lm} & - e^{2\lam}K_2^{lm}\df{}{\theta}& 
        -e^{2\lam}K_2^{lm}\df{}{\phi}\\
        \star & r\(K_5^{lm} - 2K_2^{lm}\)& 0\\
        \star & 0 & r\sin^2\theta \(K_5^{lm} - 2K_2^{lm}\)
        \end{array}\)Y_{lm}\;.
\end{align}
In addition we have the matter variables
\begin{align}
        \delta\eps \= \frac{\rho^{lm}}{r}\,Y_{lm}\label{d_eps}\\
        \delta u_r \= -\frac{e^\nu}{r}\(u_1^{lm}
        - \frac{u_2^{lm}}{r}\)Y_{lm}\\
        \(\delta u_\theta,\,\delta u_\phi\)
        \= -\frac{e^\nu}{r}u^{lm}_2
        \(\df{Y_{lm}}{\theta},\,\df{Y_{lm}}{\phi}\)\;.
\end{align}
It is only through this particular choice of expansion coefficients
that we obtain a system of equations that can numerically be
integrated in a quite straightforward and -- this is the main point --
a stable way. It is a quite well known and particularly bothersome
feature of spherical coordinates that the resulting equations usually
contain divergent terms, which exactly cancel at the origin, or
quasi-singular terms, which at the origin reduce to indefinite
expressions of the kind $0/0$. For physical reasons the latter have to
be finite. If not intuition then at least bitter experience tells us
that it can be impossible to obtain a numerical stable evolution
scheme for the ``raw'' equations without further manipulations. It is
clear that a standard numerical discretization scheme cannot fully
take into account the cancellations that occur on the analytic level,
which results in severe instabilities at the origin.

As a first step to overcome those difficulties one therefore has to
use special linear combinations of the variables to avoid the presence
of the divergent terms that have to cancel. That is the reason why we
use the particular combinations of $S$ and $T$ in the expansions of
the lapse $\alpha$ and the metric component $h_{rr}$. Furthermore, the
scaling with $r$ (e.g. $\rho^{lm}/r$ instead of just $\rho^{lm}$ in
\eqref{d_eps}) has been chosen such that the remaining terms, which
become indefinite expressions at the origin, are only terms that do not
contain any derivatives of the variables themselves. With this choice
all the coefficients but $u^{lm}_2$ have the same behavior at the
origin, namely
\begin{align*}
        Q^{lm}(t,r) \= Q^{lm}_0(t)\,r^{l+1} + Q^{lm}_1(t)\,r^{l+3} + \dots\;.
\end{align*}
The leading term of $u^{lm}_2$ is proportional to $r^l$. We will
discuss the different ways of writing the equations and the associated
numerical problems in much more detail in the section on the numerical
implementation. Besides, for the sake of clarity we will again
suppress the indices $l$ and $m$ throughout the rest of this chapter.

Our last step before writing down the equations is to replace $K_1$ by
the following quantity
\begin{align}\label{Kdef}
        r^2K \;:=&\; K_1 + 2r\(\df{K_2}{r} + \lam'K_2\) - K_5\;.
\end{align}
This is necessary in order to get rid of the last remaining singular
terms which might be a threat to the numerical evolution.

In this way, we obtain a system of 5 coupled evolution equations,
which are of first order in time but still second order in space.
There are 2 equations for the metric variables $S$ and $T$ and the
3 more for the extrinsic curvature variables $K$, $K_2$ and $K_5$:
\begin{subequations}
\label{FOsys}
\begin{align}
        \label{S}
        \df{S}{t} \= K\\
\begin{split}
\label{K}
        \df{K}{t} \= e^{2\nu-2\lam}\bigg[
        \dff{S}{r} + \(5\nu' - \lam'\)\df{S}{r}\\
        &\quad{} + \(4\(\nu'\)^2 + 5\frac{\nu'}{r} + 3\frac{\lam'}{r}
        - 2\frac{e^{2\lam} - 1}{r^2} - e^{2\lam}\frac{l(l+1)}{r^2}\)S\\
        &\quad{} + 4\(\frac{1}{r}\(\frac{\nu'}{r}\)' + 2\(\frac{\nu'}{r}\)^2
        - \frac{\lam'\nu'}{r^2}\)T\,\bigg]
\end{split}
\end{align}
\begin{align}
        \label{T}
        \df{T}{t} \= K_5\\
\begin{split}\label{K5}
        \df{K_5}{t} \= e^{2\nu-2\lam}\bigg[
        \dff{T}{r} + \(\nu' - \lam'\)\df{T}{r}\\
        &\quad{} + \(\frac{\nu'}{r} + 3\frac{\lam'}{r}
        + 2\frac{e^{2\lam} - 1}{r^2} - e^{2\lam}\frac{l(l+1)}{r^2}\)T\\
        &\quad{} + 2\(r\nu' + r\lam' - 1\)S\,\bigg]
        + 8\pi e^{2\nu}\(1 - C^2_s1\)\rho
\end{split}\\
        \label{K2}
        \df{K_2}{t} \=
        e^{2\nu-2\lam} \(r\df{S}{r} + \(2r\nu' + 1\)S
        + 2\frac{\nu'}{r}T\)\,.
\end{align}
\end{subequations}
We could easily convert this system into a first order system in time
and space by adding another two evolution equations for the first
derivative $S'$ and $T'$. However, the form of the first four
equations \eqref{S} -- \eqref{K5}, which are independent of $K_2$,
suggests to rather convert them into two coupled wave equations for
$S$ and $T$
\begin{align}
\begin{split}
        \label{waveS} \dff{S}{t} \= e^{2\nu-2\lam}\bigg[ \dff{S}{r} +
        \(5\nu' - \lam'\)\df{S}{r}\\ &{}\qquad\quad + \(4\(\nu'\)^2 +
        5\frac{\nu'}{r} + 3\frac{\lam'}{r} - 2\frac{e^{2\lam} -
        1}{r^2} - e^{2\lam}\frac{l(l+1)}{r^2}\)S\\ &{}\qquad\quad +
        4\(\frac{1}{r}\(\frac{\nu'}{r}\)' + 2\(\frac{\nu'}{r}\)^2 -
        \frac{\lam'\nu'}{r^2}\)T\,\bigg]
\end{split}\\
\begin{split}
        \label{waveT}
        \dff{T}{t} \= e^{2\nu-2\lam}\bigg[
        \dff{T}{r} + \(\nu' - \lam'\)\df{T}{r}\\
        &{}\qquad\qquad + \(\frac{\nu'}{r} + 3\frac{\lam'}{r}
        + 2\frac{e^{2\lam} - 1}{r^2} - e^{2\lam}\frac{l(l+1)}{r^2}\)T\\
        &{}\qquad\qquad + 2\(r\nu' + r\lam' - 1\)S\,\bigg]
        + 8\pi e^{2\nu}\(1 - C^2_s\)\rho\;,
\end{split}
\end{align}
which are equivalent to equations (14) and (15) of Allen et al.
\cite{AAKS98}~(Their variables $F$ and $S_{Allen}$ are related to
ours as follows: $F = T$ and $S_{Allen} = e^{2\nu}S$). As can be seen,
the wave equation for $S$ \eqref{waveS} is totally decoupled from the
fluid variable $\rho$, which only couples to the metric perturbation
$T$ in \eqref{waveT}. The equation for $K_2$ is only necessary in the
interior region, where it couples to the hydrodynamical equations,
which follow from energy conservation $D_\nu T^{\mu\nu}$, and are
given by
\begin{subequations}
\label{rho_system}
\begin{align}
\begin{split}\label{rho}
        \df{\rho}{t} \= e^{2\nu - 2\lam}\bigg[\,\df{\tilde u_1}{r} 
        - \frac{1}{r}\df{\tilde u_2}{r}\\
        &\qquad\qquad{} + \(3\nu' - \lam' + \frac{1}{r}\)\tilde u_1
        - \(3\frac{\nu'}{r} - \frac{\lam'}{r}
        + e^{2\lam}\frac{l(l+1)}{r^2}\)\tilde u_2\,\bigg]\\
        &\quad{} + \(p + \eps\)\(r\df{K_2}{r} + \(2 + r\lam'\)K_2
        - \frac{r^2}{2}K - \frac{3}{2}K_5\) + r\eps'K_2
\end{split}
\end{align}
\begin{align}
\begin{split}
        \label{u1}
        \df{\tilde u_1}{t} \= C_s^2\df{\rho}{r}
        + \(\nu'\(1 + C_s^2\) + \(C_s^2\)'\)\rho
        - \half\(p + \eps\)\(r^2\df{S}{r} + \df{T}{r} + 2rS\)
\end{split}\\
        \label{u2}
        \df{\tilde u_2}{t} \= C_s^2\rho - \half\(p + \eps\)\(r^2 S + T\)\;.
\end{align}
\end{subequations}
Here, we have defined $\tilde u_i := \(p + \eps\)u_i$. By introducing
the enthalpy perturbation
\begin{align}\label{H_def}
         H := \frac{C^2_s}{p + \eps}\rho\;,
\end{align}
the fluid equations assume a more convenient form:
\begin{subequations}
\label{H_system}
\begin{align}
\begin{split}\label{H}
        \df{H}{t} \= e^{2\nu - 2\lam}C_s^2\df{u_1}{r} 
        + C_s^2\(r\df{K_2}{r} + \(2 + r\lam'\)K_2
        - \frac{r^2}{2}K - \frac{3}{2}K_5\) - r\nu'K_2\\
        &\quad{} + e^{2\nu - 2\lam}\(C_s^2
        \(2\nu'\ - \lam'\) - \nu'\)u_1\\
        &\quad{} + e^{2\nu - 2\lam}\(C_s^2
        \(\frac{\lam'}{r} - 2\frac{\nu'}{r}
        - e^{2\lam}\frac{l(l+1)}{r^2}\) + \frac{\nu'}{r}\)u_2
\end{split}\\
        \label{uu1}
        \df{u_1}{t} \= \df{H}{r} - \half\(r^2\df{S}{r} + \df{T}{r} + 2rS\)\\
        \label{uu2}
        \df{u_2}{t} \= H - \half\(r^2S + T\)\;.
\end{align}
\end{subequations}
Interestingly, from \eqref{uu1} and \eqref{uu2} it follows that the
coefficients $u_1$ and $u_2$ are not independent of each other but
rather are related via
\begin{align}\label{dfu2}
        u_1 \= \df{u_2}{r} \;.
\end{align}
The above system \eqref{H_system}, too, can be cast into a second
order wave equation for $H$, which is equivalent to equation (16) of
Allen et al.~\cite{AAKS98}~(the different signs in the terms containing
$S$ and $T$ are correct):
\begin{align}
\begin{split}\label{waveH}
        \dff{H}{t} \= e^{2\nu - 2\lam}\bigg[\,C^2_s\dff{H}{r} 
        + \(C^2_s \(2\nu' - \lam'\) - \nu'\)\df{H}{r}\\
        &{}\qquad + \(C^2_s
        \(\frac{\nu'}{r} + 4\frac{\lam'}{r} - e^{2\lam}\frac{l(l+1)}{r^2}\)
        + 2\frac{\nu'}{r} + \frac{\lam'}{r}\)H\\
        &{}\qquad + \frac{1}{2}\nu'\(C^2_s - 1\)\(r^2\df{S}{r} - \df{T}{r}\)\\
        &{}\qquad + \(C^2_s\(\frac{7}{2}\frac{\nu'}{r} + \frac{\lam'}{r}
        - \frac{e^{2\lam} - 1}{r^2}\)- \frac{\nu'}{r}\(2r\nu' + \half\)\)
        \(r^2S + T\)\,\bigg]\;.
\end{split}
\end{align}
It thus seems that the polar oscillations of neutron stars can be
completely described by three wave equations inside the star and two
in the exterior region. However, it is even possible to further reduce
the number of equations, for we have not made use of any of the
remaining constraint equations. The Hamiltonian constraint relates the
fluid variable $\rho$ to the two metric variables $S$ and $T$:
\begin{align}
\begin{split}
\label{HC}
        8\pi e^{2\lam}\rho  \=
        - \dff{T}{r} + \lam'\df{T}{r} + r\df{S}{r}
        + \(2 - 2r\lam' +  \half e^{2\lam}l(l+1)\)S\\
        &\quad{} - \(\frac{e^{2\lam} - 1}{r^2} + 3\frac{\lam'}{r} -
        e^{2\lam}\frac{l(l+1)}{r^2}\)T\;.
\end{split}
\end{align}
It is therefore possible to eliminate $\rho$ in equation \eqref{waveT}
and thus to obtain a consistent system of equations, where both in the
interior and exterior the two spacetime variables $S$ and $T$ are used
to describe the evolution of the oscillations. In the exterior, both
$S$ and $T$ propagate with the local speed of light $e^{\nu-\mu}$, in
the interior, however, $T$ then changes its character and propagates
with the local speed of sound $e^{\nu-\mu}C_s$.

With $\rho$ being eliminated, $S$ and $T$ in the interior now become
independent variables and the Hamiltonian constraint \eqref{HC} serves
as a definition for $\rho$. In the exterior, however, $S$ and $T$ are
not independent but have to satisfy the Hamiltonian constraint with
$\rho$ set to zero. Unfortunately, we cannot use the Hamiltonian
constraint to further eliminate one of those variables, but it is
possible to combine $S$ and $T$ to form a new variable $Z$ (we use the
definition (20) of Allen et al.~\cite{AAKS98})
\begin{align}\label{Z}
        Z \= - \frac{2}{l(l+1)}
        \frac{1 - \frac{2M}{r}}{l(l+1) - 2 + \frac{6M}{r}}
        \(2rT' + \frac{2M - r\(2 + l(l+1)\)}{r - 2M}T - 2r^2S\)\;,
\end{align}
which then satisfies a single wave equation, the famous Zerilli
equation that was first derived in 1970 by F. Zerilli \cite{Zer70b} in
the context of black hole oscillations:
\begin{align}\label{Zeqn}
        \dff{Z}{t} \= \frac{\pa^2 Z}{\pa r^2_*} 
        - 2e^{2\nu}\frac{n^2(n+1)r^3 + 3n^2Mr^2 + 9nM^2r + 9M^3}
        {r^3(nr + 3M)^2}Z\;.
\end{align}
Here, we use $2n = l(l+1) - 2$, and $r_*$ is again the tortoise
coordinate from section \ref{odd}. 

The last set of equations that is still missing are the momentum
constraints:
\begin{subequations}
\begin{align}
\begin{split}
\label{MC1}
        16\pi e^{2\nu}\(p + \eps\)u_1
        \= \df{K_2}{r} - 2\df{K_5}{r}\\
        &\quad{} + rK - \(3\nu' + 3\lam' - e^{2\lam}\frac{l(l+1)}{r}\)K_2
        + 2\nu'K_5
\end{split}\\
        \label{MC2}
        16\pi e^{2\nu}\(p + \eps\)u_2
        \= r\df{K_2}{r} - r^2K + r\(\nu' + \lam'\)K_2 - 2K_5\;.
\end{align}
\end{subequations}
They do not provide us with new information since they are equivalent
to the time derivative of the Hamiltonian constraint in the following
sense. Let ${\cal H}$, ${\cal M}_1$ and ${\cal M}_2$ denote the
righthand sides of \eqref{HC}, \eqref{MC1} and \eqref{MC2},
respectively. As already pointed out in \cite{KE93b}, we then find that
the following relation holds in the exterior region:
\begin{align}
        2\df{}{t}{\cal H} \= 
        \df{}{r}\({\cal M}_1 - \frac{{\cal M}_2}{r}\)
        + \(2\nu' + \frac{1}{r}\)\({\cal M}_1 - \frac{{\cal M}_2}{r}\)
        - e^{2\mu}\frac{l(l+1)}{r^2}{\cal M}_2\;.
\end{align}
And conversely, we have for the time derivative of the momentum
constraints
\begin{subequations}
\begin{align}
        \df{}{t}{{\cal M}_1} \= 2\df{}{r}\(e^{4\nu}{\cal H}\)\\
        \df{}{t}{{\cal M}_2} \= 2e^{4\nu}\df{}{r}{\cal H}\;.
\end{align}
\end{subequations}
This is nothing else but the contracted Bianchi identities for the
Ricci tensor and can be checked by explicitly differentiating
the constraints with respect to $t$ and then making use of the
evolution equations. 

In the interior, we have already written down the connection between
the constraints since here the Bianchi identities are equivalent to
the conservation of energy-momentum by means of the field equations
and are exactly given by the fluid equations \eqref{rho_system} or
\eqref{H_system}.

From the above relations it is clear that if the Hamiltonian
constraint is satisfied for all times, so are the momentum constraints,
and vice versa.

Let us now turn to the question which system of equations we should use
for the numerical evolution. The basic idea was to use the first order
system that results from the (3+1)-split of the field equations.
However, because of the instability problems at the origin we had to
recast the first order system in such a way that it more or less
became equivalent to the system of wave equations rewritten in first
order form. This means that there is no real advantage any more in
sticking to the first order system, on the contrary, from a
computational point of view, it is much more efficient to use the wave
equations because we need fewer equations.

If we were to use the first order system \eqref{FOsys} together with
two auxiliary variables for $S'$ and $T'$ and with the three fluid
equations \eqref{rho_system}, we would have to solve 10 equations in
the interior and 6 equations in the exterior. Of course, in the
interior we could also use the Hamiltonian constraint \eqref{HC} to
eliminate $\rho$ in \eqref{waveT}. In this case, both in the interior
and the exterior, we would have to evolve 6 equations.

If we took the wave equations the maximal set would consist of only
three equations, namely \eqref{waveS}, \eqref{waveT} and
\eqref{waveH}. Here, too, we can use the Hamiltonian constraint
\eqref{HC} to eliminate the fluid variable $\rho$ in \eqref{waveT},
which leaves us with two wave equations in both the interior and the
exterior. Since this is by far the fastest way to evolve the
perturbations of neutron stars, we will use those equations for the
numerical evolution.

The equation for $S$,
\begin{subequations}
\label{wave-eqns}
\begin{align}
\begin{split}
\label{S_wave}
        \dff{S}{t} \= e^{2\nu-2\lam}\bigg[\,
        \dff{S}{r} + \(5\nu' - \lam'\)\df{S}{r}\\
        &\quad{} + \(4\(\nu'\)^2 + 5\frac{\nu'}{r} + 3\frac{\lam'}{r}
        - 2\frac{e^{2\lam} - 1}{r^2} - e^{2\lam}\frac{l(l+1)}{r^2}\)S\\
        &\quad{} + 4\(\frac{1}{r}\(\frac{\nu'}{r}\)' + 2\(\frac{\nu'}{r}\)^2
        - \frac{\lam'\nu'}{r^2}\)T\,\bigg]\;,
\end{split}
\end{align}
is valid both in the exterior and interior, whereas for $T$ we have to
distinguish the two cases. In the interior we use \eqref{waveT} with
$\rho$ replaced with the Hamiltonian constraint \eqref{HC}
\begin{align}
\begin{split}
        \label{T_wave_in}
        \dff{T}{t} \= e^{2\nu-2\lam}C^2_s\bigg[\;
        \dff{T}{r} - r\df{S}{r} - \lam'\df{T}{r}
        +\(2r\lam' - 2 - \half e^{2\lam}l(l+1)\)S\\
        &\quad{}\qquad\qquad + \(\frac{e^{2\lam} - 1}{r^2} + 3\frac{\lam'}{r}
        - e^{2\lam}\frac{l(l+1)}{r^2}\)T\;\bigg]\\
        &\quad{} + e^{2\nu-2\lam}\bigg[\;\nu'\df{T}{r} + r\df{S}{r}
        + \(2r\nu' + \half e^{2\lam}l(l+1)\)S\\
        &\quad{}\qquad\qquad
        + \(\frac{\nu'}{r} + \frac{e^{2\lam} - 1}{r^2}\)T\;\bigg]\;,
\end{split}
\end{align}
and in the exterior region we use \eqref{waveT} with $\rho$ set to zero
\begin{align}
\begin{split}
        \label{T_wave_ex}
        \dff{T}{t} \= e^{2\nu-2\lam}\bigg[\;
        \dff{T}{r} + \(\nu' - \lam'\)\df{T}{r}\\
        &\quad{} + \(\frac{\nu'}{r} + 3\frac{\lam'}{r}
        + 2\frac{e^{2\lam} - 1}{r^2} - e^{2\lam}\frac{l(l+1)}{r^2}\)T
        + 2\(r\nu' + r\lam' - 1\)S\,\bigg]\;.
\end{split}
\end{align}
\end{subequations}
If we also used the Hamiltonian constraint \eqref{HC} in the exterior,
we would obtain an equation that has lost its hyperbolic character
(just set $C_s^2 = 0$ in \eqref{T_wave_in}), which would immediately lead 
to instabilities when numerically integrated.

In the exterior we could also try to switch to the Zerilli equation
\eqref{Zeqn}, which would have the advantage of being a gauge invariant
single wave equation. In addition, for large exterior grids this would
reduce the computing time by a factor of two. From \eqref{Z} we can
compute $Z$ from $S$ and $T$, and it is also possible to invert this
expression to yield $S$ and $T$ in terms of $Z$:
\begin{subequations}\label{S_T}
\begin{align}
        T \= \(r - 2M\)Z'
        + \(\half l(l+1) - \frac{6M}{r\Lambda}\(1 - \frac{2M}{r}\)\)Z\\
\begin{split}
        S \= \(1 - \frac{2M}{r}\)Z''
        + \frac{M}{r^2}\(1 - \frac{6}{\Lambda}\(1 - \frac{2M}{r}\)\)Z'\\
        &\quad{}+ \frac{1}{r^2}\(\frac{3M}{r} - l(l+1)
        + \frac{6M}{r\Lambda}\(3 - \frac{8M}{r}\)
        - \(\frac{6M}{r\Lambda}\)^2\(1 - \frac{2M}{r}\)\)Z\;,
\end{split}
\end{align}
where
\begin{align}\label{Lambda}
        \Lambda \= l(l+1) - 2 + \frac{6M}{r}\;.
\end{align}
\end{subequations}
However, the numerical experiment shows that switching in the exterior
of the star from the variables $S$ and $T$ to the Zerilli function $Z$
and using the Zerilli equation \eqref{Zeqn} to evolve $Z$ in the
exterior causes a numerical instability. This is because $S$, $T$ and
$Z$ are not just related by some linear combination, but the relations
involve derivatives and are only valid if $S$ and $T$ satisfy the
Hamiltonian constraint, which is, of course, not strictly true in the
numerical case. For instance, if we were to numerically compute $Z$ at
grid point $i$ from $S$ and $T$ using formula \eqref{Z}, where we
approximate $T'_i$ by $(T_{i+1} - T_{i-1})/(2\Delta r)$ and then in
turn compute $S$ and $T$ at the same grid point $i$ from $Z$ using
formulas \eqref{S_T}, where again we approximate the derivatives of
$Z$ with central differences, we would see that the resulting values
could differ by quite a large amount from the original values we
started with. During the evolution, this mismatching between those
values, which would occur at the point where we switch the equations,
would rapidly amplify and spoil the whole evolution.

In \cite{Mon74a} and \cite{Mon74b}, Moncrief showed that it is
possible to construct two gauge invariant quantities $q_1$ and $q_2$,
which completely describe the stellar oscillations inside the
star. Moreover, the fluid-like quantity $q_2$ vanishes in the exterior
region by virtue of the Hamiltonian constraint and the quantity $Q :=
q_1/\Lambda$ satisfies the Zerilli function. It is even possible to
define $Q$ in the stellar interior by using the following definition
for $\Lambda$:
\begin{align}
        \Lambda \= l(l+1) - 2e^{-2\lam}\(1 + r\lam'\)\;,
\end{align}
which in the exterior agrees with \eqref{Lambda}. This would mean that
in the interior we would have one wave equation for the fluid variable
$q_1$ and another one for $Q$, which in the exterior would
automatically transform into the Zerilli equation. This therefore
would be the most efficient set of equations with the additional
advantage of being gauge invariant. Unfortunately, Moncrief does not
write down the relevant equations, which is quite understandable since
in the interior they become terribly messy.

We therefore stick to the above formulation of the perturbation
equations as a set of two coupled wave equations for both the interior
and the exterior region of the star.

\section{Boundary and junction conditions}\label{sec:bc}


There are three boundaries we have to take care of. As was already
discussed in the previous section, at the origin $r=0$ we have to
demand all variables to be regular. From Taylor expansion around $r=0$
we then can infer the analytic behavior of the various variables, which
is proportional to $r^{l+1}$ for both $S$ and $T$.

At the outer boundary far away from the star, we require the waves to
be purely outgoing.

The third boundary is the surface of the star at $r = R$, which is
formally defined by the vanishing of the total pressure $P$. Since the
perturbations will slightly deform the star, the perturbed surface
will be displaced by an amount $\xi^i$ with respect to the unperturbed
location at $r=R$. If the coordinates of the unperturbed surface are
denoted by $x^i_R$, the vanishing of the total pressure $P$ at the
displaced surface translates to $P(t,x^i_R +\xi^i) = 0$. Taylor
expansion to first order then gives
\begin{align}
        0 \;=\; P(t,x^i_R + \xi^i) \= P(t,x^i_R)
        + \xi^i\df{}{x^i}P(x^i_R)\non\\
        \= p(x^i_R) + \delta p(t,x^i_R) + \xi^i\df{}{x^i} p(x^i_R)\;.
\end{align}
In the last step, we have made use of the fact that the total pressure
$P$ is the sum of the unperturbed pressure $p$ and its Eulerian
perturbation $\delta p$. In addition, we have omitted the term that
contains the product of $\xi^i$ and $\delta p$ since it is of second
order in the perturbations. Now, the unperturbed pressure $p$ is a
function of $r$ only and it is furthermore $p(r = R) = 0$, hence we
obtain
\begin{align}
        \delta p(t,x^i_R) \= -\xi^r p'(R)\;.
\end{align}
Unfortunately, this is not a very convenient boundary condition since
we neither use the pressure perturbation nor the displacement vector
in our set of evolution equations. Therefore we must relate this
condition to the variables we use. We will try to find a condition
that gives us the time evolution of $\delta\eps$ at the stellar surface.
The first step is to use the relation $\delta\eps = dp/d\eps\,\delta
p$, which gives us
\begin{align}
        \df{}{t}\delta\eps \= - \frac{d\eps}{dp}p'\df{}{t}\xi^r 
        \;=\; -\eps' \df{}{t}\xi^r\;.
\end{align}
The time derivative of $\xi^r$ can then be related to the $r$-component of
the 4-velocity $u_r$ \cite{Ruoff96}:
\begin{align}
        \df{\xi^r}{t} \= e^{-2\lam}\(e^\nu\delta u_r - \beta_r\)\;.
\end{align}
After expansion in spherical harmonics, we finally obtain 
\begin{align}\label{BCrho}
        \df{}{t}\rho(t,R) \= \eps'\bigg[RK_2(t,R) + e^{2\nu-2\lam}\(u_1(t,R)
        - \frac{u_2(t,R)}{R}\)\bigg]\;.
\end{align}
The equivalent equation for the quantity $H$ as defined in \eqref{H_def}
reads
\begin{align}\label{BCH}
        \df{}{t}H(t,R) \= -\nu'\bigg[RK_2(t,R) + e^{2\nu-2\lam}\(u_1(t,R)
        - \frac{u_2(t,R)}{R}\)\bigg]\;.
\end{align}
Incidentally, this expression can be derived directly from the evolution
equation \eqref{H} just by setting $C_s^2$ to zero. The same is true
for the wave equation \eqref{waveH}. For polytropic equations of state
it is always $C_s^2 = 0$ at the surface of the star, hence in
\eqref{H} the boundary condition \eqref{BCH} is satisfied
automatically. For realistic equations of state the sound speed at the
surface should be that of iron, which is very small compared to the
sound speed inside the core, where it might reach almost the speed of
light for very relativistic stellar models. For practical purposes, in
those cases we just might as well set $C_s^2(r = R) = 0$.

Let us now turn to the junction conditions at the surface of the
star. We will always assume that $\eps$ and $C_s^2$ go to zero when
approaching the stellar surface $r = R$. If at the surface we had a
finite energy density (as for example in a constant density model),
this would result in discontinuities in $\lam'$, which in turn would
affect the differentiability properties of the perturbation
quantities. For polytropic equations of state, our assumptions are
always fulfilled and for realistic equations of state, both the
density and the sound speed are very small compared to their values in
the core and may thus be confidently set to zero at the surface.

The continuity of the first and second fundamental forms across the
surface ensures the continuity of the metric perturbations $S$, $T$
and $S'$. The associated extrinsic curvature variables $K$, $K_5$ and
$K'$ must be continuous as well. From \eqref{K} it follows that $S''$
is continuous, too. If in \eqref{K5} we substitute $T''$ by means of
the Hamiltonian constraint \eqref{HC}, we can see that $T'$ is
continuous. The continuity of $T''$, however, depends on the value of
$\rho$. If we let the subscripts $in$ and $ex$ represent the values
for the interior and the exterior, respectively, we have for $T''$
\begin{align}\label{T_inex}
        T_{in}'' - T_{ex}'' \= -8\pi e^{2\lam}\rho(R)\;.
\end{align}
As we shall see below, for polytropic equations of state $\rho(R)$
can either be zero, finite, or even infinite, depending on the
polytropic index $\Gamma$. This has to do with the behavior of the
derivative of the background energy density $\eps'$, which appears in
the boundary condition \eqref{BCrho}.

For polytropic equations of state, it is clear that at the surface,
both $p$ and $\eps$ vanish. The TOV equations \eqref{tov2} and \eqref{tov3} 
can be combined to yield
\begin{align}
        p' \= -\(p + \eps\)\frac{e^{2\lam}}{ r^2}
        \left(m + 4\pi r^3 p\right)\;,
\end{align}
from which we see that $p'$ vanishes at the surface. However, $\eps'$
can behave in quite different ways. From \eqref{tov3} we find
\begin{align}
        \eps' \= -\nu'\(p + \eps\)\(\frac{dp}{d\eps}\)^{-1}\;.
\end{align}
Close to the surface, it is $p \ll \eps$ and therefore
\begin{align}
        (p + \eps)\(\frac{dp}{d\eps}\)^{-1}
        &\approx\; \eps\(\frac{dp}{d\eps}\)^{-1}\;.
\end{align}
For a polytropic equation of state the square of the sound speed
$\frac{dp}{d\eps}$ is given by
\begin{align}
        \frac{dp}{d\eps} \= \kappa\Gamma\eps^{\Gamma - 1}\;,
\end{align}
hence
\begin{align}
        \eps\(\frac{dp}{d\eps}\)^{-1} \= \frac{\eps^{2 - \Gamma}}{\kappa\Gamma}
\end{align}
and
\begin{align}\label{epssurf}
        \eps' \= -\nu'\frac{\eps^{2 - \Gamma}}{\kappa\Gamma}\;.
\end{align}
Approaching the surface, $\eps \rightarrow 0$ and $\nu'$ will become
a constant. Now, from \eqref{epssurf} we see that the behavior of
$\eps'$ critically depends on the value of the polytropic index
$\Gamma$. We can distinguish three different cases.  For $\Gamma < 2$,
we have $\eps' \rightarrow 0$, for $\Gamma = 2$ we have $\eps'
\rightarrow$ const., whereas for $\Gamma > 2$ we have 
$\eps' \rightarrow -\infty$!

This is somewhat disturbing since for the boundary condition
\eqref{BCrho} this would mean that $|\rho| \rightarrow \infty$, unless
the expression in brackets vanishes. Unfortunately, this is not
automatically guaranteed! Interestingly, the boundary condition
\eqref{BCH} for $H$ is harmless for all values of $\Gamma$ since
$\nu'$ is always bounded. But, of course, if we were to compute $\rho$
from $H$ using
\begin{align}
        \rho \= (p + \eps)\(\frac{dp}{d\eps}\)^{-1}\!H
        \;\approx\;\frac{\eps^{2 - \Gamma}}{\kappa\Gamma}H
\end{align}
we would obtain an infinite value when $\Gamma > 2$ unless $H$ vanishes
at the surface. However, as in \eqref{BCrho}, \eqref{BCH} does not
guarantee the vanishing of $H$, even if $H$ is initially set to
zero. Therefore, we must ask ourselves what really happens in the case
$\Gamma > 2$, where it seems that $|\rho| \rightarrow \infty$.

First, we would like to refer to a paper of Moncrief \cite{Mon74b},
where he discusses a sufficient stability condition for the non-radial
stellar oscillations. For a polytrope he finds that for the potential
energy in the vicinity of the surface to be positive it must hold that
\begin{align}
        l(l+1) - 2 - 4\pi r^2\frac{\eps^{2 - \Gamma}}{\kappa\Gamma}
        \;&\ge\; 0\;.
\end{align}
Moncrief shows that for $l \ge 3$ this condition is always satisfied
for $6/5 < \Gamma \le 2$. For $l=2$ he obtains $6/5 < \Gamma < 4/3$.
The reason for the lower limit of $\Gamma = 6/5$ is that for smaller
$\Gamma$ there are no bounded stellar models. For $\Gamma > 2$ the
above condition will always be violated since the last term then goes
to negative infinity.

However, the above condition is only a sufficient condition. But it
cannot be a necessary condition for stability, for in that case all
stellar models with $\Gamma > 2$ would be unstable with respect to
non-radial oscillations. For $l = 2$ even the models with $\Gamma >
4/3$ would be unstable, which is not the case as mode calculations
\cite{LNS93} and the direct evolution of the perturbation equations
show.

A possibility to clarify this weird behavior of $\rho$ is to look at
the Lagrangian description of the perturbations. By definition,
Lagrangian perturbations are changes that are measured by an observer
who moves with the fluid. Hence, she would compare e.g. the energy
density at the displaced location $x^i + \xi^i$ to the original
unperturbed value at $x^i$. In mathematical terms the Lagrangian
energy perturbation reads
\begin{align}\label{Lag}
        \Delta \eps(t,x^i) \= {\cal E}(t, x^i + \xi_i(t, x^i))
        - \eps(x^i)\;.
\end{align}
Here, $\cal E$ denotes the total energy density. A similar
expression holds for the Lagrangian pressure change. To compute
$\Delta\eps$ at the surface, we set $x^i = R^i$ and obtain
\begin{align}
        \Delta \eps(t,R^i) \= {\cal E}(t, R^i + \xi^i) - \eps(R^i)\non\\
        \= {\cal E}(t, R^i + \xi^i)\;,
\end{align}
since $\eps(R^i) = 0$ for a polytropic equation of state. Furthermore,
as we already know that the total pressure $P(t, R^i + \xi^i)$ has to
vanish and since for the polytropic equation of state it is ${\cal
E}(P = 0) = 0$, we obtain $\Delta \eps(t,R^i) = 0$.

Hence, the Lagrangian energy density perturbation always vanishes
at the surface, regardless of the actual value of $\Gamma$.

What then happens to the Eulerian density perturbation? By
definition the Eulerian density perturbation $\delta \eps$ is the
difference between the perturbed energy density $\cal E$ and the
background density $\eps$ at the same location
\begin{align}
        \delta \eps(t, x^i) \= {\cal E}(t, x^i) - \eps(x^i)\;.
\end{align}
It is through Taylor expansion to linear order that we obtain the
connection between the Lagrangian and Eulerian perturbations:
\begin{align}\label{Deps}
        \Delta \eps(t,x^i) \= {\cal E}(t, x^i)
        + \xi^i(t,x^i)\df{}{x^i}{\cal E}(t, x^i) - \eps(x^i)\non\\
        \= \delta \eps(t, x^i) + \xi^i(t,x^i)\df{}{x^i}{\cal E}(t, x^i)\non\\
        \= \delta \eps(t, x^i) + \xi^i(t,x^i)\df{}{x^i}\(\eps(r)
        + \delta \eps(t, x^i)\)\non\\
        \= \delta \eps(t, x^i) + \xi^r(t,x^i)\eps'(r)   
        + \xi^i(t,x^i)\df{}{x^i} \delta \eps(t, x^i)\;.
\end{align}
The last term usually can be dropped with the argument that it is a
product ot the two infinitesimal quantities $\xi^i$ and the gradient
of $\delta\eps$ and therefore of second order in the perturbations.
However, in the $\Gamma > 2$-case this argument breaks down at the
surface, where the gradient of the background energy density $\eps'$
becomes infinite. If we were to drop the second order term, it is
clear that the Eulerian perturbation $\delta\eps$ would have to become
infinite, too, in order to compensate for the blow up of $\eps'$ and
to yield a vanishing Lagrangian perturbation $\Delta\eps$.

This shows us that in the $\Gamma > 2$-case the physically meaningful
quantity is the Lagrangian energy perturbation $\Delta \eps$, which
remains bounded everywhere and not the Eulerian energy perturbation
$\delta \eps$. However, in our case it is not possible to switch from
the Eulerian description to the Lagrangian because the latter is only
defined in the stellar interior, where we can define a fluid
displacement vector. Outside the star we have vacuum, which cannot be
displaced, hence to describe the metric perturbations, we have to rely
on the Eulerian description. Therefore, at the stellar surface we
would have to switch from the Lagrangian to the Eulerian
description. But it is right there that the Eulerian concept is
misbehaved for $\Gamma > 2$, and we would run into the same troubles.

The whole discussion seems somewhat irrelevant, for in the actual set
of equations we do not use the Eulerian perturbation of the energy
density at all. We have got rid of it by using the Hamiltonian constraint
\eqref{HC}. Also, the gradient of the background energy density does
not appear anywhere in the equations. However, we have to compute the
second derivative of $T$, which, as can be seen from the continuity
analysis at the beginning of this section, depends on the behavior of
the Eulerian energy perturbation $\rho$. From \eqref{T_inex} we infer
that in the $\Gamma > 2$-case we must have a blow up of $T''$ at the
surface. Of course, this is very troublesome for the numerical
discretization, and even for $\Gamma = 2$ we still have a
discontinuity in $T''$, which will spoil the second order convergence
of the numerical discretization scheme.

\begin{figure}[t]
\leavevmode
\epsfxsize=\textwidth
\epsfbox{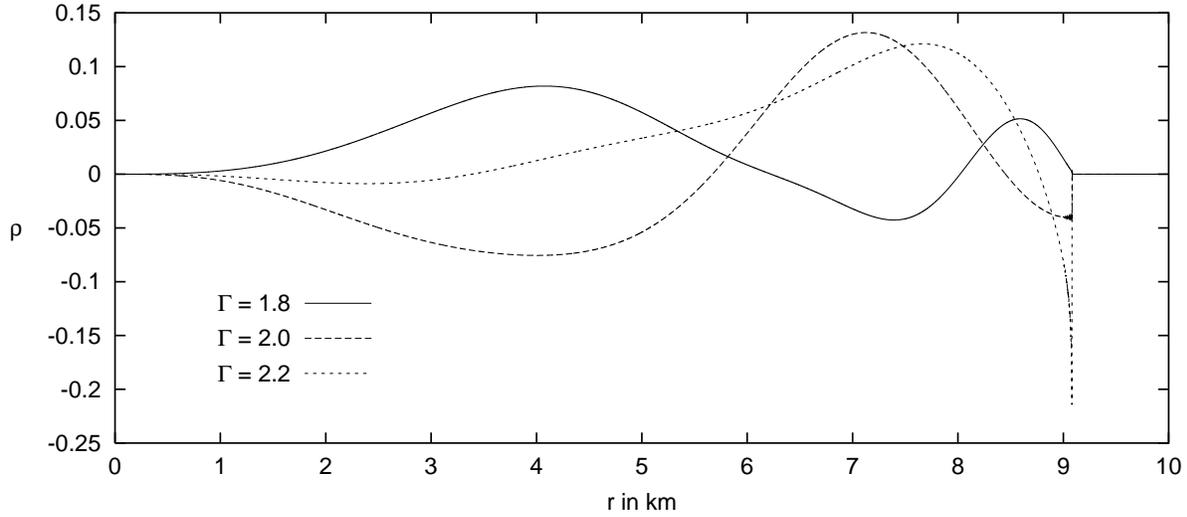}
\caption{\label{rho_poly}Snapshots of the values of $\rho$ after a certain
evolution time for polytropic stellar models with three different
polytropic indices $\Gamma = 1.8, 2.0$ and $2.2$. It is obvious that
for $\Gamma = 2.2$ $\rho$ diverges at the surface of the star.}
\end{figure}

The numerical evolutions indeed confirm the above analysis. By
computing $\rho$ with the aid of the Hamiltonian constraint \eqref{HC}
we find that for polytropic stellar models with $\Gamma > 2$, $\rho$
tends to blow up at the stellar surface, even if it was there
initially set to zero. In Fig.~\ref{rho_poly} we have plotted the
values of $\rho$ after a certain time of evolution for three different
polytropic indices, namely $\Gamma = 1.8$, $\Gamma = 2.0$ and $\Gamma
= 2.2$. The corresponding values of $\kappa$ are $0.184\,$km$^{1.6}$,
$100\,$km$^2$ and $49600\,$km$^{2.4}$ and were chosen in such a way
that for the same central density, we obtain models with the same
radius. As is evident from Fig.~\ref{rho_poly}, for $\Gamma = 1.8$,
$\rho$ vanishes at the stellar surface, for $\Gamma = 2.0$ it assumes
a constant value and for $\Gamma = 2.2$ it diverges, which is
consistent with the foregoing discussion. It should be noted that in
those numerical simulations we have used a grid size of 6400 points
inside the star in order to have a decent resolution.

\section{Numerical implementation}

It is not quite straightforward to implement a numerical
discretization scheme for the above set of equations
\eqref{wave-eqns}, and one has to worry about some severe problems
that will arise when one is not doing the right thing. First, we have
to treat the boundaries in a correct way. In the previous section we
have seen that there are in fact three boundaries one has to deal
with, namely the origin $r=0$, the surface of the star and the outer
boundary of the grid. The latter is the easiest to handle and could in
principle (if computing time does not matter) be totally ignored.

At the outer boundary we impose outgoing radiation condition, which can
be realized in the following way: Far away from the star, we know that
the asymptotic solution $\Psi(t,r)$ is an outgoing wave with a
propagation speed $c = e^{\nu-\lam}$ and an amplitude that scales with
some power of $r$:
\begin{align}
        \Psi(t,r) \= r^a\Phi(c\,t - r)\;.
\end{align}
Here $\Psi$ stands for either $S$ or $T$. Working out the different
derivatives
\begin{align}
        \df{}{t}\Psi(t,r) \= c\,r^a\Phi'(c\,t - r)\\
        \df{}{r}\Psi(t,r) \= -r^a\Phi'(c\,t - r) + a\,r^{a-1}\Phi(t,r)
        \;=\; -r^a\Phi'(c\,t - r) + \frac{a}{r}\Psi(t,r)
\end{align}
leads us to the following relationship:
\begin{align}\label{rel}
        \df{}{r}\Psi(t,r) \= -\frac{1}{c}\df{}{t}\Psi(t,r)
        + \frac{a}{r}\Psi(c\,t - r)\;.
\end{align}
Knowing this relation can help us to compute $\Psi(t,r)$ at the
boundary grid point $N$ at the next time level $n+1$. If we denote the
time step by $\Delta t$ and the spatial grid spacing by $\Delta r$, we
can discretize equation \eqref{rel} at the intermediate grid point $(N
- 1/2, n + 1/2)$ with $\Psi$, $\df{\Psi}{r}$ and $\df{\Psi}{t}$
approximated by
\begin{align}
        \Psi^{n + 1/2}_{N-1/2}
        \= \frac{1}{4}\(\Psi^{n+1}_{N-1} + \Psi^n_{N-1}
        + \Psi^{n+1}_N + \Psi^{n}_{N}\)\\
        \(\df{\Psi}{r}\)^{n + 1/2}_{N-1/2}
        \= \frac{1}{2\Delta r}\(\Psi^n_N - \Psi^n_{N-1}
        + \Psi^{n+1}_N - \Psi^{n+1}_{N-1}\)\\
        \(\df{\Psi}{t}\)^{n + 1/2}_{N-1/2}
        \= \frac{1}{2\Delta t}\(\Psi^{n+1}_{N-1} - \Psi^n_{N-1}
        + \Psi^{n+1}_N - \Psi^{n}_{N}\)\;,
\end{align}
and solve the resulting equation for the unknown value $\Psi^{n+1}_N$
\begin{align}
        \!\!\!\!\!\!\!\Psi^{n+1}_N \= \frac{1}{1 + B}\(\(1 - B\)\Psi^n_{N-1}
        - B\(\Psi^n_N + \Psi^{n+1}_{N-1}\)
        + \frac{1 - A}{1 + A}\(\Psi^{n+1}_{N-1} - \Psi^n_N\)\)\;,
\end{align}
where
\begin{align}
        A \;&:=\; \frac{\Delta x}{c\Delta t}\\
        B \;&:=\; \frac{a}{(1 - 2N)(1 + A)}\;.
\end{align}
If $a = 0$ this reduces to
\begin{align}
        \Psi^{n+1}_N \= \Psi^n_{N-1}
        + \frac{1 - A}{1 + A}\(\Psi^{n+1}_{N-1} - \Psi^n_N\)\;.
\end{align}
This is the case for $S$, whereas for $T$ we have to set $a = 1$, i.e.
the amplitude of $T$ grows linearly with $r$ as the the wave travels
outwards. As we shall see this will cause some problems when we want
to compute the Zerilli function. 

Of course, the approximation of \eqref{rel} with finite differences
will lead to a partial reflection of the outgoing wave at the grid
boundary. The reflected wave will then travel back inwards and
contaminate the numerical evolution. The effect of this contamination
depends on the choice of the initial data and in some cases can be
quite interfering. For instance we can choose initial data that will
cause a quite large outgoing initial pulse of radiation followed by a
strong ring-down. The amplitude of the final ringing of the neutron
star can then be smaller by several orders of magnitude. In this case,
even if only a very small fraction of the first pulse gets reflected,
it can strongly affect the later numerical evolution. However, we can
also construct initial data which produce oscillations with more or
less constant amplitude. In this case, the reflected part practically
does not affect the results at all. 

In any case, to be on the safe side, and if computation time does not
matter, one can always move the boundary that far afield that any
reflections will take too long to travel back to the point where the
wave signal gets extracted.

The boundary condition at the surface of the star has been described
in much detail in the previous section. It has become obvious that
for polytropic equations of state with the polytropic index $\Gamma \ge 2$
we have to deal with a discontinuity in $T''$ at the stellar surface.

This, of course, will affect the convergence of the numerical
discretization scheme, because for instance a second order scheme
converges only in second order if the second derivative is
continuous. And, indeed, by using a second order discretization
scheme that does not take care of the discontinuity, we only find
first order convergence for $\Gamma \ge 2$.

However, from a practical point of view, it does not seem necessary to
really worry about this fact. First, the use of any polytropic
equation of state with a constant polytropic index throughout the
whole star is unrealistic anyway. If we wanted to obtain more
realistic results, we would have to resort to realistic tabulated
equations of state. And for those, $\eps'$ and therewith $\rho$ is
almost zero at the surface, which results in a continuous $T''$. We
therefore discretize equations
\eqref{wave-eqns} with central differences on the whole domain.

A somewhat more tricky business is the inner boundary at $r=0$.
The reason is the fact that $r=0$ is not a physical boundary but
rather a coordinate boundary, which is only due to the choice of
spherical coordinates, and which is absent, for example, in Cartesian
coordinates. Hence, at $r=0$ we cannot impose physical boundary
conditions but we must ask for some regularity conditions the functions
have to satisfy. To understand what is going on at the origin, we
look at a simplified version of equation \eqref{waveT}, where we focus
only on the troublesome parts:
\begin{align}\label{Torg}
        \dff{T}{t} \= \dff{T}{r} - \frac{l(l+1)}{r^2}T\;.
\end{align}
Obviously the righthand side is regular if and only if $T$ has a
Taylor expansion of the kind
\begin{align}
        T(t,r) \= T^r_0(t)\,r^{l+1} + T^r_1(t)\,r^{l+3} + \dots\;.
\end{align}
The other valid solution
\begin{align}
        T(t,r) \= T^d_0(t)\,r^{-l} + T^d_1(t)\,r^{-l+2} + \dots
\end{align}
is diverging at the origin. Hence, we have to make sure, that during
the numerical evolution this solution will be suppressed. Now, there
are several possibilities to modify \eqref{Torg} by rescaling the
variable $T$. For instance, we can introduce $\widehat{T} = rT$, for
which we have the following equation:
\begin{align}\label{Twidehat}
        \dff{\widehat{T}}{t}
        \= \dff{\widehat{T}}{r} + \frac{2}{r}\df{\widehat{T}}{r}
        - \frac{l(l+1)}{r^2}\widehat{T}\;.
\end{align}
In this case $\widehat{T}$ has to be proportional to $r^l$ at the
origin. We also can get rid of the $r^l$-behavior by introducing
$\widetilde{T} = r^l\widehat{T}$. The appropriate equation then reads
\begin{align}\label{Tbar}
        \dff{\widetilde{T}}{t}
        \= \dff{\widetilde{T}}{r} + \frac{2(l+1)}{r}\df{\widetilde{T}}{r}\;.
\end{align}
Here $\widetilde{T}$ is finite at the origin and symmetric with
respect to the transformation $r \rightarrow -r$. Hence
$\widetilde{T}'(0) = 0$, but $(\widetilde{T}'/r)(0)$ is finite,
again. However, numerically, we cannot directly compute this
expression at $r=0$ since this would result in $0/0$, which, as such,
is ill-defined. However, we could use the l'H\^ospital rule to obtain
$(\widetilde{T}'/r)(0) = (\widetilde{T}''/r')(0) = \widetilde{T}''(0)$.

If we naively try to discretize any of the above equations, say, with
central differences, we will positively run into troubles. If we
ignore the divergent terms for a moment and only discretize the
simple wave equation $\ddot{T} = T''$ by means of central differences,
we find from the von Neumann stability analysis that the Courant
number $C$ is one. The Courant number $C$ determines the maximal
allowed time step size $\Delta t_{max}$ for a given spatial resolution
$\Delta r$.  Since in our case the propagation speed is one, we have
the relation $C = \Delta t_{max}/\Delta r$.

As soon as we include the divergent term $-l(l+1)T/r^2$, this is not
true any more and instead we find $C < 1$ to be a monotonically
decreasing function of $l$. For large $l$ this means that in order to
have stability we have to take very small time steps, which makes the
evolutions more and more time consuming.

We can compute the Courant condition in the same way as we did in the
section \ref{troubles}, where we discussed the influence of the dip in
the sound speed on the stability behavior of the discretized fluid
equation. We have to compute the eigenvalues of the matrix ${\bf G}$,
which acts on the vector $T^n$ of grid values at the time level $n$.
Those eigenvalues then have to be smaller than one in their moduli in
order to insure stability. Here, the only free parameter of ${\bf G}$
are $l$ and the ratio $\Delta t/\Delta r$.

\begin{figure}[t]
\leavevmode
\epsfxsize=\textwidth
\epsfbox{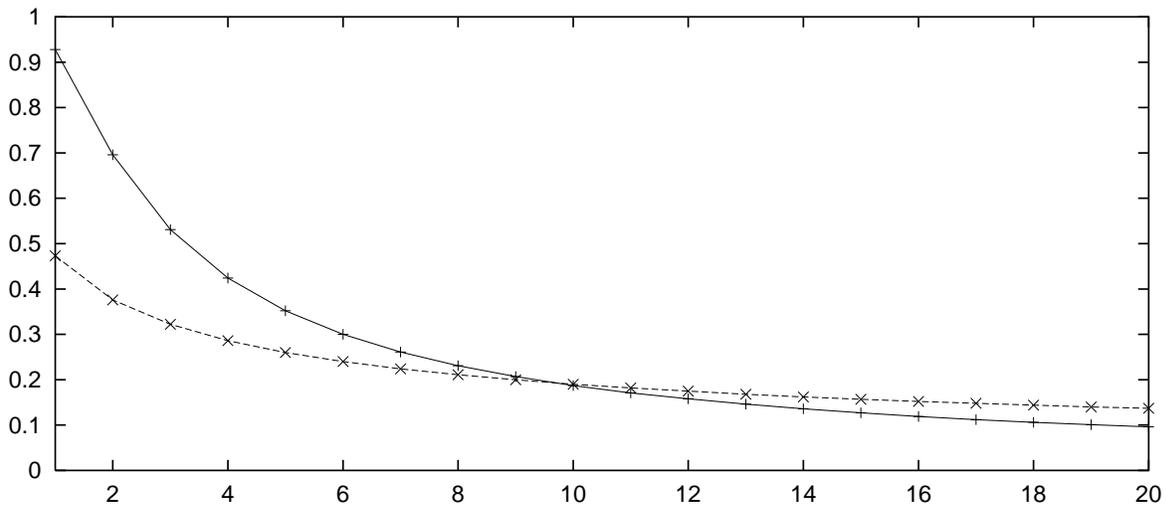}
\caption{\label{C_bc0}The Courant number $C$ as a function of $l$ for
the discretized version of equations \eqref{Torg}, \eqref{Twidehat}
(solid line) and \eqref{Tbar} (dashed line).}
\end{figure}

In Fig.~\ref{C_bc0} we show the Courant number $C$ as a function of
$l$ for the discretized versions of \eqref{Torg} and \eqref{Tbar} with
the following boundary conditions at $r = 0$: $T(0) = 0,
\widehat{T}(0) = 0$ and $\ddot{\widetilde{T}}(0) =
(2l+3)\widetilde{T}''(0)$. It turns out that $C(l)$ is the same for
the versions \eqref{Twidehat} and \eqref{Torg}. As $l$ is increased,
the allowed time step size drastically decreases. For $l = 4$ the
Courant number $C$ has shrunk by a factor of 2 compared to its value
for $l=1$. It is interesting to note that for small values of $l$
version $\eqref{Tbar}$ allows for a smaller time step size than the
other versions, but for large $l$ things get reversed. This is
probably due to the fact that in $\eqref{Tbar}$ the singular term is
only proportional to $l$, whereas in \eqref{Torg} and \eqref{Tbar} it
is proportional to $l^2$.

This reduction of the maximal allowed time step size $\Delta t_{max}$
is a quite undesirable feature, but we could live with it if we only
used small values of $l$. But why wait longer for the numerical
results than really necessary if there is a quite simple trick that
allows us to retain a Courant number of, say, $C = 0.9$ for all values
of $l$?

This trick consists in moving the boundary condition from $r_0=0$ to
some inner grid point $r_i = i\Delta r$, depending on the value of
$l$. The larger $l$ the higher the number of the boundary grid
point. Specifically, for a given $l$ we choose the boundary to be at
$r_{l-1} = (l-1)\Delta r$. Hence, for the numerical computation we
ignore all grid points $r_i$ with $i < l - 1$ and set $T(r_{l-1}) =
0$. What this basically amounts to is to cut off the bad influence of
the $l(l+1)/r^2$-term close to the origin. Of course, by doing so, we
artificially introduce some additional numerical error at the
boundary, but the numerical experiment shows that with the above
prescription this error will remain bounded and localized only at the
boundary. It thus does not have a bad influence on the evolution in
the remaining computational domain.

\begin{figure}[t]
\leavevmode
\epsfxsize=\textwidth
\epsfbox{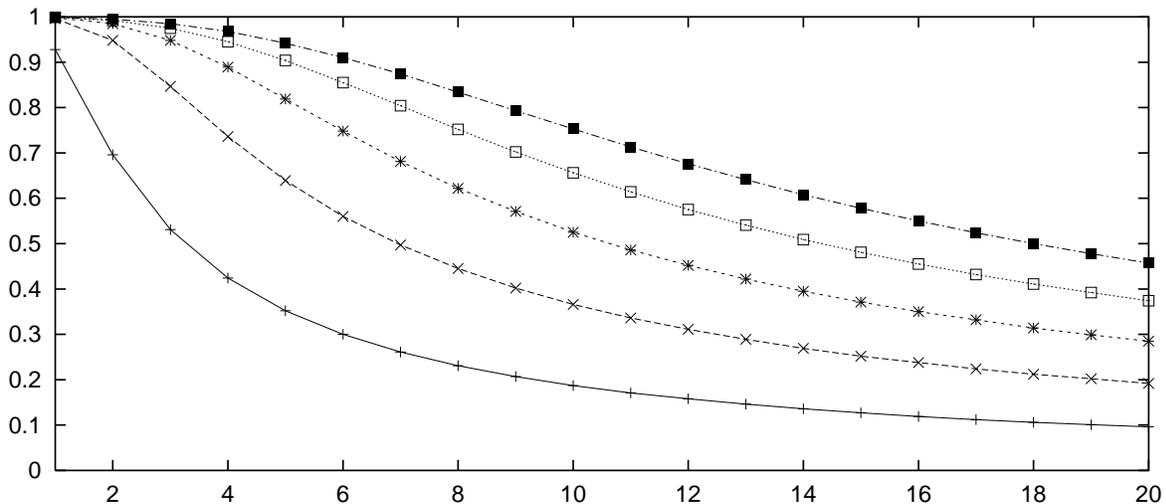}
\caption{\label{C_var_bc}The Courant number $C$ as a function of $l$ for
different boundary grid points.}
\end{figure}

In Fig.~\ref{C_var_bc} we show the Courant number $C$ as a function of $l$
for different boundary grid points. It can be clearly seen that for
fixed $l$ the Courant number $C$ increases as the boundary is moved to
a higher grid point number. Eventually, $C$ will become larger than our
desired value of 0.9.

To see whether this is indeed true for a real numerical evolution, we
evolve an analytic solution to \eqref{Torg} with periodic time
dependence: $T(t,r) = T_0(r)\cos t$, where $T_0(r)$ can be expressed
in terms of spherical Bessel functions.

\begin{figure}[p]
\begin{minipage}{\textwidth}
\leavevmode
\epsfxsize=\textwidth
\epsfbox{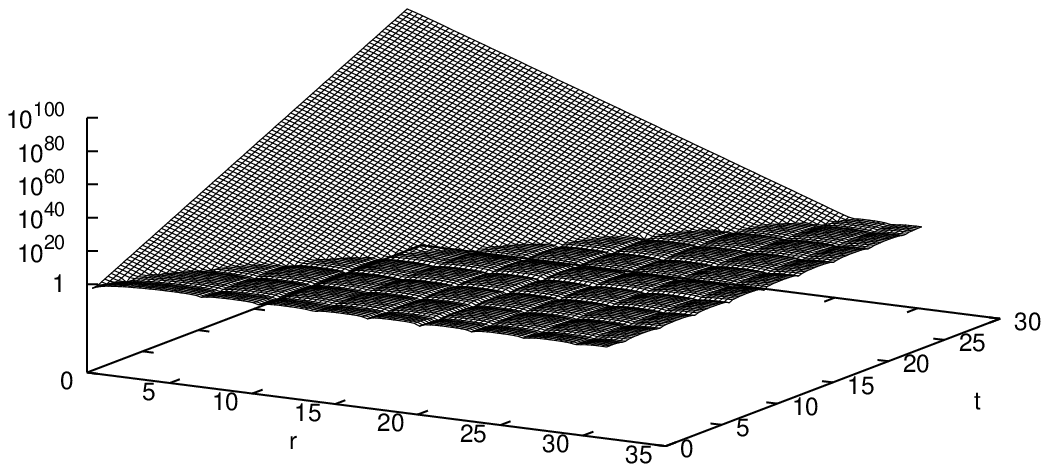}
\vspace*{-2cm}
\end{minipage}
\begin{minipage}{\textwidth}
\leavevmode
\epsfxsize=\textwidth
\epsfbox{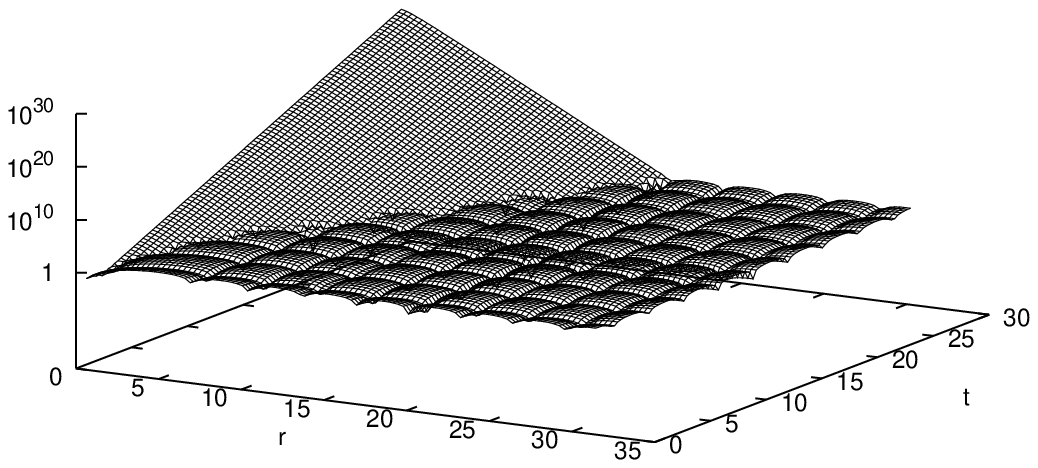}
\vspace*{-2cm}
\end{minipage}
\begin{minipage}{\textwidth}
\leavevmode
\epsfxsize=\textwidth
\epsfbox{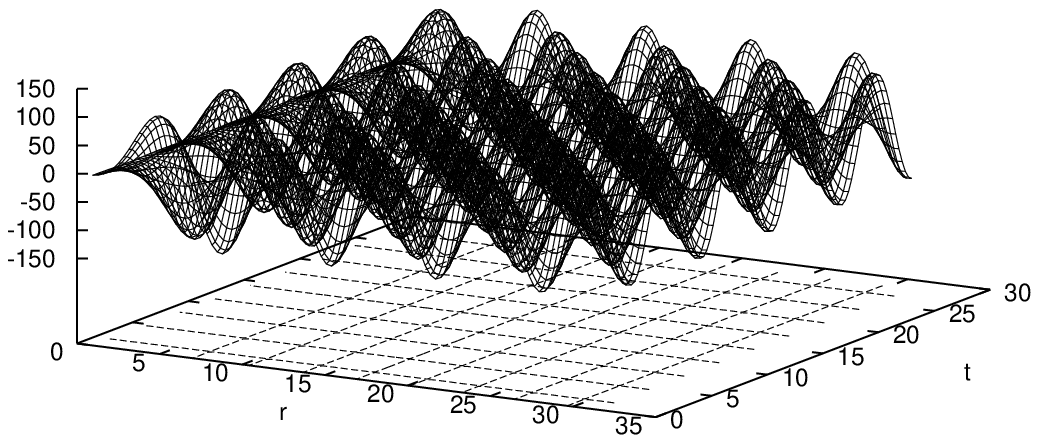}
\vspace*{-1cm}
\caption{\label{bound}Evolution of $T$ with the left boundary at different
locations. Upper panel: boundary at $i = 0$. Middle panel: at $i = 1$.
Lower panel: $i=2$.}
\end{minipage}
\end{figure}

In Fig.~\ref{bound} we show the numerical evolution of $T$ with $l=3$
for the different left boundary points $i=0$, $i=1$ and $i=2$ and $C =
0.9$. From Fig.~\ref{C_bc0} we expect the case where the boundary is
located at $i=2$ to be stable, whereas the other cases should exhibit
an instability. This is indeed what we find from the numerical
evolution. In the upper panel, where the boundary is at $i=0$, we have
a quite drastic growth at the origin. In the middle panel, we have put
the boundary at $i=1$, which reduces the strength of the instability,
and finally in the lower panel, where $i=2$, we have a stable evolution.

So far we have only discussed the numerical treatment of \eqref{Torg}.
Since the singular term in \eqref{Twidehat} is the same as in
\eqref{Torg}, the same procedure is also valid for \eqref{Twidehat}.
For practical reasons we prefer \eqref{Torg}, for we save a first
derivative. However, things change a little if we want to take
equation \eqref{Tbar}. In this case we have a different boundary
condition at the origin. Still, it is possible to move this boundary
condition farther to the right, but even by doing so we always have to
use a smaller Courant number than in the other cases. Since we would
like to have a fast numerical evolution, we therefore do not use this
version.

The reader may ask why we are presenting three different versions of
the same equation, when, at the end, we discard two of them. The
reason is that when one derives, for example, the equations for
neutron star oscillations (or any kind of (linear) evolution equations
in spherical coordinates), the ``raw'' form usually is not well suited
for the numerical treatment. One then has the freedom of rescaling the
variables. Our suggested prescription is to choose the variables in
such a way that they have a $r^{l(+1)}$-behavior at the origin. If
possible, the singular terms of the resulting equations should not
contain any derivatives. It then should be possible to cure any
instability at the origin by moving the boundary from $r = 0$ so far
to the right till the numerical scheme eventually becomes stable.

The whole discussion so far has dealt only with the second order equation.
However, it is also valid for the equivalent first order system.

\section{Convergence and the punishment for violating the constraints}

As was already discussed earlier, any valid initial data have to
satisfy the constraint equations. Once satisfied at $t=0$ the
evolution equations then guarantee, by means of the Bianchi identities,
that the constraints will be conserved for all times.

Of course, in the numerical evolution of the free evolution system,
the constraints will start to be violated due to discretization and
truncation errors. The degree of violation can then be used to monitor
the accuracy of the evolution.

In our case we use the free evolution only outside the star, whereas
inside the star we have already used up the Hamiltonian constraint to
reduce the number of equations. Therefore we will monitor the
Hamiltonian constraint \eqref{HC} only outside the star, where it is
$\rho = 0$. We do not monitor the momentum constraints \eqref{MC1} an
\eqref{MC2}, for we have already seen that they are equivalent to the
Hamiltonian constraint, and since we use the wave equations, we do not
explicitly compute the extrinsic curvature variables, anyway.

The exact evaluation of the Hamiltonian constraint should yield zero,
however, due to the numerical errors throughout the evolution the
righthand side of \eqref{HC} will deviate from zero. If the
discretization is consistent and the equations are correct, then the
violation of the Hamiltonian constraint should converge to zero as the
resolution is increased. With a second order discretization scheme
the error should decrease by a factor of four if one doubles the
resolution.

To check the convergence of our code, we will monitor the Hamiltonian
constraint as a function of time at some arbitrary location outside
the star. In the following we will use polytropic stellar models with
$\Gamma = 2$ and $\kappa = 100\,$km$^2$. The advantage of this choice
is that it produces smooth functions for $p$, $\eps$, and $C_s$, but
the disadvantage is the discontinuity in the second derivative of $T$.

\begin{figure}[t]
\leavevmode
\epsfxsize=\textwidth
\epsfbox{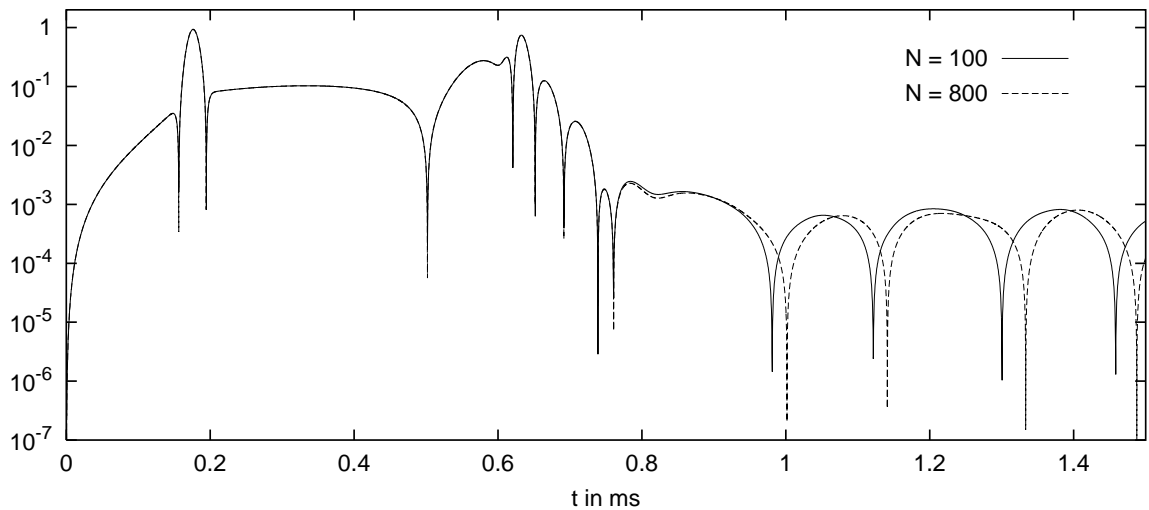}
\caption{\label{pulse}Wave form of $T$ for two runs with $N = 100$ and
$N = 800$. The main difference shows up in the fluid ringing after
$t=1\,$ms, where there is an obvious phase shift between the wave forms
that are obtained with the different resolutions.}
\end{figure}

As initial data, let us choose a time symmetric gravitational wave
pulse that travels towards the star and gets scattered back. In
Fig.~\ref{pulse} we first show the evolution of the variable $T$ at a
fixed location outside the star for the two different resolutions of
$N = 100$ and $N = 800$ grid points inside the star. As the amplitude
of the wave signal changes by some orders of magnitude, we use a
logarithmic scale and show the modulus of $T$.

The signal consists of three characteristic features. Because of the
time symmetry of the initial data the initial pulse splits into one
outgoing and one ingoing pulse. Since the observation point is
situated further outside than the initial pulse, we first see the
outgoing part of the pulse. This pulse is followed by the scattered
and distorted ingoing pulse and finally by some oscillatory ringing,
which consists of the strongly damped $w$-modes and the fluid $f$- and
$p$-modes, which have much smaller amplitudes.

Up to the point where the fluid ringing starts there is basically no
difference in the signal for the different resolutions. However, in
the fluid ringing there is an obvious phase shift for the different
resolutions. For low resolutions the frequencies of the fluid modes
tend to be larger than their actual values. Also, the amplitudes of
the higher modes are underestimated in this case.

\begin{figure}[t]
\leavevmode
\epsfxsize=\textwidth
\epsfbox{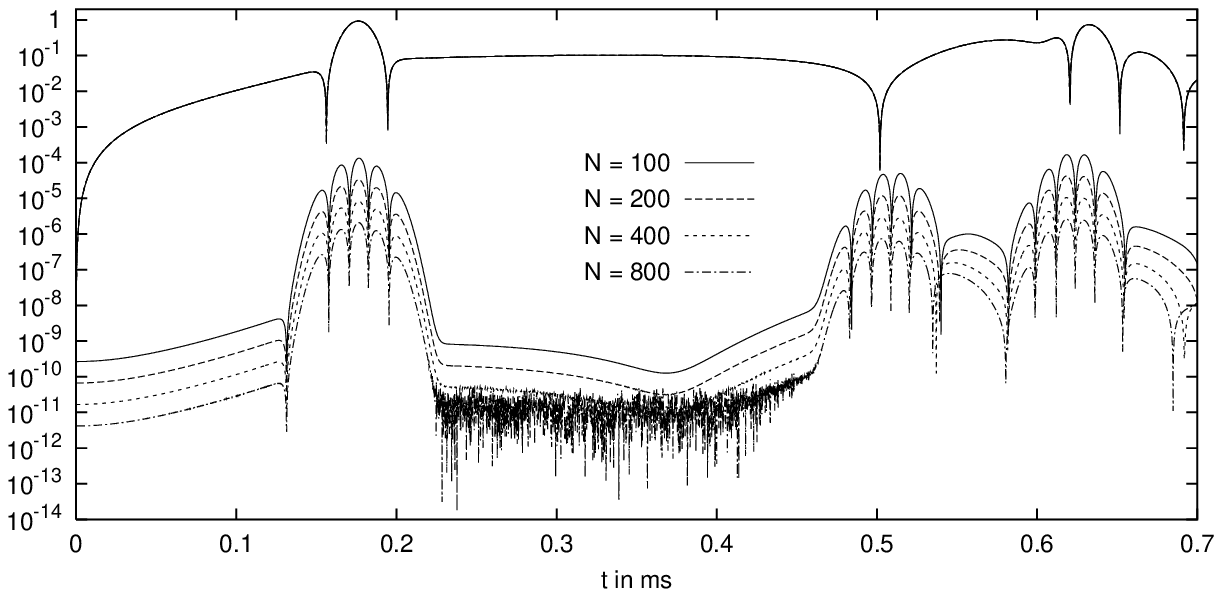}
\epsfxsize=\textwidth
\epsfbox{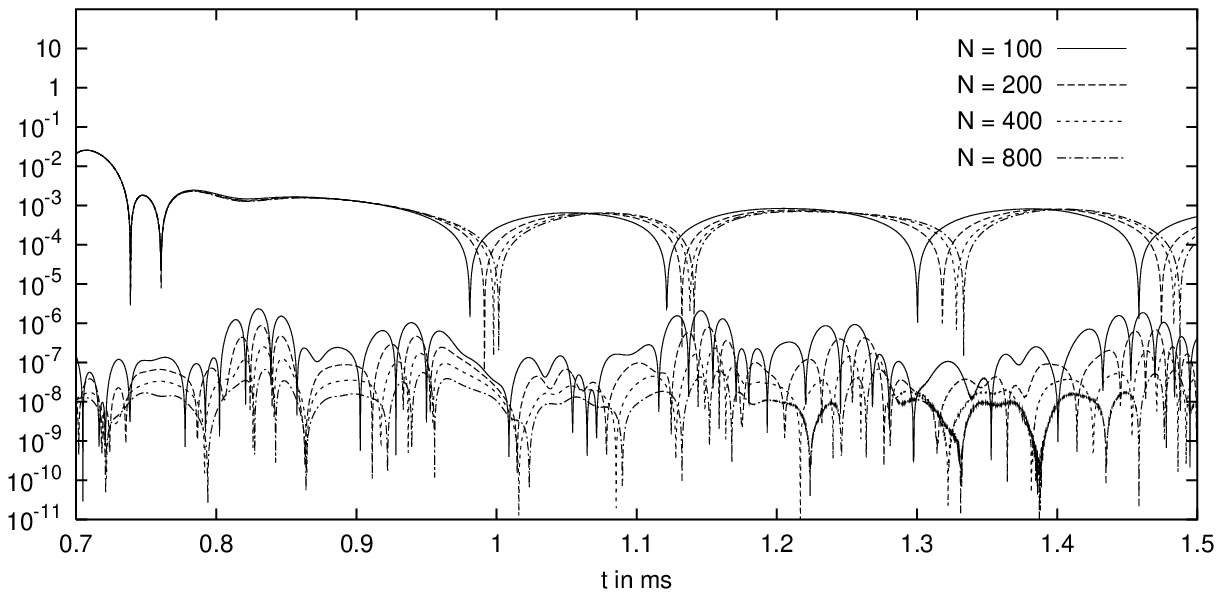}
\caption{\label{conv}Same evolution as in Fig.~\ref{pulse} for four
different resolutions. In addition to the variable $T$, we show the
violation of the Hamiltonian constraint \eqref{HC}. For more details
see text.}
\end{figure}

\begin{figure}[t]
\leavevmode
\epsfxsize=\textwidth
\epsfbox{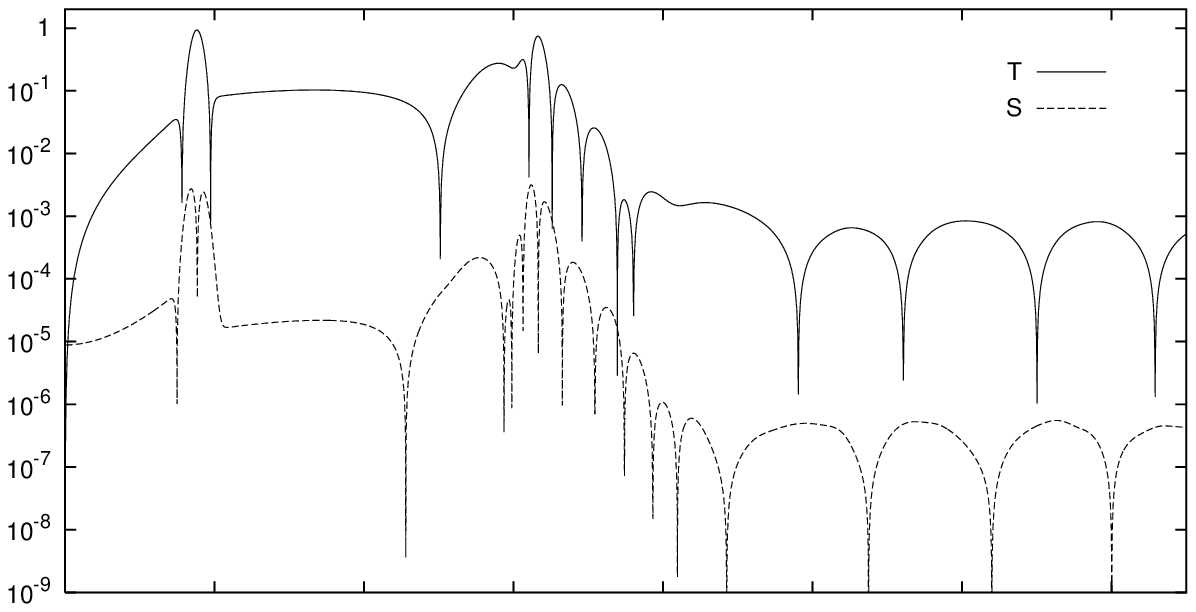}
\epsfxsize=\textwidth
\epsfbox{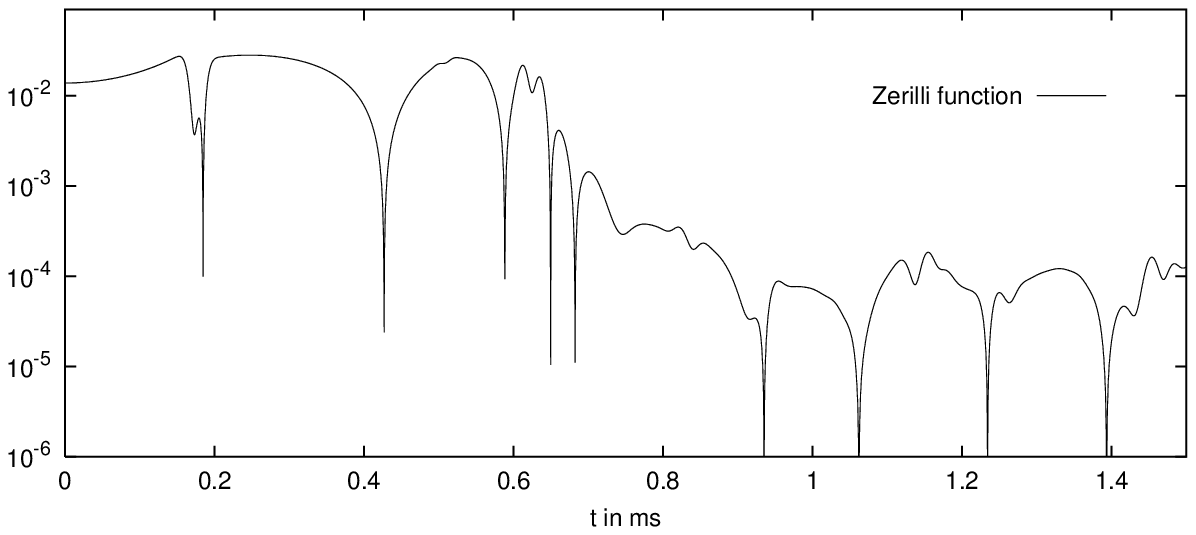}
\caption{\label{res100}Evolution for a resolution of $N = 100$ grid points
inside the star. The upper panel shows $S$ and $T$, in the lower panel
we show the Zerilli function $Z$. Note that even that $S$ and $T$ seem
to be quite smooth, the Zerilli function looks quite jagged.}
\end{figure}

Let us now turn to the Hamiltonian constraint. In Fig.~\ref{conv} we
show the violation of the Hamiltonian constraint for the same
evolution as in Fig.~\ref{pulse}, this time for four different
resolutions starting with $N = 100$ and doubling each time. For
comparison we also include the evolution of $T$. The evolution of the
constraint is similar to the evolution of $T$, but it also has some
distinctive features. First of all, the oscillation frequencies are
much higher, which is what we would expect since the violation of the
constraint is mainly due to the residual error in the approximation of
the derivatives by finite differences, which is proportional to some
higher derivatives. Secondly, in addition to the outgoing and
scattered pulses, there is a third one in between, which stems from a
partial reflection of the ingoing pulse right at the surface of the
star. For the very high resolutions the violation of the constraint
has a lower limit due to the finite machine precision, which manifests
itself in the noise that is present for $N = 800$.

\begin{figure}[t]
\leavevmode
\epsfxsize=\textwidth
\epsfbox{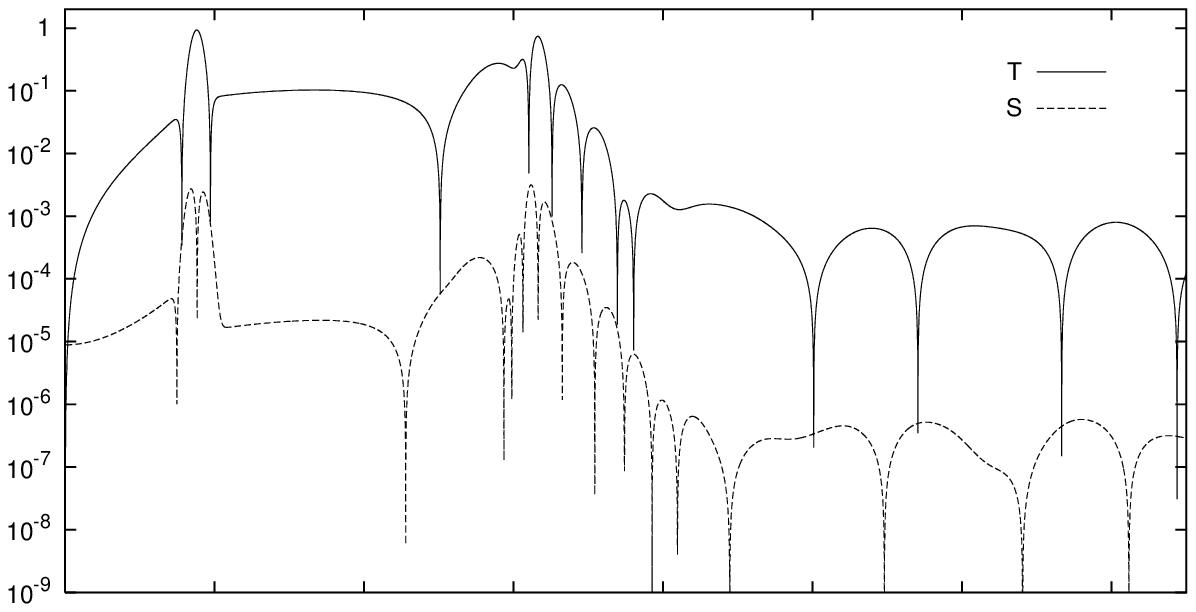}
\epsfxsize=\textwidth
\epsfbox{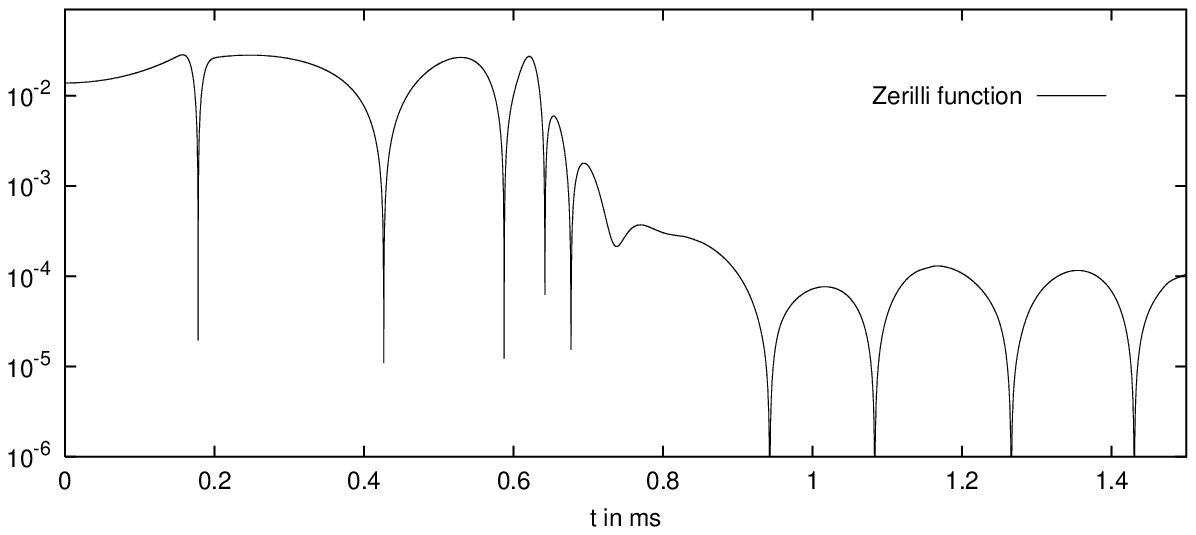}
\caption{\label{res800}Same evolution as in Fig.~\ref{res100}, this time
with a resolution of $N = 800$. Again, we show $S$ and $T$ in the upper
panel, which appear quite similar to Fig.~\ref{res100}. In contrast to
the jagged appearance of the low resolution Zerilli function in
Fig.~\ref{res100} the high resolution Zerilli function in the lower
panel is very smooth.}
\end{figure}

\abcfig
\begin{figure}[t]
\leavevmode
\epsfxsize=\textwidth
\epsfbox{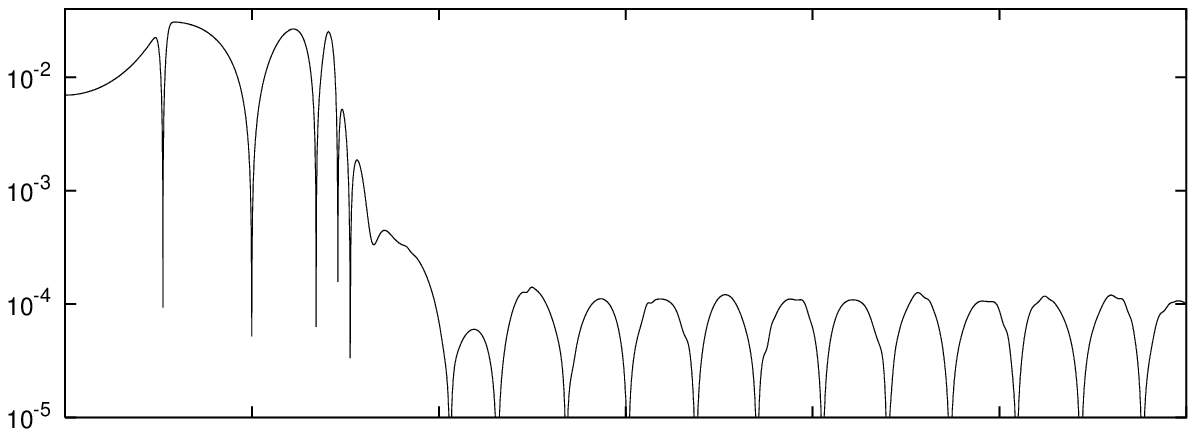}
\epsfxsize=\textwidth
\epsfbox{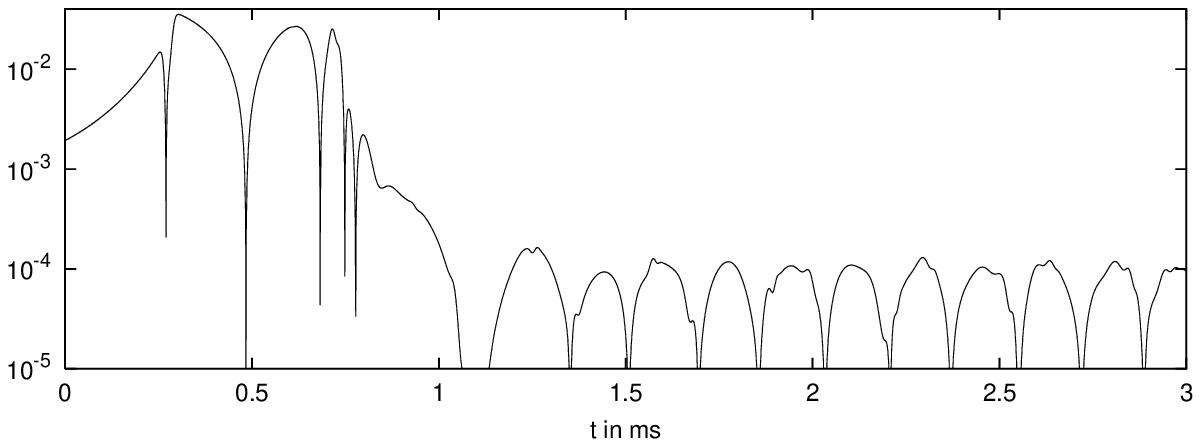}
\caption{\label{Za}Zerilli function evaluated at $r = 125\,$km
(upper panel) and $r = 250\,$ km (lower panel).}
\end{figure}

As one can clearly see, in the logarithmic plot of Fig.~\ref{conv}
the spacing between the curves for the different resolutions is
equidistant, which means that we can translate each curve to match its
adjacent one by multiplying it with a certain factor. For the upper
part of Fig.~\ref{conv} this factor turns out to be exactly $4$, which
tells us that, indeed, we have second order convergence. However, in
the ringing phase, which is shown in the lower panel of Fig.~\ref{conv},
this is not true any more. Here we can see that the behavior of the
Hamiltonian constraint is quite different for the various
resolutions. The different curves would not match if we tried to
superpose them onto each other by translating them in the appropriate
way. This is because the oscillation frequencies of the fluid modes
are slightly different for different resolutions. Of course, they
start to converge for increasing resolution, but unfortunately the
convergence is only of first order. As was already explained in the
previous section, this fact is due to the stellar surface, where the
second derivative of $T$ is discontinuous. A second order
discretization scheme can only yield second order convergence if the
second derivatives are continuous, which is not the case.

Another somewhat nasty feature that goes hand in hand with the
violation of the constraint is the sensitivity in the computation of
the Zerilli function. As was already mentioned in the section 4.2,
the Zerilli function $Z$ is the single gauge invariant function that
fully describes the polar perturbations of the Schwarzschild
spacetime. It can be constructed from $S$ and $T$ by means of formula
\eqref{Z}, however, it is crucial that $S$ and $T$ satisfy the
Hamiltonian constraint. Unfortunately, even small violations of the
constraint can lead to quite crummy Zerilli functions. This can be
seen in Figs. \ref{res100} and \ref{res800}, where we show the
evolution of $S$ and $T$ together with the resulting Zerilli function
$Z$ for the two different resolutions of $N = 100$ and $N = 800$.
Whereas the wave forms of $S$ and $T$ are smooth and look almost alike
for the different resolutions, the Zerilli function $Z$ is quite rough
for the low resolution, because of the violation of the constraint.
For $N = 800$ this violation has decreased enough to produce a smooth
Zerilli function.

\begin{figure}[t]
\leavevmode
\epsfxsize=\textwidth
\epsfbox{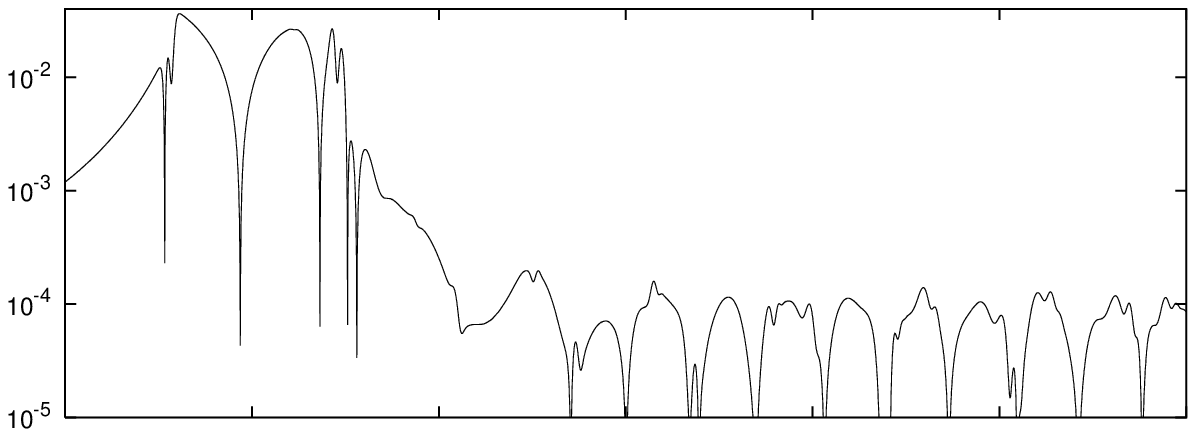}
\epsfxsize=\textwidth
\epsfbox{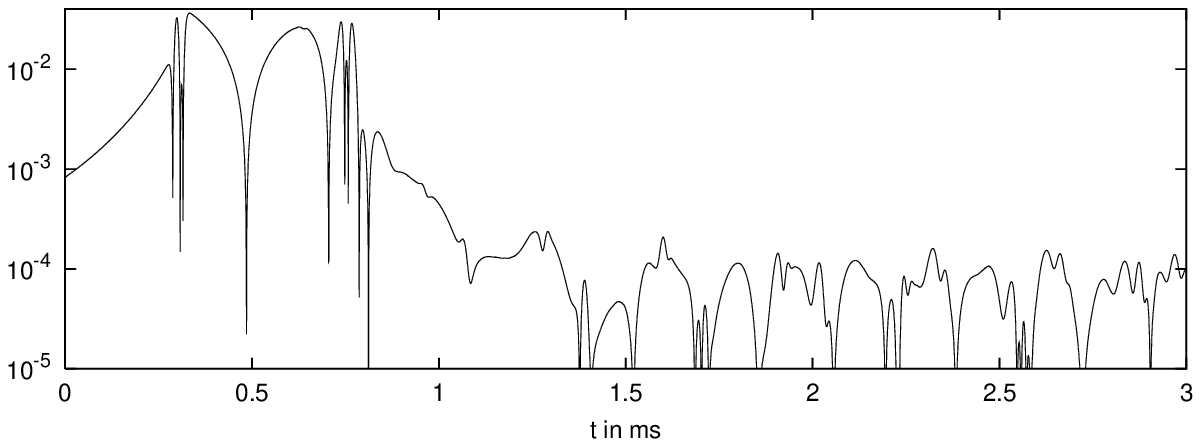}
\caption{\label{Zb}Zerilli function evaluated at at $r = 500\,$km
(upper panel) and $r = 1000\,$ km (lower panel).}
\end{figure}
\resetfig

Unfortunately, the error in the Zerilli function increases as a
function of $r$. This means that even if the Zerilli function computed
for $N = 800$ looks very smooth, it can again become quite distorted
if we move to larger $r$. We then would have to further increase the
resolution in order to obtain accurate enough results.

This is quite unfortunate since if one is interested in the radiation
power of the oscillations, one has to compute the Zerilli function at
spatial infinity. Of course, this is numerically not feasible, hence
one would like to extract the Zerilli function as far away from the
star as possible. But there is the caveat. In the wave zone far away
from the star, the amplitude of the variable $S$ remains bounded as
the wave propagates towards infinity, whereas the amplitude of $T$
grows linearly with $r$. The amplitude of the Zerilli function, of
course, has to remain bounded, since it is related to the radiated
power (see appendix C), which has a finite asymptotic value. It is
therefore clear that in the computation of $Z$ by means of \eqref{Z}
the growing behavior of $T$ somehow must exactly cancel. But, again,
this happens if and only if $S$ and $T$ exactly satisfy the
Hamiltonian constraint. If there is only a slight violation of the
constraint, the cancellation cannot occur at a hundred percent level
and thus the Zerilli function will eventually start to grow for large
$r$. This growth is particularly dominant for the high frequency
components in the wave form. They get much more amplified than the low
frequency modes as we move towards infinity. Thus, in a power spectrum
of the Zerilli function, which is recorded very far away from the star
we will have a bias towards the high frequency components, which is a
pure numerical artefact.

To illustrate the above discussion, in Figs.~\ref{Za} and \ref{Zb} we
show the Zerilli function $Z$ constructed from $S$ and $T$ at the four
different locations of $r=$ 125, 250, 500, and $1000\,$km for a given
resolution of 400 grid points inside the star. At $r=125\,$km, which
is the upper panel of Fig.~\ref{Za}, the Zerilli function appears quite
smooth. At $r=250\,$km, the panel below, some small bumps in the fluid
ringing appear, which get enhanced in the upper panel of Fig.~\ref{Zb},
where $Z$ is extracted at $r=500\,$km. Finally, in the lower panel
of Fig.~\ref{Zb} panel at $r=1000\,$km, we can see that the high
frequency components have totally distorted the shape of the Zerilli
function.

We have already discussed in section \ref{polar} that it is not
possible to switch from the variables $S$ and $T$ to the Zerilli
function in the exterior region, because of the numerical instability
associated with this procedure. The most practical way to obtain a
quite reliable Zerilli function, therefore, is not to completely switch
to the Zerilli function in the exterior region but to additionally
evolve $Z$ together with $S$ and $T$. That is, close to the surface of
the star we construct $Z$ from $S$ and $T$ by means of formula
\eqref{Z} and then use the Zerilli equation \eqref{Zeqn} to
independently evolve $Z$ parallel to $S$ and $T$. Of course, this
amounts to the additional computational expenditure of evolving an
extra wave equation, but we get rewarded by obtaining much nicer
results.

\begin{figure}[t]
\leavevmode
\epsfxsize=\textwidth
\epsfbox{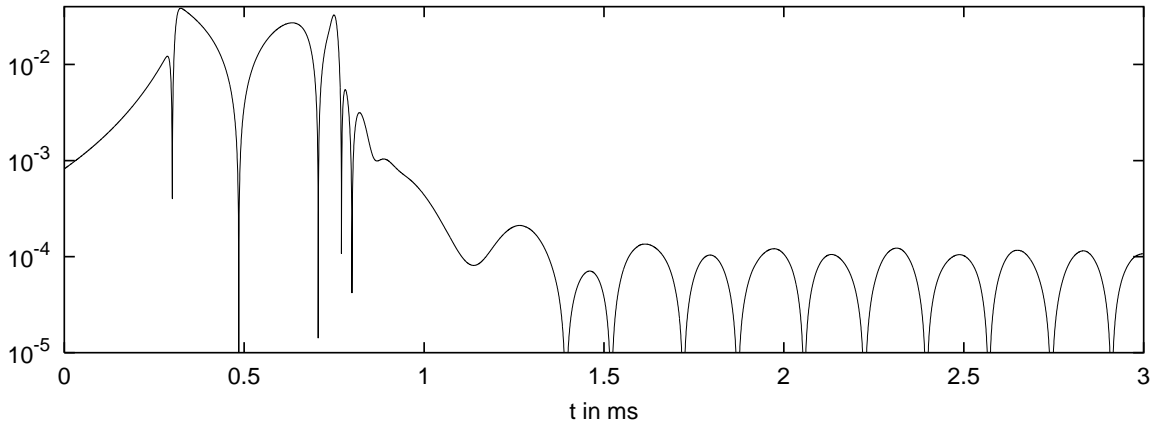}
\caption{\label{Zeq(r)}Zerilli function at $r=1000\,$km evolved with
the Zerilli equation. Compare with bottom panel of Fig.~\ref{Zb}.}
\end{figure}

In Fig.~\ref{Zeq(r)} we show the Zerilli function evolved with the
Zerilli equation and then extracted at $r=1000\,$km. The difference to
the Zerilli function in the lower panel of Fig.~\ref{Zb}, which should
be the same, is quite striking but plausible. There is no
amplification of any high frequency components, and thus the evolved
Zerilli function in Fig.~\ref{Zeq(r)} remains smooth, even at large
radii.

In the following section, where we present the results for various
stellar models and initial data, we always evolve the Zerilli function
together with $S$ and $T$.

\section{Results for polytropic equations of state}

If not mentioned otherwise, throughout this section we always use
polytropic models with $\Gamma = 2$ and $\kappa = 100\,$km$^2$.

Before presenting any results we have to discuss the various ways of
constructing initial data, which boils down to the question of how to
solve the constraints. Basically, there are only two different classes
of initial conditions we can construct.

First, we could excite the neutron star by hitting it with a
gravitational wave. This means that in the star we initially set all
the perturbation variables to zero. In the exterior we can at some
finite domain arbitrarily prescribe one of the metric perturbations
and use the Hamiltonian constraint to solve for the remaining one. For
the sake of simplicity, we prescribe $T$ to be a narrow Gaussian
located sufficiently far outside the star. The Hamiltonian constraint
then reduces to a first order ordinary differential equation for
$S$. Or, we could prescribe the Zerilli function and use \eqref{S_T}
to compute the appropriate values of $S$ and $T$. In this case we
would not have to solve any differential equation since by
construction, $S$ and $T$ satisfy the Hamiltonian constraint.

The other type of initial data is an initial perturbation of the fluid
with the associated perturbation of the metric. This means that we
arbitrarily prescribe $\rho$ and use the Hamiltonian constraint to
solve for the metric variables $S$ and $T$. However, this cannot be
done uniquely, since we have two variables to solve for but only one
equation. The two easiest possibilities are to set one variable to
zero and solve for the remaining one. As we shall see, it can make a
huge difference in the resulting wave form, whether for the same
$\rho$, we choose $S$ to be zero and solve for $T$ or, vice versa, set
$T$ to zero and solve for $S$.

Of course, all the above choices of prescribing initial data are pure
ad hoc prescriptions and have no astrophysical meaning whatsoever. As
a first attempt to construct initial data by means of a more physical
prescription, Allen et al.~\cite{AAKLR99} have tried to apply the very
successful method of constructing initial data for black hole
collisions in the close limit to the case of the final stage
of two merging neutron stars. Still, from an astrophysical point of
view, this procedure, which works surprisingly well in the black hole
case is somewhat questionable in its applicability for neutron stars.

The construction of physically realistic initial data is still a very
open field and involves a deep understanding of the physics behind
core collapses or neutron star quakes (glitches), which one assumes to
be the most important mechanisms to produce significant oscillations
of the neutron star. That is why we have to stick to our ad hoc data.
Nevertheless, they still can give us important insight into what kind
of modes can be excited, and what generic wave forms look like.

Moreover, for the sake of simplicity, we stick to time symmetric initial
data. In this case, the momentum constraints are trivially satisfied
since all extrinsic curvature variables are zero. In
\cite{AKLPS99} Andersson et al.~extend the close limit approach of Allen
et al.~\cite{AAKLR99} to the head-on collision of boosted neutron
stars. Here the initial data are not time symmetric any more and
therefore they have to explicitly solve the momentum constraints for
the extrinsic curvature variables. Unfortunately we cannot use their
initial data in a straightforward way since their data do not conform
to the Regge-Wheeler gauge. From section \ref{polar} we know that in
the Regge-Wheeler gauge the extrinsic curvature variable
$\widehat{K}_4$ has to vanish, otherwise $\widehat{T}_1$, which has to
vanish, too, would assume nonzero values. The initial data of Allen et
al.~have non-vanishing $\widehat{K}_4$.

Let us now turn to the first choice of initial data, the impinging
gravitational wave. In previous simulations \cite{AAKS98} it was
shown that for a reasonable neutron star model in general a
gravitational wave will excite the first $w$-mode, the $f$-mode and
several $p$-modes. However, the authors confined themselves to only
one polytropic model with \hbox{$\eps_0 = 3\!\cdot\!10^{15}$g/cm$^3$} and
considered only the quadrupole ($l=2$) case.

Here we would like to present a much more exhaustive survey of wave
forms for a whole series of polytropic stellar models, ranging from
very low mass up to ultra-compact models. Of course, both the low mass
and the ultra-relativistic stellar models are quite unrealistic;
especially the latter ones, for they are unstable with respect to
radial oscillations. Nevertheless, it is quite interesting to see the
change in the wave forms as one moves along the series of different
models. In addition to the quadrupole case, we also pay some attention
to $l=3$ and $l=4$. 

\begin{table}[t]
\begin{center}
\begin{tabular}{|c||c|c|c|c|}\hline
\multicolumn{5}{|c|}{\rule[-2.5mm]{0mm}{7.5mm}Polytropic stellar models
($\Gamma = 2, \kappa = 100\,$km$^2$)}\\\hline
\rule[-2.5mm]{0mm}{7.5mm}Model & $\eps_0\;[$g/cm$^3]$ & $M\;[M_\odot]$ & $R\;[$km$]$ & R/M\\
\hline\hline
\rule[0mm]{0mm}{5mm}1 & $5.0\!\cdot\! 10^{14}$ & 0.495 & 11.58 & 15.843\\
2 & $1.0\!\cdot\! 10^{15}$ & 0.802 & 10.81 & 9.137\\
3 & $2.0\!\cdot\! 10^{15}$ & 1.126 & 9.673 & 5.819\\
4 & $3.0\!\cdot\! 10^{15}$ & 1.266 & 8.862 & 4.740\\
5 & $5.0\!\cdot\! 10^{15}$ & 1.348 & 7.787 & 3.912\\
6 & $7.0\!\cdot\! 10^{15}$ & 1.343 & 7.120 & 3.585\\
7 & $1.0\!\cdot\! 10^{16}$ & 1.300 & 6.466 & 3.369\\
8 & $3.0\!\cdot\! 10^{16}$ &1.097 & 5.211 & 3.217\\
9 & $5.0\!\cdot\! 10^{16}$ &1.031 & 4.992 & 3.280\\
\hline
\end{tabular}
\caption{\label{models}List of the used polytropic stellar models
and their physical parameters.}
\end{center}
\end{table}

Because $S$ and $T$ are gauge dependent quantities and as such not
very meaningful, in the plots we therefore always show the Zerilli
function $Z$ obtained by the prescription of the last section.

As initial data we choose a narrow Gaussian in $T$ centered around $r
= 50\,$km. For the runs with $l=2$, the resolution is 500 grid points
inside the star. For $l=3$ we have to increase the resolution to 2000
grid points inside the star in order to obtain reasonable results. The
observation point is located at $r = 100\,$km and the outer boundary
has been moved far enough to the right in order to prevent any
contamination from reaching the observation point before the evolution
stops. The physical properties of the used stellar models are given in
table \ref{models}.

\begin{figure}[p]
\leavevmode
\epsfxsize=\textwidth
\epsfbox{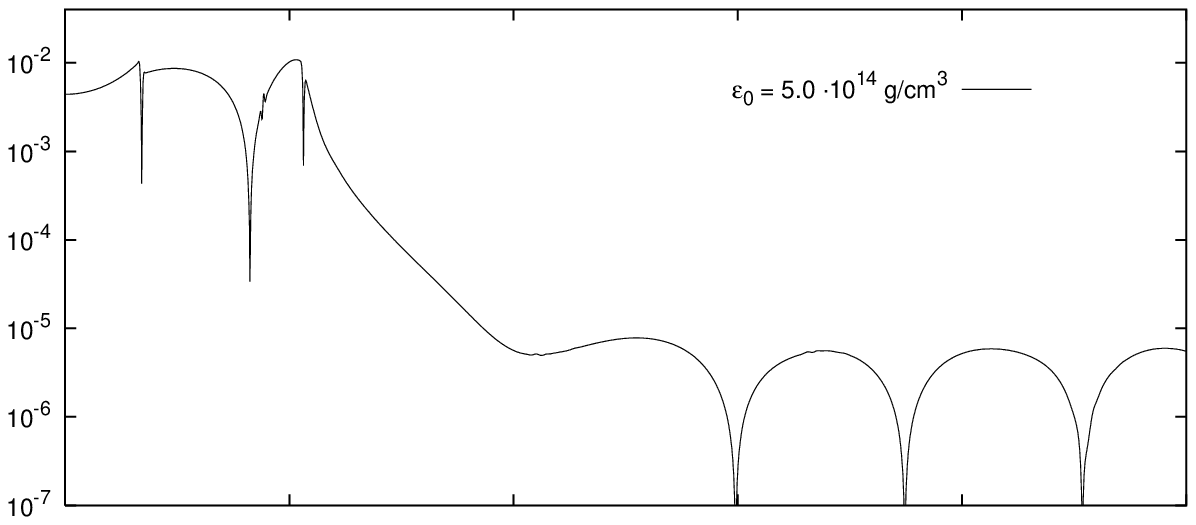}
\epsfxsize=\textwidth
\epsfbox{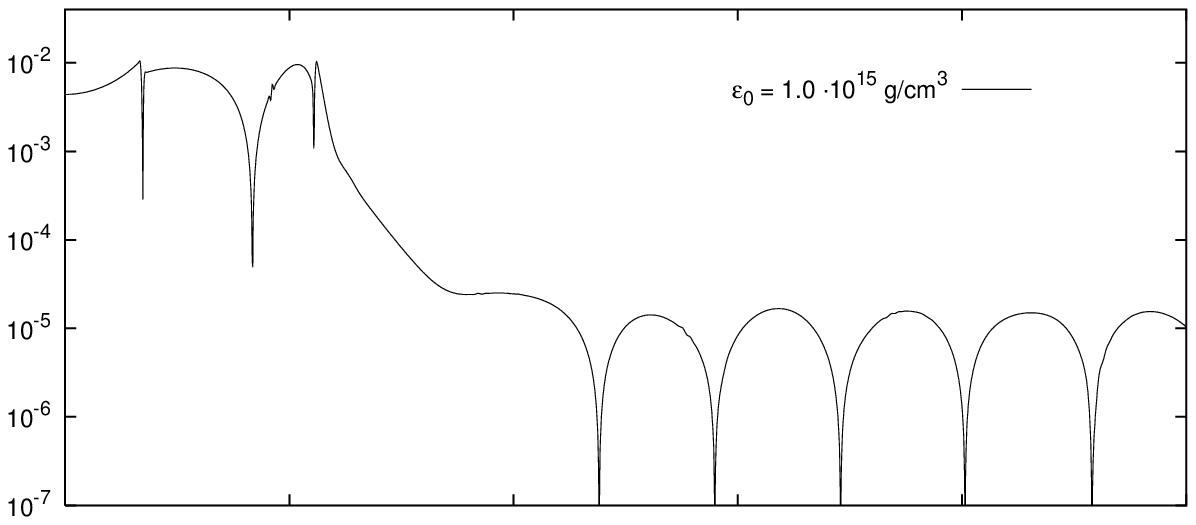}
\epsfxsize=\textwidth
\epsfbox{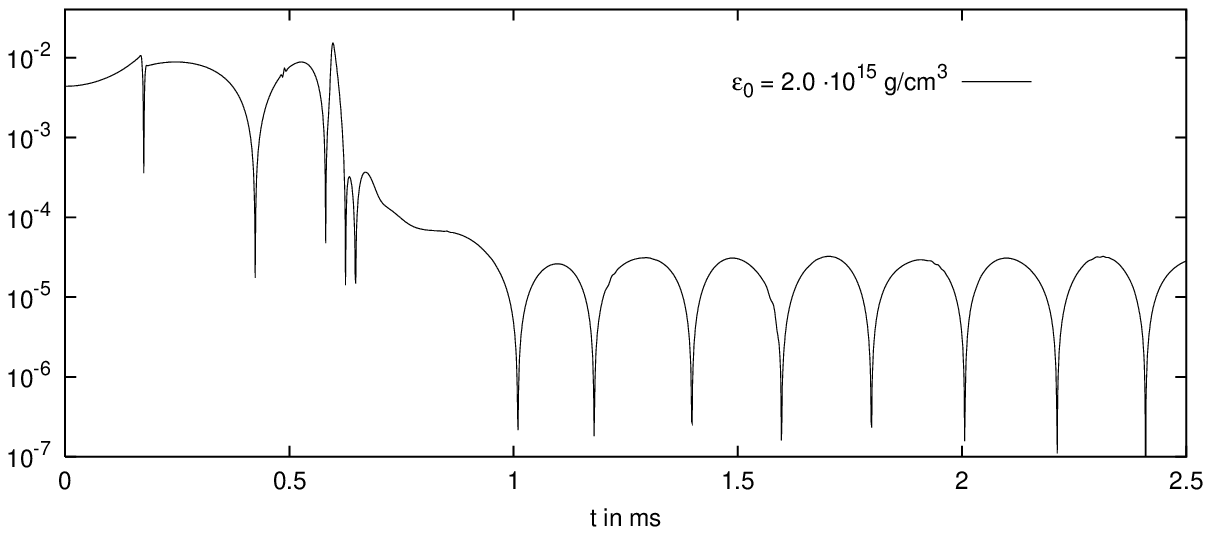}
\caption{\label{M1-3_l=2}$l=2$ wave forms for the low mass models
M1 -- M3.}
\end{figure}

Figure \ref{M1-3_l=2} shows the evolution of the modulus of the
Zerilli function for the stellar models M1 -- M3 for $l=2$ on a
logarithmic scale. The wave forms of the low mass models M1 and M2 are
quite similar: After the reflected wave pulse has crossed the
observation point around $t = 0.5\,$ms, the amplitude of the wave
signal drops rapidly by several orders of magnitude until the final
fluid ringing starts, which is dominated by the $f$-mode. The less
relativistic the model the lower the amplitude of the fluid
ringing. Also, the frequency of the $f$-mode decreases as the models
become less relativistic. Obviously, in all three cases, there is no
$w$-mode presence at all!  This is somewhat surprising since mode
calculations show that there exist $w$-modes for those
models. Instead, the wave forms show a tail-like fall-off that is
characteristic for the late time behavior of black hole oscillations.
For black holes, the wave form consists of the exponentially damped
quasi-normal modes and an additional ring-down, which obeys a power law,
and which results from the backscattering of the gravitational waves
at the curved background spacetime. After the quasi-normal modes have
damped away, it is this tail that dominates the late-time evolution.

In the case of those less relativistic stellar models M1 and M2, the
damping of the $w$-modes is quite strong and we therefore can see a
tail-like fall-off before the fluid ringing starts.

\begin{figure}[p]
\leavevmode
\epsfxsize=\textwidth
\epsfbox{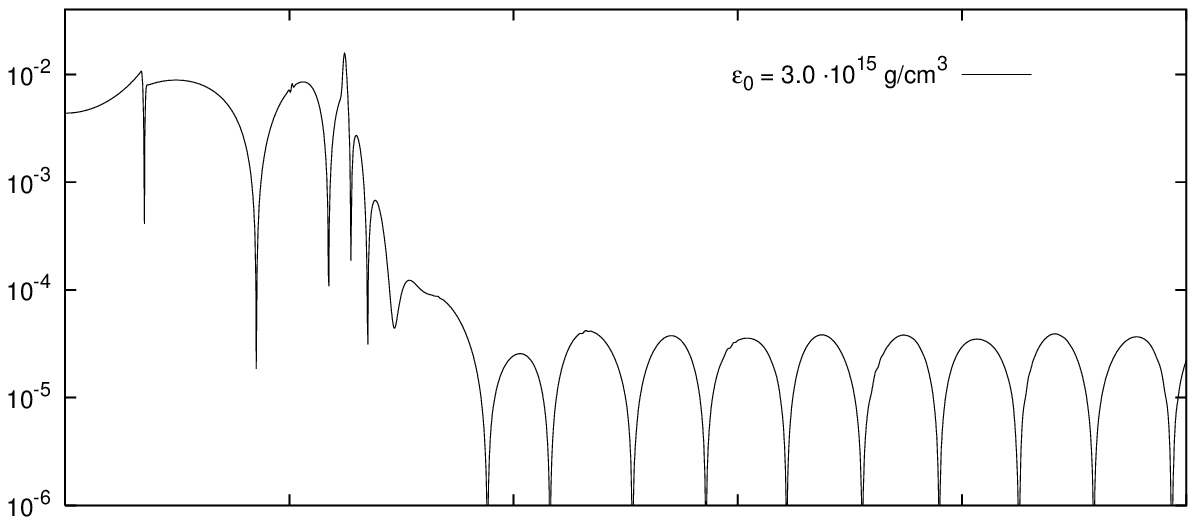}
\epsfxsize=\textwidth
\epsfbox{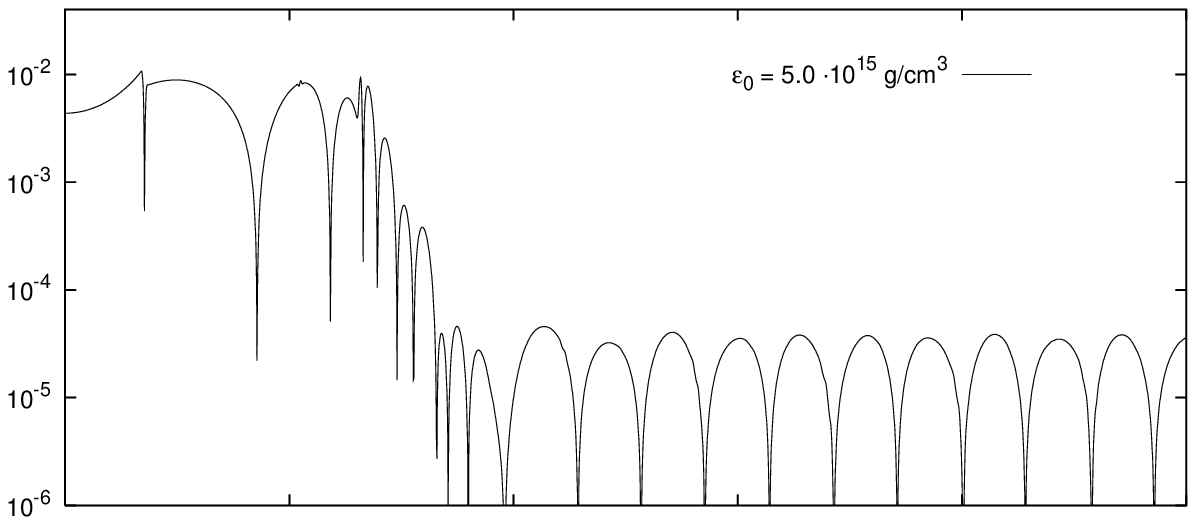}
\epsfxsize=\textwidth
\epsfbox{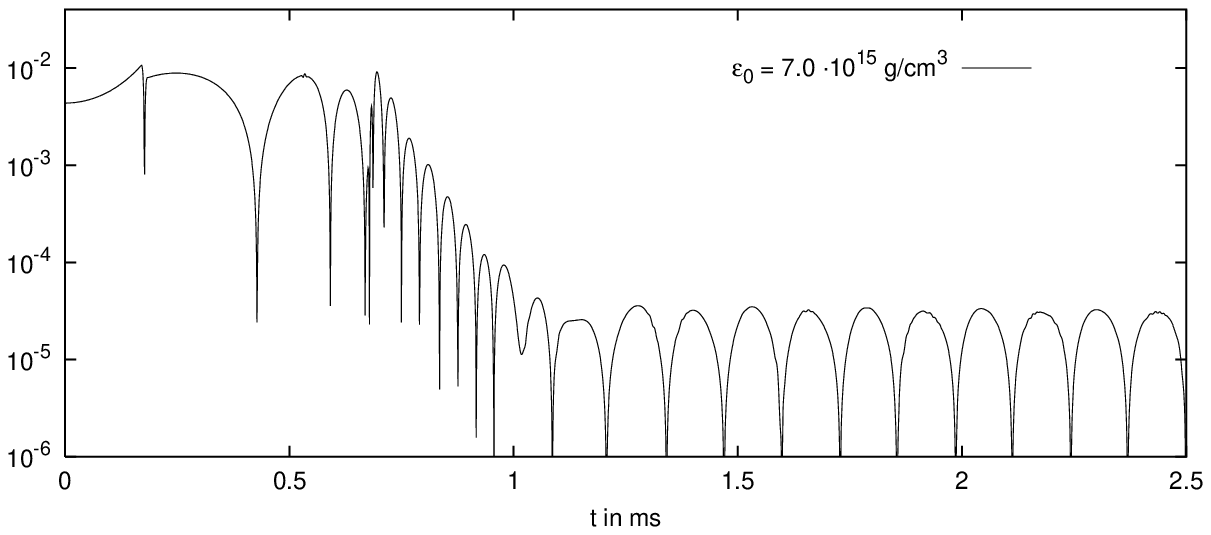}
\caption{\label{M4-6_l=2}$l=2$ wave forms for the intermediate mass
models M4 -- M6.}
\end{figure}

\begin{figure}[p]
\leavevmode
\epsfxsize=\textwidth
\epsfbox{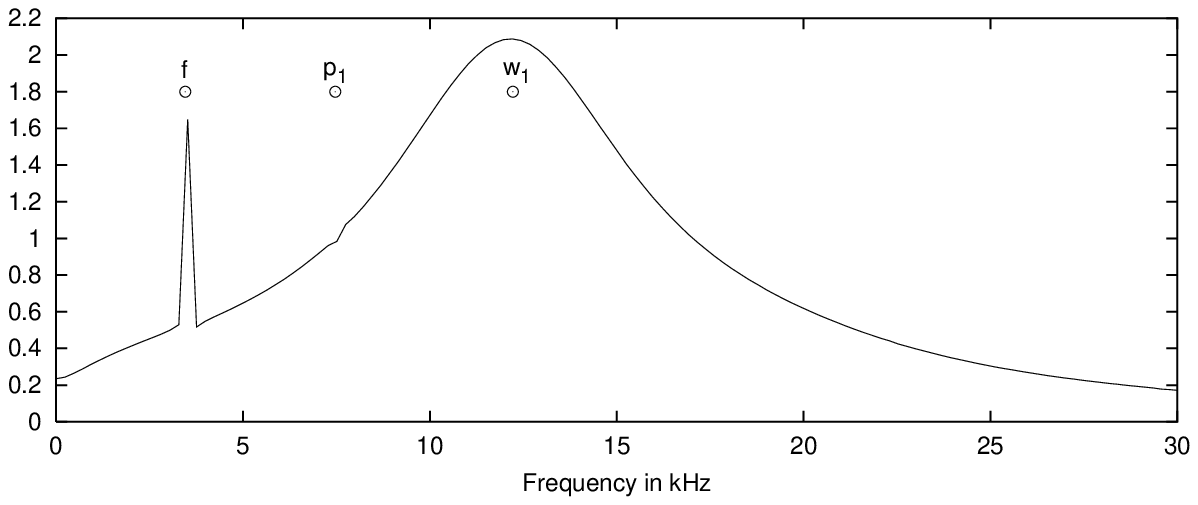}
\caption{\label{fft_l=2_M6}Power spectrum for model M6. The $f$-mode,
the first $p$-mode and the first $w$-mode are clearly present.}
\epsfxsize=\textwidth
\epsfbox{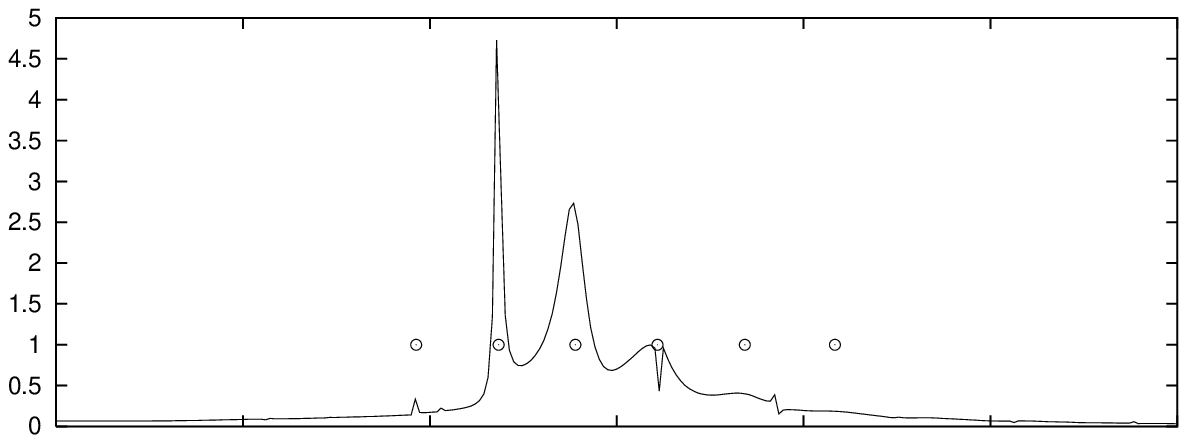}
\epsfxsize=\textwidth
\epsfbox{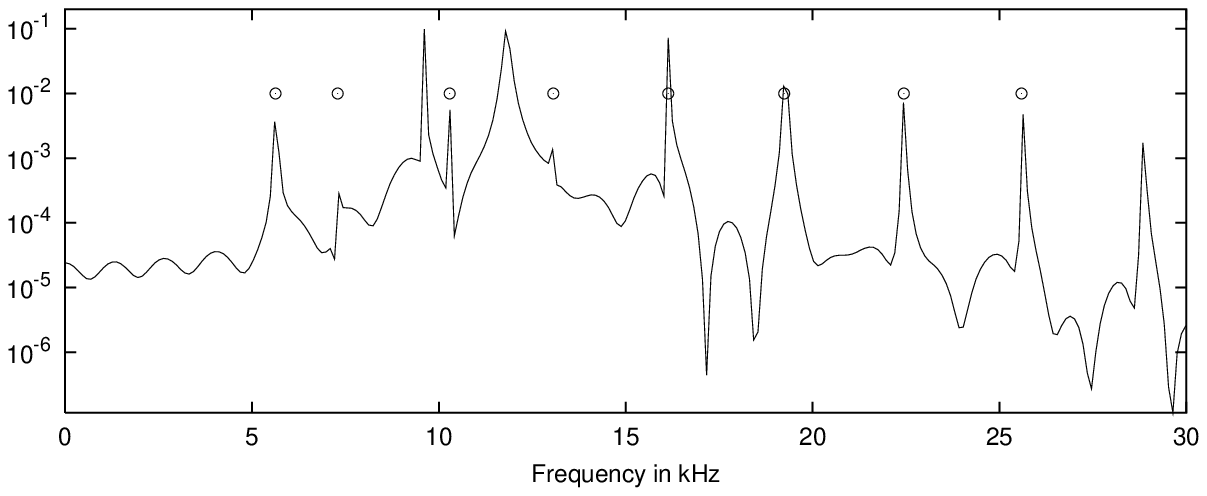}
\caption{\label{fft_l=2_M9}Power spectra of the evolution for Model M9
with $l=2$. In the upper panel the Fourier transformation is taken for an
early starting time and shows the presence of the modest damped
$w$-modes. The lower spectrum is for a later time, where most of the
$w$-modes have damped away and the very long-lived $p$-modes prevail.}
\end{figure}

It is only with model M3 that things start to change. Here we can see
some timid oscillations before the amplitude again starts dropping
down to give way for the final fluid ringing. In the wave form of
model M4, which is shown in Fig.~\ref{M4-6_l=2}, a strongly damped
oscillation, which corresponds to the first $w$-mode, is clearly
visible. In models M5 and M6, with the latter being right above the
stability limit, it is even more pronounced. In Fig.~\ref{fft_l=2_M6}
we show the power spectrum of the wave form of model M6. Also included
are the frequencies of the $f$-mode, the first $p$-mode and the first
$w$-mode that were obtained by a mode calculation with a program of
M.~Leins \cite{Leins94}. The agreement is excellent.

\begin{figure}[p]
\leavevmode
\epsfxsize=\textwidth
\epsfbox{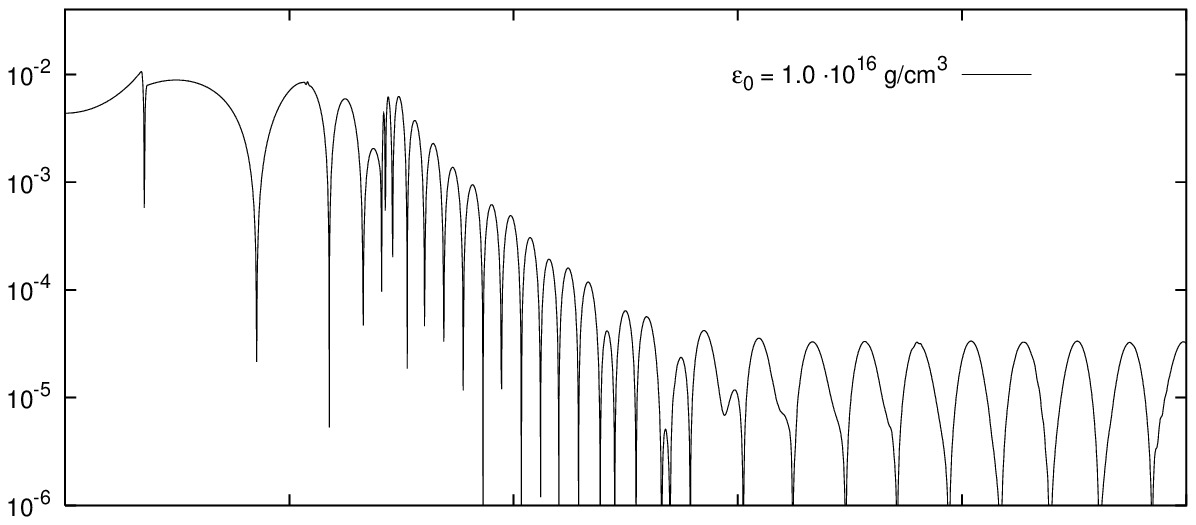}
\epsfxsize=\textwidth
\epsfbox{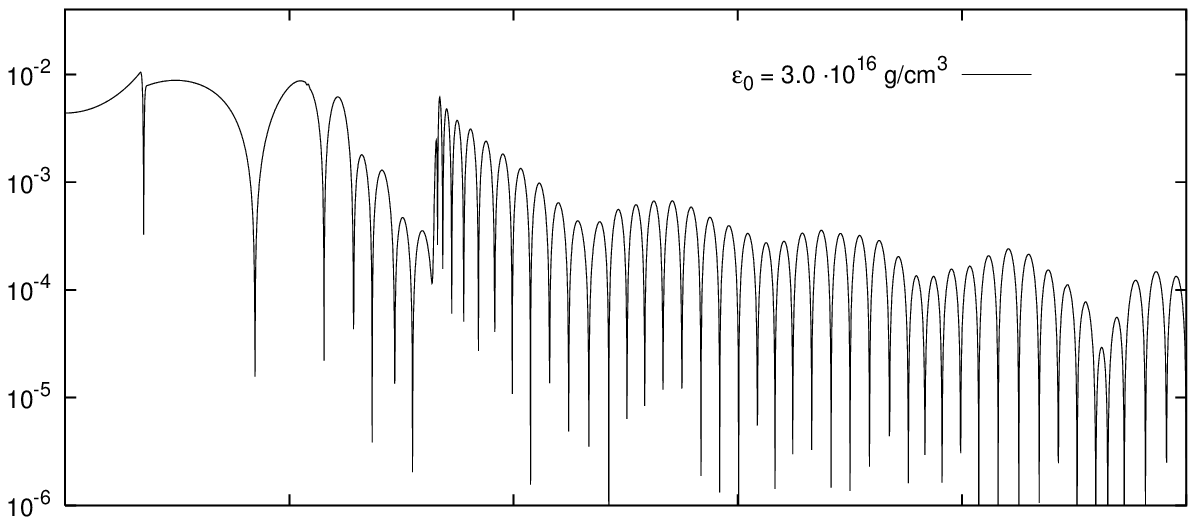}
\epsfxsize=\textwidth
\epsfbox{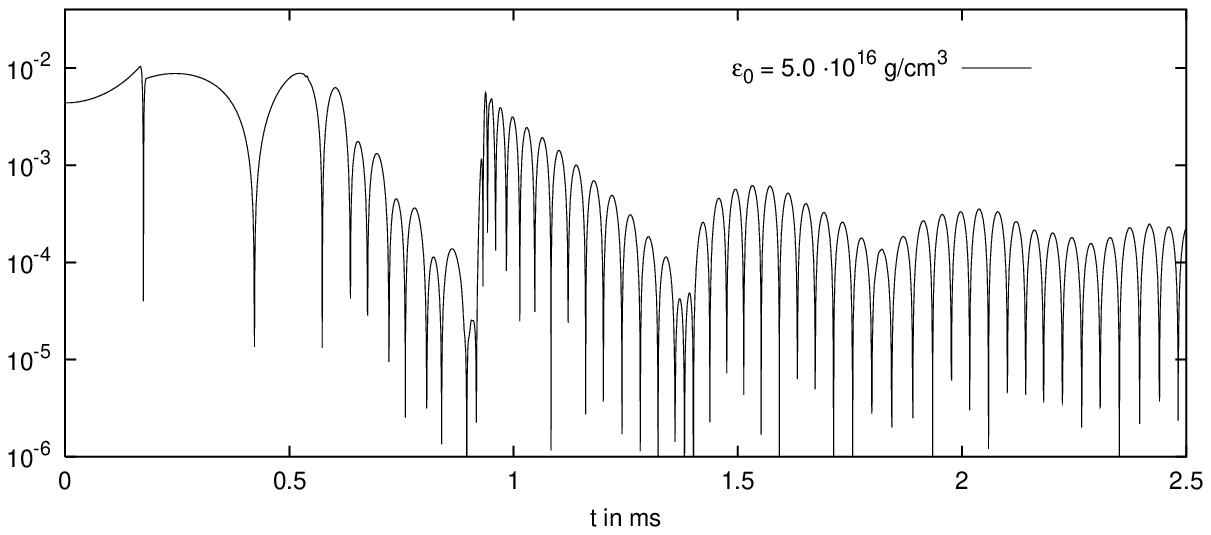}
\caption{\label{M7-9_l=2}$l=2$ wave forms for the ultra-relativistic
models M7 -- M9.}
\end{figure}

\begin{figure}[p]
\leavevmode
\epsfxsize=\textwidth
\epsfbox{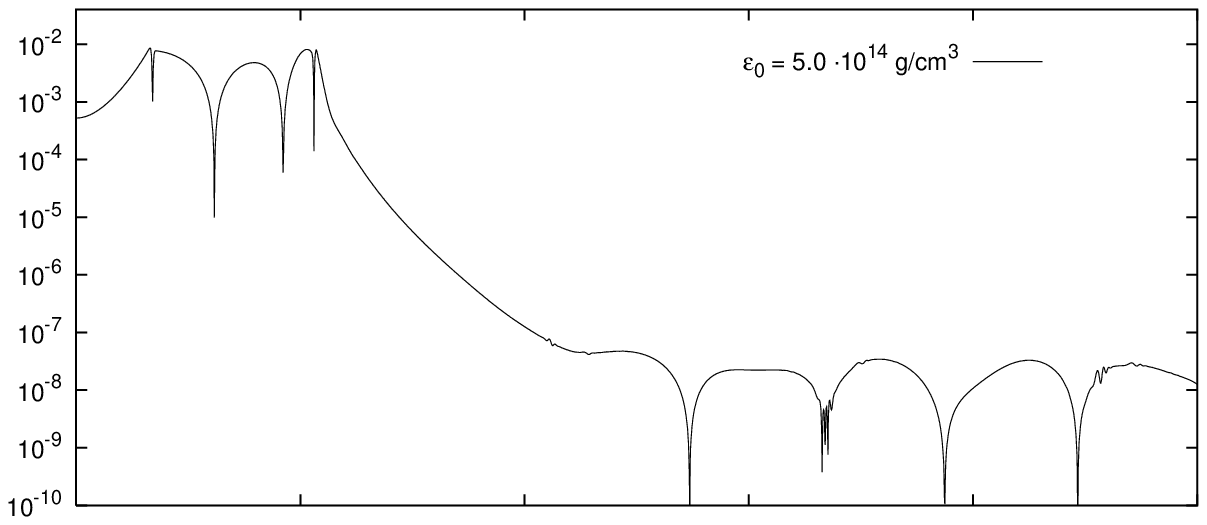}
\epsfxsize=\textwidth
\epsfbox{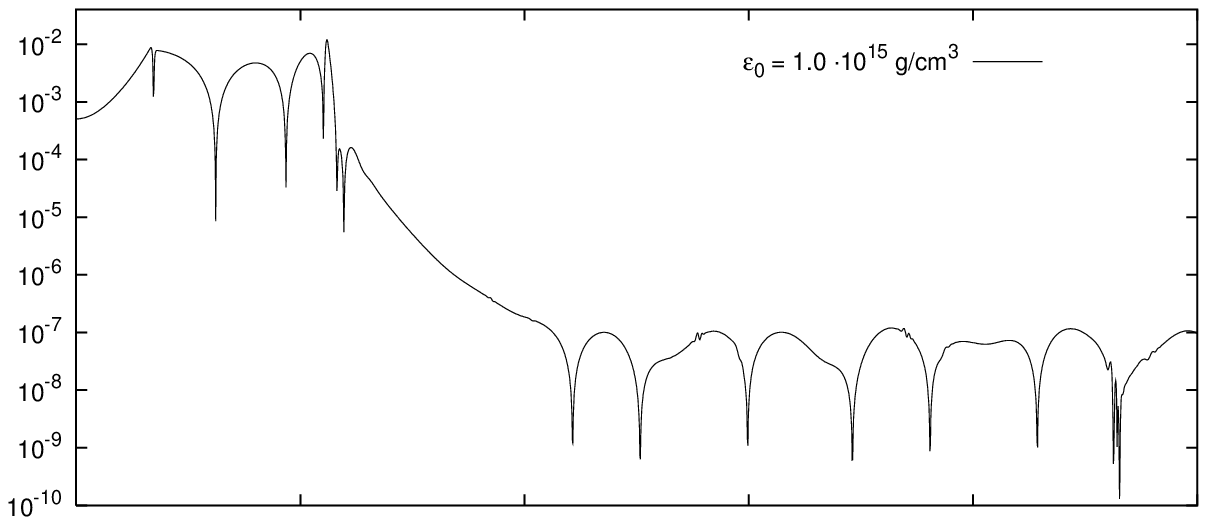}
\epsfxsize=\textwidth
\epsfbox{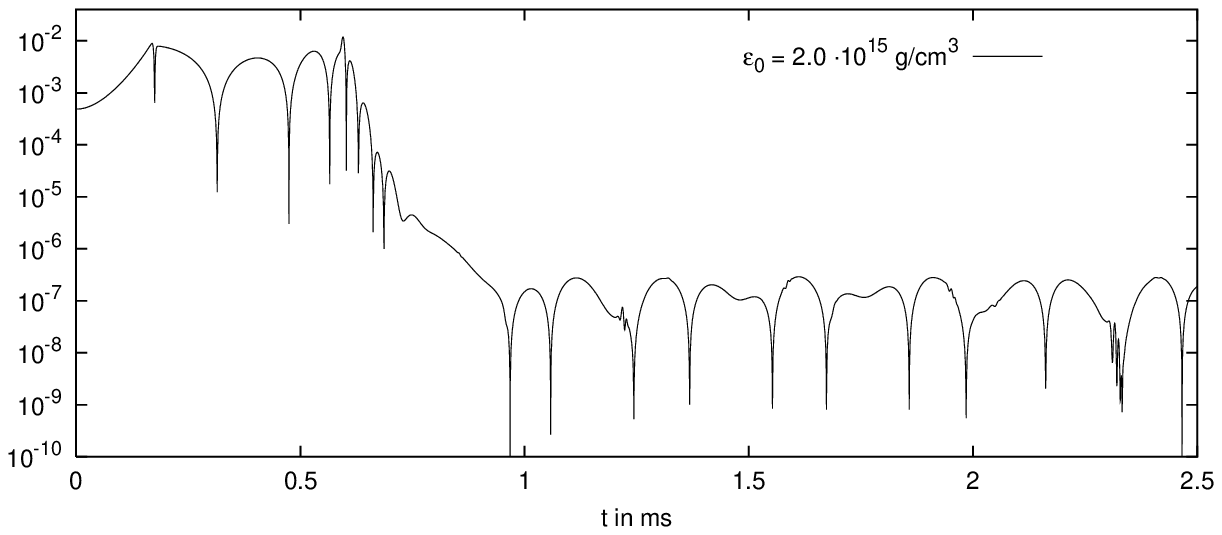}
\caption{\label{M1-3_l=3}$l=3$ wave forms for the low mass models
M1 -- M3.}
\end{figure}

\begin{figure}[p]
\leavevmode
\epsfxsize=\textwidth
\epsfbox{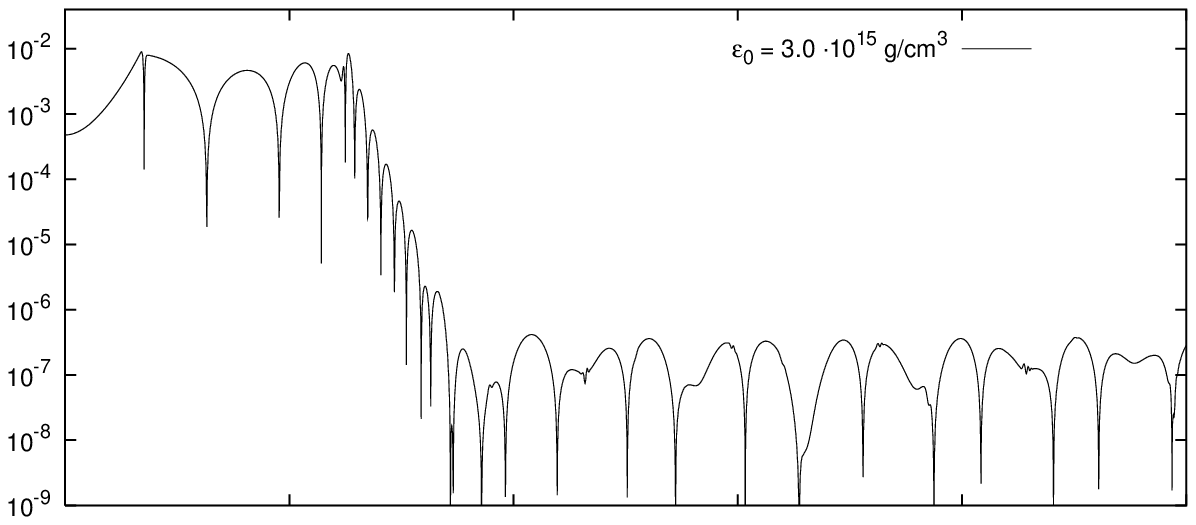}
\epsfxsize=\textwidth
\epsfbox{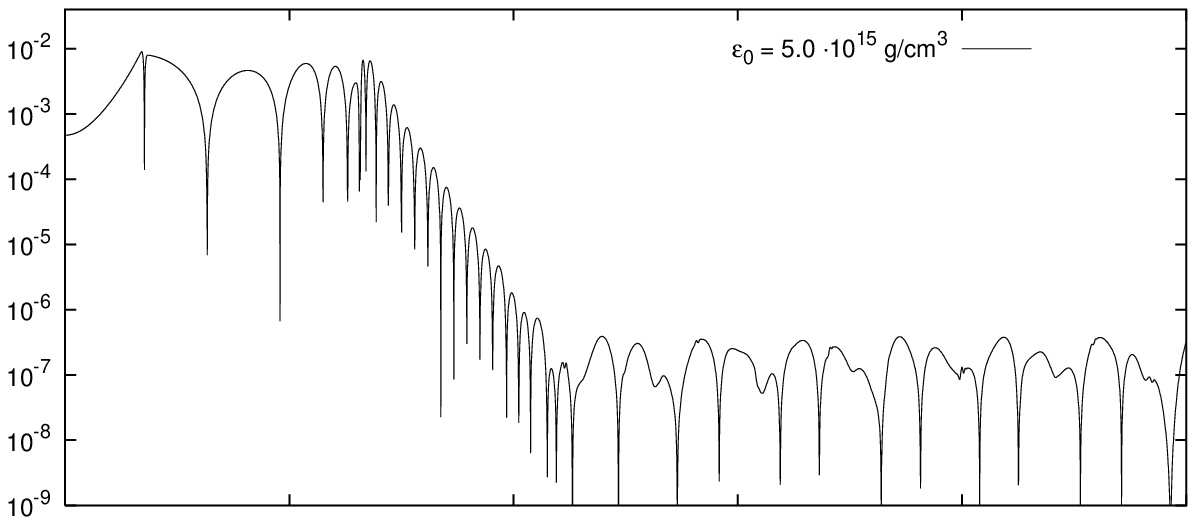}
\epsfxsize=\textwidth
\epsfbox{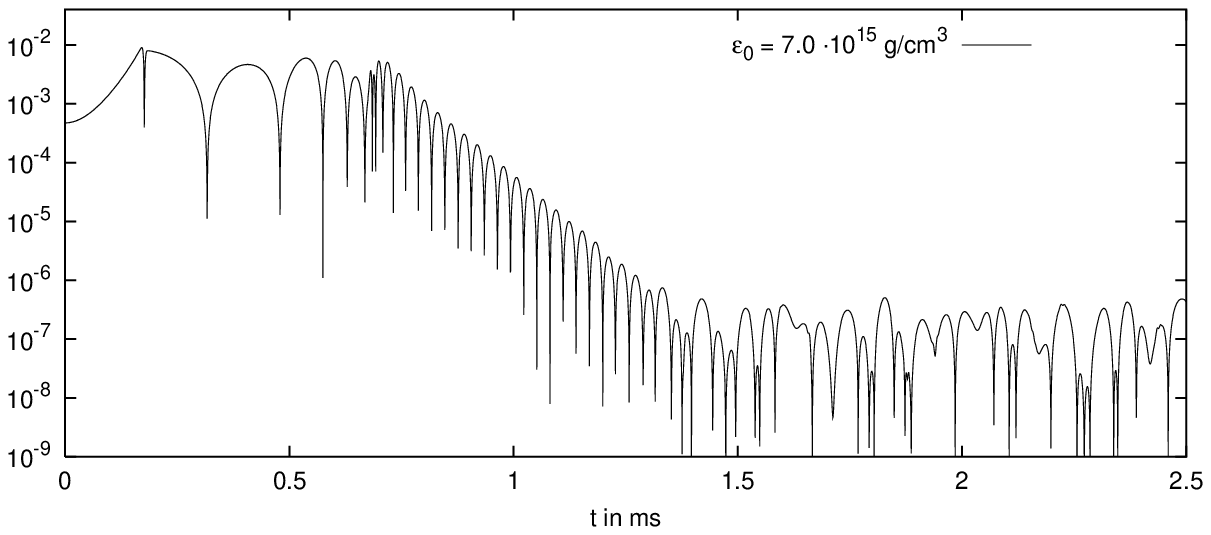}
\caption{\label{M4-6_l=3}$l=3$ wave forms for the intermediate mass
models M4 -- M6.}
\end{figure}

\begin{figure}[p]
\leavevmode
\epsfxsize=\textwidth
\epsfbox{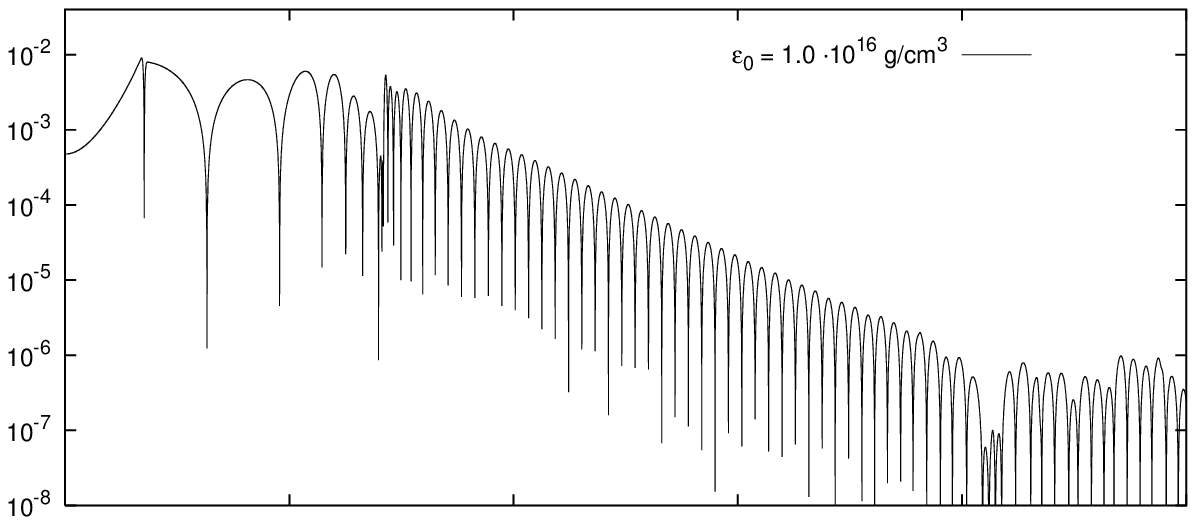}
\epsfxsize=\textwidth
\epsfbox{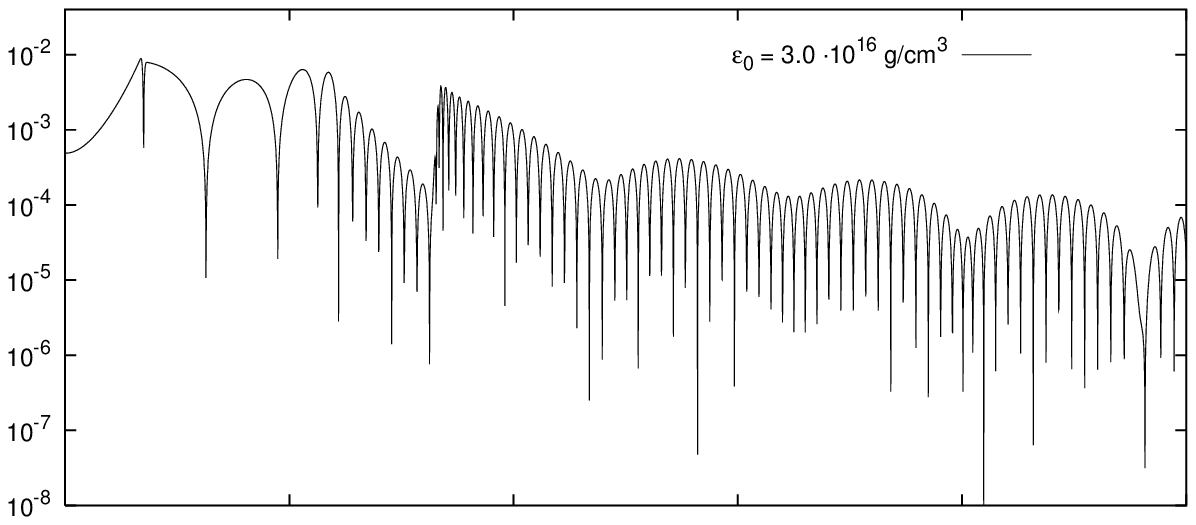}
\epsfxsize=\textwidth
\epsfbox{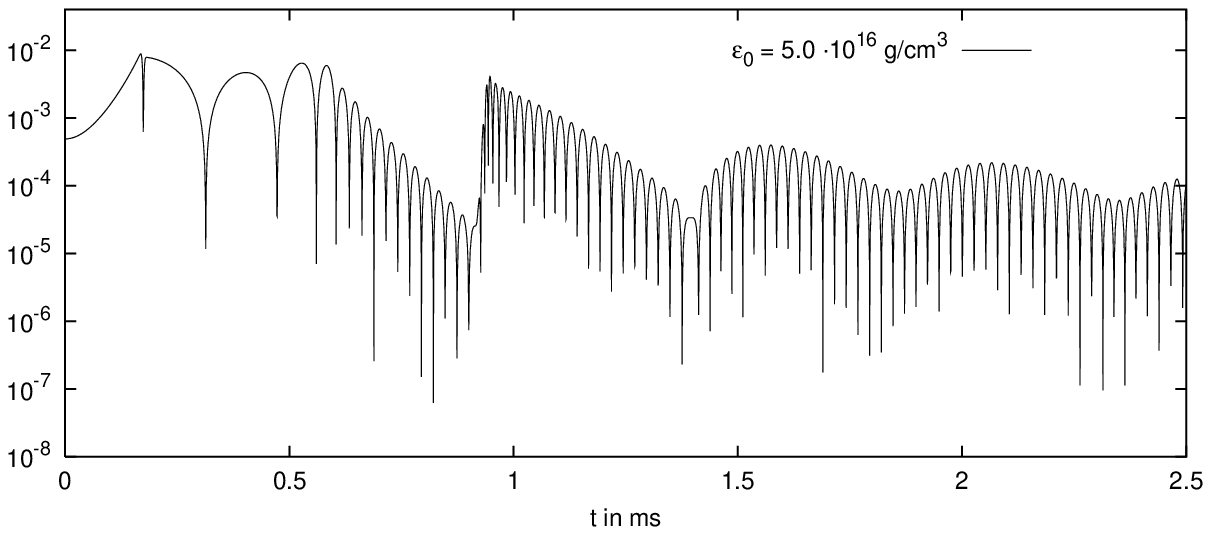}
\caption{\label{M7-9_l=3}$l=3$ wave forms for the ultra-relativistic
models M7 -- M9.}
\end{figure}

From comparing the wave forms of models M4 -- M7 we can discern that
the $w$-modes get more and more long-lived as the neutron stars become
more and more relativistic, which is confirmed by explicit mode
calculations. Finally, for the ultra-relativistic models M8 and M9 in
Fig.~\ref{M7-9_l=2} the wave forms again change their shapes quite
drastically, which hints at the appearance of a new effect. In those
cases there exists another family of very long-lived modes, which are
called trapped modes, and were first found by Chandrasekhar \& Ferrari
\cite{CF91b} for the axial perturbations of ultra-relativistic
stars. As was discussed in section \ref{odd} the axial modes can be
described both within and outside the star by a single wave equation
with a potential term $V$, whose shape depends on the stellar
model. For less relativistic models $V$ is dominated by the
centrifugal term $l(l+1)/r^2$ and is therefore a monotonically
decreasing function of $r$. As the central density increases, the
stellar models get more compact and the remaining terms become more
important. Above a certain central density, $V$ starts to develop a
local minimum inside the star, which then gives rise to the existence
of the new family of trapped modes. For our polytropic equation of
state this happens at a central density of about $1.1\!\cdot\!
10^{16}\,$g/cm$^3$.

For the polar case we have two coupled equations and it is therefore
not possible to write down an effective potential. But there also
exist polar trapped modes for stellar models with central densities
above a certain value, which is about the same where the axial trapped
modes start to appear.

In Fig.~\ref{fft_l=2_M9} we show two power spectra of the wave form of
model M9, one taken from the time $t=1\,$ms, where the trapped modes
dominate, and one taken from a much later time, where they have mostly
damped away and the fluid modes dominate. Here, too, the agreement
with the modes obtained by a direct mode calculation is evident.

\begin{figure}[t]
\leavevmode
\epsfxsize=\textwidth
\epsfbox{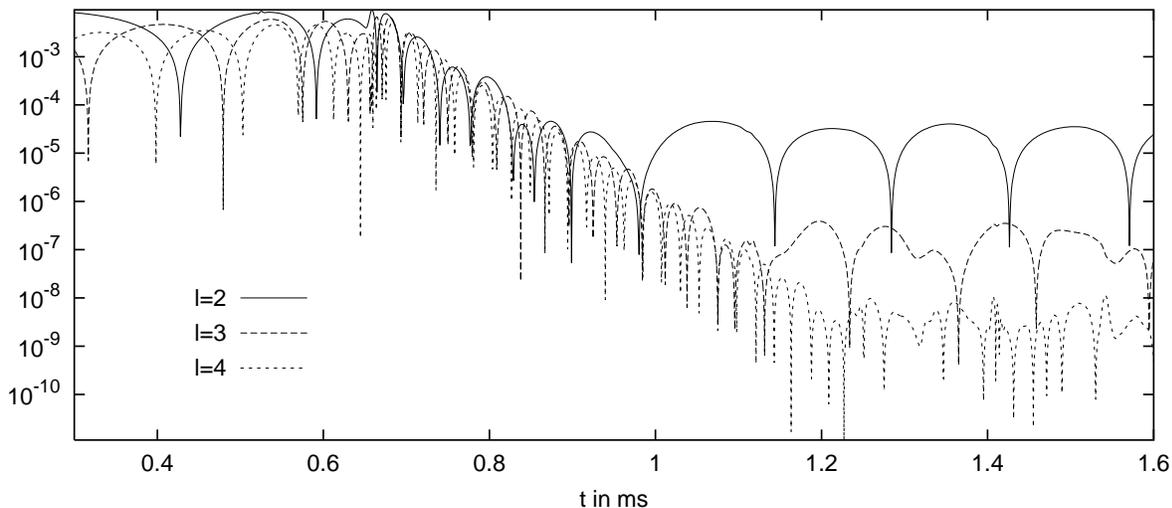}
\caption{\label{l=234}Wave forms for $l=2,3,4$ for model M6. The higher
$l$ the less energy gets transfered to the fluid.}
\end{figure}

In Fig.~\ref{l=234} we compare the wave forms of model M6 for $l=2$, 3,
and 4. It is obvious that the higher $l$ is, the less energy the impinging
gravitational wave can transfer to the fluid. The amplitudes of the
fluid modes decrease by about two orders of magnitude for increasing $l$.

There is another interesting feature in the wave forms for the
ultra-compact stars. By comparing the three panels of
Fig.~\ref{M7-9_l=2} we can see that the reflection of the impinging
wave packet with the following quasi-normal ringing gets more and more
delayed as the star becomes more and more compact. This feature is
also evident in Fig.~\ref{M7-9_l=3}, which is the $l=3$ case.

For model M7 the reflection occurs at around $t=0.7\,$ms, for model M8
after $t=0.8\,$ms, and for model M9 it is at $t=0.9\,$ms. For all
models the wave forms up to the time of about $t=0.6\,$ms are quite
similar. But for the ultra-relativistic models a gap starts to develop
between $t=0.6\,$ms and the point where the reflected wave packet
crosses the observer and the quasi-normal ringing starts. This is due
to the fact that the more compact the model the slower the local
(coordinate) speed of light (which is given by $e^{\nu-\mu}$) inside
the star and the longer the time the wave packet takes to penetrate the
star, get reflected and leave the star again. In the time during which
the wave packet is ``trapped'' inside the neutron star, there is an
exponentially damped oscillation with a characteristic frequency that
depends almost only on $l$ and $M$ and that starts in all cases
considered at about $t=0.6\,$ms.

The surprise is that the frequency and damping time of this ring-down
do not correspond to any of the quasi-normal modes of the particular
stellar model. They rather correspond to the (complex) quasi-normal
frequency of a black hole with the same mass as the star! The
correspondence is not quite perfect, but the frequencies agree within
less than one percent, only the damping of the black hole is slightly
larger than that of the neutron star. Still, this is a rather
unexpected result and, to our knowledge, has not been reported before.

\begin{figure}[t]
\leavevmode
\epsfxsize=\textwidth
\epsfbox{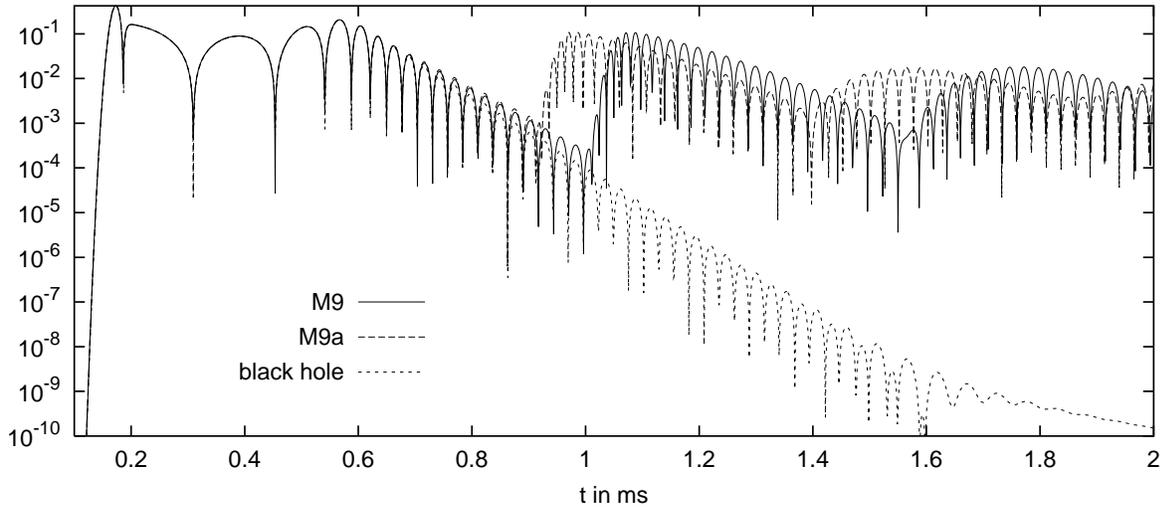}
\caption{\label{odd_comp}Evolution of the same initial data for two
neutron star models (M9 and M9a) and a black hole with the same mass.}
\end{figure}

This peculiar feature is also present in the axial case, which is
apparent in Fig.~\ref{odd_comp}, where we evolve the same initial data
for two different stellar models. The first stellar model is model M9
of table \ref{models}, the other one, let us call it M9a, is a
polytropic star with $\Gamma = 2.2$ and $\kappa = 28911.95$. This
choice yields the same total mass for the same central energy density
as for model M9. The radius, however, is somewhat smaller and given by
$R = 4.58\,$km. In addition, we also evolve the initial data for a
black hole with the same mass as the two stellar models.

Up to $t = 0.6\,$ms all three wave forms are identical since the
exterior is given by the same Schwarzschild metric for all three
cases. As soon as the wave packet hits the surface of the star, things
start to change slightly. It is evident that the first ring-down phase
is not exactly the same in the neutron star and the black hole case. Even
for the two different stellar models it is not identical. But this
should not cause any wonder since we are dealing with quite different
objects, on the contrary, it is remarkable how close the wave forms
are up to the point where the wave packet gets reflected in the
neutron star case. The most obvious difference is that the damping for
the black hole is somewhat stronger than for the neutron stars. The
frequencies are almost the same.

For a black hole the normalized $(2M = 1)$ complex frequencies of the
least damped quasi-normal modes for $l=2$ and $l=3$ are given both for
the polar and axial case as
\begin{align*}
        l = 2: w_1 \= 0.74734 + 0.17792\I\\
        l = 3: w_1 \= 1.19888 + 0.18540\I\;.
\end{align*}

Since the potentials inside the star are somewhat different for the
two stellar models and quite different from the black hole, it must be
the outer parts of the potentials, which are the same in all three
cases, that are mainly responsible for the first ring down. Concerning
the stars it is clear that the inner part of the potential is
important, too, for it is responsible for the stellar quasi-normal
modes, especially the trapped modes, which do not exist in the black
hole case. But it seems that in the black hole case the parameters of
the least damped mode, which is the most important one, are
predominantly determined by the outer parts of the Regge-Wheeler
potential in the axial case, or the Zerilli potential in the polar
case.

To corroborate our presumption, we look at initial data inside the
star, now again for the polar case. That is we prescribe a fluid
perturbation in $\rho$ and use the Hamiltonian constraint to solve for
the metric variables $S$ and $T$. As was mentioned earlier, this
cannot be done in a unique way, hence we consider the two extreme
cases. For given $\rho$ we can either set $T=0$ and solve for $S$, or
vice versa, set $S=0$ and solve for $T$.

\begin{figure}[p]
\leavevmode
\epsfxsize=\textwidth
\epsfbox{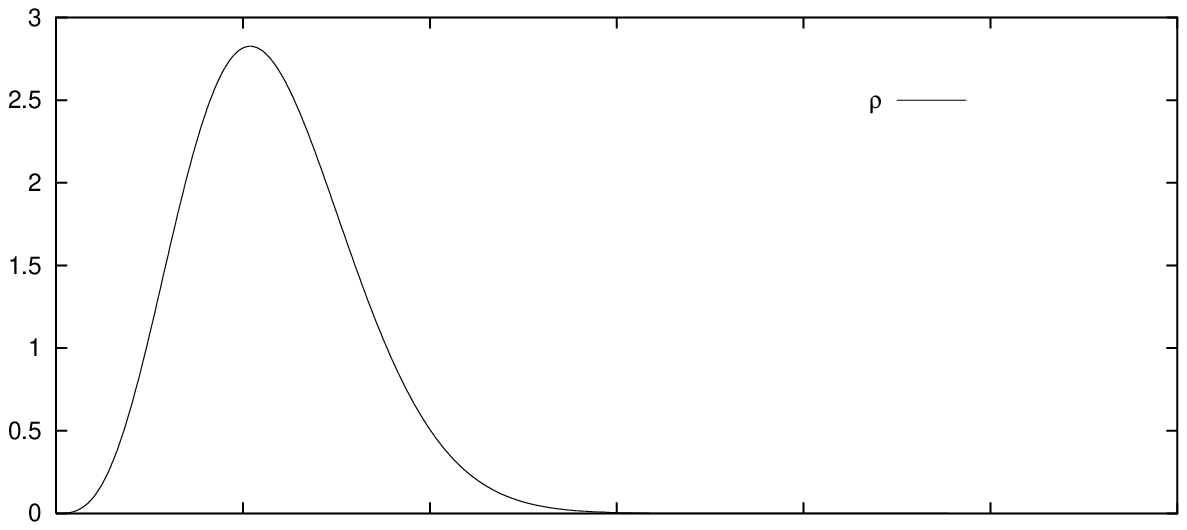}
\epsfxsize=\textwidth
\epsfbox{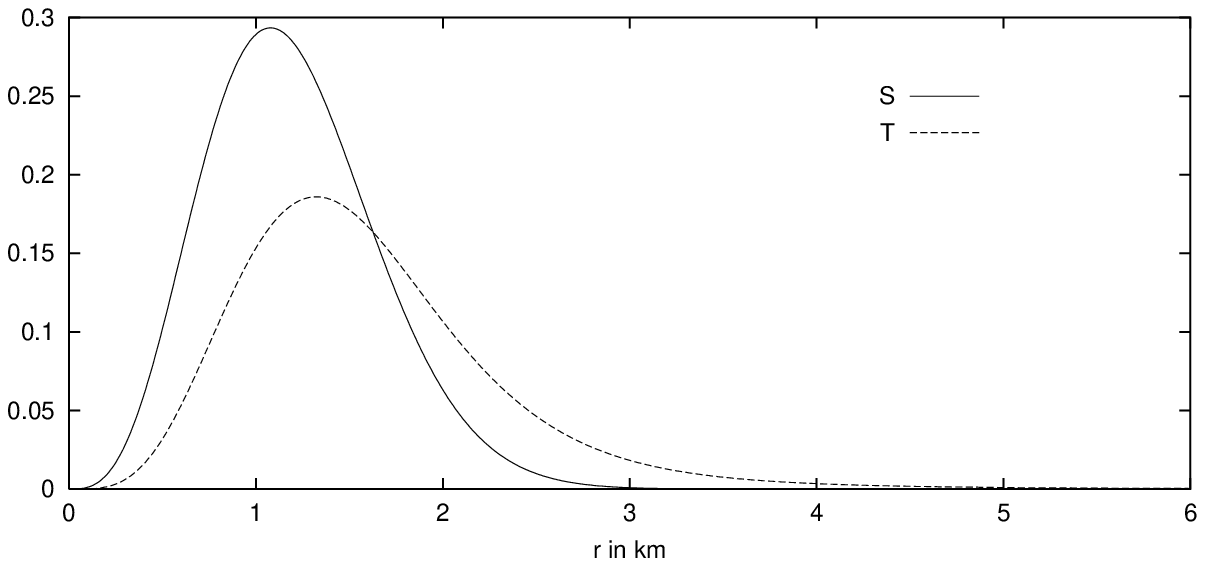}
\epsfxsize=\textwidth
\epsfbox{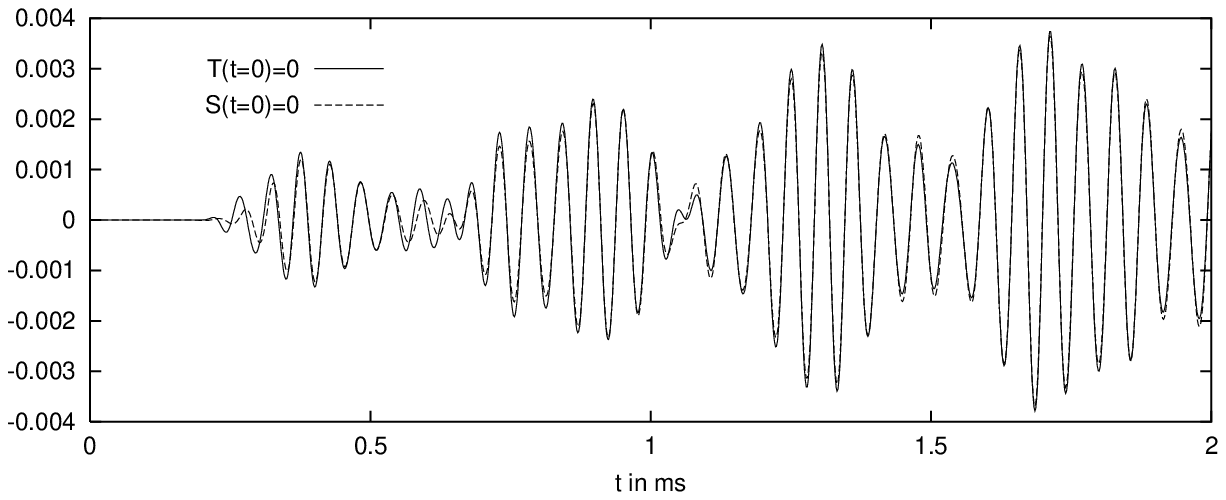}
\caption{\label{b=0}Evolution of the initial data which are located
inside the potential. The upper panel shows the initial fluid
perturbation. In the middle we show the two possible solutions of the
Hamiltonian constraints for the metric variable $S$ with $T=0$ and for
$T$ with $S=0$. The lower panel shows the evolution of both kinds of
initial data.}
\end{figure}

\begin{figure}[p]
\leavevmode
\epsfxsize=\textwidth
\epsfbox{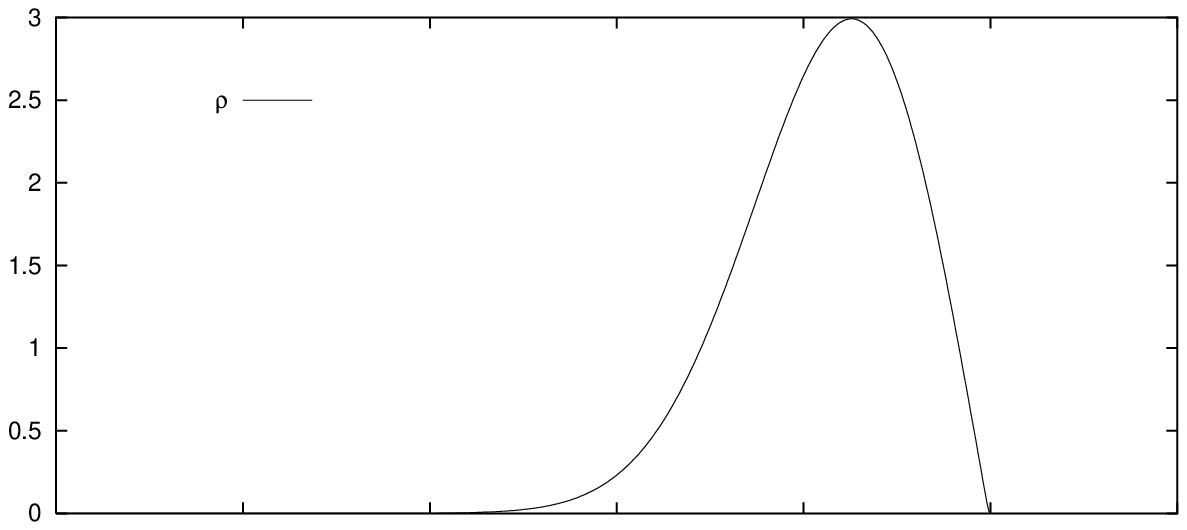}
\epsfxsize=\textwidth
\epsfbox{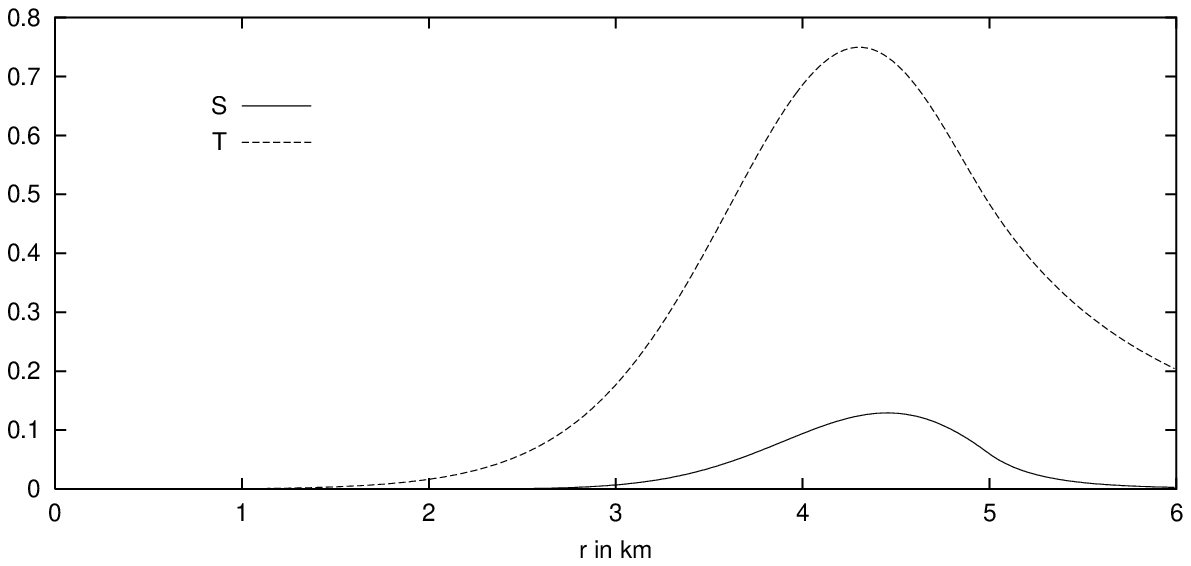}
\epsfxsize=\textwidth
\epsfbox{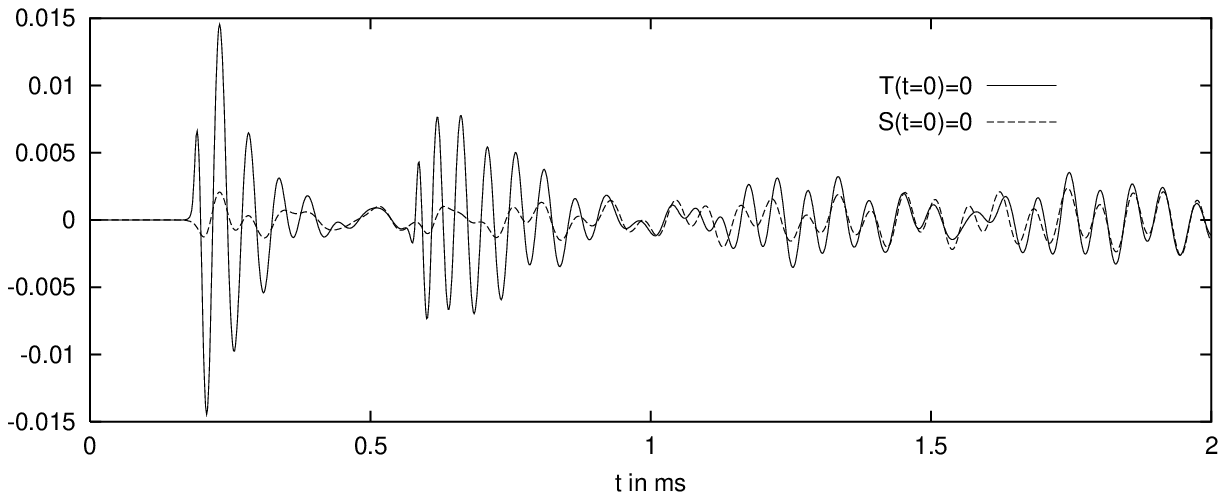}
\caption{\label{b=5}Evolution of the initial data which are located
outside the dip of the potential. As in Fig.\ref{b=0} the upper panel
shows the initial fluid perturbation. In the middle we show the two
possible solutions of the Hamiltonian constraints for the metric
variable $S$ with $T=0$ and for $T$ with $S=0$. The lower panel shows
the evolution of both kinds of initial data.}
\end{figure}

Furthermore, we can choose $\rho$ to have its major contribution close
to the stellar surface, that is at the peak of the potential, or
closer to the stellar core, that is right inside the dip of the
potential. In Figs.~\ref{b=0} and \ref{b=5} we show the initial data
and the evolution for both cases. 

It is interesting to see that in the first case the wave forms for
the two kinds of initial data, where we either solve for $S$ or for $T$,
are almost identical. Furthermore, it is evident that there is no
sign at all of a black-hole-like ring-down phase. The spectral analysis
of the resulting wave forms shows that there are predominately fluid
modes and only some of the first long-lived trapped modes present.

In the second case, things have changed dramatically. Here the wave
forms for the two kinds of initial data are totally different. Those
where we initially set $T=0$ and solve for $S$ lead to a wave form that
is somewhat similar to the case of an impinging gravitational
wave. The frequency of the ring-down after the first pulse is, indeed,
close to the first quasi-normal mode of an equal mass black hole. It is
only after the second pulse that we find the characteristic
quasi-normal modes of the neutron star. The spectral analysis reveals
that both the trapped and fluid modes are excited.

However, if we now initially set $S=0$ and solve for $T$, the
resulting wave form is quite different. There is no trace of any black
hole frequency whatsoever, and in the spectrum we scarcely find any of
the higher trapped modes. All in all the wave form is quite similar to
the case where the fluid perturbation is located near the center of
the star.

\begin{figure}[t]
\leavevmode
\epsfxsize=\textwidth
\epsfbox{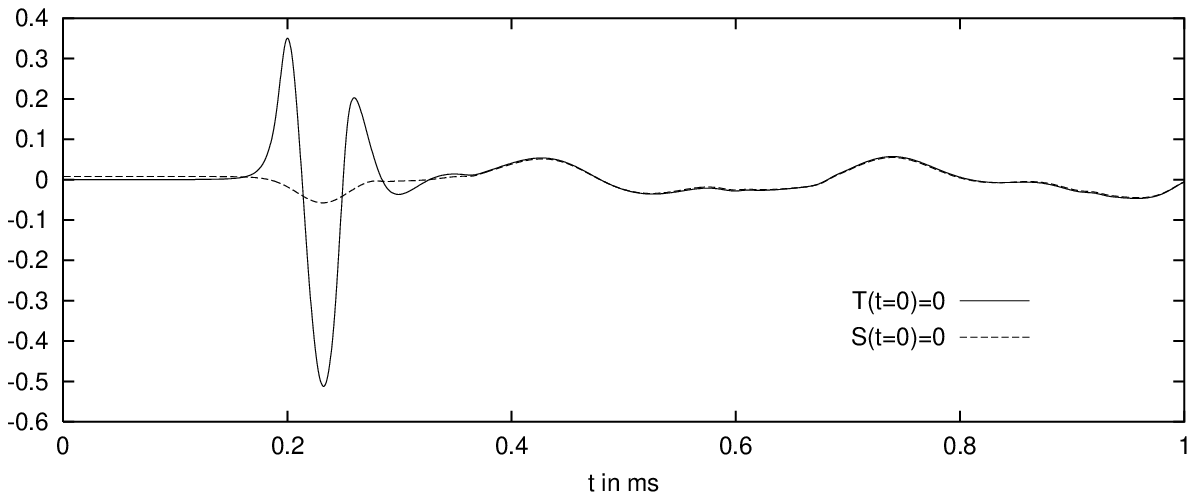}
\caption{\label{ST_3e15}Evolution of two sets of initial data 
with the same initial fluid perturbation, but in one case it is
$S(t=0) = 0$, and in the other it is $T(t=0) = 0$.}
\end{figure}

It should be mentioned that a long term evolution shows that both wave
forms start to look alike. This can be explained in such a way that
for both kinds of initial data the fluid modes are excited to the same
extent since we prescribe the same initial fluid perturbation $\rho$,
but in the case where $S$ differs from zero and $T=0$, we have an
additional excitation of the $w$-modes, which because of their strong
damping will quickly fade away. Eventually, both wave forms contain
only the long-lived modes and therefore will have the same time
dependence.

Thus, it seems that in the interior it is the variable $S$ which is
mainly responsible for the metric modes, and $T$ is responsible for
the fluid modes, even if $T$ is actually a metric perturbation.
However, it has already become clear that in the interior $T$ more or
less assumes the role of the fluid variable.

We demonstrate this effect more clearly for a less relativistic
stellar model, which does not have any trapped modes. Hence, here any
excited $w$-modes will immediately damp away and give way to the
long-lived fluid modes. In Fig.~\ref{ST_3e15} we show the evolution for
a given initial fluid perturbation $\rho$, where we either solve the
Hamiltonian constraint for $T$ or for $S$. In the former case the
resulting wave form is composed of fluid modes exclusively, whereas in
the latter case we see a short burst of radiation, which then fades
away very quickly. The remaining fluid oscillations are then almost
identical in both cases.

It thus seems that the prescription of $\rho$ determines more or
less exclusively how much energy we initially put into the fluid
modes. In addition, the value of $S$ determines how much energy is put
into the $w$-modes. The remaining variable $T$ is then fixed by the
Hamiltonian constraint. If we set $S=0$, we do not obtain any $w$-modes
at all. If we increase $S$, more and more energy is released through
$w$-modes without significantly changing the energy of the fluid
modes because of the weak coupling between gravity and matter.

Of course, we cannot choose arbitrarily high values for $S$ without
affecting the energy of the fluid modes. If we increase $S$ by too
large an amount, the coupling will transfer the energy of the metric
into the fluid. Hence, in this case, the excitation of the fluid modes
after the $w$-modes have radiated away will be much stronger than for
small values of $S$.

It thus seems that the metric variable $S$ has a quite important
influence on the form of the gravitational waves that are emitted by
neutron stars. If we assume the oscillations of a neutron star to be
excited by either some star quakes or by the precursing core collapse
in the stage of the birth of the neutron star, and not by an impinging
gravitational wave, then the emitted wave form will strongly depend on
whether or not this particular physical process was able to produce
some non-vanishing $S$. From the decomposition of the spatial metric
perturbations \eqref{p_metric} we see that $T$ is similar to a
conformal factor, for if we set $S = 0$ the perturbed metric
\eqref{p_metric} is conformally flat. The value of $S$ is therefore a
measure of the deviation from the conformal flatness. Hence, if the
physical process somehow conserves the conformal flatness of the
metric to some degree, it is clear that one cannot expect to have a
significant excitation of $w$-modes.

Andersson \& Kokkotas \cite{AK98b} have shown that by extracting the
frequencies and damping times of the first $w$-mode and the $f$-mode
in a wave signal one can reveal important stellar parameters such as
mass and radius, which then can be used to restrict the set of
possible equations of state. But this procedure stands and falls with
the presence of $w$-modes in the signal. It is therefore important to
investigate the processes that lead to oscillations of neutron stars
in greater detail on their ability to excite $w$-modes in a
significant manner.

It is also clear that any construction of numerical initial data that
rely on conformal flatness as in \cite{AAKLR99,AKLPS99} will
suppress the presence of $w$-modes.

\section{Using realistic equations of state}

When we try to switch to a realistic equation of state, we will run
into the same numerical troubles as in the radial case in chapter 3,
i.e. we will have instabilities that are associated with the dip in
the sound speed at the neutron drip point in conjunction with too low
a resolution.

This is because the structure of the fluid equation \eqref{waveH} is
of the same kind as in the radial case. Here we have found a convenient way
to get rid of those problems by introducing a new ``hydrodynamical
tortoise coordinate'' which stretches those parts of the neutron star
where the sound speed assumes low values.

Of course, this transformation is only defined in the stellar interior
for the fluid equations, for it is only there that the propagation
speed is the speed of sound (apart from the factor $e^{\nu-\lam}$).
If we still wanted to use the system of equations \eqref{wave-eqns},
where in the interior $T$ plays the role of the fluid, we would have
to switch from the $x$-grid in the interior to the $r$-grid in the
exterior, which is somewhat inconvenient. It is therefore much more
natural to explicitly include the fluid equation \eqref{waveH}, which
is defined in the interior only. Thence, it is only \eqref{waveH}
which will be transformed according to \eqref{transf}, whereas we keep
the wave equations for $S$ and $T$ as they are given in \eqref{waveS}
and
\eqref{waveT}.

However, this means that in the interior we have to simultaneously
evolve the fluid variable $H$ on a different grid than the metric
variables $S$ and $T$. Because of the coupling at each time step we
have to interpolate $H$ from the $x$-grid onto the $r$-grid in order
to update equation \eqref{waveT}, and, vice versa, both $S$ and $T$
from the $r$-grid onto the $x$-grid in order to update equation
\eqref{waveH}, which can easily be done using spline interpolation.

\begin{figure}[t]
\leavevmode
\epsfxsize=\textwidth
\epsfbox{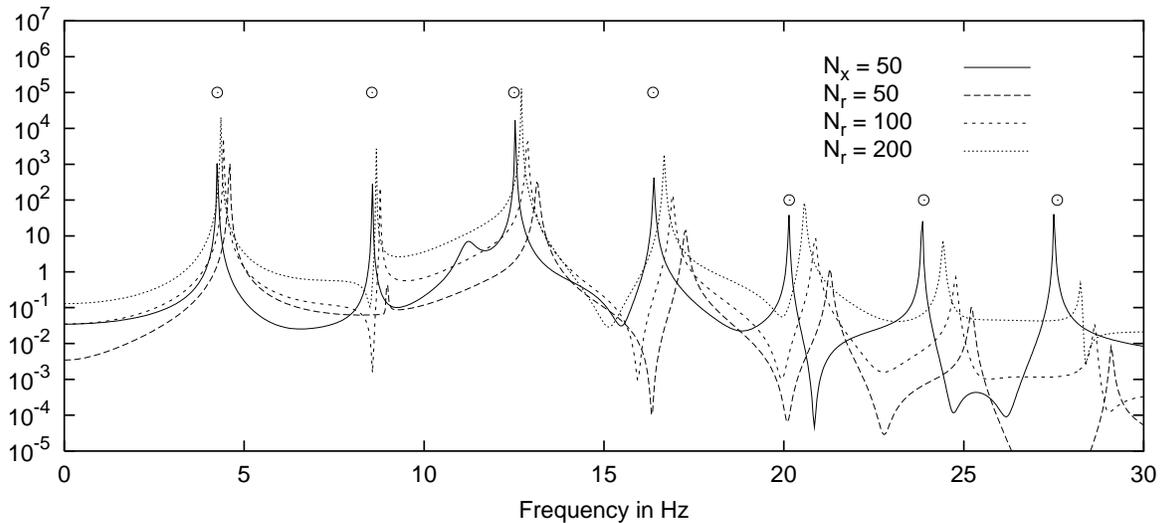}
\caption{\label{power_x_r}Comparison of the power spectra for a resolution
of $N_x = 50$ grid points on the $x$-grid with different resolutions on
the $r$-grid.}
\end{figure}

The transformed fluid part of the fluid equation \eqref{waveH} reads
\begin{align}
\begin{split}
\label{Hwave}
        \dff{H}{t} \= e^{2\nu - 2\lam}\bigg[\,\dff{H}{x}
         + \(\(2\nu_{,x} + \lam_{,x}\) - \frac{\nu_{,x}}{C_s^2}
        - \frac{C_{s,x}}{C_s}\)\df{H}{x}\\
        &{}\mbox{\hspace*{2cm}} + \(C_s\(\frac{\nu_{,x}}{r}
        + 4\frac{\lam_{,x}}{r} - C_se^{2\lam}\frac{l(l+1)}{r^2}\)
        + \frac{\lam_{,x}}{rC_s} + 2\frac{\nu_{,x}}{rC_s}\)H\,\bigg]\;.
\end{split}
\end{align}
The subscript $x$ denotes a derivative with respect to $x$. As in the
radial case, the boundary condition for $H$ at the surface of the star
simply translates into
\begin{align}
        \df{H}{x}|_{x(R)} \= 0\;.
\end{align}
Apparently, it is the region close to the surface, which is mainly
responsible for the fluid modes. To obtain the right mode frequencies
in the power spectrum, it is important to have high resolution close to
the surface of the star and this is what can be accomplished by our
new coordinate $x$. In Fig.~\ref{power_x_r} we show the power
spectra of wave forms obtained from runs with different
resolutions. The initial data always were some fluid excitations at the
center of the star together with $S=0$. In this case we suppress any
$w$-mode contribution.

It is apparent that even for the quite moderate resolution of 50 grid
points on the $x$-grid inside the star we obtain very accurate
frequencies for the first couple of fluid modes. On the $r$-grid,
however, a resolution of 200 grid points is still not enough to obtain
the same accuracy, and the peaks of the higher $p$-modes in the
spectrum are still quite far off their true values. Of course, those
results were obtained with a polytropic equation of state, since
otherwise we would not have been able to do the evolution on the
$r$-grid because of the occurring instability.

After all those pleasant features of our new $x$-coordinate, we should
also mention one drawback that comes along with the transformation of
the fluid equation, and which affects the use of polytropic equations
of state with $\Gamma \ge 2$. In this case, we already know from the
discussion of the boundary condition that the perturbed Eulerian
energy density at the stellar surface can either be finite ($\Gamma =
2$) or even infinite ($\Gamma > 2$). This then goes hand in hand with
a discontinuity or an infinite jump in $T''$.

\begin{figure}[p]
\leavevmode
\epsfxsize=\textwidth
\epsfbox{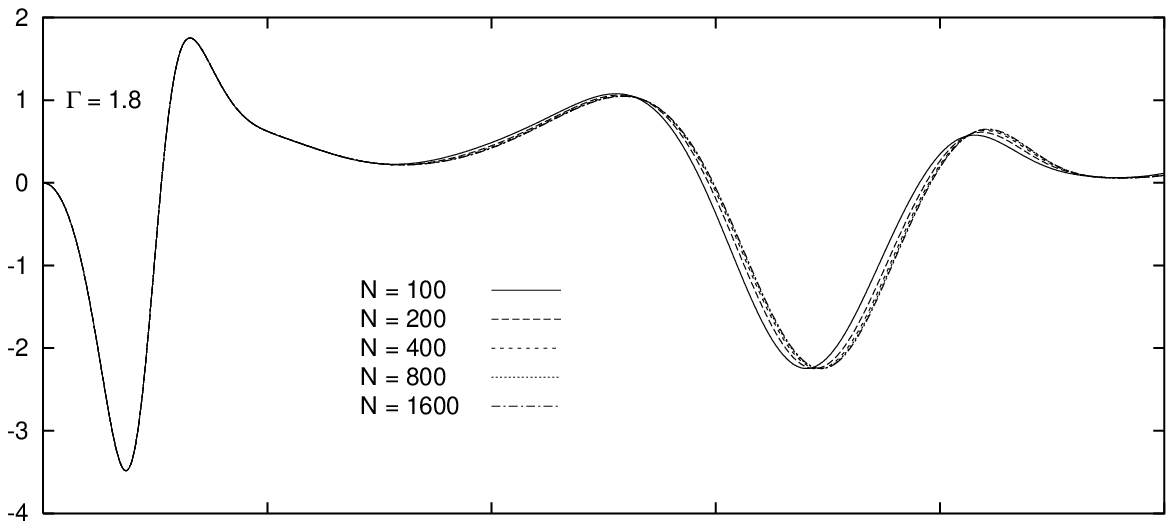}
\epsfxsize=\textwidth
\epsfbox{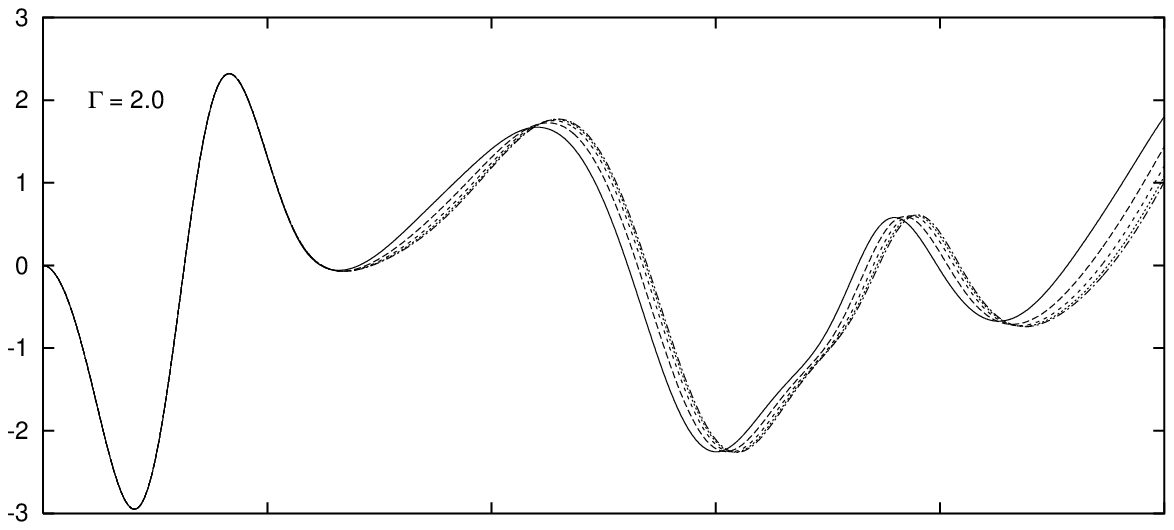}
\epsfxsize=\textwidth
\epsfbox{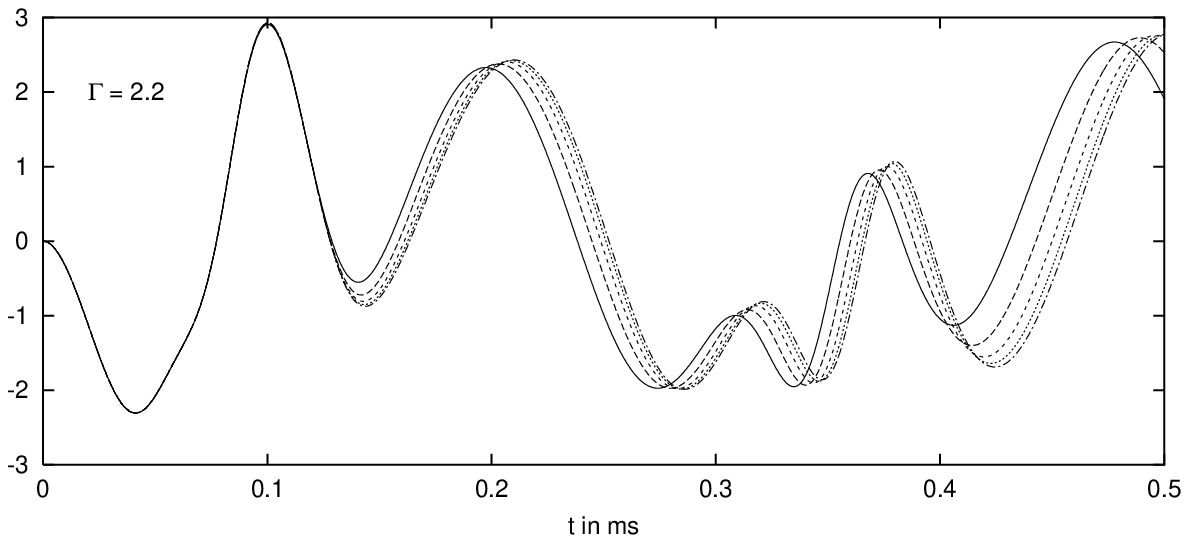}
\caption{\label{r_conv}Evolutions on the $r$-grid for $\Gamma = 1.8$, 2.0
and 2.2, each with different resolutions. For $\Gamma = 1.8$ the
convergence is much faster than for $\Gamma = 2.0$ and 2.2.}
\end{figure}

\begin{figure}[p]
\leavevmode
\epsfxsize=\textwidth
\epsfbox{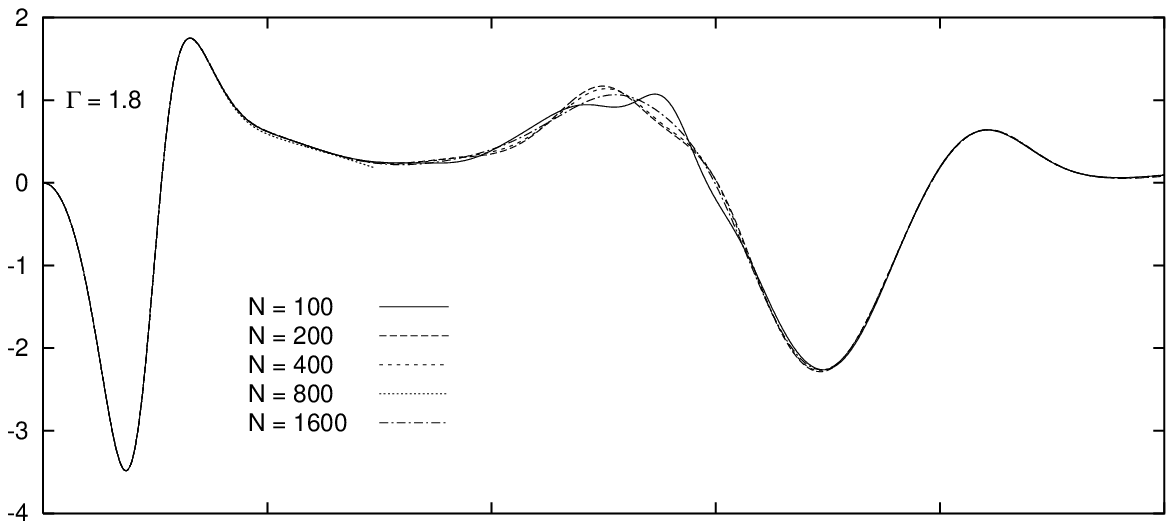}
\epsfxsize=\textwidth
\epsfbox{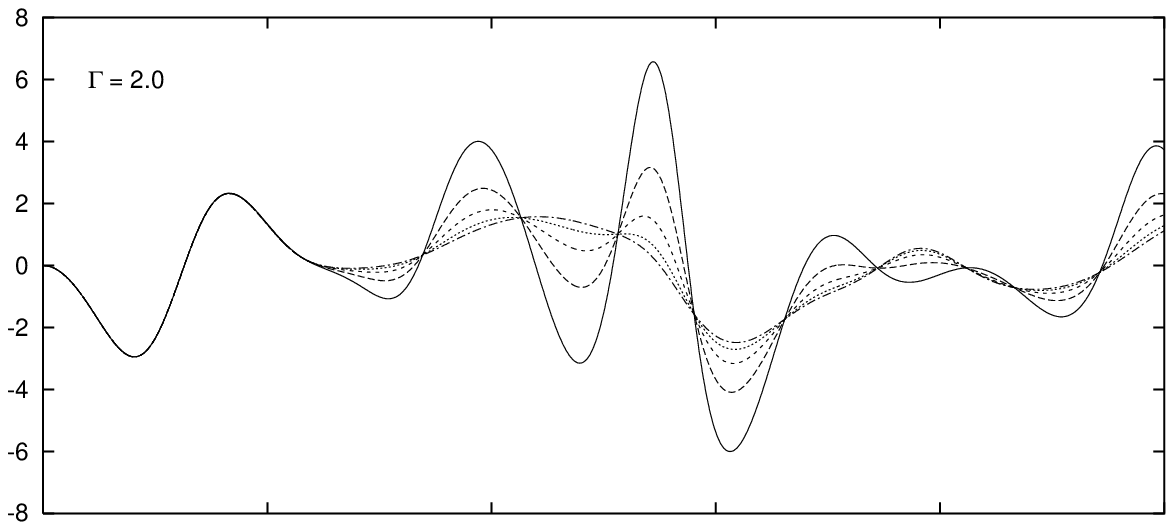}
\epsfxsize=\textwidth
\epsfbox{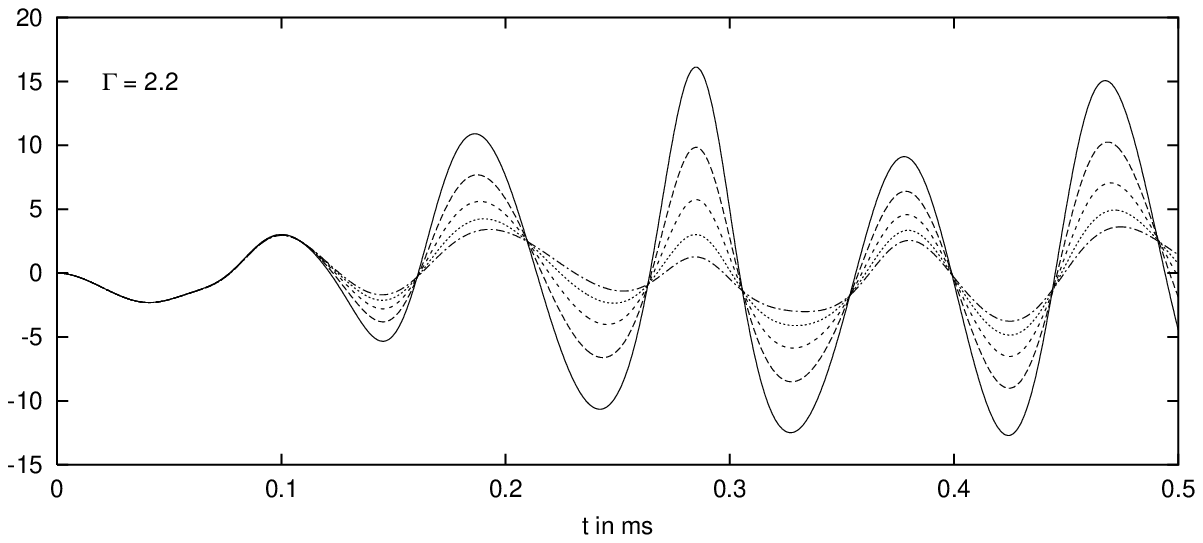}
\caption{\label{x_conv}Same evolutions as in Fig.~\ref{r_conv}, but 
this time on the $x$-grid for $\Gamma = 1.8$, 2.0
and 2.2, each with different resolutions. Again, for $\Gamma = 1.8$ the
convergence is the best, whereas for $\Gamma = 2.2$ it is very bad.}
\end{figure}

\begin{figure}[p]
\leavevmode
\epsfxsize=\textwidth
\epsfbox{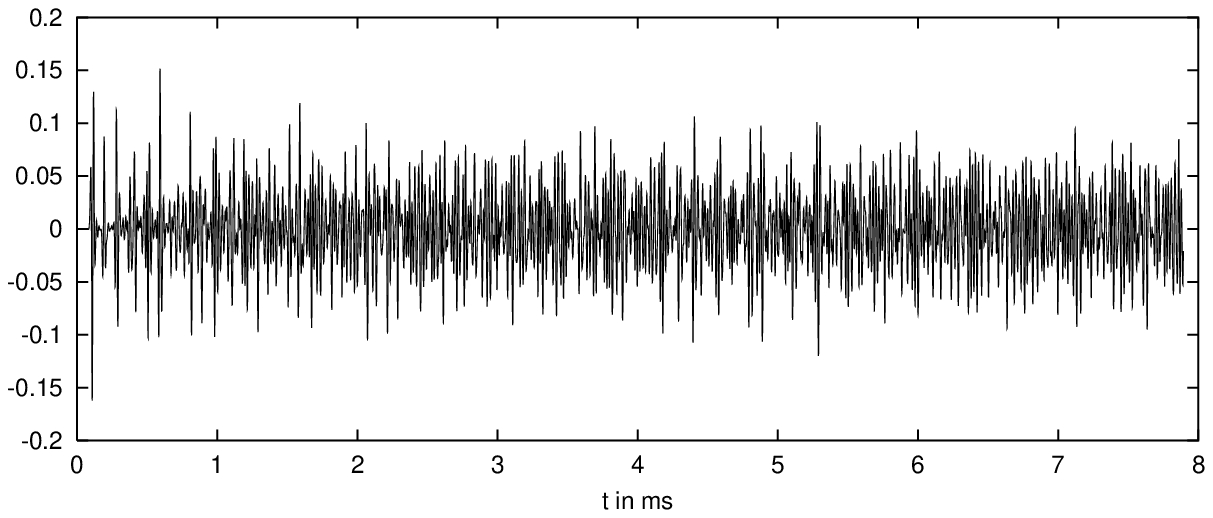}
\epsfxsize=\textwidth
\epsfbox{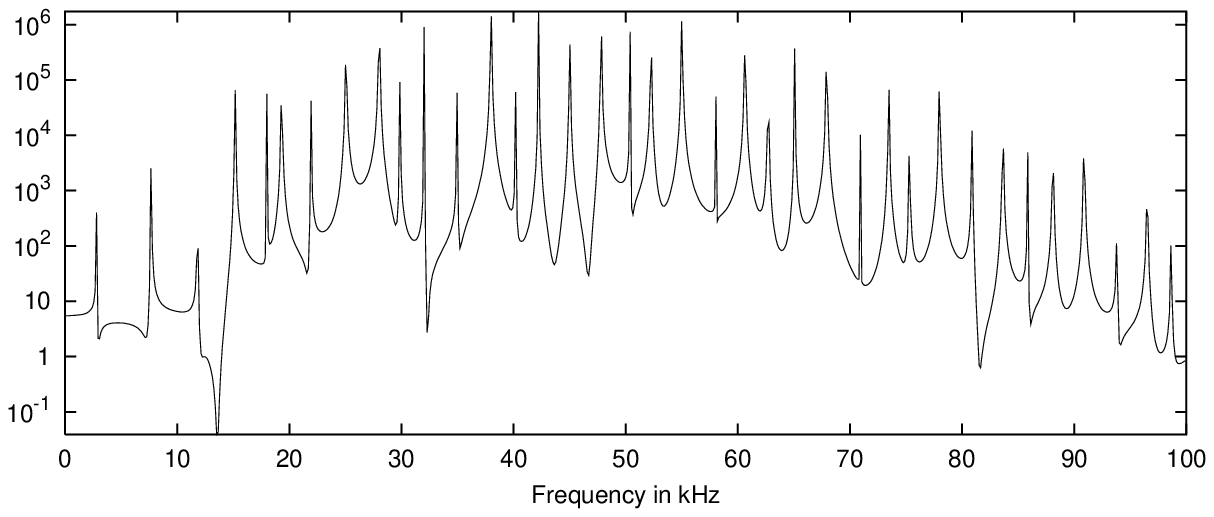}
\epsfxsize=\textwidth
\epsfbox{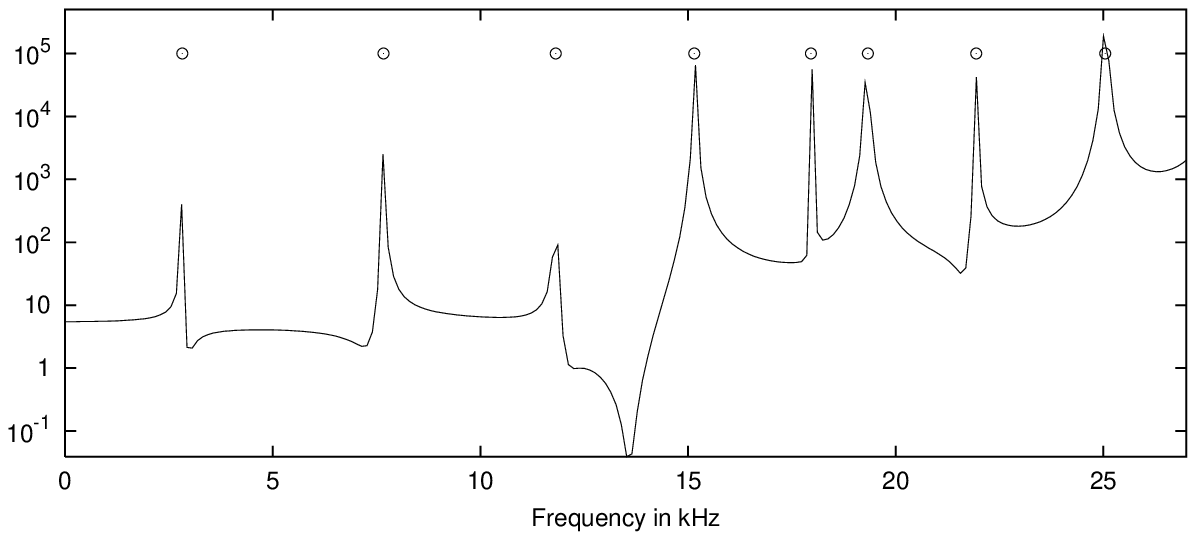}
\caption{\label{MPA}Upper panel: Evolution of a sharp initial fluid
pulse using the MPA equation of state. The power spectrum in the
middle panel reveals that dozens of $p$-modes are excited. In the lower
panel we include the modes computed with explicit mode calculation.}
\end{figure}

In the discretization of the wave equations \eqref{wave-eqns} we more
or less ignored this fact and computed $T''$ everywhere, the stellar
surface included, by means of central differences with second order
accuracy. Of course, for $\Gamma \ge 2$ the neglect of the appropriate
treatment of the discontinuity was punished by the reduction of the
order of convergence from second down to first. But otherwise we did
not have any serious disadvantages.

Unfortunately, with the new coordinate things get much worse. Because
of the better resolution of the fluid at the surface, the
discontinuity in $T''$ will be much more pronounced. In our case this
then leads to some artificial reflections at the surface, which in the
case of $\Gamma > 2$ are truly severe. In Figs.~\ref{r_conv} and
\ref{x_conv} we show the wave forms of $S$ right at the stellar
surface for different polytropes with $\Gamma = $1.8, 2.0 and 2.2.
Figure \ref{r_conv} shows the evolutions on the $r$-grid and
Fig.~\ref{x_conv} on the $x$-grid.

Whereas for the conventional discretization in $r$ the main inaccuracy
of low resolutions is the phasing, we can see that using the
$x$-coordinate we get some ugly additional bumps in the wave function,
which start to vanish as the resolution is increased. For $\Gamma =
1.8$, $T''$ is continuous at the surface, thus there is only a small
reflection for the lowest resolution of 100 grid points inside the
star. For resolutions of 400 or more, the wave form seems to have
more or less converged to its proper shape. For $\Gamma = 2$, the bumps
for low resolutions are much higher and it is only for the quite large
number of 1600 grid points inside the star that the wave form
coincides with the one from the $r$-grid. But for $\Gamma = 2.2$ things
seem to be quite hopeless. Even for 1600 grid points the shape of the
wave form still has almost nothing to do with the corresponding one of
the $r$-grid. One would have to double the resolution some more
times before one could obtain an acceptable wave form.

This seems to be quite a nasty feature of the new set of equations and
it might be worth trying to implement an appropriate handling of the
discontinuity. However, for our purposes this is not necessary. The main
motivation for transforming the equations was to get rid of the
instability that occurs when we switch to realistic equations of
state. And this can be accomplished with the new version of the
equations. Moreover, for realistic equations of state, the
perturbation of the energy density $\rho$ is almost zero at the
surface, and thus, there is only a negligible discontinuity in
$T''$. Thus, we will not encounter any of those interfering
reflections, and it is possible to obtain reasonable wave forms without
having to rely on unacceptably high resolutions as it would be the
case with the old discretization in order to overcome the instability.

In Fig.~\ref{MPA} we demonstrate the usefulness of our coordinate
transformation. We show the evolution of a sharp peak in the fluid
perturbation $\rho$, which leads to the excitation of a multitude of
fluid modes. This is convincingly confirmed by the power spectrum
which shows 37 modes in the interval up to $100\,$kHz. And they agree
with the modes which are obtained by the direct mode computation.


\chapter{Shooting particles at the neutron star}

In the previous chapter we demonstrated that it is possible to excite
the oscillations of a neutron star and that the resulting
gravitational wave signal contains the expected quasi-normal
modes. However, all the chosen initial data were purely arbitrary and
void of any astrophysical meaning. This arbitrariness will now be
removed by simulating a physical process that could excite the modes
of a neutron star. We will use the gravitational field of a moving
mass $\mu$ to perturb the neutron star, hoping that this will lead to
excitations of some of its eigenmodes. Since this investigation takes
place in the framework of perturbation theory, we have to require that
the mass $\mu$ is much smaller than the mass $M$ of the neutron star,
or $\mu/M \ll 1$. Furthermore, we will treat the orbiting mass $\mu$
as a point mass. In this particle limit, the gravitational field of
the mass $\mu$ can be considered as a perturbation of the spherical
background field that is due to the neutron star. The particle will
move on a geodesic in the background metric since deviations thereof
due to gravitational radiation are effects of second order and will
therefore be neglected.

We will not consider collisions of particles with the neutron star
\cite{Bor97} because it is not clear at all how to treat the impact
and subsequent merge of the particle with the neutron star. Therefore
we will only focus on circular and scattering orbits. Previous studies
of excitations of neutron stars by orbiting particles were performed
in the frequency domain \cite{FGB99,TS99,AP99}. Since we already have
the evolution code, we will include the particle and do the explicit
time integration. As we shall show, considering the dynamics in the
time domain has as a consequence that we are forced to ``smooth out''
the particle; that is, the $\delta$-functions in the sources of the
equations due to the presence of the particle are approximated by
Gaussians. We show, however, that our model of the particle is
self-consistent and convergent. Another important issue is the
prescription of appropriate initial data that satisfy the constraints.
We will face the same problems as in the previous chapter, since the
constraints do not provide us with a unique way to find valid initial
data. However, by resorting to the flat space case we can find analytic
initial data which can serve as a good approximation in the case that the
particle is initially far away from the neutron star.

We now proceed how to mathematically describe the particle and how
to incorporate it into the perturbation equations.
\newpage
\section{Adding a particle}
The energy-momentum tensor for a point particle with mass $\mu$ moving
on a geodesic $X^\lam(\tau)$ in a Schwarzschild metric is given by
\begin{align}\label{part_tmn}
        {\cal T}^{\mu\nu} \= \mu\int\delta^{(4)}(x^\lam-X^\lam(\tau))U^\mu
        U^\nu{\rm d}\tau\non\\
        \= \mu\frac{U^\mu U^\nu}{U^t r^2}
        \delta(r-R(t))\,\delta(\cos\theta\ - \cos\Theta(t))\,
        \delta(\phi-\Phi(t))\;,
\end{align}
where $U^\mu$ is the particle's 4-velocity
\begin{align}
        U^\mu \= \frac{dX^\mu}{d\tau}\;,
\end{align}
and $\tau$ the particle's proper time along its trajectory. We now
orient the coordinate system in such a way that the particle's orbit
coincides with the equatorial plain of the neutron star ($\Theta =
\pi/2$). Besides, the particle's coordinate time $T$ is identical with
the time coordinate $t$ of the spacetime in which it moves. Therefore
we will use $t$ to parametrize the path of the particle $X^\lam(t) =
[t,\, R(t),\,\pi/2,\,\Phi(t)]$.

From the geodesic equations $\frac{DU^\mu}{d\tau} \equiv U^\nu D_\nu
U^\mu = 0$ we find:
\begin{subequations}\label{geo_eqs}
\begin{align}
        \label{geo1}
        \frac{dt}{d\tau} \= e^{2\lam}E\\
        \label{geo2}
        \(\frac{dR}{d\tau}\)^2 \= E^2 - e^{2\nu}\(1 + \frac{L^2}{R^2}\)\\
        \label{geo3}
        \frac{d\Phi}{d\tau} \= \frac{L}{R^2}\;,
\end{align}
\end{subequations}
where $E$ and $L$ are the energy and angular momentum per unit mass of
the particle, respectively. We also recall that for the Schwarzschild
metric we have
\begin{align}
        e^{2\nu} \= e^{-2\lam} \;=\; 1 - \frac{2M}{r}\;.
\end{align}
We can use \eqref{geo1} to eliminate the proper time $\tau$ from
equations \eqref{geo2} and \eqref{geo3}:
\begin{subequations}
\begin{align}
        \label{geo2t}
        \(\frac{dR}{dt}\)^2 \= e^{4\nu}\(1
        - \frac{e^{2\nu}}{E^2}\(1 + \frac{L^2}{R^2}\)\)\\
        \label{geo3t}
        \frac{d\Phi}{dt} \= e^{2\nu}\frac{L}{R^2E}\;.
\end{align}
\end{subequations}
On the one hand those equations can be used to replace the quantities
$v_R \equiv \frac{dR}{dt}$ and $v_\Phi \equiv \frac{d\Phi}{dt}$ in the
source terms of the particle. But on the other hand we also have to
explicitly solve them for the particle's trajectory coordinates $R(t)$
and $\Phi(t)$, since we need those coordinates in the
$\delta$-functions $\delta(r-R(t))$ and $\delta(\phi-\Phi(t))$.

The perturbation equations for the isolated neutron star are given by
\eqref{hij} and \eqref{kij}. In the vacuum region, the matter terms in
\eqref{kij} vanish, of course. However, when we add the particle, we
have to include its energy-momentum tensor in \eqref{kij}, which then
reads
\begin{align}\label{kij_part}
\begin{split}
        \d_t k_{ij} \= - \d_i \d_j \alpha +
        \Gamma ^k_{\phantom{i}ij}\d_k\alpha + \delta \Gamma
        ^k_{\phantom{i}ij}\d _k e^\nu\\
        &{}\quad + \alpha R_{ij}
        + e^\nu\delta R_{ij}
        - 8\pi e^\nu\({\cal T}_{ij} + \half\,g^{\mu\nu}
        {\cal T}_{\mu\nu}\)\;.
\end{split}
\end{align}
To obtain the initial data, we also have to modify the perturbed
constraint equations \eqref{hhc} and \eqref{mmc}, which then read
\begin{align}
        g^{ij}\delta R_{ij} - h^{ij} R_{ij} \= 16 \pi
        e^{-2\nu}{\cal T}_{00}\label{hcc}\\
        g^{jk}\(\d _i k_{jk} - \d _j k_{ik}
        - \Gamma^l_{\phantom{i}ik} k_{jl} + \Gamma^l_{\phantom{i}jk} k_{il}\)
        \= 8\pi e^{-\nu}{\cal T}_{0i}\;.\label{mcc}
\end{align}
In writing these constraint equations, we have assumed that initially 
the fluid of the neutron star is unperturbed.

To simplify the above set of equations, we again want to get rid of
the angular dependence by expanding the equations in Regge-Wheeler
harmonics $\lbrace[\rw^A_{lm}]_{\mu\nu}(\theta,\phi)\rbrace_{A=1,\dots,10}$. We
then obtain equations for the expansion coefficients, which only depend
on $r$ and $t$. For the metric part of the equations the decomposition
can be done in exactly the same way as in chapter 2, and the
energy-momentum tensor of the particle can be written as
\begin{align}\label{exp_tmn}
        {\cal T}_{\mu \nu}(t,r,\theta,\phi) \= \sum_{l = 0}^{\infty}
        \sum_{m = -l}^l \sum_{A = 1}^{10}
        t_A^{lm}(t,r)\; [\rw^A_{lm}]_{\mu\nu}(\theta,\phi)\;.
\end{align}
Unfortunately, the Regge-Wheeler harmonics $\rw^{lm}_A$ do not form an
orthonormal set, and it is thus not possible to directly obtain the 
coefficients $t_A^{lm}(t,r)$ by means of the orthogonality relation.
We rather have to take some detour. We have to construct an orthonormal
set of tensor harmonics $\mz^A_{lm}$, which can be expressed as some linear
combination of the Regge-Wheeler harmonics $\rw^A_{lm}$:
\begin{align}
        \mz^A_{lm} \= \sum_{B = 1}^{10}C_{AB}
        \rw^B_{lm}\:.
\end{align}
We then compute the coefficients ${\hat t}_A^{lm}(t,r)$
corresponding to the orthonormal set $\mz^A_{lm}$ by using the
orthogonality relation
\begin{align}
        {\hat t}_A^{lm} \= \int_{S^2}(\mz^A_{lm})^{*\mu\nu}
        {\cal T}_{\mu\nu}\,{\rm d\Omega}\;.
\end{align}
Finally, the desired coefficients $t_A^{lm}(t,r)$ can be obtained
by
\begin{align}
        t_A^{lm} \= \sum_{B = 1}^{10}C_{BA}{\hat t}_{B}^{lm}\;,
\end{align}
which then will be plugged in our equations.

One convenient orthonormal set $\mz^A_{lm}$ can be found in Zerilli
\cite{Zer70c}. In Appendix A we will list the complete set and
explicitly show the relation to the Regge-Wheeler harmonics. In
Appendix B, we then demonstrate how to derive the coefficients
$t_A^{lm}(t,r)$.

If we want to include the particle terms in the equations for the
extrinsic curvature \eqref{kij_part}, we have to use some caution. In
chapter 2 we saw that the expansion of \eqref{kij_part} into tensor
harmonics lead to evolution equations for the six coefficients
$\widehat{K}^{lm}_i$. By choosing the Regge-Wheeler gauge together
with the appropriate initial data we could reduce the evolution
equation for $\widehat{K}^{lm}_4$ to the trivial case
$\df{}{t}\widehat{K}^{lm}_4 = 0$. This then ensured the vanishing of
$\widehat{K}^{lm}_4$ for all times.

However, in the presence of the particle this is not true anymore.
Instead of a zero on the righthand side, we have a source term
\begin{align}\label{K4_part}
        \df{}{t}\widehat{K}^{lm}_4 \= -8\pi t^{lm}_8\;.
\end{align}
But this means that during the evolution $\widehat{K}^{lm}_4$ will start
to differ from zero, which, because of \eqref{T1}, in turn would make
$\widehat{T}^{lm}_1$ non-vanishing. This is somewhat unfortunate but it
can be remedied by choosing a different lapse function $\alpha$. If we
pick
\begin{align}
        \alpha \= -\half e^{\nu}\(\frac{T^{lm}}{r} + rS^{lm}
        + 16\pi t_8^{lm}\)Y^{lm}\;,
\end{align}
the last term exactly cancels the righthand side of \eqref{K4_part}
and the vanishing of both $\widehat{T}_1$ and $\widehat{K}_4$ can be
guaranteed.

Again, we have to distinguish between the axial and the polar
perturbations. It is only the polar part which will be studied in this
chapter, nevertheless, for the sake of completeness and because the
inclusion of the particle terms in \eqref{axial} -- \eqref{oddMC}
is straightforward, we also present the axial equations for the particle:
\begin{subequations}
\label{axial_part}
\begin{align}
        \df{V_4}{t} &= e^{4\nu}\(\df{K_6}{r}
         + 2\(\nu' - \frac{1}{r}\)K_6\) - e^{2\nu}K_3\\
        \df{K_3}{t} &= \frac{l(l+1) - 2}{r^2}V_4
        - 16\pi e^{2\nu}t_7\\
        \df{K_6}{t} &= \df{V_4}{r} - 8\pi t_{10}\;.
\end{align}
\end{subequations}
The momentum constraint relates the extrinsic curvature coefficients
to the particle's source term via
\begin{align}\label{odd2}
        \df{K_3}{r} + \frac{2}{r}K_3
        - \frac{l(l+1) - 2}{r^2}K_6
        &= 16\pi e^{2\lambda} t_4\;.
\end{align}
The relevant source terms of the particle read:
\begin{subequations}
\begin{align}
        t_4 &= e^{2\nu}\frac{\mu L}{r^2 l(l+1)}\delta(r-R(t))
        \df{}{\Theta}Y^*_{lm}\\
        \label{t7}
        t_7 &= e^{2\lambda}\frac{\mu L}{r^2 l(l+1)}v(t)
        \delta(r-R(t))\df{}{\Theta}Y^*_{lm}\\
        \label{t10}
        t_{10} &= -e^{2\nu}\frac{2\mbox{i}m\mu L^2}
        {r^2 El(l+1)(l-1)(l+2)}\delta(r-R(t))
        \df{}{\Theta}Y^*_{lm}\;.
\end{align}
\end{subequations}
For radial infall, it is $L = 0$ and all source terms vanish. Hence, in
this case the radiation is of even parity only.

We can now use the system of equations of chapter 4, and in
\eqref{K} and \eqref{K5} we just have to add the following source
terms:
\begin{subequations}
\begin{align}
\begin{split}\label{K+source}
        \df{K}{t} \= \dots + 16\pi \frac{e^{4\nu}}{r}
        \bigg[t_8'' - 2 t_6' + \(5\nu' - \frac{3}{r}\)t_8' + t_5
        - 2\(3\nu' - \frac{1}{r}\)t_6\\
        &{}\hspace*{3cm}{}+ 2\((\nu')^2 - 4\frac{\nu'}{r}
        + \frac{3 - e^{2\lambda}}{r^2}\)t_8 - \frac{e^{2\lambda}}{r^2}t_9\,
        \bigg]
\end{split}\\
        \df{K_5}{t} \= \dots + 16\pi e^{4\nu}\bigg[t_8' - 2t_6
        + 2\(\nu' - \frac{1}{r}\)t_8 + \frac{e^{2\lambda}}{r}t_9
        - \frac{r}{2}e^{2\lambda}t\,\bigg]\;.\label{K5+source}
\end{align}
\end{subequations}
The appropriate source terms are derived in Appendix B and are given by
\begin{subequations}
\begin{align}
        t_5 \= e^{6\lambda}\frac{\mu E}{r^2}v^2(t)\delta(r-R(t))
        Y^*_{lm}\\
        t_6 \= -e^{2\lambda}\frac{\mbox{i} m\mu L}{r^2l(l+1)}v(t)
        \delta(r-R(t))Y^*_{lm}\\
        t_8 \= e^{2\nu}\frac{\mu L^2\(l(l+1) - 2m^2\)}{r^2El(l+1)(l-1)(l+2)}
        \delta(r-R(t))Y^*_{lm}\\
        t_9 \= e^{2\nu}\frac{\mu L^2\(l(l+1) - m^2 - 1\)}{r^2E(l-1)(l+2)}
        \delta(r-R(t))Y^*_{lm}\\
        t \= -e^{2\nu}\frac{\mu}{r^2E}\delta(r-R(t))Y^*_{lm}\;.
\end{align}
\end{subequations}
With the explicit form of the source terms, \eqref{K+source} and
\eqref{K5+source} read:
\begin{align}
\begin{split}
        \df{K}{t} \= \dots + 8\pi e^{2\nu}\frac{\mu}{r^3}\bigg[
        2e^{4\lam}Ev^2\delta + \mbox{i}m \frac{vL}{r(n+1)}\bigg(2r\delta'
        + (e^{2\lam} - 7)\delta\bigg)\\
        & + \frac{e^{2\nu}L^2}{r^2En(n+1)}\bigg(
        \(m^2(n + 27) - (n+1)(2n + 27)\)\delta\\
        &{} + (n + 1 - m^2)\(e^{2\nu}r^2\delta''
        + \frac{r}{2}(9 - 23e^{2\nu})\delta'
        + 3\(e^{2\lam} + 13e^{2\nu}\)\delta\)\bigg)\bigg]Y^*_{lm}
\end{split}
\end{align}
\begin{align}
\begin{split}
        \df{K_5}{t} \= \dots + 8\pi e^{4\nu}\frac{\mu}{r}\bigg[
        \(\frac{1}{E} + 2\mbox{i}m e^{2\mu}\frac{vL}{r(n+1)}\)\delta\\
        &{}\qquad\qquad\qquad\qquad + \frac{L^2}{r^2En(n+1)}
        \bigg(e^{2\nu}(n+1 - m^2)\(r\delta' - 6\delta\)\\
        &\qquad\qquad\qquad\qquad\qquad
        + \((n+1)(2n+3) - m^2(n+3)\)\delta\bigg)\bigg]Y^*_{lm}\;,
\end{split}
\end{align}
where we have defined
\begin{align}
        n \;\equiv\; \half l(l+1) - 1\:.
\end{align}
For the Zerilli equation \eqref{Zeqn} we obtain the following
source term:
\begin{align}
\begin{split}
        \dff{Z}{t} \= \dots - 16\pi e^{4\nu}\frac{\mu}{r^2E\Lambda(n+1)}
        \bigg[e^{2\nu}(L^2 + r^2)\delta'\\
        & + \bigg( \frac{L^2}{2rn}\(m^2(2n+3) - 4n(n+1) - 3
        + e^{2\nu}(3 - 2n - 3m^2)\)\\
        &\qquad - 2\mbox{i}m e^{2\mu}vLE
        + 12\frac{ME^2}{\Lambda} - r(n+1) + 3M
        \bigg)\delta\bigg]Y^*_{lm}\;.
\end{split}
\end{align}
The Hamiltonian constraint \eqref{hcc} with the particle term reads
\begin{align}
\begin{split}\label{hc}
        &T'' + \nu'T' - rS'
        - \(\frac{\nu'}{r} + e^{2\lambda}\frac{l(l+1)}{r^2}\)T
        - \(2r\nu' + 2 + \half e^{2\lambda}l(l+1)\)S\\
        \= -8\pi re^{4\lambda}t_1\;,
\end{split}
\end{align}
and the momentum constraints \eqref{mcc} lead to the following
two equations
\begin{subequations}\label{mcpart}
\begin{align}\label{mc1}
        rK'_5 - \frac{1}{2}e^{2\lambda}l(l+1)K_2
        - r^2K - \(r\nu' + 1\)K_5
        \= -8\pi r^2t_2\\
        rK'_2 - r^2K - 2K_5\label{mc2}
        \= 16\pi rt_3\;.
\end{align}
\end{subequations}
Here, the appropriate particle terms are
\begin{subequations}
\begin{align}
        t_1 \= e^{2\nu}\frac{\mu E}{r^2}\delta(r-R(t))
        Y^*_{lm}\\
        t_2 \= -e^{2\lambda}\frac{\mu E}{r^2}
        v(t)\delta(r-R(t))Y^*_{lm}\\
        t_3 \= e^{2\nu}\frac{\mbox{i} m\mu L}{r^2l(l+1)}\delta(r-R(t))
        Y^*_{lm}\;.
\end{align}
\end{subequations}
In the momentum constraints \eqref{mcpart} the quantity $K_2$ still
appears, which, again, is not used in the evolution equations.
Therefore, we will get rid of it by differentiating \eqref{mc1} with
respect to $r$ and using \eqref{mc1} and \eqref{mc2} to eliminate
$K_2$ and $K_2'$. The resulting equation is then second order in $K_5$
and reads
\begin{align}\label{MC2nd}
\begin{split}
        \!\!\!\!\!K_5''& + \nu'K_5' - \(\frac{\nu'}{r}
        + e^{2\lambda}\frac{l(l+1)}{r^2}\)K_5
        - rK' - \(2r\nu' + 2 + \half e^{2\lambda}l(l+1)\)K\\
        \= -8\pi\(rt_2' + 2\(r\nu' + 1\)t_2 
        - e^{2\lambda}\frac{l(l+1)}{r}t_3\)\;.
\end{split}
\end{align}
It should be noted that this equation also follows from taking the
time derivative of the Hamiltonian constraint \eqref{hc} and taking
into account that $\dot{S} = K$ and $\dot{T} = K_5$, which is consistent
with the discussion in chapter 4.

We can actually verify that the following relations have to hold:
\begin{align}
        \dot{t}_1 \= e^{4\nu}\bigg[t_2' + 2\(\nu' + \frac{1}{r}\)t_2
        - e^{2\mu}\frac{l(l+1)}{r^2}t_3\bigg]\\
        \dot{t}_2 \= e^{4\nu}\bigg[t_5' - e^{4\mu}\nu't_1 + \(3\nu' + \frac{2}{r}\)t_5
        - \frac{e^{2\mu}}{r^3}\bigg(l(l+1)(rt_6 - t_8) + 2t_9\bigg)\bigg]\\
        \dot{t}_3 \= e^{4\nu}\bigg[t_6' + 2\(\nu' + \frac{1}{r}\)t_6 
        - \frac{e^{2\mu}}{r^2}\bigg((1 - l(l+1))t_8 + t_9\bigg)\bigg]
\end{align}

The constraint equations serve as initial value equations that have to
be solved on the initial slice $t=0$ in order to obtain valid initial
data. Unfortunately, there is no unique way to solve those
equations. This is due to the fact that to a particular solution of
the inhomogeneous equations, we can always add a solution of the
homogeneous equation, which would correspond to adding some arbitrary
gravitational waves. The problem of finding the ``right'' initial data
that represent only the perturbations which are due to the presence of
the particle and contain no additional radiation will be discussed in
more detail in section \ref{sec:init}.

For a particle falling from rest we would have $v(t = 0) = 0$ and $L =
0$, and therefore $t_2 = t_3 = 0$. Thus the momentum constraint
\eqref{MC2nd} can be trivially satisfied by setting the extrinsic
curvature variables to zero, which corresponds to time symmetric
initial data. We then are left with solving the Hamiltonian constraint
\eqref{hc}. Of course, a particle falling from rest would fall
radially towards the neutron star and eventually hit its surface.
Since we want to avoid such an impact, we have to give the particle
some angular momentum. We also will give it some radial velocity in
order to decrease the time it takes to come close to the neutron
star. However, this means that we have to solve the momentum
constraint \eqref{MC2nd}, too.

\section{Numerical implementation}
\label{sec:num}
In order to solve the equations on the computer, we have to use the
explicit forms of the spherical harmonics $Y_{lm}$. The perturbation
equations without particle are degenerate with respect to $m$, since
the background metric is spherically symmetric. However, the presence
of the particle breaks this symmetry and we have to consider the
various $m$-cases. Fortunately, we do not have to consider all
possible values of $m$ for a given value of $l$, since for negative
$m$ the spherical harmonics just undergo sign change and phase shift
($Y^*_{lm} = (-1)^m Y_{l,-m}$). The advantage of putting the particle
in the equatorial plain ($\Theta = \frac{\pi}{2}$) is that in the
polar case we only have to deal with multipoles with $m = l,\,
l-2,\,\dots$; the remaining ones with $m = l-1,\,l-3,\,\dots$ are
axial multipoles. Since the evolution code can only handle real valued
perturbations, we have to treat the real and imaginary parts of the
spherical harmonics separately. Finally, all the equations will be
solved on a finite grid, hence we have to approximate the
$\delta$-function by a narrow Gaussian
\begin{align*}
        \delta(r-R(t)) \;&\approx\; \frac{1}{\sigma\sqrt{2\pi}}
        e^{-\frac{(r - R(t))^2}{2\sigma^2}}\;,\qquad\mbox{$\sigma$ small}\;.
\end{align*}
We also need first and second derivatives, which can be expressed as
\begin{align*}
        \delta'(r-R(t)) \;&\approx\;
        \frac{R(t) - r}{\sigma^2}\delta(r-R(t))\\
        \delta''(r-R(t)) \;&\approx\; \frac{1}{\sigma^2}
        \(\frac{(R(t) - r)^2}{\sigma^2} - 1\)\delta(r-R(t))\;.
\end{align*}
For this approximation to be valid, we have to ensure the convergence
of the solution for $\sigma \rightarrow 0$. This can be done in two
different ways. On the one hand we can look at the convergence of the
waveforms that are obtained in the evolution, and on the other hand
we can monitor the violation of the constraints. A possible way do to
so is to monitor the following quantity
\begin{align}\label{I}
        I \= \frac{1}{8\pi\mu EY^*_{lm}}\int{re^{2\nu}
        \(\mbox{rhs of \eqref{hc}}\)dr}\;,
\end{align}
where the domain of integration is the region outside the neutron
star. In the limit $\sigma\rightarrow 0$ it is $I = 1$ throughout the
whole evolution, which is, of course, not true in the discretized
form. Numerically, we cannot take this limit with a fixed grid size,
since eventually we cannot sufficiently resolve the Gaussian.
Therefore, by performing this limit, we also have to decrease the grid
spacing in order to resolve the narrowing Gaussian. We do so by
keeping the ratio $\sigma/\Delta x$ constant throughout the sequence.
From the numerical data we find second order convergence. In the
following, we will use $\sigma/\Delta x \approx 0.15$, which gives us
a decent resolution of the Gaussian and its derivatives.

There is still a subtle point. In deriving the source terms (see
Appendix C), we have tacitly transformed the particle coordinate $R$
into the spacetime coordinate $r$ since the presence of the
$\delta$-function makes a distinction unnecessary. However, in the
evolution equations we have to take derivatives of the source terms
with respect to $r$ but not with respect to $R$, and therefore we
would obtain different source terms if we had not changed the $R$'s
into $r$'s. As an example consider the following two source terms
\begin{align}
        S_1(r) &= f(r)\delta(r - R)
\end{align}
and
\begin{align}
        S_2(r) &= f(R)\delta(r - R)\;,
\end{align}
which are equivalent because of the presence of the $\delta$-function.
But if we now differentiate $S_1$ and $S_2$ with respect to $r$ we
obtain for $S_2$ just the derivative of the $\delta$-function, whereas
for $S_1$ we also have to differentiate $f$. Analytically this does
not make a difference, but if we approximate the $\delta$-function by
a Gaussian, then the two expressions for $S_1$ and $S_2$ and their
respective derivatives are different. To gain accuracy, we should have
kept e.g.~an expression like $r^2d\phi/d\tau\delta(r - R)$ as
$Lr^2/R^2\delta(r - R)$ and not just as $L\delta(r - R)$. However,
this is rather cumbersome, and for the numerical evolutions the actual
difference is negligible, so we have assumed the source terms to be of
the form of $S_1$.

\section{Setting up the initial conditions}
\label{sec:init}
\begin{figure}[t]
\leavevmode
\epsfxsize=\textwidth
\epsfbox{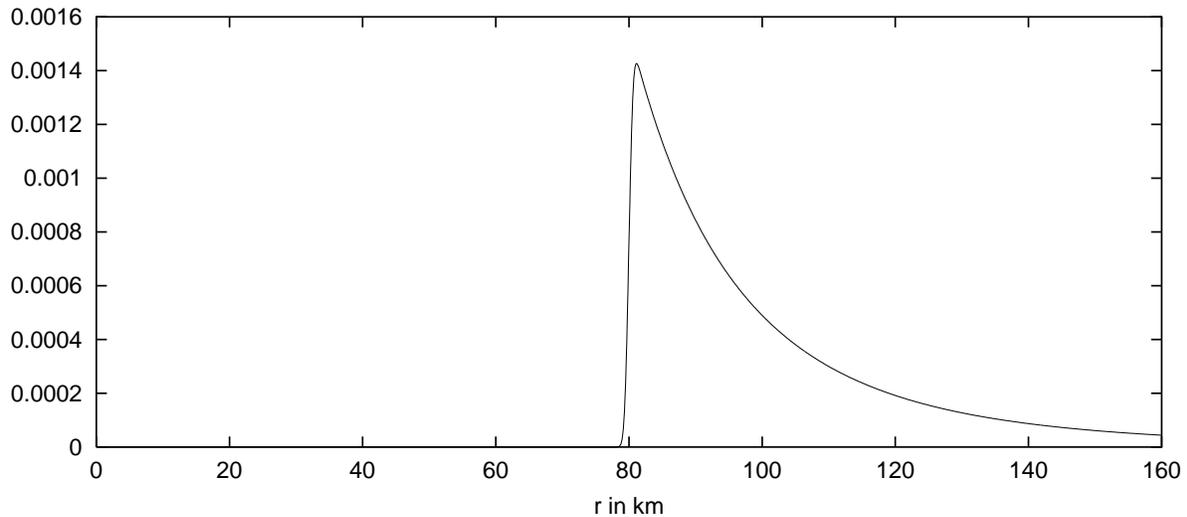}
\caption{\label{Sinit}Profile of $S_0$ with $T_0 = 0$. Note that
        $S_0$ exhibits a discontinuity at the particle's location.}
\end{figure}

\begin{figure}[t]
\epsfxsize=\textwidth
\epsfbox{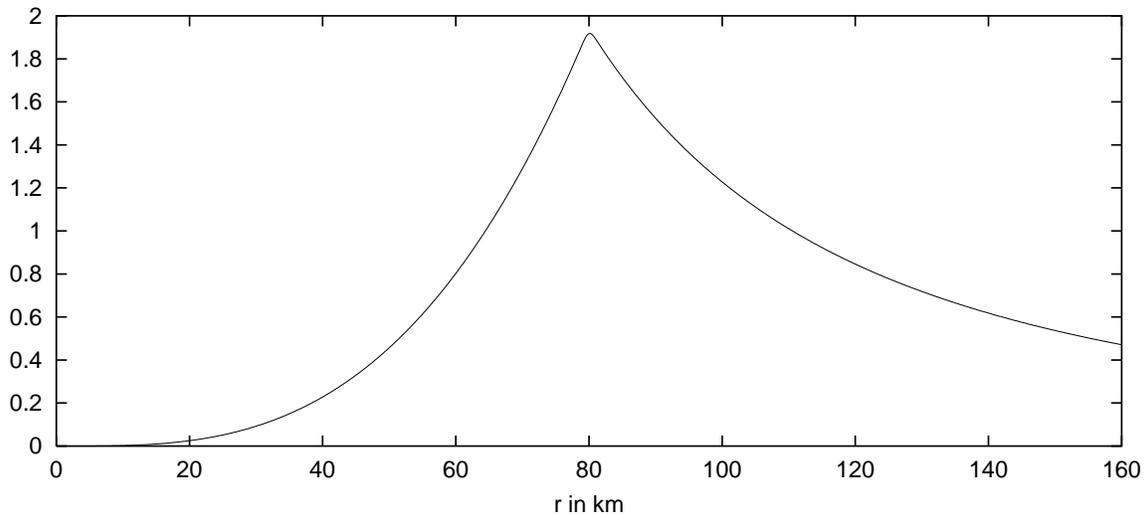}
\caption{\label{Tinit}Profile of $T_0$ with $S_0 = 0$. Note that $T_0$
        is continuous in contrast to the initial profile of $S_0$ in
        Fig.~\ref{Sinit}}
\end{figure}

As we already know from chapter 4, any construction of initial data
for a particle falling from rest or starting with some initial
velocity involves solving the Hamiltonian constraint \eqref{hc}.
However, \eqref{hc} is one equation for the two metric quantities $S$
and $T$, which means that we have more or less the same freedom in
choosing the initial values as in chapter 4. Let us denote the initial
values of $S$ and $T$ at $t = 0$ by $S_0$ and $T_0$, respectively. We
then can, for example, either choose $T_0 = 0$ and solve for $S_0$, or
do it the other way round and set $S_0 = 0$ and solve for $T_0$. In the
former case, we would have to solve a first order equation for $S_0$,
the latter case would lead to a second order equation for $T_0$. In
Figs.~\ref{Sinit} and \ref{Tinit} we show those two possibilities. By
comparing the different shapes of $S_0$ and $T_0$ we might already
deduce that the case in which we set $T_0 = 0$ and solve for $S_0$
(Fig.~\ref{Sinit}) is less favorable since $S_0$ exhibits a
discontinuity at the location of the particle. In the other case
(Fig.~\ref{Tinit}) $T_0$ still exhibits a kink, but it is continuous.

To assess which choice is the more natural, we consider a particle
initially at rest in flat spacetime. Of course, the particle will
remain at rest since there is no matter around that could attract the
particle. Thus, the perturbation of the spacetime that is created by
the particle will be stationary. Hence the equations of motion for the
metric perturbations $S$ and $T$ will read $\df{S}{t} = 0$ and
$\df{T}{t} = 0$. From this conditions and from the Hamiltonian
constraint it then follows that $S$ has to vanish.

Of course, in the presence of the neutron star, those arguments do not
hold any more, and $S_0=0$ will not be the right choice of initial
data, but if the particle initially is far enough away from the
neutron star, the error in setting $S_0=0$ should be very small. This
error actually corresponds to an introduction of an extra amount of
gravitational radiation, which is not at all related to the radiation
that is emitted when the particle moves through the spacetime. This
extra amount will propagate during the evolution and eventually excite
oscillations of the neutron star. However, if we put the particle far
enough away from the neutron star, the strength of the induced
oscillations should be small compared to the ones excited when the
particle comes close to the neutron star, which can be confirmed by
the numerical evolutions. Close to the neutron star, where the
gravitational field is strongest, setting $S_0=0$ will cause $S$ to
bulge up and send a wave towards infinity. Again, this amount of
radiation is by far smaller than the radiation that will come directly
from the particle itself.

\begin{figure}[p]
\leavevmode
\vspace*{-5mm}
\epsfxsize=\textwidth
\epsfbox{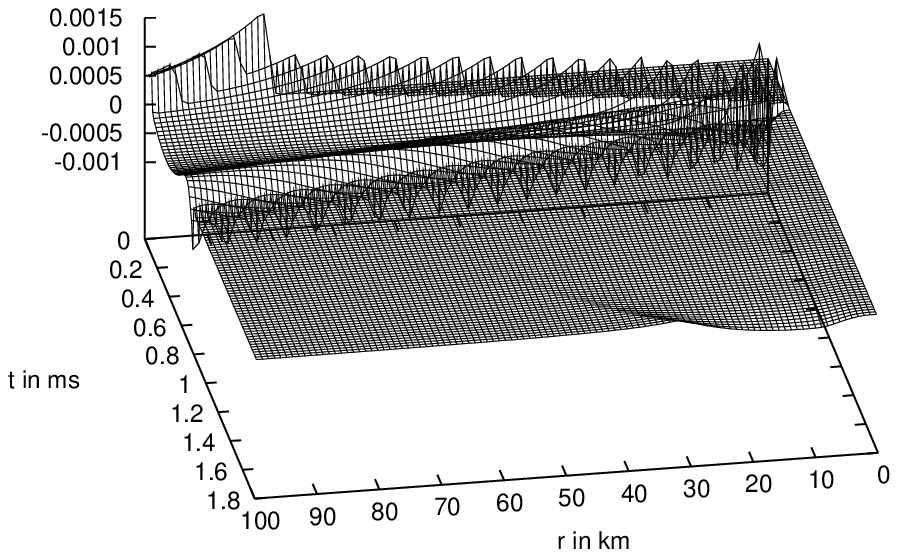}
\caption{\label{S3d_1}Evolution of $S$ for $T_0 = 0$ and $S_0$ 
        from Fig.~\ref{Sinit}. We can see two bursts of gravitational
        waves that propagate both in and outwards. The ingoing pulse
        gets reflected at the origin and travels back outwards again.}
\epsfxsize=\textwidth
\epsfbox{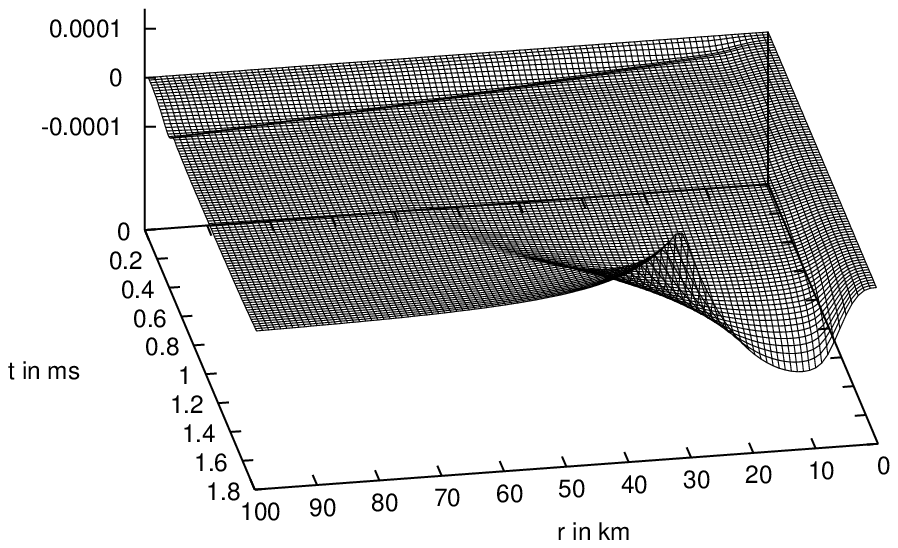}
\vspace*{-1cm}
\caption{\label{S3d_2}Evolution of $S$ for $S_0 = 0$ and $T_0$
        from Fig.~\ref{Tinit}. Here, we can see a wave emerging from
        the vicinity of the neutron star and travelling outwards.
        However, the amplitude is much smaller than in
        Fig.~\ref{S3d_1}.}
\end{figure}

\begin{figure}[p]
\leavevmode
\vspace*{-5mm}
\epsfxsize=\textwidth
\epsfbox{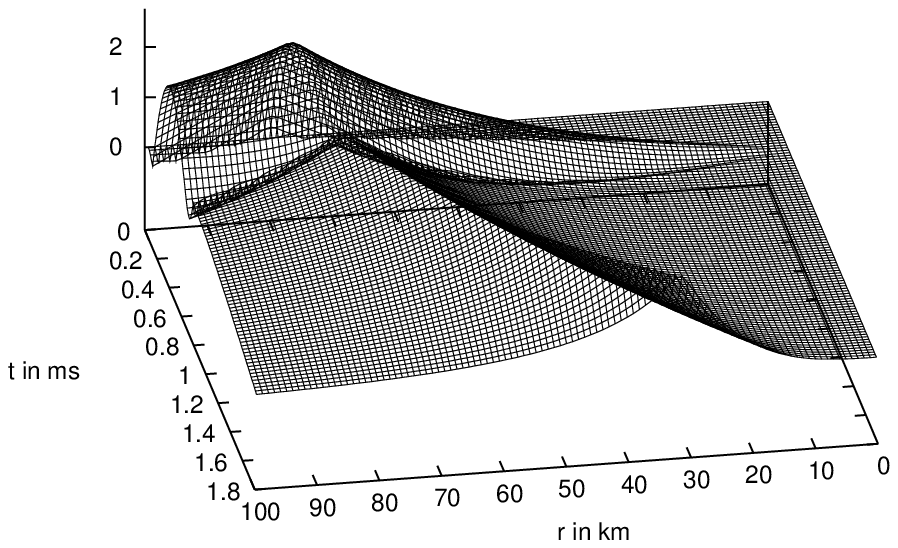}
\caption{\label{T3d_1} Evolution of $T$ for $T_0 = 0$ and $S_0$ from 
        Fig.~\ref{Sinit}. We can see the build-up of $T$, which is
        disturbed by the reflected part of the wave that was sent out
        by the particle.}
\epsfxsize=\textwidth
\epsfbox{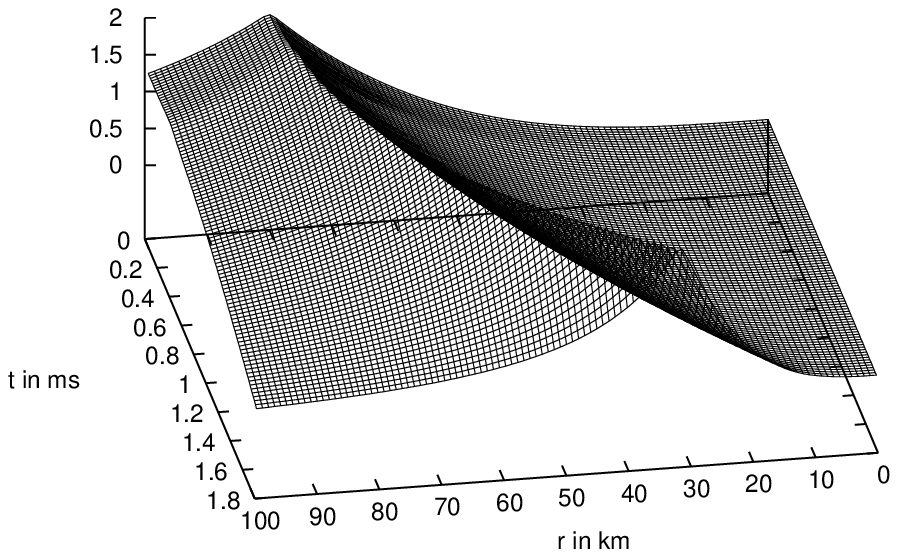}
\vspace*{-1cm}
\caption{\label{T3d_2} Evolution of $T$ for $S_0 = 0 $ and $T_0$ from 
        Fig.~\ref{Tinit}. It is evident that $T$ basically has its
        ``right'' shape right from the beginning.}
\end{figure}

In Figs.~\ref{S3d_1} through \ref{T3d_2} we show the evolution of the
two possible choices of initial data for a particle falling from
rest. In Fig.~\ref{S3d_1} and Fig.~\ref{T3d_1} we show the evolution
of $S$ and $T$ for the initial data of Fig.~\ref{Sinit}, and in
Fig.~\ref{S3d_2} and Fig.~\ref{T3d_2} the evolution of $S$ and $T$ for
the initial data of Fig.~\ref{Tinit}. The differences are
obvious. The initial shape of $T$ in the latter case is almost
unchanged during the evolution, whereas $S$ starts to aquire its
``right'' shape. In the other case we can see a huge burst of
radiation propagating in both directions. In the same time $T$ is
acquiring its ``right'' shape. We also include the evolution of the
Hamiltonian constraint, where we plot the righthand side of
\eqref{hc}, which monitors the ``path'' of the particle. In both cases
the graphs agree as it should be, of course, for the Hamiltonian
constraint should only monitor the energy density of the particle,
independent of any gravitational waves. Lastly, we also show the
metric function $S$ and $T$ after a certain time $t$, which clearly
shows that regardless of the chosen initial data they will adjust
themselves to their proper values after having radiated away the
superfluous initial wave content.

\begin{figure}[p]
\leavevmode
\vspace*{-5mm}
\epsfxsize=\textwidth
\epsfbox{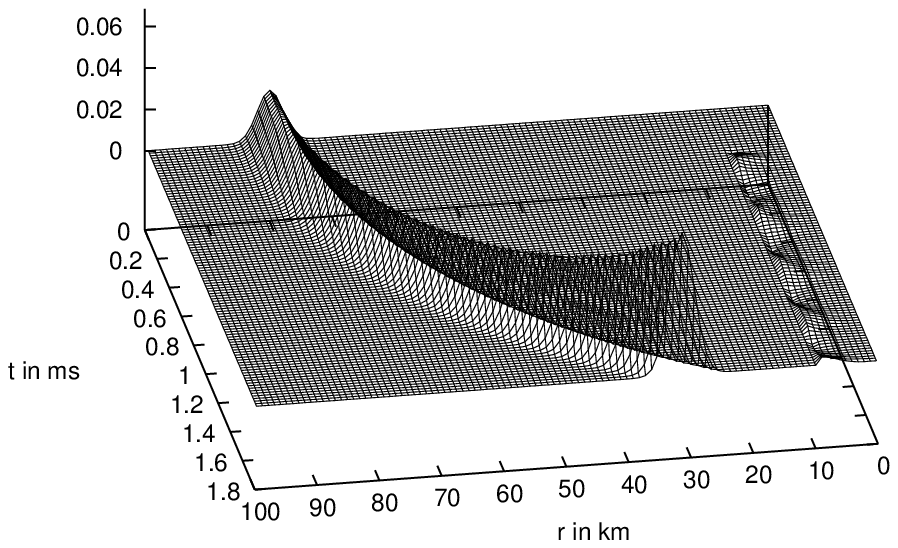}
\caption{\label{HC3d_1} Evaluation of the Hamiltonian constraint during the 
        evolution for $T_0 = 0$. The Gaussian shape of the particle
        is clearly visible and has been chosen to be very broad for a
        better visualization. The particle initially is at rest and
        starts falling towards the neutron star. On the very right side
        we can see the matter oscillations of the neutron star.}
\epsfxsize=\textwidth
\epsfbox{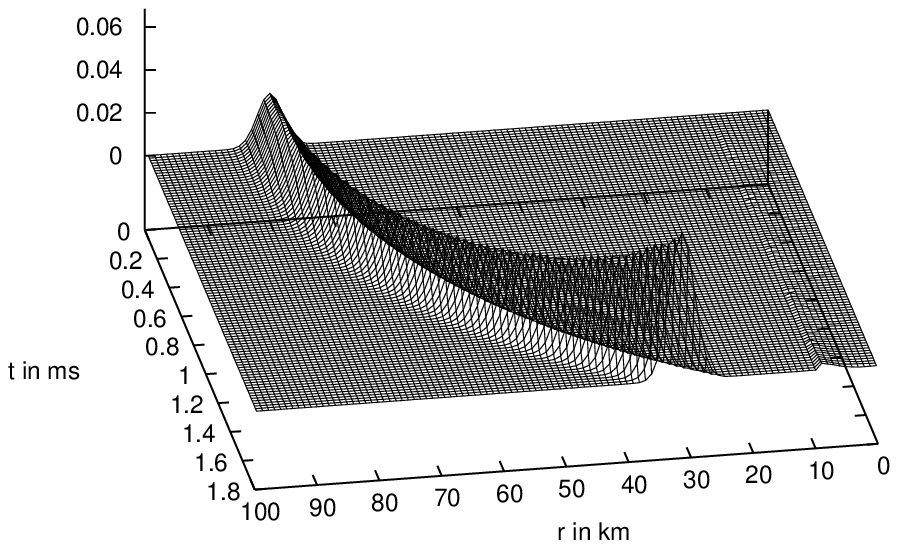}
\caption{\label{HC3d_2} Evaluation of the Hamiltonian constraint during the 
        evolution for $S_0 = 0$. As expected, it basically does not
        differ from Fig.~\ref{HC3d_1}, for the gravitational waves do
        not give any contribution. Note, however, that the matter
        oscillations of the neutron star are much smaller than in
        Fig.~\ref{HC3d_1}.}
\end{figure}

\begin{figure}[t]
\leavevmode
\epsfxsize=\textwidth
\epsfbox{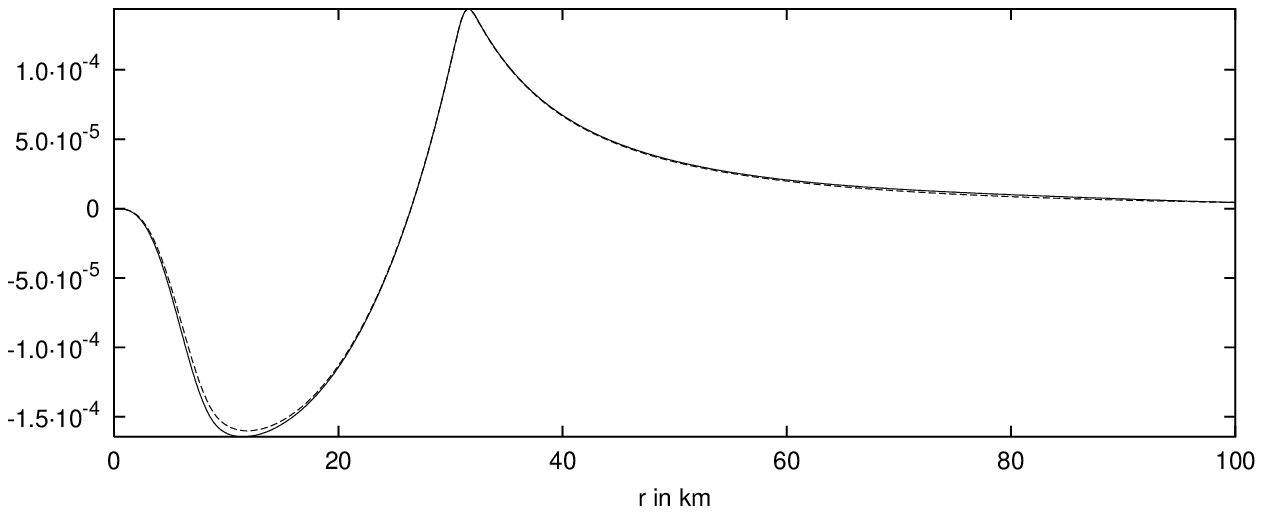}
\caption{\label{STfinal1} Plot of the variables $S$ at the end of the
        evolution of the different initial data. The artificial
        radiation of the initial data has been radiated away and $S$
        has assumed its ``right'' profile that is independent of the
        initial data. The difference between the different profiles is
        due to the fact that the initial data with $T_0 = 0$
        contain much more radiation, which excites the neutron stars to
        pulsations, which in turn disturb the profile of $S$.}
\epsfxsize=\textwidth
\epsfbox{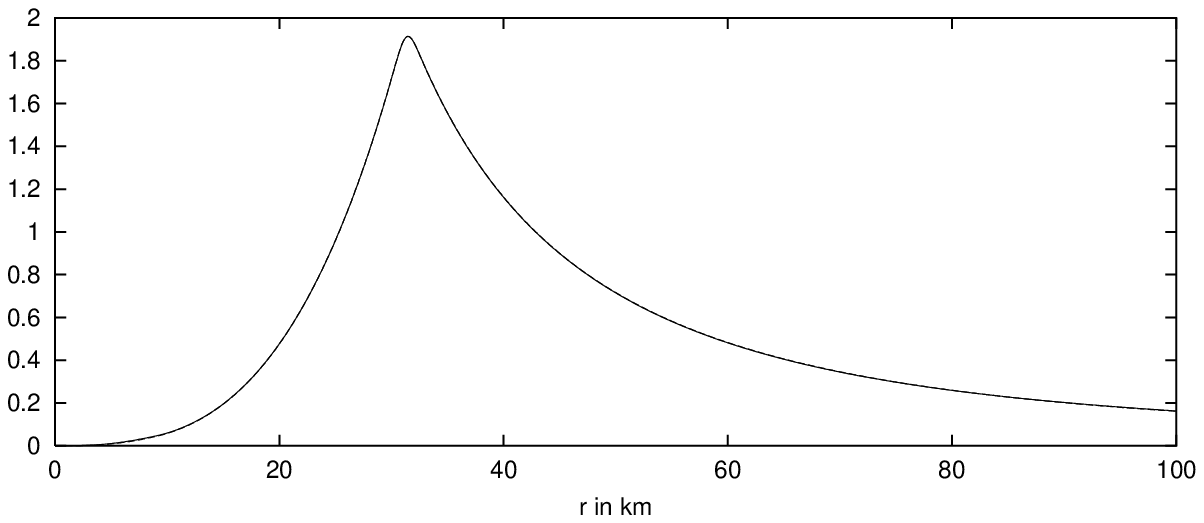}
\caption{\label{STfinal2} Plot of the variables $T$ at the end of the
        evolution of the different initial data. Here, too, the
        artificial radiation of the initial data has been radiated
        away and $T$ has assumed its ``right'' profile, which is
        independent of the initial data. In contrast to $S$ there is
        almost no difference in the two profiles.}
\end{figure}

We should note that we cannot escape the whole ambiguity of how to
choose the variables $S$ and $T$ by using the Zerilli formalism
instead. There, any regular initial data are valid initial data since
by construction the Zerilli function always satisfies the constraint
equations. Hence, we cannot see a priori whether ot not the chosen
initial data will have additional radiation content.

If the particle initially is not at rest, in addition to solving the
Hamiltonian constraint \eqref{hc} we also have to solve the momentum
constraints \eqref{mc1} and \eqref{mc2}. Of course, as
with the Hamiltonian constraint we again are faced with the same kinds
of ambiguity in solving that equation.

Now, for a particle initially being at rest or very slow, or for
circular orbits, we do not have to care about what kind of initial
data we choose, for we do not have to worry about the initial
gravitational wave pulses, since they travel with the speed of light
and will long be gone when the particle, which is much slower, comes
close to the star. However, if the particle's initial velocity is
close to the speed of light, then the particle ``rides'' on its own
wave pulse and it will be not clear any more whether the excitation of
some particular modes of the neutron star is due the the particle
itself or due to the initial burst, which comes from the inappropriate
initial data. If the particle is slow enough, those two effects can
clearly be distinguished. For very fast particles this is not possible
any more and this is particularly bothersome, insofar it is known that
a pulse of gravitational waves will predominately excite
$w$-modes. Now, if we have a wave signal from a particle that grazes a
neutron star with almost the speed of light, and we find that, indeed,
there are some traces of a $w$-mode, how then can we make sure that
this is a ``real'' signal and not an artefact due to the inappropriate
initial data?

To obtain an approximate answer, we again turn to the flat space case.
In this limit, the equations governing the evolution of $S$ and $T$
reduce to two simple coupled wave equations with a source term
that takes the presence of the particle into account:
\begin{subequations}\label{eqs_flat}
\begin{align}
        \dff{S}{t} \= \dff{S}{r} - \frac{l(l+1)}{r^2}S
        + 16\pi\mu E\frac{v^2}{r^3}\delta(r - R(t))Y^*_{lm}\label{S_flat}\\
        \dff{T}{t} \= \dff{T}{r} - \frac{l(l+1)}{r^2}T + 4S
        - 8\pi\frac{\mu}{rE}\delta(r - R(t))Y^*_{lm}\;.\label{T_flat}
\end{align}
\end{subequations}
In flat space the particle moves on a straight line with constant
velocity $v$, hence it is
\begin{align}
        R(t) \= r_0 + vt\;,
\end{align}
with $r_0$ being the initial location of the particle. Furthermore,
the normalized energy $E$ is just the Lorentz factor
\begin{align}
        E \= \frac{1}{\sqrt{1 - v^2}}\;.
\end{align}
It is interesting to note that the wave equation for $S$
\eqref{S_flat} is totally decoupled from the one for $T$. However, the
solutions of \eqref{S_flat} and \eqref{T_flat} have to satisfy the
flat space Hamiltonian constraint, which reads
\begin{align}\label{HCflat}
        \dff{T}{r} - \frac{l(l+1)}{r^2}T - r\df{S}{r}
        - \(2 + \half l(l+1)\)S
        \= - 8\pi\frac{\mu E}{r}\delta(r - R(t))Y^*_{lm}\;.
\end{align}
We now seek for an exact solution of \eqref{S_flat} that obeys the right
boundary conditions at the origin and at infinity. Once it is found,
we may use \eqref{HCflat} to numerically compute the appropriate
$T$. We state that a solution for \eqref{S_flat} is given by the following
series ansatz:
\begin{align}\label{Sansatz}
\begin{split}
        S(t,r) \= A\(\sum_{i = 0}^{\infty}a_i \frac{r^{2i+l+1}}{(r_0 +
        vt)^{2i+l+3}}\Theta(r_0 + vt - r) \right.\\
        &{}\hspace*{1cm}\left. + \sum_{i = 0}^{\infty}b_i\frac{(r_0 +
        vt)^{l-2i-2}} {r^{l-2i}}\Theta(r - r_0 - vt)\)\;,
\end{split}
\end{align}
where $\Theta$ is the Heaviside function, which satisfies
\begin{align}
        \Theta(x) \= \begin{cases}
        0, & x < 0\\
        1, & x \ge 0\\ \end{cases}\;.
\end{align}
Continuity at $r = r_0 + vt$ requires that
\begin{align}\label{ab}
        \sum_{i = 0}^{\infty}a_i \= \sum_{i = 0}^{\infty}b_i\;.
\end{align}
The overall amplitude will be determined by $A$, hence we deliberately 
may set
\begin{align}
        \sum_{i = 0}^{\infty}a_i \= \sum_{i = 0}^{\infty}b_i \,=\, 1\;.
\end{align}
The amplitude $A$ and the coefficients $a_i$ and $b_i$ can be found
by plugging \eqref{Sansatz} into \eqref{S_flat}. For $A$ we find
\begin{align}\label{S0}
        A \= \frac{16\pi\mu E^3 v^2Y^*_{lm}}{2l + 1
        + 2\sum_{i = 0}i(a_i - b_i)}\;,
\end{align}
and the coefficients $a_i$ and $b_i$ are determined by the following
recursion relations
\begin{align}
        a_{i+1} \= a_iv^2\frac{(2i + l + 3)(2i + l + 4)}
        {2(i + 1)(2i + 2l + 3)}\\
        b_{i+1} \= b_iv^2\frac{(l - 2i - 2)(l - 2i - 3)}
        {2(i + 1)(2i - 2l + 1)}\;.
\end{align}
It is interesting to note that whereas the series in $a_i$ will extend
to infinity, the series in $b_i$ will always terminate because one of
the two factors in the numerator will become zero for some $i$. The
series in $a_i$ will converge if and only if $|v| < 1$.

\begin{figure}[p]
\leavevmode
\epsfxsize=\textwidth
\epsfbox{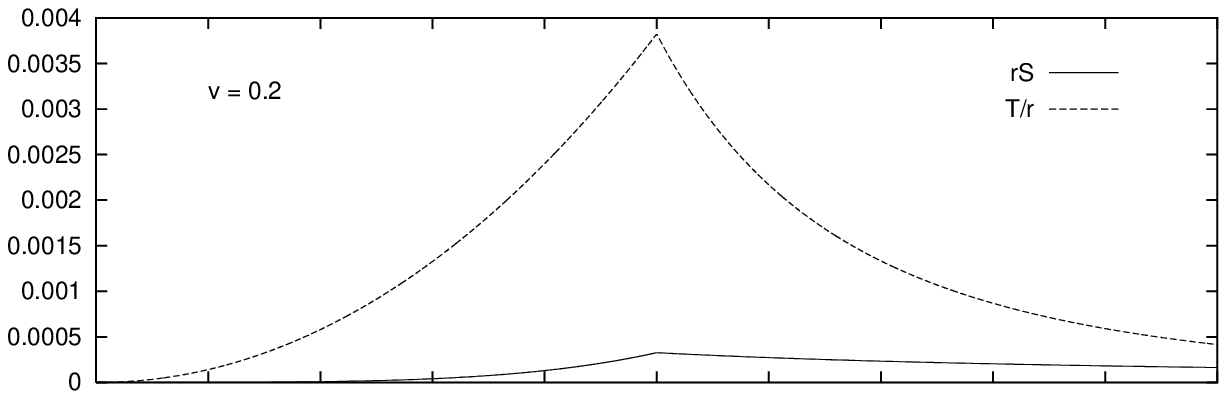}
\epsfxsize=\textwidth
\epsfbox{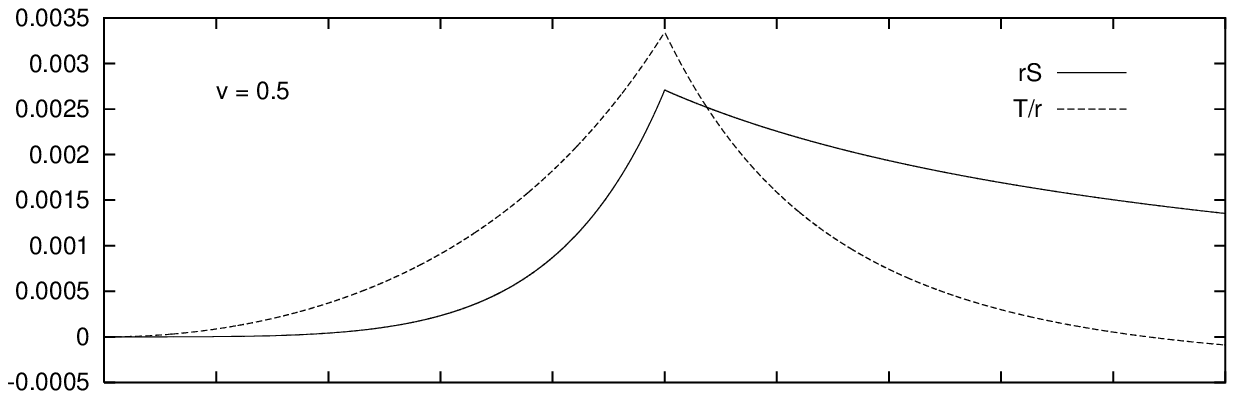}
\epsfxsize=\textwidth
\epsfbox{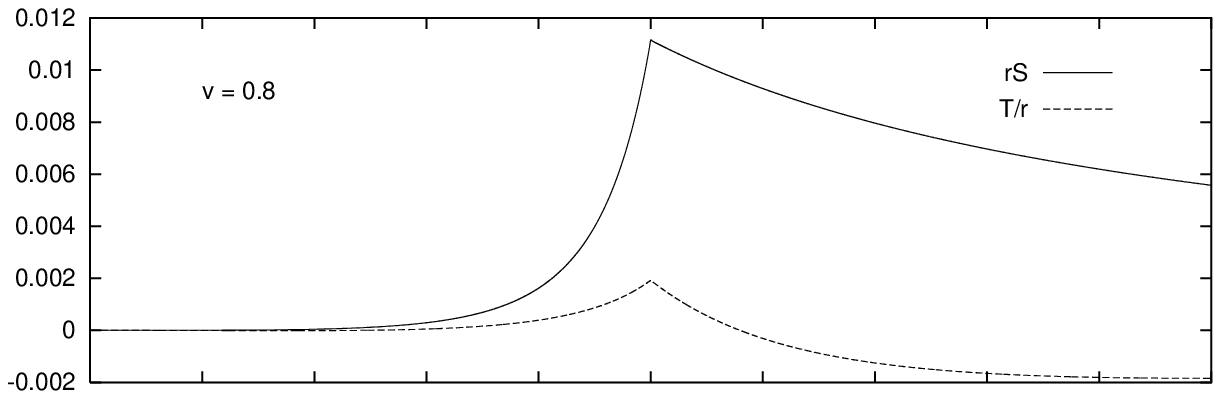}
\epsfxsize=\textwidth
\epsfbox{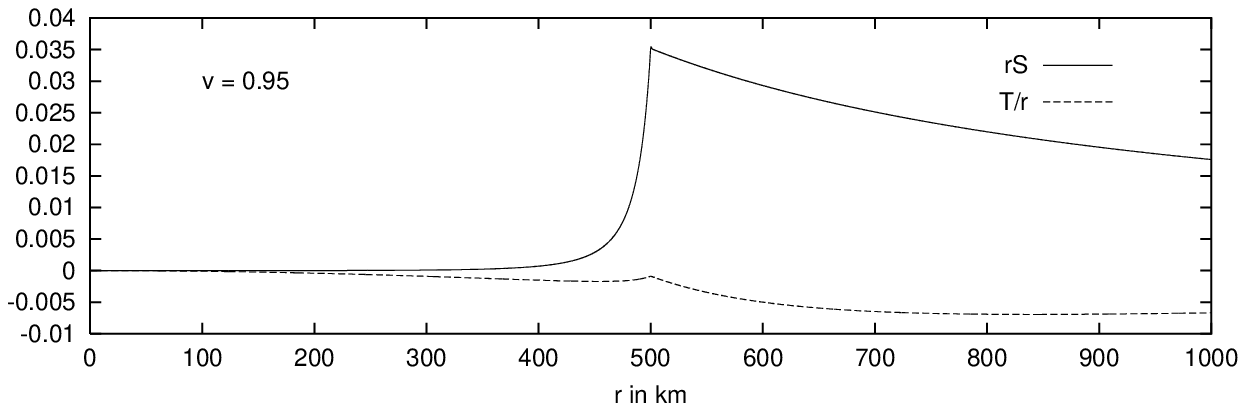}
\caption{\label{analytic}Initial data for a particle with different initial
radial velocities $v$, where $S$ is given by \eqref{Sansatz} and $T$ is
obtained by solving \eqref{HCflat}. For small velocities, $T/r$
dominates $rS$, whereas for ultra-relativistic velocities, $rS$
dominates $T/r$.}
\end{figure}

In Fig.~\ref{analytic} we show the initial data obtained from
\eqref{Sansatz} for a particle that is located at $r = 500\,$km with
different initial velocities. Here, we plot $rS$ and $T/r$ since only
those quantities can be meaningfully compared with each other. For
$v=0$ it is $rS=0$ but the amplitude of $rS$ grows rapidly when the
particle's velocity approaches the speed of light, whereas the peak of
$T/r$ slightly decreases. In the ultra-relativistic limit $rS$ totally
dominates over $T/r$.

Of course, the above solutions for $S$ and $T$ are not valid any more
if we consider the particle in curved spacetime, because they would
violate the constraints. However, we can still use \eqref{Sansatz} as
a prescription for $S$ and then use the curved space Hamiltonian
constraint \eqref{hc} to solve for $T$. Furthermore, we can compute $K$
from $K = dS/dt$ and then use the momentum constraint \eqref{MC2nd} to
compute $K_5$. As long as the particle does not have any angular
momentum and is far away from the neutron star, the thus obtained
initial values should be a good approximation for a boosted particle
on a Schwarzschild background. However, if the particle has a large
angular momentum, there will be additional source terms in the
evolution equations and our approximation should break down. However,
we are mainly interested in trajectories which come very close to the
neutron star, and hence have only small angular momentum. In this
case the above prescription still yields good initial data.

\section{Numerical results}

To compare the code with known results, we consider a particle in a
circular orbit with radius $r_0$ around the neutron star and compute
the radiated energy at infinity. Numerically, this can be accomplished
by evaluating the Zerilli function $Z$, which can be computed from $S$
and $T$ by means of formula \eqref{Zeqn}, at some large distances.

\begin{figure}[t]
\leavevmode
\epsfxsize=\textwidth
\epsfbox{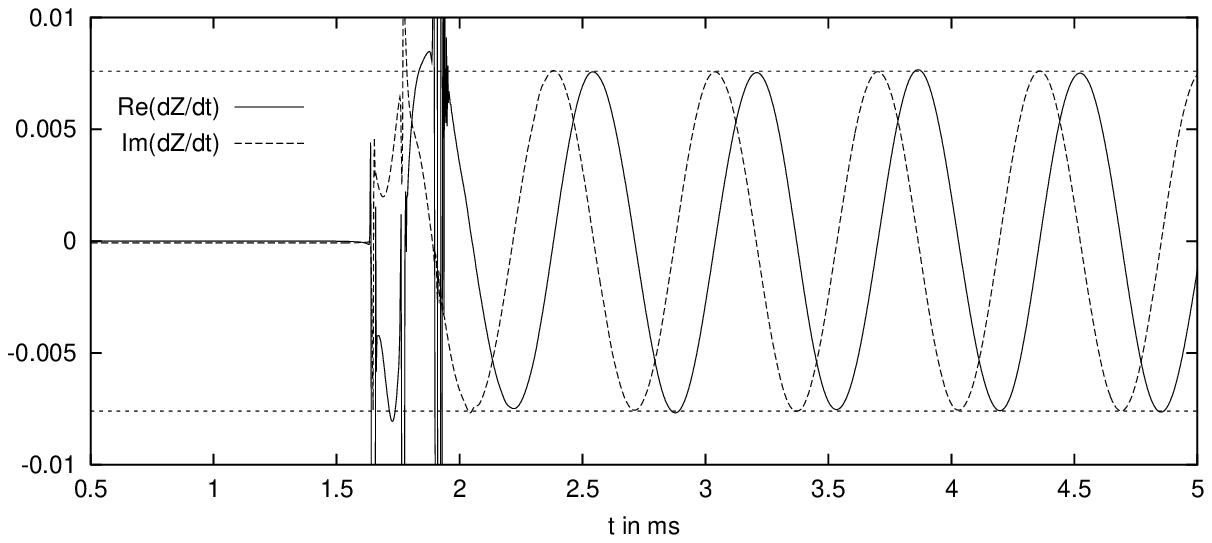}
\caption{\label{circ}Evolution of the real and imaginary parts of
$\dot{Z}_{22}$ at $r = 500\,$km. After the initial radiation
bursts the wave forms show periodical oscillations with twice the
orbital frequency. The amplitude is about $0.0076$, which corresponds
to a radiation power of $(M/\mu)^2dE_{22}/dt = 5.46\!\cdot\!
10^{-5}$.}
\end{figure}

\noindent
The radiated power for some particular values of $l$ and $m$ can then
be computed from
\begin{align}
        \frac{dE_{lm}}{dt} \= \frac{1}{64\pi}\frac{(l+2)!}{(l-2)!}\,
        |\dot{Z}_{lm}|^2\;.
\end{align}
(For a derivation, see Appendix C). Cutler et al.~\cite{CFPS} have
numerically computed the normalized gravitational power
$(M/\mu)^2dE_{lm}/dt$ that is radiated by a particle orbiting a black
hole. In table II they compile the multipole components for $r_0/M =
10$. To compare the output of our code to those results, we choose the
mass of the neutron star to be M = $1.99\,$km and the orbit of the
particle to be at $r_0 = 19.9\,$km in order to obtain the ratio of
$r_0/M = 10$. The mass of the particle is set to $\mu = 1\,$km. In
Fig.~\eqref{circ} we show $\dot{Z}_{lm}$ as a function of time
extracted at $r = 500\,$km for $l=m=2$. After some wave burst that
comes from the inappropriate choice of initial data, we see that the
signal is periodic with a frequency of twice the orbital frequency of
the particle. The amplitude is about $0.0076$, which corresponds to a
radiated power of $(M/\mu)^2dE_{22}/dt = 5.46\!\cdot\! 10^{-5}$, which
is in excellent agreement with Cutler et al.~,who obtain a value of
$(M/\mu)^2dE_{22}/dt = 5.388\!\cdot\!10^{-5}$. The slightly higher
value of our result may be due to the fact that at $r = 500\,$km the
Zerilli function is still somewhat off its asymptotic value, which
will be a little bit smaller. Also, a black hole absorbs some of the
radiation, whereas a neutron star will re-emit all of the incoming
radiation. We find that we agree with all the polar modes compiled in
table II of Cutler et al.~within a few percent, only for the case
$l=3$ and $m=1$ we find the radiated power to be
$5.9\!\cdot\!10^{-10}$ instead of their value of
$5.71\!\cdot\!10^{-8}$.

It is clear that a particle on a circular orbit will not excite the
eigenmodes of the neutron star in a significant manner, unless the
orbital frequency is close to the frequency of a stellar quasi-normal
mode. In this case, we can have a resonant excitation of this
particular mode and the radiated energy flux can drastically increase
\cite{Koj87}. It is clear, however, that only the low-frequency modes
like the $f$-mode can be excited by this mechanism; the frequencies of
the $w$-modes are much too high to lie in the frequency range of the
orbiting particle.

Therefore the only way to possibly excite $w$-modes are very
eccentric orbits, where the periastron is very close to the surface of
the star, or scattering processes with very small impact parameters.
The investigation of the latter is the main objective of this work.

Before presenting the results for scattering orbits, we should
comment on some difficulties that are related with the use of our
evolution code.

As we have learned in the previous chapter, the computation of the
Zerilli function is somewhat troublesome, for the violation of the
Hamiltonian constraint during the numerical evolution prevents the
exact cancellation of the growth in $T$. However, we managed to go
around this problem by extracting the Zerilli function $Z$ very close
to the star and then using the Zerilli equation itself to propagate
$Z$ towards infinity. This lead to very reliable results.

There should be nothing that prevents us from doing the same thing for
the particle. Of course, we would have to include the appropriate
source term which takes care of the presence of the particle.
Unfortunately, the inclusion of this term is not straightforward, but
for our purposes it is not really necessary either.

Our concern is not so much to obtain the quantitative amount of energy
that gets radiated away in a scattering process. We are much more
interested in some qualitative statements concerning the relative
excitations of the various neutron star modes, and especially whether
or not the particle can excite the $w$-modes. For this purpose it is
enough to look at the waveforms of $S$ and $T$, which also contain all
the relevant information.

Of course, it is not totally impossible to compute a quite accurate
Zerilli function as we have already demonstrated for the circular
orbits. Here, the problems with the amplification of the high
frequency components are not present, since the orbital frequency of
the particle is comparably low. Still, we were forced to resort to
quite high resolution in order to obtain the desired accuracy.

Before we go on with the discussion we should clarify one point that
might give rise to some confusion. We have decomposed the particle,
which is represented by a 3-dimensional $\delta$-function, into its
various multipole components, which are labeled by $l$ and $m$. Each
of those multipoles represents an infinitely thin matter shell,
whose surface density is proportional to the spherical harmonic
$Y_{lm}(\theta,\phi)$. It is only by the summation over all the shells, i.e.
over all $l$ and $m$, that the particle gets localized at the
particular point ($R$, $\Theta$, $\Phi$) in space. In the following
we focus on $l = m = 2$, but we will still use the term ``particle''
though we better had to speak of a ``quadrupole shell''.

Ideally, we should extract the wave form at $r = \infty$, which is
clearly not possible on a finite grid. We have to record the signal
at some finite distance from the star. But if the particle moves on an
unbounded orbit, it means that it will eventually cross the location
of the observer.

Now, the particle is always slower than the propagation speed of the
gravitational waves, which, of course, propagate with the speed of
light. Hence, the observer will usually see the wave signal before he
is hit by the particle (better: quadrupole shell). However, when the
particle is very fast and the observer is not far enough away from the
neutron star, not enough time has elapsed for the interesting part of
the signal to cross the observer before the particle arrives. That
means that the signal will be a superposition of the ``real''
gravitational wave signal and the influence of the gravitational field
of the particle itself.

\begin{figure}[t]
\leavevmode
\epsfxsize=\textwidth
\epsfbox{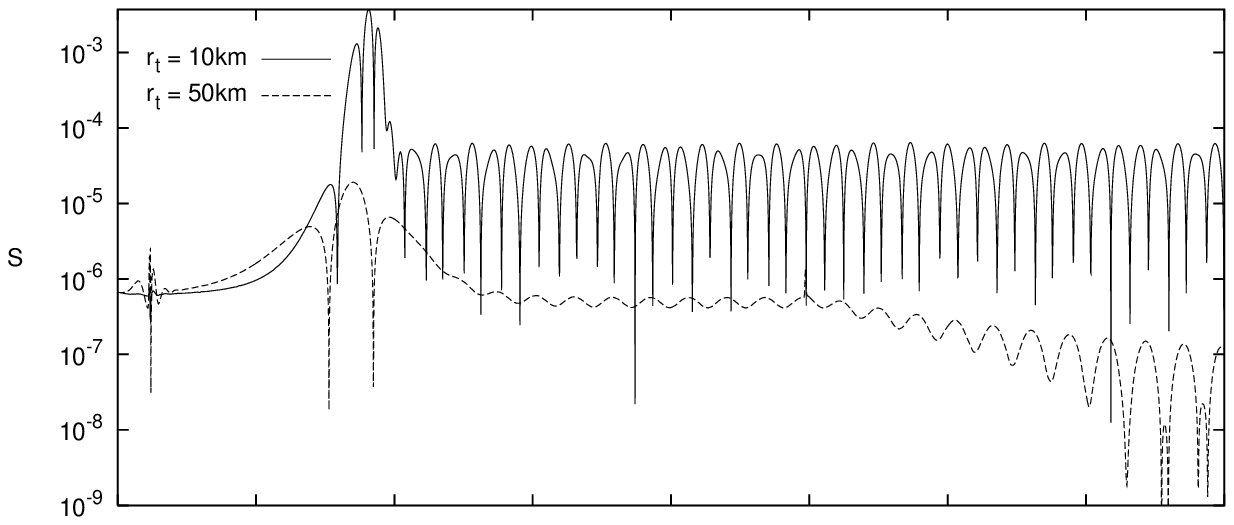}
\epsfxsize=\textwidth
\epsfbox{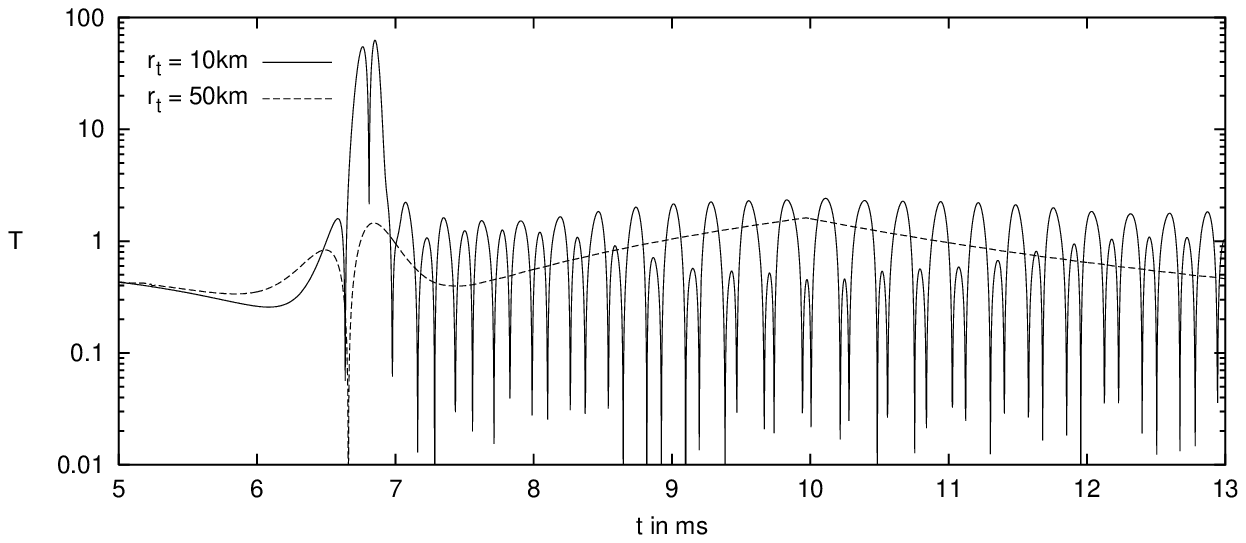}
\caption{\label{ST10_50}It is shown the difference of excitation strengths
for the two different grazing radii $r_t = 10\,$km (solid line) and
$r_t = 50\,$km (dashed line). The waveform of $S$ (upper panel) for
$r_t = 10\,$km is totally unaffected by the presence of the particle,
whereas for $r_t = 50\,$km the gravitational field shifts the
amplitude of the signal to higher values. For $T$ (lower panel) this
effect is much more pronounced and can already be detected for $r_t =
10\,$km. For $r_t = 50\,$km, the oscillations of the neutron star are
totally buried in the gravitational field of the particle. At about $t
= 10\,$ms the particle crosses the observer who is located at $r_{obs}
= 1000\,$km.}
\end{figure}

\begin{figure}[t]
\leavevmode
\epsfxsize=\textwidth
\epsfbox{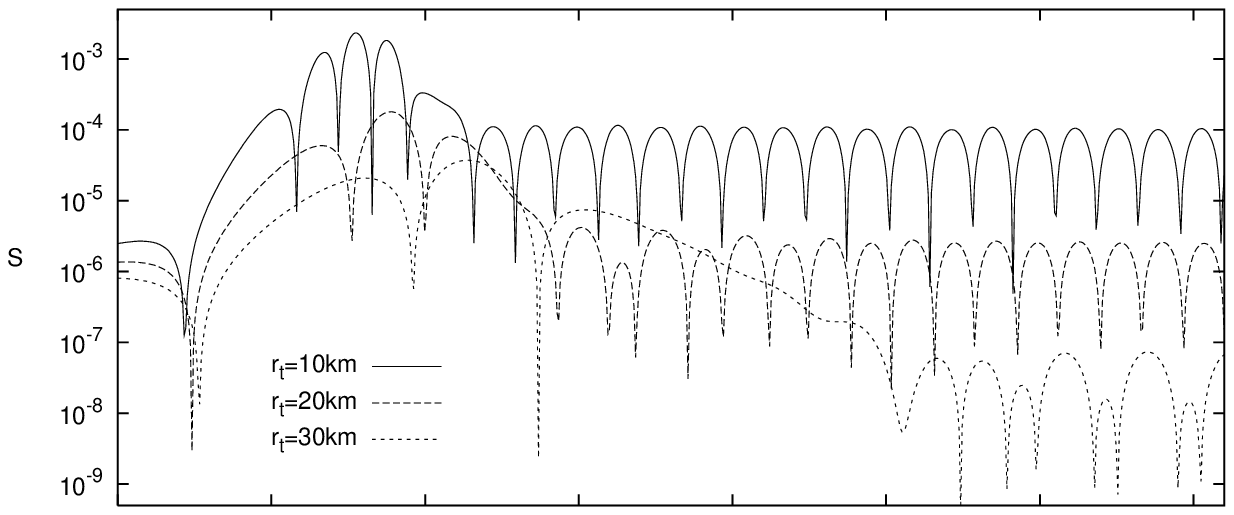}
\epsfxsize=\textwidth
\epsfbox{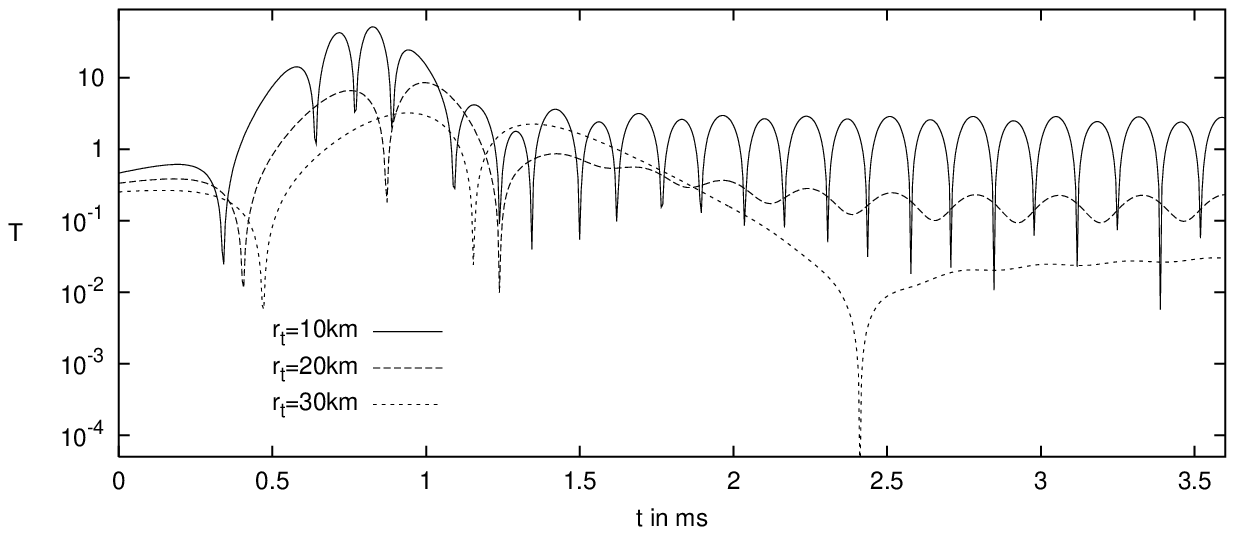}
\caption{\label{ST_v0=0.1}Wave forms of $S$ and $T$
for the three different grazing radii $r_t = 10$km, $r_t = 20$km and
$r_t = 30$km. The initial velocity of the particle is $v_0 = 0.1$.}
\end{figure}

\begin{figure}[t]
\leavevmode
\epsfxsize=\textwidth
\epsfbox{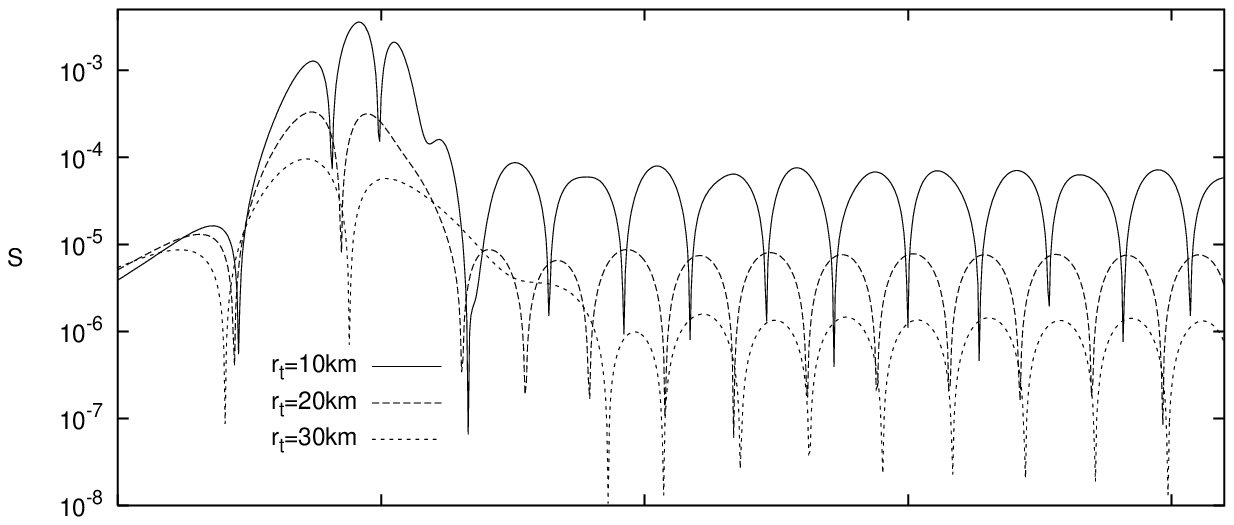}
\epsfxsize=\textwidth
\epsfbox{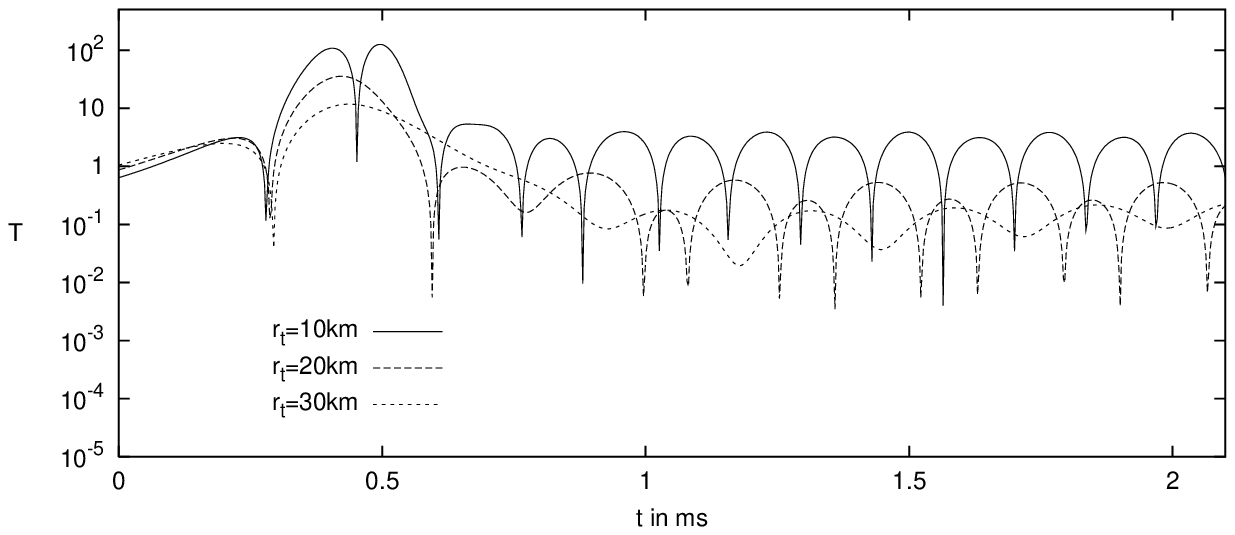}
\caption{\label{ST_v0=0.5}Wave forms of $S$ and $T$
for the three different grazing radii $r_t = 10$km, $r_t = 20$km and
$r_t = 30$km. The initial velocity of the particle is $v_0 = 0.5$.}
\end{figure}

\begin{figure}[t]
\leavevmode
\epsfxsize=\textwidth
\epsfbox{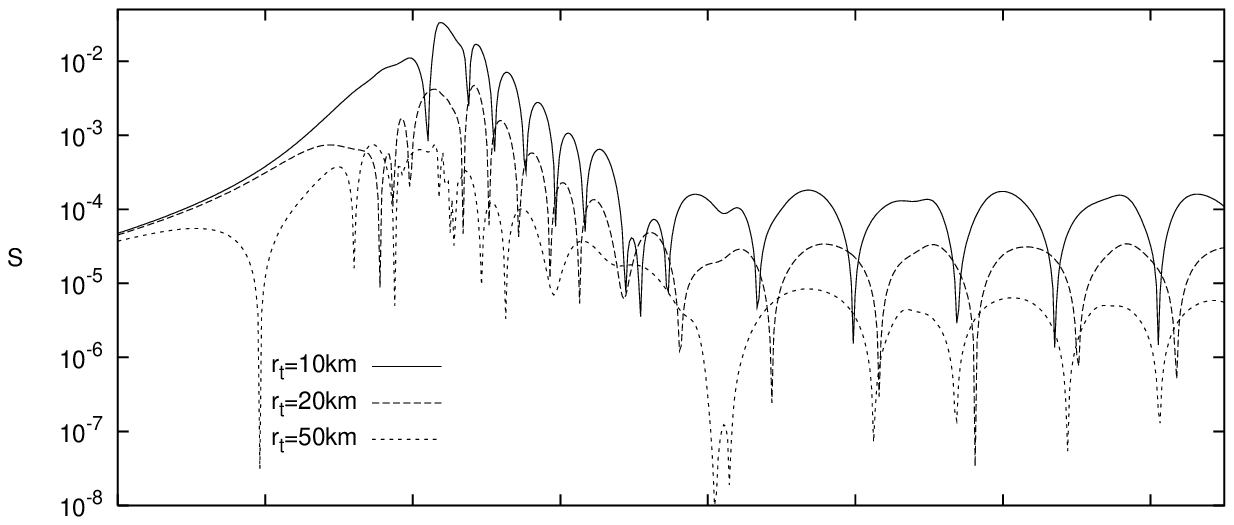}
\epsfxsize=\textwidth
\epsfbox{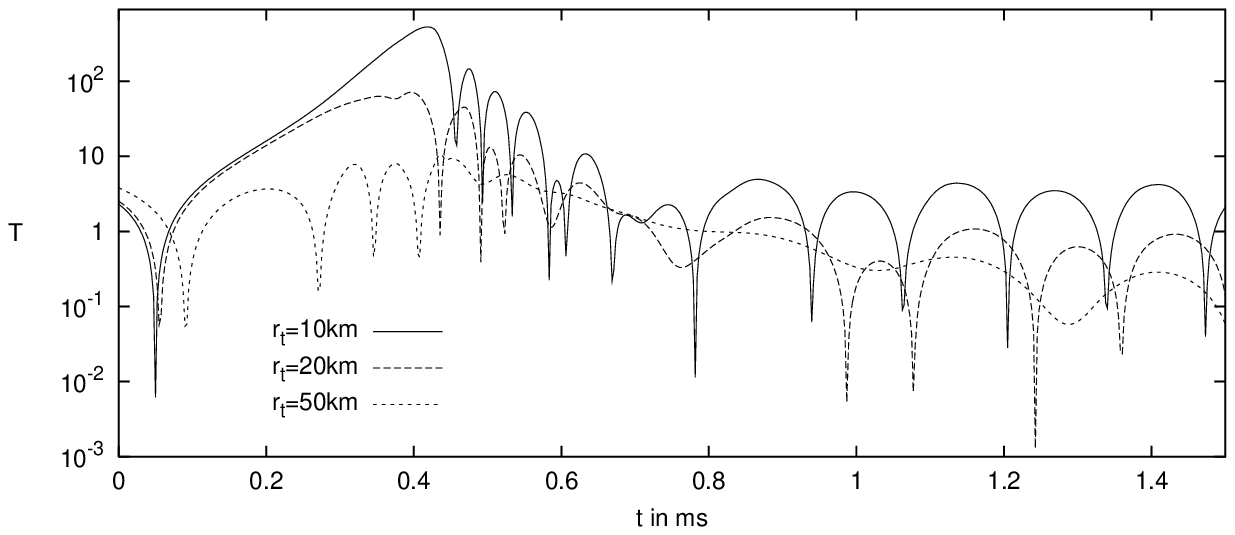}
\caption{\label{ST_v0=0.97}Wave forms of $S$ and $T$
for the three different grazing radii $r_t = 10$km, $r_t = 20$km and
$r_t = 50$km. The initial velocity of the particle is $v_0 = 0.97$.}
\end{figure}

The farther the observer moves outwards, the better the separation of
the two components in the signal can be made. Furthermore, the
influence of the particle, i.e. the strengths of the source terms in
the equations for $S$ and $T$ decrease with $1/r^3$ for $S$ and with
$1/r$ for $T$ (see equations \eqref{eqs_flat}), whereas the amplitude
of the outgoing signal remains constant for $S$ and grows linearly for
$T$. But this means that the presence of the particle affects the wave
form of $T$ much stronger than that of $S$. 

Of course, the effective interfering influence of the particle
strongly depends on the actual excitation strength of the neutron star
oscillations. The latter in turn depends on the value of the particle's
turning point $r_t$, which is the smallest distance between the
particle and the neutron star. The closer the particle approaches the
neutron star, i.e. the smaller $r_t$ is, the higher is the amplitude
of the induced oscillations. For large $r_t$ the induced fluid
oscillations are so weak that they will totally be buried within the
gravitational field of the particle. This is depicted in
Fig.~\ref{ST10_50}, where we show the wave forms of $S$ and $T$ for
two different orbits with turning points of $r_t = 10\,$km and $r_t =
50\,$km. The influence of the gravitational field on the wave forms,
especially on $T$, is obvious.

\begin{figure}[t]
\leavevmode
\epsfxsize=\textwidth
\epsfbox{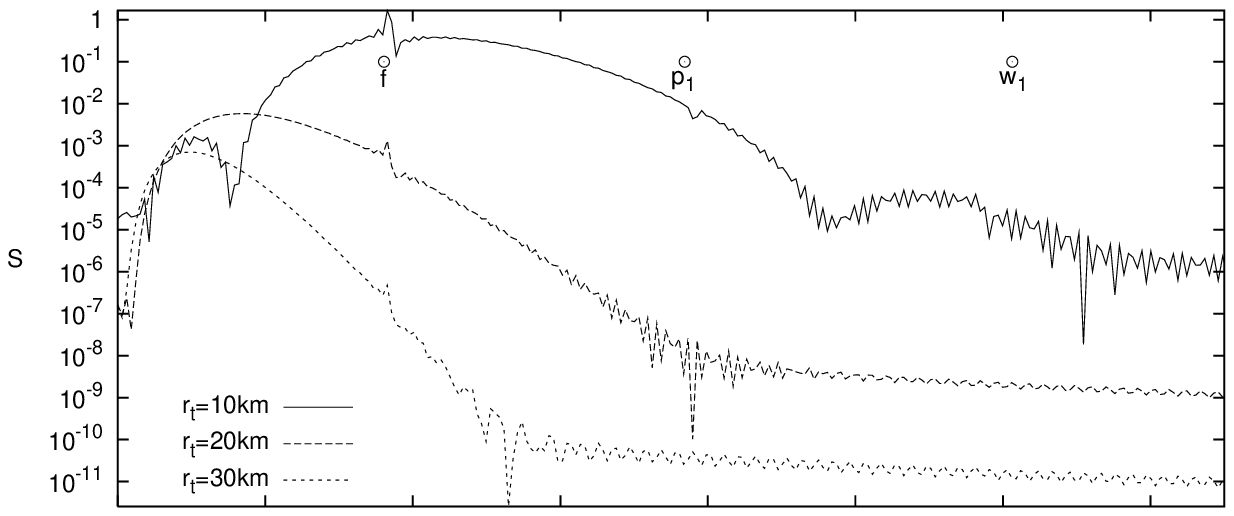}
\epsfxsize=\textwidth
\epsfbox{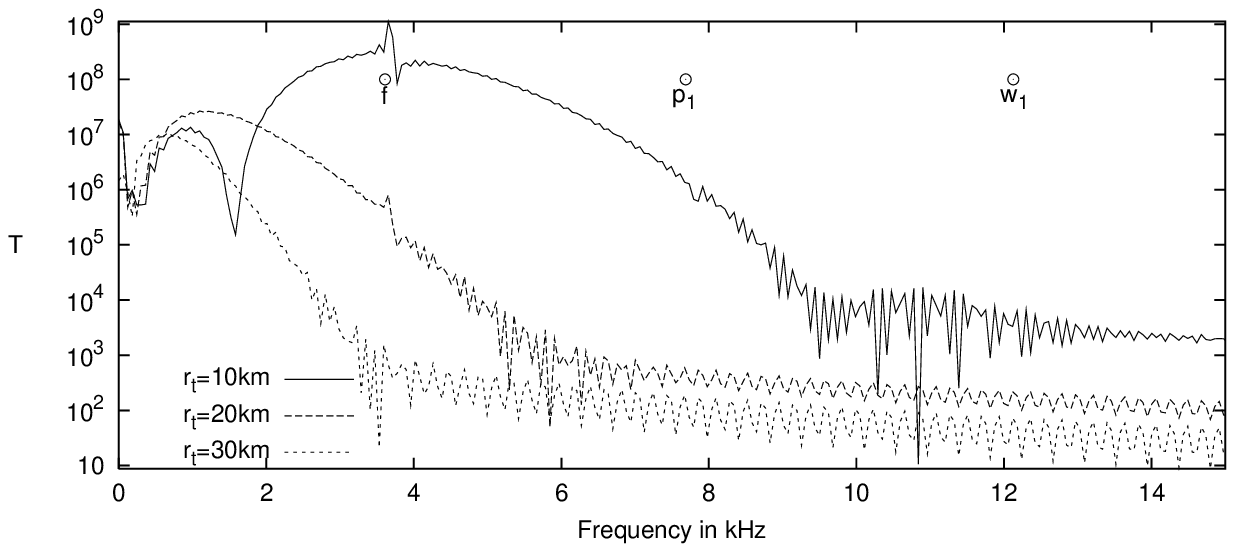}
\caption{\label{ST_v0=0.1fft}Fourier transformation of the wave forms for
$v_0 = 0.1$.}
\end{figure}

\begin{figure}[t]
\leavevmode
\epsfxsize=\textwidth
\epsfbox{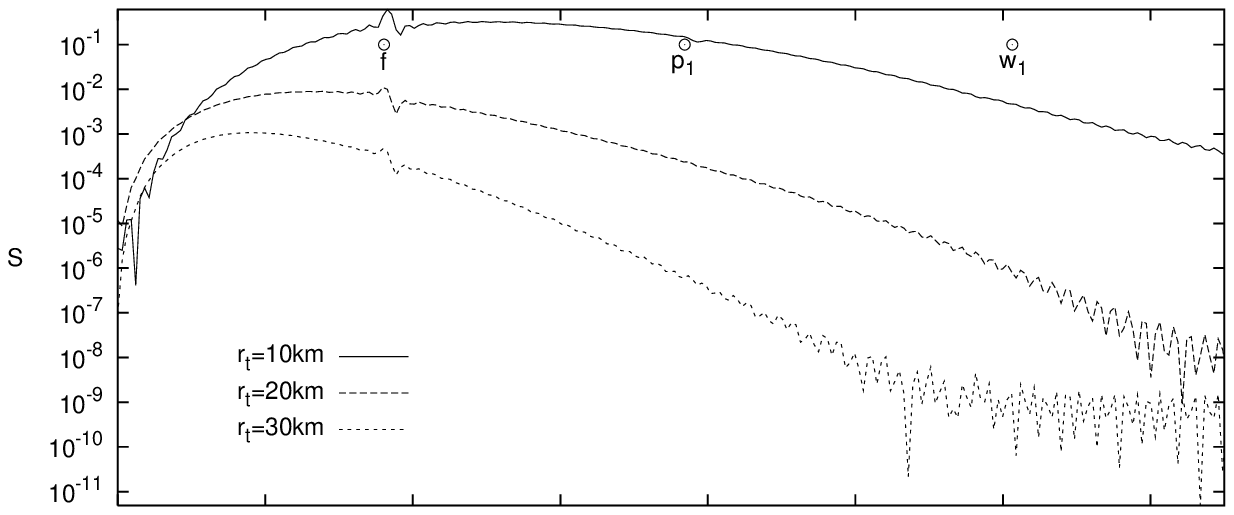}
\epsfxsize=\textwidth
\epsfbox{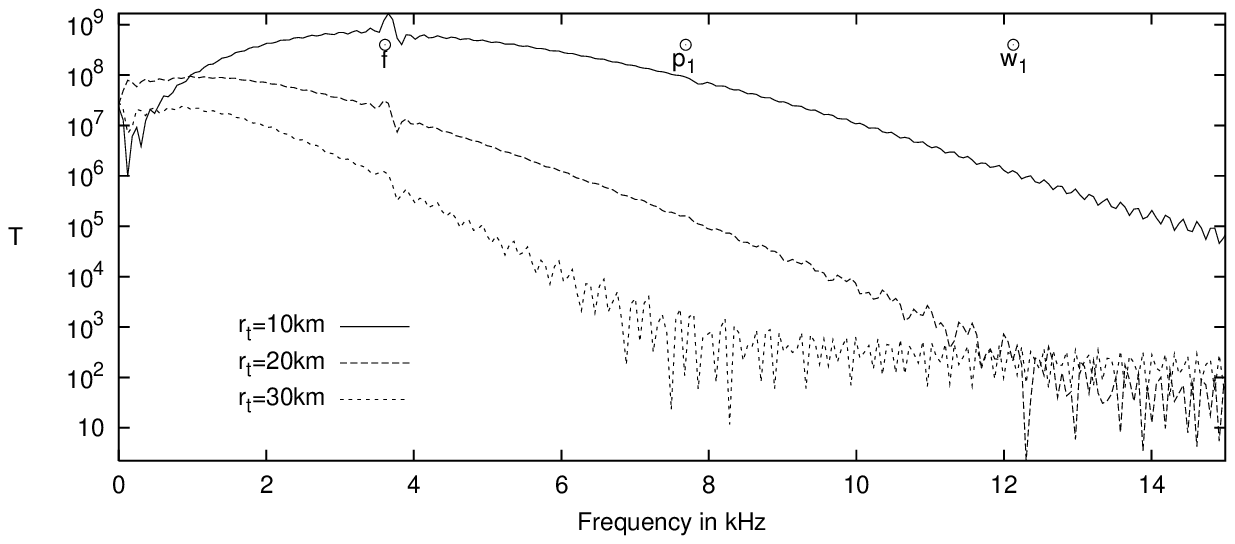}
\caption{\label{ST_v0=0.5fft}Fourier transformation of the wave forms for
$v_0 = 0.5$.}
\end{figure}

\begin{figure}[t]
\leavevmode
\epsfxsize=\textwidth
\epsfbox{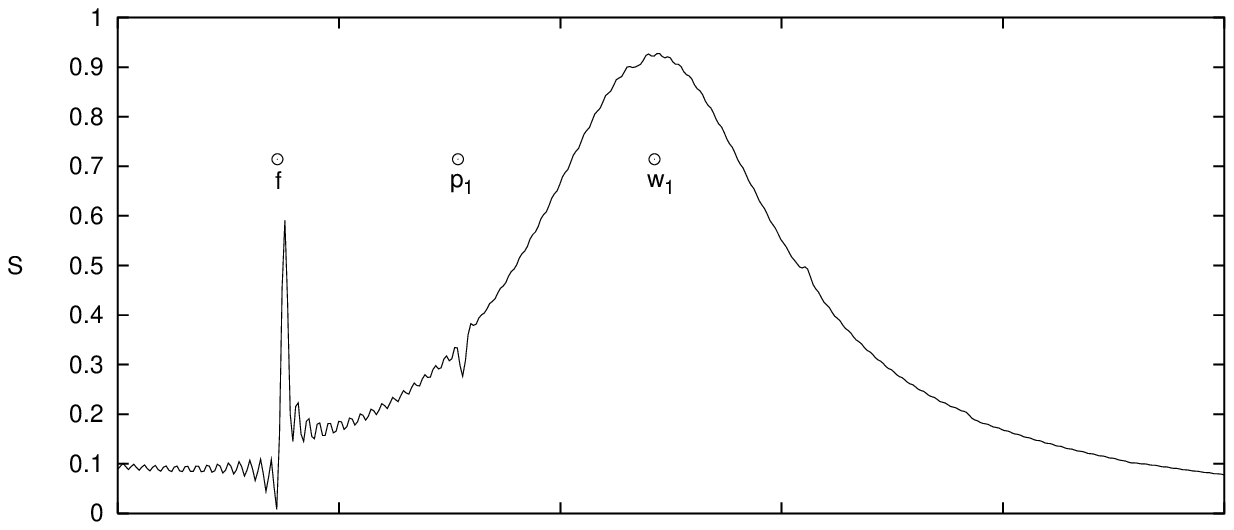}
\epsfxsize=\textwidth
\epsfbox{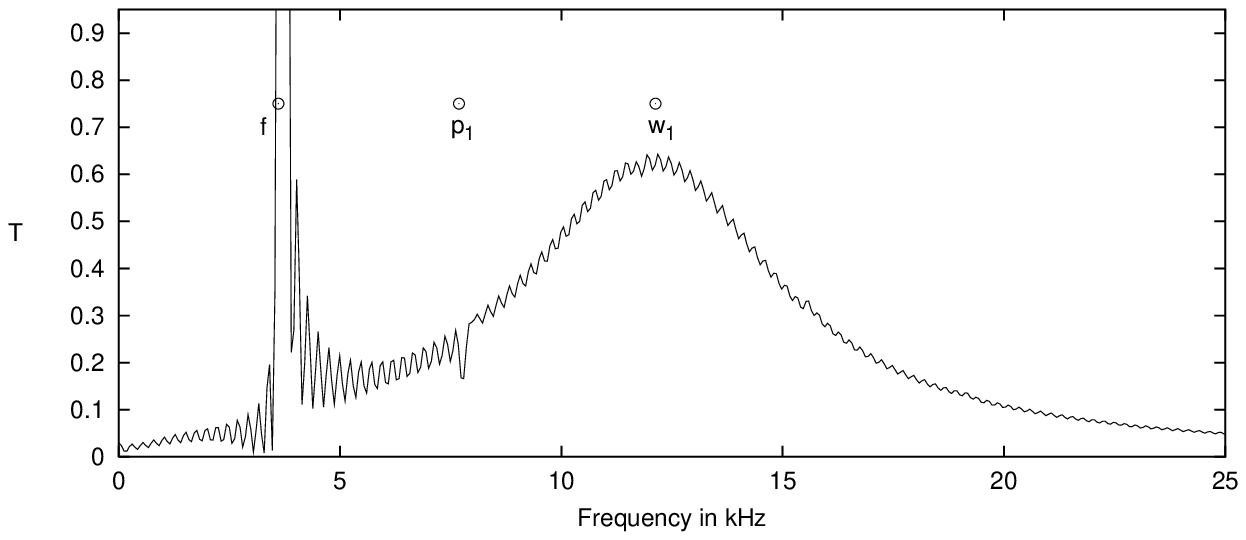}
\caption{\label{ST_v0=0.97fft}Fourier transformation of the wave form for
$r_t = 10$km and $v_0 = 0.97$. The scales are arbitrary. The presence
of the first $w$-mode is clearly visible.}
\end{figure}

In Figs.~\ref{ST_v0=0.1} through \ref{ST_v0=0.97} we show the wave
forms of $S$ and $T$ for different initial radial velocities $v_0$ and
different parameters of the turning point $r_t$. Figures \ref{ST_v0=0.1fft}
through \ref{ST_v0=0.97fft} show the corresponding Fourier transformations
for the orbit with $r_t = 10\,$km.

From Figs.~\ref{ST_v0=0.1} and \ref{ST_v0=0.5} and their
corresponding power spectra Figs.~\ref{ST_v0=0.1fft} and
\ref{ST_v0=0.5fft} it is clear that for ``small'' initial velocities
$v_0 \le 0.5$ there is no hint of a $w$-mode excitation in the
signal. Instead the signal consists of a first pulse, which comes from
the part of the particle's orbit close to the turning point, where the
particle is closest to the neutron star and radiates the most
strongly. The power spectrum of this part of the signal should peak at
about twice the angular velocity $\omega = d\Phi/dt$ of the particle
at the turning point $r_t$, since the particle's source term is
proportional to $\cos(m\omega t)$. For $v_0 = 0.5$ and $r_t = 10\,$km
the peak is around $5.5\,$kHz.

As soon as the particle leaves the star, its radiation
strongly decreases and the fluid modes of the excited neutron star
take over. Here, we find that almost all the energy is in the
$f$-mode, only a tiny fraction is in the first $p$-mode.

It is only for very high initial values of $v_0$ and very small values
of $r_t$ that there is indeed a significant excitation of
$w$-modes. By sampling different initial velocities we find that for
$v_0 \approx 0.7$ we can spot a trace of the first $w$-mode. And for
$v_0$ approaching 1, we can obtain a quite strong excitation of the
first $w$-mode, which can be seen in Figs.~\ref{ST_v0=0.97} and
\ref{ST_v0=0.97fft}.

\section{Discussion}

With the particle we have a physical even though probably a quite
unrealistic process, which induces oscillations in a neutron star. By
unrealistic we mean that this mechanism is unlikely to generate
gravitational waves that are strong enough to be detectable on Earth.

Whereas all the considered kinds of initial data of Chapter 4 could
excite the $f$-mode and some of the first $p$-modes, this was not the
case for the $w$-modes. There, only a special class of initial data
was able to excite those modes.

Andersson and Kokkotas \cite{AK98b} have shown how to extract the
physical parameters of a neutron star, like mass $M$ and radius $R$
and possibly the equation of state, if the complex frequencies of the
$f$-mode and the first $w$-mode are known. Of course, it is only
possible to detect the first $w$-mode of a neutron star if it gets
excited by some physical mechanism.

We used an orbiting mass $\mu$ in the particle limit \hbox{$\mu\ll M$}
to assess whether or not it is possible to excite the first
$w$-mode. We found that in order to excite a significant amount of
$w$-modes the particle's velocity at infinity must be much more than
70 percent of the speed of light. Moreover, the excitation strength of
the modes rapidly decreases as the particle's turning point $r_t$
increases. This is in agreement with the results of Ferrari et
al.~\cite{FGB99}, who find that no significant $w$-mode excitation is
observable. However, they do not consider the cases where the particle
has an energy $E\gg 1$, which is necessary if it were to excite any
$w$-modes.

The question now arises: Can we infer anything about what happens in
an astrophysical event? Our results show that the particle initially
has to be incredibly fast in order to excite a significant amount of
$w$-mode content. It is clear that there do not exist any
astrophysical events at all that could accelerate the particle (which,
of course does not represent an elementary particle, but some heavy
extended object, like a planet or an asteroid or even another neutron
star) up to more than half the speed of light. Hence, we might safely
conclude that any astrophysical scenario that can be simulated by a
particle orbiting a neutron star will not produce any detectable
amount of $w$-modes.

But what if the object collides with the neutron star? It has been
suggested \cite{FGB99} that in this case there might be some
significant excitation of $w$-modes. In particular, simulations of a
particle that radially falls onto a neutron star were done by Borelli
\cite{Bor97} who found that, indeed, both fluid and $w$-modes were
excited to a significant level. However, one serious drawback of those
calculations is that as soon as the particle hits the surface of the
star, the calculation is stopped. This leads to an overestimation of
the high frequency components in the power spectrum. Besides, in this
study only axial modes were investigated, it is therefore clear
that the $w$-modes will always show up in the spectrum because they
will not be screened by the presence of fluid modes, which can happen
in the even parity case. Now, even in the particle limit, it is not
possible to simulate a collision since it not clear at all what
happens when the particle hits the surface. A first step might be,
though, to let the particle go right through the neutron star without
being affected by the presence of the neutron star matter. In case
there is a significant amount of $w$-modes one might conclude that in
a realistic scenario, where the physics of the impact is included,
there still might be $w$-modes present, maybe they are even more
strongly excited than in the unphysical ``tunnelling'' case. However,
our view is that even in the impact case the particle still must have
an initial velocity that is a significant fraction of the speed of
light. Since for scattering orbits with initial speeds less than 50
percent of the speed of light there is no hint of any $w$-mode
presence even for the closest trajectories, we believe that this
should not change very much even if the particle's turning point lies
inside the neutron star. Of course, in this case it seems reasonable
to infer that the $p$-modes of the fluid would be much more strongly
excited.

Our results in chapter 4 indicate that the presence of the $w$-modes
in the signal is closely related with the value of $S$. This
particular role of $S$ seems also to be confirmed in our particle
simulations, for we have seen that for ultra-relativistic particles
the value of $rS$ greatly dominates the value of $T/r$. And it is only
there that we find a significant amount of $w$-modes.

Still, the question remains as to what happens in a realistic
astrophysical scenario. We have excluded the possibility of $w$-modes
excitation by means of a realistic scattering process. However, it is
not possible to confer our results to the merger process of a binary
neutron star system, for the latter cannot be adequately described
within perturbation theory. Only if the final object does not
immediately collapse to a black hole, it might wildly oscillate and
emit some significant radiation through $w$-modes.

\renewcommand{\chaptermark}[1]{\markboth{\appendixname\ \thechapter.\
    #1}{}}
\begin{appendix}
\chapter{Relations between the Regge-Wheeler and
Mathews-Zerilli harmonics}

The evolution equations (\ref{kij_part}) will be expanded using the
Regge-Wheeler harmonics $[\rw^A_{lm}]_{\mu\nu}$, which are defined
in \cite{RW57}, and which do not form an orthonormal set. However, in
order to perform the expansion of the energy-momentum tensor
(\ref{part_tmn}), we need an orthonormal set. This is given by the
Mathews-Zerilli harmonics $[\mz^A_{lm}]_{\mu\nu}$
\cite{Math62,Zer70a}, which are orthonormal with respect to the
following inner product
\begin{align}\label{prod}
        \int_{S^2}[\mz^A_{lm}]^{*\mu\nu}
        [\mz^{A'}_{l'm'}]_{\mu\nu}\,{\rm d\Omega} \=
        s_A\delta_{AA'}\delta_{ll'}\delta_{mm'}\;,
\end{align}
where the asterisk denotes complex conjugation and 
\begin{align}
        [\mz^A_{lm}]^{\mu\nu} \= \eta^{\mu\kappa}\eta^{\nu\sigma}
        [\mz^A_{lm}]_{\kappa\sigma}\;.
\end{align}
Because of the use of the inverse Minkowski metric $\eta^{\mu\nu}$ to raise
the indices, the inner product (\ref{prod}) is not positive definite
and it is
\begin{align}
        s_A \= \begin{cases}
        +1,& A = 1, 5,\dots, 10\\
        -1,& A = 2,3,4\\
        \end{cases}\;.
\end{align}
For the sake of completeness we list the whole set of the 10 tensor
harmonics $\mz^A_{lm}$, where we correct some errors in \cite{Zer70c}:
\begin{align*}
        [\mz^1_{lm}]_{\mu\nu} \=
        \( \begin{array}{cccc}
        1 & 0 & 0 & 0 \\
        0 & 0 & 0 & 0 \\
        0 & 0 & 0 & 0 \\
        0 & 0 & 0 & 0 \\
        \end{array} \) Y_{lm} \\
        {}[\mz^2_{lm}]_{\mu\nu} \=
        \frac{\mbox{i}}{\sqrt{2}}\( \begin{array}{cccc}
        0 & 1 & 0 & 0 \\
        1 & 0 & 0 & 0 \\
        0 & 0 & 0 & 0 \\
        0 & 0 & 0 & 0 \\
        \end{array} \) Y_{lm}
\end{align*}
\begin{align*}
        {}[\mz^3_{lm}]_{\mu\nu} \= 
        \frac{\mbox{i} r}{\sqrt{2l(l+1)}}
        \( \begin{array}{cccc}
        0 & 0 & \df{}{\theta} & \df{}{\phi} \\
        0 & 0 & 0 & 0 \\
        \df{}{\theta} & 0 & 0 & 0 \\
        \df{}{\phi} & 0 & 0 & 0 \\
        \end{array} \) Y_{lm} \\ 
        {}[\mz^4_{lm}]_{\mu\nu} \= 
        \frac{r}{\sqrt{2l(l+1)}}
        \( \begin{array}{cccc}
        0 & 0 & \frac{1}{\sin\theta} \df{}{\phi} & 
        -\sin\theta\df{}{\theta} \\
        0 & 0 & 0 & 0 \\
        \frac{1}{\sin\theta} \df{}{\phi} & 0 & 0 & 0 \\
        -\sin\theta\df{}{\theta} & 0 & 0 & 0 \\
        \end{array} \) Y_{lm}\\
        {}[\mz^5_{lm}]_{\mu\nu} \=
        \( \begin{array}{cccc}
        0 & 0 & 0 & 0 \\
        0 & 1 & 0 & 0 \\
        0 & 0 & 0 & 0 \\
        0 & 0 & 0 & 0 \\
        \end{array} \) Y_{lm} \\ 
        {}[\mz^6_{lm}]_{\mu\nu} \=
        \frac{r}{\sqrt{2l(l+1)}}
        \( \begin{array}{cccc}
        0 & 0 & 0 & 0 \\
        0 & 0 & \df{}{\theta} & \df{}{\phi} \\
        0 & \df{}{\theta}& 0 & 0 \\
        0 & \df{}{\phi} & 0 & 0 \\
        \end{array} \) Y_{lm} \\ 
        {}[\mz^7_{lm}]_{\mu\nu} \= 
        \frac{\mbox{i} r}{\sqrt{2l(l+1)}}
        \( \begin{array}{cccc}
        0 & 0 & 0 & 0 \\
        0 & 0 & \frac{1}{\sin\theta} \df{}{\phi} & 
        -\sin\theta\df{}{\theta} \\
        0 & \frac{1}{\sin\theta} \df{}{\phi} & 0 & 0 \\
        0 & -\sin\theta\df{}{\theta} & 0 & 0 \\
        \end{array} \) Y_{lm} \\ 
        {}[\mz^8_{lm}]_{\mu\nu} \=
        \frac{r^2}{\sqrt{2l(l+1)(l-1)(l+2)}}
        \( \begin{array}{cccc}
        0 & 0 & 0 & 0 \\
        0 & 0 & 0 & 0 \\
        0 & 0 & W_{lm} & X_{lm} \\
        0 & 0 & X_{lm} & -\sin^2\theta\,W_{lm} \\
        \end{array} \) \;,\\ 
        {}[\mz^9_{lm}]_{\mu\nu} \=
        \frac{r^2}{\sqrt{2}}\( \begin{array}{cccc}
        0 & 0 & 0 & 0 \\
        0 & 0 & 0 & 0 \\
        0 & 0 & 1 & 0 \\
        0 & 0 & 0 & \sin^2\theta\\
        \end{array} \) Y_{lm} \\ 
        {}[\mz^{10}_{lm}]_{\mu\nu} \=
        \frac{\mbox{i} r^2}{\sqrt{2l(l+1)(l-1)(l+2)}}
        \( \begin{array}{cccc}
        0 & 0 & 0 & 0 \\
        0 & 0 & 0 & 0 \\
        0 & 0 & \frac{1}{\sin\theta}\,X_{lm} & -\sin\theta\,W_{lm} \\
        0 & 0 & -\sin\theta\,W_{lm} & -\sin\theta\,X_{lm} \\
        \end{array} \) \\ 
\end{align*}
with 
\begin{align}
        \label{Xdef} X_{lm}
        \;&:=\; 2\(\df{}{\theta} - \cot\theta\)\df{}{\phi} Y_{lm}\\
        \label{Wdef} W_{lm}
        \;&:=\; \(\dff{}{\theta} - \cot\theta\df{}{\theta}
        - \frac{1}{\sin^2\theta}\dff{}{\phi}\) Y_{lm}
        \;=\;\(l(l+1) + 2\dff{}{\theta}\)Y_{lm}\;,
\end{align}
and $Y_{lm}$ are the ordinary scalar harmonics. There are three odd
parity harmonics $\mz^4_{lm}, \mz^7_{lm}$ and $\mz^{10}_{lm}$, the
remaining seven ones have even parity.

In terms of the Mathews-Zerilli harmonics $\mz^A_{lm}$ the
Regge-Wheeler harmonics $\rw^A_{lm}$ read:
\enlargethispage{1cm}
\begin{align*}
        \rw^1_{lm} \= \mz^1_{lm}\\
        \rw^2_{lm} \= -\mbox{i}\sqrt{2}\,\mz^2_{lm}\\
        \rw^3_{lm} \= -\frac{\mbox{i}}{r}\sqrt{2l(l+1)}\,\mz^3_{lm}\\
        \rw^4_{lm} \= -\frac{1}{r}\sqrt{2l(l+1)}\,\mz^4_{lm}\\
        \rw^5_{lm} \= \mz^5_{lm}\\
        \rw^6_{lm} \= \frac{1}{r}\sqrt{2l(l+1)\,}\mz^6_{lm}\\
        \rw^7_{lm} \= \frac{\mbox{i}}{r}\sqrt{2l(l+1)}\,\mz^7_{lm}\\
        \rw^8_{lm} \= \frac{\sqrt{2}}{2r^2}l(l+1)
        \(\sqrt{\frac{(l-1)(l+2)}{l(l+1)}}\,
        \mz^8_{lm} - \mz^9_{lm}\)\\
        \rw^9_{lm} \= \frac{\sqrt{2}}{r^2}\,\mz^{9}_{lm}\\
        \rw^{10}_{lm} \= \frac{i}{2r^2}\sqrt{2l(l+1)(l-1)(l+2)}\,
        \mz^{10}_{lm}\;.
\end{align*}
The inverse relations are given by
\begin{align*}
        {}\mz^1_{lm} \= \rw^1_{lm}\\
        {}\mz^2_{lm} \= \frac{\mbox{i}}{\sqrt{2}}\,\rw^2_{lm}\\
        {}\mz^3_{lm} \= \frac{\mbox{i} r}{\sqrt{2l(l+1)}}\,\rw^3_{lm}\\
        {}\mz^4_{lm} \= -\frac{r}{\sqrt{2l(l+1)}}\,\rw^4_{lm}\\
        {}\mz^5_{lm} \= \rw^5_{lm}\\
        {}\mz^6_{lm} \= \frac{r}{\sqrt{2l(l+1)}}\,\rw^6_{lm}\\
        {}\mz^7_{lm} \= -\frac{\mbox{i} r}{\sqrt{2l(l+1)}}\,\rw^7_{lm}\\
        {}\mz^8_{lm} \= \frac{r^2}{\sqrt{2l(l+1)(l-1)(l+2)}}
        \(l(l+1)\,\rw^9_{lm} + 2\,\rw^8_{lm}\)\\
        {}\mz^9_{lm} \= \frac{r^2}{\sqrt{2}}\,\rw^9_{lm}\\
%
        {}\mz^{10}_{lm} \= -\frac{2\mbox{i} r^2}
        {\sqrt{2l(l+1)(l-1)(l+2)}}\,\rw^{10}_{lm}\;.
\end{align*}

\chapter{The source terms of the particle}

The particle's energy-momentum tensor \eqref{part_tmn} will be
expanded using the orthonormal Mathews-Zerilli harmonics defined in
appendix A:
\begin{align}
	\label{expan}
	{\cal T}_{\mu\nu} \= \sum_{l = 0}^{\infty}
	\sum_{m = -l}^l \sum_{A = 1}^{10}
	\hat{t}^{lm}_{A} [\mz^A_{lm}]_{\mu\nu}\;.
\end{align}
By using the orthonormality condition \eqref{prod} we can compute the 
coefficients $\hat{t}^{lm}_A$ through
\begin{align}\label{coeff}
        \hat{t}_A^{lm} \= \int_{S^2}[\mz^A_{lm}]^{*\mu\nu}{\cal T}_{\mu\nu}
        \,{\rm d\Omega}\;.
\end{align}
We thus obtain the following set:
\begin{align*}
	\hat{t}_{1}^{lm} \= e^{4\nu}\frac{\mu}{r^2}\frac{dt}{d\tau}
	\delta(r-R(t))Y^*_{lm}\\
	\hat{t}_{2}^{lm} \= \mbox{i}\sqrt{2}\frac{\mu}{r^2}\frac{dR}{d\tau}
	\delta(r-R(t))Y^*_{lm}\\
	\hat{t}_{3}^{lm} \= e^{2\nu}\frac{2\mbox{i}\mu}{r\sqrt{2l(l+1)}}
	\delta(r-R(t))\frac{d}{d\tau}Y^*_{lm}\\
	\hat{t}_{4}^{lm} \= e^{2\nu}\frac{2\mu}{r\sqrt{2l(l+1)}}
	\delta(r-R(t))\(\frac{1}{\sin\Theta}\frac{d\Theta}{d\tau}
	\df{}{\Phi} - \sin\Theta\frac{d\Phi}{d\tau}\df{}{\Theta}\)
	Y^*_{lm}\\
	\hat{t}_{5}^{lm} \= e^{4\lambda}\frac{\mu}{r^2}
	\frac{d\tau}{dt}\(\frac{dR}{d\tau}\)^2
	\delta(r-R(t))Y^*_{lm}\\
	\hat{t}_{6}^{lm} \= e^{2\lambda}\frac{2\mu}{r\sqrt{2l(l+1)}}
	\frac{d\tau}{dt}\frac{dR}{d\tau}\delta(r-R(t))
	\frac{d}{d\tau}Y^*_{lm}\\
	\hat{t}_{7}^{lm} \= e^{2\lambda}\frac{2\mbox{i}\mu}{r\sqrt{2l(l+1)}}
	\frac{d\tau}{dt}\frac{dR}{d\tau}\delta(r-R(t))
	\(\sin\Theta\frac{d\Phi}{d\tau}\df{}{\Theta}-
	\frac{1}{\sin\Theta}\frac{d\Theta}{d\tau}\df{}{\Phi}
	\)Y^*_{lm}
\end{align*}
\begin{align*}
\begin{split}
	\hat{t}_{8}^{lm} \= \frac{\mu}{\sqrt{2l(l+1)(l-1)(l+2)}}
	\frac{d\tau}{dt}\delta(r-R(t))\\
	&\quad \left[\(\(\frac{d\Theta}{d\tau}\)^2
	- \sin^2\Theta\(\frac{d\Phi}{d\tau}\)^2\)
	W^*_{lm}
	+ 2\frac{d\Theta}{d\tau}\frac{d\Phi}{d\tau}
	X^*_{lm}\right]
\end{split}\\
	\hat{t}_{9}^{lm} \= \frac{\mu}{\sqrt{2}}\frac{d\tau}{dt}
	\delta(r-R(t))\(\(\frac{d\Theta}{d\tau}\)^2
	+ \sin^2\Theta\(\frac{d\Phi}{d\tau}\)^2\)
	Y^*_{lm}\\
\begin{split}
	\hat{t}_{10}^{lm} \= \frac{\mbox{i}\mu}{\sqrt{2l(l+1)(l-1)(l+2)}}
	\frac{d\tau}{dt}\delta(r-R(t))\sin\Theta\\
	&\quad \left[\(\(\frac{d\Phi}{d\tau}\)^2
	- \frac{1}{\sin^2\Theta}\(\frac{d\Theta}{d\tau}\)^2\)
	X^*_{lm}
	+ 2\frac{d\Theta}{d\tau}\frac{d\Phi}{d\tau}
	W^*_{lm}\right]\;.
\end{split}
\end{align*}
Here, $Y^*_{lm}, W^*_{lm}$ and $X^*_{lm}$ are functions of the
particle's angular position $\Theta$ and $\Phi$ parametized by the
coordinate time $t$, therefore all derivatives with respect to proper
time $\tau$ are to be understood as
\begin{align}
	\frac{d}{d\tau} = \frac{dt}{d\tau}\frac{d}{dt}\;.
\end{align}
Since for the decomposition of the evolution equations we use the
Regge-Wheeler harmonics $\rw^A_{lm}$, which do not form an orthonormal
set, we have to expand the energy-momentum tensor \eqref{part_tmn} in
Regge-Wheeler harmonics as well:
\begin{align}
	\label{expanRW}
	{\cal T}_{\mu\nu} \= \sum_{l = 0}^{\infty}
	\sum_{m = -l}^l \sum_{A = 1}^{10}
	t^{lm}_A [\rw^A_{lm}]_{\mu\nu}\;.
\end{align}
Of course, we cannot obtain the Regge-Wheeler coefficients $t^{lm}_A$
by means of a formula similar to \eqref{coeff}, but we can construct
them from the Mathews-Zerilli coefficients $\hat{t}_A^{lm}$. Using the
relationship between the different sets of tensor harmonics given in
Appendix A, we find:
\begin{align*}
	t_1^{lm} \= \hat{t}_{1}^{lm}\\
	t_2^{lm} \= \frac{\mbox{i}}{\sqrt{2}}\hat{t}_{2}^{lm}\\
	t_3^{lm} \= \frac{\mbox{i} r}{\sqrt{2l(l+1)}}\hat{t}_{3}^{lm}\\
	t_4^{lm} \= -\frac{r}{\sqrt{2l(l+1)}}\hat{t}_{4}^{lm}\\
	t_5^{lm} \= \hat{t}_{5}^{lm}\\
	t_6^{lm} \= \frac{r}{\sqrt{2l(l+1)}}\hat{t}_{6}^{lm}\\
	t_7^{lm} \= -\frac{\mbox{i} r}{\sqrt{2l(l+1)}}\hat{t}_{7}^{lm}
\end{align*}
\begin{align*}
	t_8^{lm} \= \frac{2r^2}
	{\sqrt{2l(l+1)(l-1)(l+2)}}\hat{t}_{8}^{lm}\\
	t_9^{lm} \= \frac{r^2}{\sqrt{2}}\(\hat{t}_{9}^{lm} + \sqrt{\frac{
	l(l+1)}{(l-1)(l+2)}}\hat{t}_{8}^{lm}\)\\
	t_{10}^{lm} \= -\frac{2\mbox{i} r^2}
	{\sqrt{2l(l+1)(l-1)(l+2)}}\hat{t}_{10}^{lm}\;.
\end{align*}
We now restrict the motion of the particle to the equatorial plane
$\Theta = \frac{\pi}{2}$. In this case it is $\frac{d\Theta}{d\tau} =
0$ and $\sin\Theta = 1$, and we can use the geodesic equations
\eqref{geo1} and \eqref{geo3} to substitute all expressions containing
derivatives with respect to proper time $\tau$. Furthermore, we can
obtain quite simple relations for the derivatives of $Y^*_{lm}$:
\begin{align}
	\df{}{\Phi}Y^*_{lm} \= -\mbox{i} mY^*_{lm}\\
	\dff{}{\Theta}Y^*_{lm} \= \(m^2 - l(l+1)\)Y^*_{lm}\;.
\end{align}
This then gives us a somewhat simpler set of coefficients:
%
\begin{align*}
	t_1^{lm} \= e^{2\nu}\frac{\mu E}{r^2}\delta(r-R(t))
	Y^*_{lm}\\
	t_2^{lm} \= -e^{2\lambda}\frac{\mu E}{r^2}
	v(t)\delta(r-R(t))Y^*_{lm}\\
	t_3^{lm} \= e^{2\nu}\frac{\mbox{i} m\mu L}{r^2l(l+1)}\delta(r-R(t))
	Y^*_{lm}\\
	t_4^{lm} \= e^{2\nu}\frac{\mu L}{r^2l(l+1)}\delta(r-R(t))
	\df{}{\Theta}Y^*_{lm}\\
	t_5^{lm} \= e^{6\lambda}\frac{\mu E}{r^2}v^2(t)\delta(r-R(t))
	Y^*_{lm}\\
	t_6^{lm} \= -e^{2\lambda}\frac{\mbox{i} m\mu L}{r^2l(l+1)}v(t)
	\delta(r-R(t))Y^*_{lm}\\
	t_7^{lm} \= e^{2\lambda}\frac{\mu L}{r^2l(l+1)}v(t)
	\delta(r-R(t))\df{}{\Theta}Y^*_{lm}\\
	t_8^{lm} \= e^{2\nu}\frac{\mu L^2(l(l+1) - 2m^2)}
	{r^2El(l+1)(l-1)(l+2)}\delta(r-R(t))Y^*_{lm}\\
	t_9^{lm} \= e^{2\nu}\frac{\mu L^2(l(l+1) - m^2 - 1)}{r^2E(l-1)(l+2)}
	\delta(r-R(t))Y^*_{lm}\\
	t_{10}^{lm} \= -e^{2\nu}\frac{2\mbox{i} m\mu L^2}
	{r^2El(l+1)(l-1)(l+2)}\delta(r-R(t))
	\df{}{\Theta}Y^*_{lm}\;,
\end{align*}
where $v(t) = \frac{dR}{dt}$ is the radial velocity of the particle.
The field equations also require the computation of the trace 
\begin{align}
	{\cal T} \= g^{\mu\nu}{\cal T}_{\mu\nu} \;=\; \sum_{l,m}t^{lm}Y_{lm}\;,
\end{align}
with 
\begin{align}
	t^{lm} \= -e^{2\lambda}t_1^{lm}+ e^{-2\lambda}t_5^{lm}
	- \frac{l(l+1)}{r^2}t_8^{lm} + \frac{2}{r^2}t_9^{lm}\;.
\end{align}
Using the explicit forms of the coefficients, we obtain
\begin{align}
	t^{lm} \= \frac{\mu}{r^2}\delta(r-R(t))
	\(Ev^2(t)e^{4\lambda} - E + e^{2\nu}\frac{L^2}{r^2E}\)
	Y^*_{lm}\;,
\end{align}
and by making use of the geodesic equation \eqref{geo2} we can reduce
this expression to
\begin{align}
        t^{lm} \= -e^{2\nu}\frac{\mu}{r^2E}\delta(r-R(t))
        Y^*_{lm}\;.
\end{align}

\chapter{Derivation of the radiation extraction formula}
The radiated energy emitted per unit time and unit angle is given by 
\cite{LL}
\begin{align}
	\frac{d^2E}{dt\,d\Omega} &= \frac{r^2}{16\pi}\left[
	\(\df{}{t}h_{[\theta][\phi]}\)^2
	+ \frac{1}{4}\(\df{}{t}h_{[\theta][\theta]}	
	- \df{}{t}h_{[\phi][\phi]}\)^2\right]\;,
\end{align}
where 
\begin{align}
	h_{[i][j]} \;&:=\; \sqrt{g^{ii}g^{jj}}\,h_{ij}\;,
\end{align}
or specifically
\begin{align}
	h_{[\theta][\phi]} &= \frac{1}{r^2\sin\theta}h_{\theta\phi}\\
	h_{[\theta][\theta]} &= \frac{1}{r^2}h_{\theta\theta}\\
	h_{[\phi][\phi]} &= \frac{1}{r^2\sin^2\theta}h_{\phi\phi}\;.
\end{align}
The metric components $h_{[i][j]}$ have to be in the radiation gauge,
which asserts that only
\begin{align}
	h_{[\theta][\phi]} &= {\cal O}(1/r)\\
	h_{[\theta][\theta]} - h_{[\phi][\phi]} &= {\cal O}(1/r)\;,
\end{align}
whereas all other components must fall off faster than
${\cal O}(1/r)$. Furthermore, it must hold that $h_{[\theta][\theta]} =
-h_{[\phi][\phi]} + {\cal O}(1/r^2)$. We now expand the angular parts of the
metric into the orthonormal Mathews-Zerilli tensor harmonics given in
Appendix A. Since we focus on the polar perturbations, we only need to
consider $\mz_8^{lm}$ and $\mz_9^{lm}$, and we have
\begin{align}
	h_{[i][j]} &= \sum_{l,m}
	\( \begin{array}{cc}
	\(\frac{1}{\sqrt{2}}{\wh K}_{lm}Y_{lm}
	+ \frac{1}{N}{\wh G}_{lm}W_{lm}\)
	& \frac{1}{N\sin\theta}{\wh G}_{lm}X_{lm}\\
	\frac{1}{N\sin\theta}{\wh G}_{lm}X_{lm}
	& \(\frac{1}{\sqrt{2}}{\wh K}_{lm}Y_{lm}
	- \frac{1}{N}{\wh G}_{lm}W_{lm}\)\\
	\end{array} \)\;,
\end{align}
where we use
\begin{align}
	N := \sqrt{2l(l+1)(l-1)(l+2)}\;.
\end{align}
This gives us
\begin{align}
	\frac{d^2E}{dtd\Omega} &= \frac{r^2}{16\pi}\(
	\abs{\sum_{l,m}\frac{\dot{\wh G}_{lm}}{N^2\sin^2\theta}X_{lm}}^2
	+ \abs{\sum_{l,m}\frac{\dot{\wh G}_{lm}}{N^2}W_{lm}}^2\)\;,
\end{align}
where we have to take the square moduli because we are dealing with
complex metric coefficients. To obtain the total energy flux, we have
to integrate over a 2-sphere
\begin{align}
	\frac{dE}{dt} &= \frac{r^2}{16\pi}
	\sum_{l,m,l',m'}\frac{1}{N^2}\dot{\wh G}_{lm}\dot{\wh G}_{*l'm'}
	\int_{S^2}\(\frac{X_{lm}X^*_{l'm'}}{\sin^2\theta}
	+ W_{lm}W^*_{l'm'}\)d\Omega\;.
\end{align}
Now, because of the orthonormality relation the integral is just
\begin{align}
	\int_{S^2}\(\frac{X_{lm}X^*_{l'm'}}{\sin^2\theta}
	+ W_{lm}W^*_{l'm'}\)d\Omega &= \frac{1}{2}N^2
	\delta_{ll'}\delta_{mm'}\;,
\end{align}
and we obtain
\begin{align}\label{E1}
	\frac{dE}{dt} &= \frac{r^2}{32\pi}\sum_{l,m}|\dot{\wh G}_{lm}|^2\;.
\end{align}
Since the actual metric perturbations are derived using the expansion
into the Regge-Wheeler harmonics, we have to express \eqref{E1} in
terms of those variables. The relation between $\wh G_{lm}$ and the
Regge-Wheeler variable $G_{lm}$ is easily derived to be
\begin{align}
	\wh G_{lm} &= \frac{N}{2}G_{lm}\;,
\end{align}
which leads to the following radiation formula
\begin{align}\label{E2}
	\frac{dE}{dt} &= \frac{r^2}{64\pi}\sum_{l,m}l(l+1)(l-1)(l+2)
	|\dot{G}_{lm}|^2\;.
\end{align}
Note, however, that $G_{lm}$ is still in the radiation gauge and not
in the Regge-Wheeler gauge. In the Regge-Wheeler gauge we have $G_{lm}
= 0$, and the above formula would not make any sense. We now have to
devise a way which relates $G_{lm}$ to the variables in the
Regge-Wheeler gauge. This can be accomplished as follows. Following
Moncrief \cite{Mon74a}, we construct some gauge invariant quantities, in
particular the Zerilli function, and examine their asymptotic
behavior. We then will show that the Zerilli function has the same
asymptotic behavior as $G_{lm}$ in the radiation gauge. Hence, the
radiated power is proportional to the time derivative of the Zerilli
function, which can be easily constructed from the metric variables in
the Regge-Wheeler gauge.

Let us write the general expansion of the even parity perturbations of
the spacial part of the metric in terms of Regge-Wheeler harmonics
(Here we use the notation of Moncrief and focus on fixed values $l$
and $m$:)
\begin{align}
	h_{rr} &= e^{2\lambda}H_2Y_{lm}\\
	h_{r\theta} &= h_1\df{Y_{lm}}{\theta},\quad
	h_{r\phi} \;=\; h_1\df{Y_{lm}}{\phi}\\
	h_{\theta\theta} &= r^2\(KY_{lm} + G\dff{Y_{lm}}{\theta}\)\\
	h_{\theta\phi} &= \frac{r^2}{2}G X_{lm}\\
	h_{\phi\phi} &= r^2\sin^2\theta\(KY_{lm}
	- G\(l(l+1)Y_{lm} + \dff{Y_{lm}}{\theta}\)\)\;.
\end{align}
We can then construct the following gauge invariant quantity
\begin{align}
	q_1 := 4re^{-4\lambda}k_2 + rl(l+1)k_1\;,
\end{align}
where 
\begin{align}
	k_1 &= K + e^{-2\lambda}\(rG' - \frac{2}{r}h_1\)
\end{align}
and
\begin{align}
	k_2 &= \frac{1}{2}\(e^{2\lambda}H_2 - e^{\lambda}\(re^{\lambda}
	K\)'\)\;.
\end{align}
We now have to determine the asymptotic behavior of $q_1$. To do so, we
have to know the asymptotic behavior of the metric coefficients $h_1$,
$H_2$, $K$ and $G$, which, of course, will depend on the chosen
gauge. However, because of its gauge invariance, $q_1$ will always
have the same asymptotic behavior, regardless in what gauge the metric
variables are.

Therefore we assume our metric variables to be in the radiation
gauge, where we know the asymptotic behavior. Since
$h_{[\theta][\theta]} = -h_{[\phi][\phi]} + {\cal O}(1/r^2)$, the
sum $h_{[\theta][\theta]} + h_{[\phi][\phi]}$ has to be of
order ${\cal O}(1/r^2)$, i.e.
\begin{align}
	h_{[\theta][\theta]} + h_{[\phi][\phi]} &= \(2K - l(l+1)G\)Y_{lm}
	\;=\;{\cal O}(1/r^2)\;.
\end{align}
But this can only hold if
\begin{align}
	2K &= l(l+1)G + {\cal O}(1/r^2)
\end{align}
for the leading order. Using this relation we find for $k_1$ and $k_2$
\begin{align}
	k_1 \;&\sim\; K + rG' \;\sim\; \frac{1}{2} l(l+1)G + rG'\\
	k_2 \;&\sim\; -\frac{1}{2}K - \frac{r}{2}K'
	\;\sim\; -\frac{1}{4}l(l+1)\(G + rG'\)\;,
\end{align}
where ``a $\sim$ b'' means $a$ equals $b$ plus higher order
terms. This leaves us with
\begin{align}
       	q_1 \;&\sim\; \frac{r}{2}l(l+1)\(l(l+1) - 2\)G\;.
\end{align}
Moncrief finally defines the quantity
\begin{align}
	Q \;&:=\; \frac{q_1}{l(l+1) - 2 + \frac{2M}{r}}\;,
\end{align}
which satisfies the Zerilli equation. The leading order for $Q$ is then
\begin{align}
	Q \;&\sim\; \frac{r}{2}l(l+1)G\;,
\end{align}
which is equivalent to 
\begin{align}\label{QG}
	G \;&\sim\; \frac{2}{rl(l+1)}Q\;.
\end{align}
This relation only holds if $G$ is given in the radiation
gauge. But in the radiation formula \eqref{E2} this is the case and
we may substitute $G$ by means of \eqref{QG}, which yields
\begin{align}\label{E3}
	\frac{dE}{dt} &= \frac{1}{16\pi}\sum_{l,m}\frac{(l-1)(l+2)}{l(l+1)}
	|\dot{Q}_{lm}|^2\;.
\end{align}
We now have to express $Q$ in terms of our actual perturbation
variables $S$ and $T$, which are in the Regge-Wheeler gauge. We first
recall that in the Regge-Wheeler gauge, it is 
\begin{align}
	h_1 &= 0\\
	H_2 &= \frac{T}{r} + rS\\
	K &= \frac{T}{r}\\
	G &= 0\;.
\end{align}
Then we compute $k_1$ and $k_2$, which is readily done:
\begin{align}
	k_1 &= \frac{T}{r}\\
	k_2 &= \frac{1}{2}\(e^{2\lambda}\(\frac{T}{r} + rS\)
	 - e^{\lambda}\(e^{\lambda}T\)'\)\;.
\end{align}
Next is $q_1$
\begin{align}
	q_1 &= 2re^{-4\lambda}\(e^{2\lambda}\(\frac{T}{r} + rS\)
	- e^{\lambda}\(e^{\lambda}T\)'\) + l(l+1)T\\
	&= 2e^{-2\lambda}\(\(1 - r\lambda'\)T + r^2S - T'\) + l(l+1)T\;.
\end{align}
Using
\begin{align}
	\lambda' = \frac{M}{r\(2M - r\)}
\end{align}
and collecting terms yields
\begin{align}
	q_1 &= -\(1 - \frac{2M}{r}\)\(2rT'
	+ \frac{2M - r\(2 + l(l+1)\)}{r - 2M}T - 2r^2S\)\;.
\end{align}
If we furthermore define $Z$ to be 
\begin{align}
	Z &= \frac{2}{l(l+1)}Q \;=\; \frac{2}{l(l+1)}
	\frac{q_1}{l(l+1) - 2 + \frac{2M}{r}}\;,
\end{align}
we find from \eqref{QG} that
\begin{align}
	G \;&\sim\; \frac{Z}{r}\;,
\end{align}
and the radiation formula \eqref{E3} finally reads
\begin{align}
	\frac{dE}{dt} &= \frac{1}{64\pi}\sum_{l,m} l(l+1)(l-1)(l+2)
	|\dot{Z}_{lm}|^2 = \frac{1}{64\pi}\sum_{l,m}\frac{(l+2)!}{(l-2)!}
	|\dot{Z}_{lm}|^2\;,
\end{align}
with 
\begin{align}
	\!\!\!\!\!Z &= -\frac{2}{l(l+1)}
	\frac{1 - \frac{2M}{r}}{l(l+1) - 2 + \frac{6M}{r}}
	\(2rT' + \frac{2M - r\(2 + l(l+1)\)}{r - 2M}T - 2r^2S\)\;,
\end{align}
which agrees with our definition of the Zerilli function in \eqref{Z}.

\end{appendix}
\newpage
\appendix
\addcontentsline{toc}{chapter}{Bibliography}
\hypertarget{bib}{}

\newpage
\thispagestyle{plain}
\addcontentsline{toc}{chapter}{Danksagung}
\hypertarget{dank}{}
\section*{Danksagung}

\sloppy
Ich danke allen Institutsmitgliedern, voran unseren Systemies, die
bereit waren, sich f\"ur meine Computer- und Druckerprobleme Zeit zu
nehmen und mir dabei weiterzuhelfen.

Ganz herzlich m\"ochte ich mich bei Prof.~Dr.~Ruder bedanken, der es
mir nicht nur durch eine Stelle beim SFB 382 erm\"oglicht hat, diese
Doktorarbeit anzufertigen, sondern auch nie einen Dienstreisewunsch
abgeschlagen hat, und sei es bis nach Indien gewesen. Auch danke ich
Priv.~Doz.~Dr.~Harald Riffert, der ein hilfsbereiter Ansprechpartner
war und stets ein Ohr f\"ur Probleme meinerseits hatte. Des weiteren
sei Dr.~Hans-Peter Nollert erw\"ahnt, der geduldig die verschiedenen
Versionen dieser Arbeit Korrektur gelesen und mit seinen Tipps und
Kommentaren sehr zu ihrer Vollendung beigetragen hat.

Leider kann ich mich nicht mehr pers\"onlich bei meinem anf\"anglichen
Betreuer Prof.~Dr.~Heinz Herold bedanken, zu dem ich, bevor er seiner
langen Krankheit erlag, jederzeit mit meinen Anliegen kommen konnte,
und der mir in vielen Dingen weitergeholfen hat. Heinz, Du fehlst!

Besonderer Dank gilt dem DAAD, da er mir durch ein HSPIII-Stipendium
einen Forschungs\-aufenthalt in den USA an der Penn State University
erm\"oglicht hat.

I thank Pablo Laguna for being my adviser at Penn State University,
and for always being open for my questions even at the morning hours,
where nobody was supposed to disturb him.

Without having had Michael Sipior as a reliable drinking buddy I would
have saved lots of money and would have had much more evenings to do
some useless work. Prost Mike!

I also wish to thank Kostas Kokkotas who transformed my suggstion for
a short business trip to Greece into a two year Postdoc fellowship,
which will give me the possibility to delve into the investigation of
oscillations of rotating neutron stars.
$\mathnormal{E\hspace*{-1pt}v\hspace*{-1pt}\chi\hspace*{-1pt}
\alpha\hspace*{-1pt}\rho\hspace*{-1pt}\iota\hspace*{-1pt}\sigma
\hspace*{-1pt}\tau\hspace*{-1pt}\acute{\omega}
\;\pi \hspace*{-1pt}o\hspace*{-1pt}\lambda\hspace*{-1pt}\acute{v}}.$
I am looking forward to this new collaboration.

Mein herzlichster Dank geht an meine allerliebste Freundin Sibylla,
die die Hochs und Tiefs der letzten Jahre, die nicht nur mit dieser
Doktorarbeit verbunden waren, geduldig ertragen hat, und die in
guten wie in schlechten Zeiten stets zu mir gehalten hat.

Ein weiteres Dankesch\"on geht an den Ex-Tatler Uwe Fischer, der
meine CD-Sammlung um einige gute Aufnahmen bereichert hat.

Zu guterletzt m\"ochte ich mich bei meinen Eltern bedanken, die mir
ein recht sorgenfreies Physikstudium samt einj\"ahrigem
USA-Aufenthalt in Oregon erm\"oglicht haben. Besonders danke ich
meinem Vater, der sicherlich nicht geglaubt hat, da\ss er in seinem
wohlverdienten Ruhestand nochmals eine Englischarbeit zu korrigieren
h\"atte.


\newpage
\end{document}


\begin{center}
{\Large \bf Papers on Radial Oscillations of Neutron Stars}\\[1cm]
\end{center}

\normalsize{\bf Dynamical instability of gaseous masses approaching
the \ss~limit in \gr.} --- S. Chandrasekhar, \prl{12}{114}{1964}\\

\normalsize{\bf The dynamical instability of gaseous masses approaching
the \ss~limit in \gr.} --- S. Chandrasekhar, \apj{140}{417}{1964}

\abs In this paper the theory of the infinitesimal, baryon-number
conserving, adiabatic, radial oscillations of a gas sphere is
developed in the framework of \gr. A variational base for determining
the characteristic \freqs~of \oscs~is established. It provides a
convenient method for obtaining sufficient conditions for the
occurence of dynamical instability. The principal result of the
analysis is the demonstration that the Newtonian lower limit 4/3, for
the ratio of the specific heats $\gamma$, for insuring dynamical
stability is increased by effects arising from \gr; indeed, is
increased to an extent that, so long as $\gamma$ is finite, dynamical
instability will intervene before a mass contracts to the limiting
radius $(\ge 2.25 GM/c^2)$ compatible with hydrostatic
equilibrium. Moreover, if $\gamma$ should exceed 4/3 only by a small
amount, then dynamical instability will occur if the mass should
contract to the radius
\begin{eqnarray*}
	R_c = \frac{K}{\gamma - 4/3}\frac{2GM}{c^2}
	\qquad (\gamma \rightarrow 4/3)
\end{eqnarray*}
where $K$ is a constant depending, principally, on the density distribution
in the configuration. The value of the constant $K$ is explicetly 
evaluated for the homogeneous sphere of constant energy density and
the polytropes of indices $n = 1,2,$ and 3.\\[5mm]

\normalsize{\bf A catalogue of methods for studying the normal modes
of radial \pul~of general-\rel~\sms} --- James M. Bardeen, Kip. S. Thorne
and David W. Meltzer, \apj{145}{508}{1966}

\abs A catalogue is presented of various methods now available for
studying the normal modes of radial \pul~of stars in \gr. The methods
are classified according to the normal-mode properties which they
investigate; and the relative merits of the methods are
discussed.\\[5mm]

\normalsize{\bf Normal modes of radial \pul~of stars at the end point
of thermonuclear evolution} --- David W. Meltzer and Kip. S. Thorne,
\apj{145}{514}{1966}

\abs The lowest three normal modes of radial pulsation are studied
for stellar configurations at the end point of thermonuclear
evolution.  Two alternative \eoss~are assumed -- the Harrison-Wheeler
\eos, and the Skyrme-Cameron-Saakyan \eos. The adiabatic index is
discussed for each \eos, taking into account the extend to which
period changes in nuclear composition are produced by stellar
\pul. The periods and $e$-folding times of the lowest three normal
modes of a wide range of configurations are calculated to an accuracy
of $\pm1$ per cent. The eigenfunctions of some of the normal modes
and the \pul~energy which they can store are also calculated; and a
discussion is presented of the damping of neutron-star \puls~and of
the consequent heating of the star. [...]\\[5mm]

\normalsize{\bf Radial \oscs~of zero-temperature white dwarfs and
\nss~below nuclear densities} --- G. Chanmugam, \apj{217}{799}{1977}

\abs The radial \oscs~of zero-temperature degenerate stars with subnuclear
densities have been calculated for the fundamental and first two excited
modes using recent \eoss. The \calcs~were made under the assumption that
$(a)$ the dynamical time scale $\tau \gg \tau_N$, where $\tau_N$ is the 
nuclear relaxation time, in which case the adiabatic index $\gamma =
\gamma_E$ (calculated along the \eos) and $(b)$ $\tau \ll \tau_N$ with
$\gamma = \gamma_C$ calculated at constant composition. For case $a$, it
is shown that white dwarfs are dynamically stable for $\rho_c < 1.2\times
10^9$g cm$^{-3}$ while \nss~are dynamically stable for $\rho_c > 10^{14}$
g cm$^{-3}$. The instability for white dwarfs in case $a$ arises from
principally from the phase transitions than from the effects of \gr,
confirming the conjecture of Baym et al. Low-density \nss~have for the
fundamental (unstable) mode ``$e$-folding'' times with a factor 100 less
than the typically expected dynamical time scale. It is shown that this
is consistent with the unusual decay of the radial eigenfunction, a 
qualitative explanation of which is given. For denstities below the
neutron drip point, it is shown that $\gamma_C = \gamma_E$ except at
a finite number of phase transition points. For densities just above
the neutron drip point, it is shown analytically that $\gamma_C$ must
be greater that 4/3 in the critical region where $\gamma_E$ drops
below 4/3. For case $b$, white dwarfs are dynamically stable for
$\rho_c < 4\times 10^{10}$ g cm$^{-3}$, while \nss~ are dynamically
stable for $\rho_c > 7\times 10^{12}$ g cm$^{-3}$ as had been suggested
previously by the author. Case $b$ is shown to be more appropriate for
infinitesimal \perts. It is also pointed out that arguments given earlier
to indicate that high-density \wds~and and low-density \nss~may not exist,
because of instabilities over long time scales, are lacking rigor.\\[5mm]

\normalsize{\bf The radial \oscs~of \nss} --- Edward N. Glass and
Lee Lindblom, \apj{53}{93}{1983} + {\bf Erratum},\apjs{71}{173}{1989}

\abs We continue our investigation of the influence of the \eos~of matter
(in the nuclear density regime) on the dynamical behavior of \nss.
The properties of the two lowest \freq~radial \osc~modes of \nss,
constructed from 13 \eoss, are presented. We find that the \freqs~ of
both radial \osc~mods studied here are generally higher than the
frequency of the corresponding \quad~$f$ mode.\\[5mm]

\normalsize{\bf Radial \oscs~of \nss~and strange stars} --- H.M. V\"ath and
G. Chanmugam, \aa{260}{250}{1992}

\abs We have determined the two lowest frequency radial \osc~modes 
constructed for several \eoss. The results for \nss~differ from those
of Glass \& Lindblom (1983). We find that the central density where
$\omega^2_0 = 0$ agrees in our calculations with that where the star
reaches its maximal mass, as expected, but contrary to their calculations.
The \osc~\freqs~of strange stars have qualitatively a different dependence
on the central density compared with the case of \nss, as the periods
of all modes go the zero when the central density of the strange star
approaches its smallest possible value.\\[5mm]

\normalsize{\bf Radial modes of rotating \nss~in the Chandrasekhar-Friedman 
formalism} --- B. Datta, S.S. Hasan, P.K. Sahu, A.R. Prasanna,
\ijmpd{7}{49}{1998}

\abs Eigen\freqs~of radial \puls~of ``slowly'' rotating \nss~are
calculated in a general \rel~formalism given by Chandrasekhar \& Friedman.
It is found that the square of the \freqs~ are always a decreasing 
function of the cental density of the neutron star. The decrease of the
squared \freq~is sensitive to the \eos~of \ns~matter, and is 
illustrated using realistic models.\\[5mm]

\normalsize{\bf Radial \puls~and stability of protoneutron stars}
--- D. Gondek, P. Haensel, and J.L. Zdunik, \aa{325}{217}{1998}

\abs Radial \puls~of newborn \nss~(protoneutron stars) are studied
for a range of internal temperatures and entropies per baryon predicted
by the existing numerical simulations. Protoneutron star models are
constructed using a realistic \eos~of hot dense matter, and under
various assumptions concerning stellar interior (large trapped lepton
number, zero trapped lepton number, isentropic, isothermal). Under
prevailing conditions, linear \oscs~of a protoneutron star can be treated
as adiabatic, and evolutionary effects can be neglected on dynamic
timescale. The dynamic behavior is governed by the adiabatic index, which
in turn depends on the physical state of the stellar interior. The
eigenfrequencies of the lowest radial modes of linear, adiabatic pulsations
are calculated. Stability of protoneutron stars with respect to small
radial perturbations is studied, and the validity of the static criteria
is discussed.\\[5mm]

\begin{center}
{\Large \bf Papers on Dipole Oscillations of Neutron Stars}\\[1cm]
\end{center}

{\normalsize \bf Non-Radial pulsation of general-\rel~\sms.
V. Analytic analysis for $l = 1$}
--- Alfonso Campolattaro, Kip S. Thorne, \apj{159}{847}{1970}

\abs The theory of small, adiabatic, dipole \perts~of a star away from
hydrostatic equilibrium is developed within the framework of \gr.  The
analysis is linearized in the \pert~amplitudes. The ``\op''
\perts~describe differential rotation, while the ``\ep''
\perts~describe \pul. In the \pul~case there are two degrees of freedom
associated with the fluid motion and {\em no} degrees of freedom in
the \grav~field (no dipole \gws). During the \pul~the star's external
\grav~field is completely unperturbed, to first order in the
amplitude, despite the fact that its surface performs a finite
back-and-forth motion. In both the \pul~and the rotational cases, the
Einstein field \eqs~are put into simple forms suitable for numerical
integration.\\[5mm]

\normalsize{\bf A \vp~and a stability criterion for the dipole modes
of \puls~of \sms~in \gr}
--- Steven L. Detweiler, \apj{201}{440}{1975}

\abs A \vp~is found for the \freqs~of the dipole modes of \puls~of
spherical \sms~in \gr. This principle reveals that if the
\rel~\ss~discriminant, $S$, is not positive definite, then a
\sm~is unstable to dipole \perts. In addition, if $S$ is positive
definite the eigen\freqs~of all of the normal modes are real.\\[5mm]

\normalsize{\bf The dipole \oscs~of general \rel~\nss}
--- Lee Lindblom, Randall J. Splinter, \apj{345}{925}{1989}

\abs We investigate the theory of the dipole \oscs~of fully \rel~\nss.
The \eqs~governing these modes are reduced to a third-order system;
and the \vp~for the \freqs~of these modes is reduced to a form involving
only two independent functions. The \eqs~that determine the damping
of these modes, which is due to viscous dissipation, are presented.
The \freqs~and viscous damping times for a range of realistic \ns~models
based on a number of \eoss~and a range of masses are determined by
solving these dipole \oscs~\eqs numerically.\\[5mm]

\normalsize{\bf The accuracy of the \rel~Cowling \app}
--- Lee Lindblom, Randall J. Splinter, \apj{348}{198}{1990}

\abs We investigate the accuracy of two versions of the \rel~Cowling
\app~for the \oscs~of general \rel~\sms: one by McDermott, Van Horn
and Scholl, and a refinement by Finn. The \osc~\freqs~and the
viscous damping time scales of several of the lowest \freq~dipole
$p$-modes in the \apps~are compared to the values for the quantities.
We find, as expected, that the accuracy of the Cowling \app~improves
as the number of nodes in the eigenfunctions of the mode increases.
Furthermore, the McDermott, Van Horn and Scholl version of the Cowling
\app~is more accurate than Finn's refined version for the low-order
$p$-modes studied here.\\[5mm]

\begin{center}
{\Large \bf Papers on Nonradial Oscillations of Neutron Stars}\\[1cm]
\end{center}

\normalsize{\bf Non-radial \pul~of general-\rel~\sms.
I. Analytic analysis for $l \ge 2$} --- Kip S. Thorne, Alfonso
Campolattaro, \apj{149}{591}{1967} + Erratum \apj{152}{673}{1968}.

\abs The theory of small, adiabatic, \nr~perturbations of a star away
from hydrostatic equilibrium is developed within the framework of
general relativity. The unperturbed equilibrium configuration is an
arbitrary, non-rotating, general-\rel~\sm. The departures from
equilibrium are analyzed into tensor spherical harmonics and then into
complex normal modes with various mixtures of incoming and outgoing
\gws. A discussion is given of the expansion of real, physical \puls~with
purely outgoing \grav~radiation in terms of the complex normal
modes. Criteria are developed for stability against \nr~\puls; and
methods are devised for computing numerically the \pul~\freqs,
eigenfunction, and \grav-\rad~damping times of the stable, real
\qnms~of \pul.\\[5mm]

\normalsize{\bf Gravitational \rad~damping}
--- Kip S. Thorne,\prl{21}{320}{1968}

\abs An analysis in presented of the emission of \gws~by
a star in \nr~\pul. This analysis, rigorous to first order
in the \pul~amplitude and to all orders in $G$ and $c$, exhibits
damping of the \puls~as the waves are emitted (\grav~radiation
reaction).\\[5mm]

\normalsize{\bf Non-radial \pul~of general-\rel~\sms.
II. Properties of the \gws}
--- Richard Price, Kip S. Thorne, \apj{155}{163}{1969}

\abs An analysis in presented of the properties of the
\gws~emitted by a fully \rel, non-rotating stellar
model, which is pulsating in a \nr~manner. Attention is focused
on {\it quasi-normal-mode \puls~} in which the star is excited at
some initial moment of time, it radiates \gws~thereafter, and
these waves gradually damp its nearly periodic \puls. For such
\qnms~a discussion is presented of the polarization of
the \gws~and of the energy, momentum, and angular momentum
which the waves transport.\\[5mm]

\normalsize{\bf Non-radial \pul~of general-\rel~\sms.
III. Analytic and numerical results for \nss}
--- Kip S. Thorne, \apj{158}{1}{1969}

\abs Numerical techniques are developed for calculating the
properties of the \nr, quasi-normal \puls~of fully \rel~stellar
models. These techniques are then applied to calculate, for four
realistic neutron-star models, the following characteristics of the
lowest few \quad~$p$-type modes: the \pul~periods ($\sim 10^{-3}$
sec), the \grav-radiation damping times ($\sim 1$sec), the
\pul~energies ($\sim 10^{52}$ ergs), the power emitted in \grav~waves
($\sim 10^{52}$ ergs sec$^{-1}$), and the shapes of the fluid
motions. Analytic techniques are used to show that many, and perhaps
all, of the $g$-type modes of a \ns~have zero \freq, if thermal
pressures are negligible compared with nuclear pressures.\\[5mm]

\normalsize{\bf Non-radial \pul~of general-\rel~\sms.
IV. The weak-field limit}
--- Kip S. Thorne, \apj{158}{997}{1969}

\abs In previous papers of this series techniques were developed for
calculating the properties of the \nr, \qn~\puls~of fully \rel~\sms.
In this paper those techniques are reexamined in the weak-field,
nearly Newtonian limit, and are found to predict an emission of
\grav~\rad at a rate which agrees with the standard weak-field \app~to
\gr. The present analysis also exhibits explicitly the
\rad-reaction forces inside the star and shows that they sap \pul~energy
from the star at the same rate as the waves carry it off. Following a
train of ideas due to William L. Burke, the analysis is generalized to
any weak-field, slow-motion system; and the relationship of the
analysis to the post-Newtonian expansion schemes of Chandrasekhar and
Fock is discussed. It is shown that the dominant \rad-reaction forces
associated with each changing multipole moment can be incorporated
into Newtonian theory by a simple change in the \bc~for the Newtonian
potential at radial infinity.\\[5mm]

\normalsize{\bf Non-radial \pul~of general-\rel~\sms.
VI. Corrections}
--- James R. Ipser, Kip S. Thorne, \apj{181}{181}{1973}

\abs The \eqs~governing the \nr~\puls~of general-\rel~\sms are shown
to form a fourth-order system, rather than a fifth-order system as
was believed previously. Incorrect statements about the application
of \bcs~at the stellar surface are also rectified. The errors corrected
here do not invalidate any of the final results obtained in previous
papers in this series.\\[5mm]

\normalsize{\bf A \vp~and a stability criterion for the
\nr~modes of \puls~of \sms~in \gr}
--- Steven L. Detweiler, James R. Ipser, \apj{185}{685}{1973}

\abs Small \nr~perturbations of spherical \sms~in general relativity
are studied. A \vp~for the normal modes of \nr~\puls with $l \ge 2$ is
derived. In this principle, an expression for the squared \freq~is
stationary with respect to arbitrary variations of three independent
dynamical variables if and only if the variations are carried out
about the eigenfunctions of a normal mode. The stationary value of the
\freq~is the associated eigen\freq. For normal modes with purely
outgoing radiation, the \vp~is modified to yield an approximate scheme
for computing first the real part and then the imaginary part of an
outgoing eigen\freq~with small imaginary part. Associated with the
\vp~is an energy-like quantity that obeys a conservation law. It is
shown that a model has no neutral, or zero-\freq, modes if the density
decreases outwards monotonically and if the \rel~\ss~discriminant is
nonzero. If the discriminant is actually positive, the outgoing normal
modes are stable.\\[5mm]

\normalsize{\bf Gravitational \perts~of spherically symmetric systems. I. 
the exterior problem} --- Vincent Moncrief, \aop{88}{323}{1974}

\abs Gravitational \perts~of the \ss~metric are treated from
a point of view which is adapted, in a natural way, to the gauge group
of the perturbed Einstein \eqs. The metric \perts~are explicitly
decomposed into their gauge invariant, gauge dependent and constrained
parts and a \vp~for the \pert~\eqs~is derived. The Regge-Wheeler and
Zerilli \eqs~are rederived and shown to have a gauge invariant 
significance. The Hamiltonian for the \perts~is constructed and
used to discuss the stability properties of the \ss~black
hole.\\[5mm]

\normalsize{\bf Gravitational \perts~of spherically symmetric systems. II. 
Perfect fluid interiors} --- Vincent Moncrief, \aop{88}{343}{1974}

\abs Methods developed in a previous paper on \perts~of the
\ss~metric are here extended to the treatment of \perts~of
\pf~\sms. The \perts~of a \pf~sphere are explicitly decomposed
into their gauge invariant and gauge dependent parts and a \vp~for the
\pert~equations is derived. The Hamiltonian for the \perts~is constructed
and a sufficient condition for stability against \nr, radiative
\perts~is derived from it. The stability criterion is applied to two
interesting classes of \sms, polytropic white dwarfs models and
high-density \ns~cores with pressure proportional to energy
density.\\[5mm]

\normalsize{\bf A variational \calc~of the fundamental \freqs~of
\quad~\puls~of fluid spheres in \gr}
--- Steven L. Detweiler, \apj{197}{203}{1975}

\abs A \vp~is used to calculate, in a relatively simple manner, the
\freqs~of \quad~\oscs~of \rel~\ns~models for a variety of \eoss~and 
central densities. Typical periods and damping times are on the order
of milliseconds and seconds, respectively, in agreement with earlier
estimates obtained by Thorne in a different manner. For spheres of
incompressible fluids, the analysis indicates that the fundamental
\quad~mode becomes unstable as $2M/R$ attains its limiting value
of $8/9$.\\[5mm]

\normalsize{\bf Torsional \oscs~of \nss}
--- Bonny L. Schumaker, Kip S.Thorne, \mnras{203}{457}{1983}

\abs Motivated by the possibility that torsional \oscs~of \nss~may be 
observable in the timing of pulsar subpulses and/or in future
\grav-wave detectors, this paper develops the detailed mathematical
theory of such torsional \oscs~and of the \gws~they emit. The
\oscs~are analysed using the formulation of first-order \perts~of a
fully general \rel~spherical \sm. All sources of damping are ignored
except \grav~\rad~reaction. The \perts~are resolved into spherical
harmonics, which decouple from each other. For each harmonic this
paper presents \eqs~of motion, an action principle, an energy
conservation law and a Liapunow-type proof that the \oscs~are always
stable. Each harmonic is then resolved into normal modes with outgoing
\gws~(time dependence $e^{i\omega t}$ with $\omega$ complex) and an
eigenvalue problem as posed for the eigenfunctions and the
eigenfrequencies $\omega$.  Five methods of solving the eigenvalue
problem are presented; three methods are valid in general (the method
of resonances, the variational method and the method of energy
conservation); one is valid on the slow-motion \app~(wavelength of
waves large compared to star) and one is valid in the weak-gravity
\app. For \sms~with weak gravity and with radially constant density
and shear modulus the eigenvalue problem is solved analytically. An
appendix develops a general theory of action principles for systems
with radiative boundary conditions -- a theory which is then used to
derive the action principles in the body of the paper and which could
be useful for a variety of other problems involving physical systems
coupled to radiation.\\[5mm]

\normalsize{\bf The \quad~\oscs~of~\nss}
--- Lee Lindblom and Steven L. Detweiler, \apjs{53}{73}{1983}

\abs We investigate here the influence of the \eos~of matter (in the
nuclear density regime) on the dynamical behavior of \nss. The
properties of the \quad~($f$ mode) \oscs~of \nss~constructed from 13
\eoss~are presented. An efficient algorithm for computing the complex
eigenfrequencies of the \nr~($p$ and $f$ mode) \oscs~of \nss~is
presented. The results of our computations are compared to relevant
astrophysical observations.\\[5mm]

\normalsize{\bf On the nonradial \pul~of general \rel~\sms}
--- Steven L. Detweiler and Lee Lindblom, \apj{292}{12}{1985}

\abs We present a nonsingular fourth-order system of equations
which describe the nonradial \puls~of \gr~\sms. These equations
represent an improvement over previous discussions in the 
literature which presented fifth-order systems or singular
fourth-order systems to describe these \puls. We also present
the nonsingular power-series solutions to these \eqs near $r=0$.\\[5mm]

\normalsize{\bf Nonradial $g$-mode \oscs~of warm \nss}
--- P.N. McDermott, H.M. Van Horn, J.F. Scholl, \apj{268}{837}{1983}

\abs We have computed adiabatic $l=1$ and 2, $p$-, $f$-, and $g$-mode
fluid \oscs~of finite-temperature \ns~models. We use the ``relativistic
Cowling \app,'' defined in this paper, in which all \perts~of the
\grav~potentials are neglected. This \app~is expected to be adequate
for the $g$-modes and even gives reasonable results for the periods of
$p$- and $f$-modes. [...] For the fluid $g$-mode \oscs~we have
studied, however, \grav~\rad~damping is exceedingly weak, typical
values of $\tau$ greatly exceeding the age of the universe. These very
long \grav~\rad~damping times are a consequence of the much longer
$g$-mode periods, which greatly reduce the relative power radiated,
and of concentration of the oscillatory motions to the outermost
several meters of the neutron star, which reduces the effective
\quad~moment of the \oscs. Damping of \ns~$g$-mode \oscs~will thus be
dominated by other processes. The confinement of the $g$-mode
eigenfunctions to the crust suggests that the effects of the nonzero
shear modulus (which we have not included) may be quite
important.\\[5mm]

\normalsize{\bf The \nr~\osc~spectra of \nss}
--- P.N. McDermott, C.J. Hansen, H.M. Van Horn, R. Buland, \apj{297}{L37}{1985}

\abs We have computed \nr~\osc~spectra for a number of realistic,
finite-temperature, nonrotating, nonmagnetic \nss~containing solid
crusts.  The spectra consist of $p$-modes (with periods $\Pi < 0.2$
ms), $g$-modes trapped in the fluid ``ocean'' layer ($\Pi > 100$ ms),
$g$-modes trapped in the fluid core ($\Pi > 10$ s), shear ($s$-) and
toroidal ($t$-) modes confined to the solid crust ($\Pi < 1$ ms), and
interfacial ($i$-)modes associated with the stellar surface ($f$-mode:
$\Pi\approx 0.2-0.4$ ms), the ``ocean''/crust interface ($\Pi\approx
10-100$ ms), and the crust/core interface ($\Pi\approx 10-100$ ms).\\[5mm]

\normalsize{\bf The accuracy of the \quad~\app~for the \grav~\rad~from
pulsating stars}
--- Eugene Balbinski, Steven Detweiler, Lee Lindblom, Bernard F. Schutz,
\mnras{213}{553}{1985}

\abs The methods for computing the damping times of the \nr~\oscs~of
stars due to \grav~\rad~emission are compared: (i) solving the
complete linearized \eqs~of \gr~and (ii) using an approximate method
based on the \quad~formula. It is shown that the results of the fully
\rel~calculation approach the results based on the \quad~for
sufficiently non-\rel~stars (i.e. $GM/c^2R$ sufficiently small).  the
accuracy of the \app~method is investigated by considering
\poly~stars with a range of \poly~indices. The \app~is worse
for larger \poly~indices, i.e. for greater condensation. We conclude
the when Newtonian models are used to approximate \rel~ones, models
with the same value of $GM/c^2R$ (rather than, say, central density)
should be compared.\\[5mm]

\normalsize{\bf Nonradial \oscs~of \nss}
--- P.N. McDermott, H.M. Van Horn, C.J. Hansen, \apj{325}{725}{1988}

\abs Finite-temperature \ns~models with a fluid core, solid crust,
and thin surface fluid ``ocean'' are capable of sustaining a broad diversity
of normal modes of \oscs. We have performed linear, adiabatic, Newtonian,
\nr~pulsation analyses of such stars, including the elastic
effects of the \ns~crust. [...] Damping mechanisms we have investigated
include \grav~\rad~damping, neutrino emission damping, electromagnetic
\rad~from an oscillating stellar magnetic field, nonadiabatic effects,
and internal friction and viscosity. Which of these mechanisms dominates
depends upon the type of mode, the spherical harmonic index $l$, the
equilibrium model, and other parameters. Finally, we suggest that surface
$g$-modes may be excited during the thermonuclear outbursts in X-ray
bursters and may lead to observable quasi-periodic \oscs.\\[5mm]

\normalsize{\bf Normal modes of a model radiating system}
--- Kostas. D. Kokkotas and Bernard F. Schutz, \grg{18}{913}{1986}

\abs In order to gain insight into normal modes of realistic radiating
systems, we study the simple model problem of a finite string and a
semi-infinite string coupled by a spring. As expected there is a family
of modes which are basically the modes of the finite string slowly
damped by the ``radiation'' of energy to infinity on the semi-infinite
string. But we also study another family of modes, found by Dyson in a
different model problem, which are strongly damped modes of the 
semi-infinite string itself. These may be analogous to the modes of
black holes, and they are likely to be present in relativistic stars
as well. The question of whether the instability in these modes which
Dyson found is present in realistic stars remains open.\\[5mm]

\normalsize{\bf Stellar resonant \oscs~coupled to \gws}
--- Yasufumi Kojima, \ptp{77}{297}{1987}

\abs Gravitational waves emitted by a particle moving in a circular 
orbit around a \sm~are examined. First-order \pert~\eqs~are solved
both inside and outside the star. A resonant \osc~of the star occurs
when the \freq~of the \gws~produced by the particle coincides
with the \qnm~of the star. The amplitude of the \gw~at the resonance
exhibits a sharp intrinsic peak depending on the \eos~used in the
equilibrium configuration.\\[5mm]

\normalsize{\bf Two families of normal modes in \rel~stars}
--- Yasufumi Kojima, \ptp{79}{665}{1988}

\abs Normal modes of \rel~stars are examined in order to study the
systems consisting of a star and \gws. It is found that there are two
classes of normal modes in such a radiating system. They are extremely
different in the decay rates of the complex eigen\freqs. One class is
a well-known family of weakly damped modes and the other is a family
of strongly damped modes, which are discovered for the first time in
numerical calculations for the \rel~stellar \oscs. The weakly damped
modes represent the slow decay of the stellar \oscs~due to \grav~\rad.
As the system becomes more \rel, the emission rate of \gws~increases.
As a result the decay rates of these modes increase. On the other
hand, the strongly damped modes essentially represent rapid decay of
the \gws~penetrated into the star because displacements of matter
inside the star are found to be very small. The decay rates of these
modes decrease as the system becomes more \rel.\\[5mm]

\normalsize{\bf The effect of viscosity on \ns~\oscs}
--- Curt Cutler, Lee Lindblom, \apj{314}{234}{1987}

\abs In this paper we compute the rate at which \ns~\oscs~are damped
by the presence of viscosity and thermal conductivity in the neutron
matter. In our computation we use fully \rel~\eqs~to describe the
\ns~\oscs, we use the best available expressions for the dissipation
coefficients in neutron matter, and we approximate the effects of
superfluidity on the dissipation mechanisms. We compute damping times
for a range of \ns~masses and temperatures, and we use a variety of
\eoss~to describe the highest density portion of the neutron matter.The
time scales computed here are used to improve previous estimates of
the maximum rotation rate of \nss.\\[5mm]

\normalsize{\bf Non-radial \puls~of \nss~with a crust}
--- Lee Samuel Finn, \mnras{245}{82}{1990}

\abs A fully \rel~treatment of \nr~pulsational modes of spherical
stars with solid crusts is presented. This work complements Schumaker
\& Thorne's \rel~treatment of torsional modes. The stain tensor for
arbitrary \ep~elastic deformations of a spherical star is derived.
A Hookean law is assumed to relate the stress to the strain, and the 
pulsational \eqs~and their \bcs~are derived from the Einstein field
\eqs~and the \eqs~of motion. The Einstein field \eqs~and \eqs~of motion
are then used to find the Lagrangian density and an action principle
for the pulsational problem. This action principle may prove useful
in future studies of the stability of \nr~pulsational modes
for stars with solid crust.\\[5mm]

\normalsize{\bf On the \nr~\oscs~of a star}
--- Subrahmanyan Chandrasekhar, Valeria Ferrari,\prsla{432}{247}{1991}

\abs A complete theory of the \nr~\oscs~of a static spherically
symmetric distribution of matter, described in terms of an energy
density and an isotropic pressure, is developed, {\it ab initio}, on
the premise that the \oscs~are excited by incident \gws. The \eqs, as
formulated, enable the decoupling of the \eqs~governing the \perts~of
the metric of the space-time from the \eqs~governing the 
hydrodynamical variables. This decoupling of the \eqs~reduces the
problem of determining the complex characteristic \freqs~of the 
\qnms~of the \nr~\oscs~to a problem in the scattering of incident
\gws~by the curvature of the space-time and the matter content of
the source acting as a potential. The present paper is restricted [..] 
to the case when the underlying \eos~is barotropic. The algorism
developed for the determination of the \qnms~is directly confirmed by
comparison with an independent evaluation by the extant alternative
algorism. Both polar and axial \perts~are considered. Dipole
\oscs~[...] are also treated as a particular simple special case.
[...]\\[5mm]

\normalsize{\bf On the \nr~\oscs~of a star. II Further amplifications}
--- Subrahmanyan Chandrasekhar, Valeria Ferrari, Roland Winston
\prsla{434}{635}{1991}

\abs The two algorisms that one uses for determining the complex 
characteristic \freqs, $\sigma = \sigma_0 + $i$\sigma_i$, belonging
to the \qnms~of \oscs~of a star, are shown to be equivalent. In the
first algorism, one searches directly in the complex $\sigma$-plane
to satisfy the requirements that at infinity there are only outgoing
waves (and no incoming waves). In the second algorism, one restricts
oneself to real $\sigma$s and selects the solution with the minimum
flux of \grav~\rad~at infinity.\\[5mm]

\normalsize{\bf On the \nr~\oscs~of a star. III A reconsideration of the
axial modes}
--- Subrahmanyan Chandrasekhar, Valeria Ferrari,\prsla{434}{449}{1991}

\abs It is shown that for stars with radii in the range $2.25 GM/c^2 <
R <$ ca. $3GM/c^2$, \qn~axial modes of \oscs~are possible. These modes
are explicitly evaluated for \sms~of uniform density.\\[5mm]

\normalsize{\bf On the \nr~\oscs~of a star. IV An application of the 
theory of Regge poles}
--- Subrahmanyan Chandrasekhar, Valeria Ferrari,\prsla{437}{133}{1992}

\abs It is shown how the flow of energy in the form of \grav~\rad, 
through a star in \nr~\oscs, can be determined by a suitable adaption
of Regge's theory of potential scattering in the quantum theory. While
the resonant scattering of axial \gws~can be treated by Regge's theory,
essentially, in its original form, the treatment of resonant scattering
of polar \gws~requires a generalization of the theory beyond its
standard range and the incorporation of a 'flux integral' derived
from independent considerations. Illustrative examples of the 
applications of the theory are provided.\\[5mm]

\normalsize{\bf Nonradial \puls~of \sms~in \gr}
--- James R. Ipser, Richard H. Price \prd{43}{1768}{1991}

\abs In the Newtonian theory, a method was developed recently that enables
one to solve the previously intractable problem of obtaining solutions
to the eigen\eqs~for the normal modes of rotating \sms. In this paper,
a closely related method is used to treat the \ep~\nr~modes of spherical
\sms~in \gr. In this treatment of generally nonbarotropic \perts, the
Lagrangian displacement is eliminated by solving for it in terms of a
certain fluid \pert~quantity $\bar\delta h$ (the enthalpy \pert~in the
barotropic case) and the Regge-Wheeler metric \perts~$K$ and $H_0$.
One is then led to an explicetly fourth-order system, consisting of
two second-order \eqs~for either pair $K$ and $\bar\delta h$ or $K$
and $H_0$. The fourth-order system for $K$ and $H_0$, from which all
fluid \pert~variables have been eliminated, is relatively simple and
provides a new way to compute \qn~\nr~modes and to analyze the interaction
of a \sm~with incident \gw. The analysis provide insight into the
connection between the Newtonian and \rel~theories of stellar \puls.
The relationship of the analyses to the recent work of Chandrasekhar
and Ferrari is discussed.\\[5mm]

\normalsize{\bf Relation of gauge formalisms for \puls~of
general-relativistic stellar models}
--- Richard H. Price, James R. Ipser,\prd{44}{307}{1991}

\abs 

\normalsize{\bf Initial-boundary value problem for the spherically
symmetric Einstein \eqs~for a perfect fluid}
--- Saskia Kind, J\"urgen Ehlers, \cqg{10}{2123}{1993}

\abs It is shown that for a given spherically symmetric distribution
of a perfect fluid on a spacelike hypersurface with boundary and a
given, time-dependent boundary pressure, there exists a unique,
local-in-time solution of the Einstein \eqs. A \ss~spacetime
can be attached to the fluid body if and only if the boundary pressure
vanishes. We assume a smooth \eos~for which the density and the speed
of sound remain positive for vanishing pressure.\\[5mm]

\normalsize{\bf Relativistic stellar \oscs~treated as an initial value
problem}
--- Saskia Kind, J\"urgen Ehlers, Bernd G. Schmidt, \cqg{10}{2137}{1993}

\abs The linearized Einstein equations for a static, spherically symmetric
fluid ball and its empty surroundings are considered. It is shown that,
given initial data obeying the constraints, there exists a unique
solution, which describes the motion of the perturbed fluid and the
\grav~waves propagating inside and outside the fluid ball. The physical
junction conditions for the boundary of the ball suffice to determine
the evolution inside and outside of the ball in terms of initial values.
The \eos~is assumed smooth and such that the density and the speed
of sound remain positive for vanishing pressure.\\[5mm]

\normalsize{\bf $W$-modes: a new family of normal modes of pulsating
\rel~stars}
--- Kostas D. Kokkotas, Bernard F. Schutz, \mnras{255}{119}{1992}

\abs We demonstrate explicitly the existence of a new family of
outgoing-wave normal modes of pulsating \rel~stars, the first such
family known that has {\it no analogue} in Newtonian stars. These
modes were discovered by the authors in a toy model, where they were
called strongly damped normal modes. Kojima then found the first
examples of these modes in realistic spherical \poly~\sms. Here
we give a number of arguments that demonstrate the existence of this
family unequivocally, and we calculate a large number of eigen\freqs.
Physically, the modes arise directly from the coupling of the fluid
\oscs~of the star to the \gw~\oscs of the space-time metric. Previously
studied modes of \rel~stars have been close analogues of modes of
Newtonian stars, where the coupling to \gws~mainly generated a small
imaginary part to the frequency. Such modes can be classified using
the Newtonian classes: $f$-, $p$-, $g$- and $r$-modes. The present
modes, by contrast, have {\it no} Newtonian analogue. They are
primarily \oscs~of the space-time metric in which the fluid hardly
vibrates at all. We christen them $w$-{\it modes} (\grav-$w$ave {\it
modes}). These modes are strongly damped, being characterized by
complex \freqs~with unusually large imaginary parts, comparable to
their real parts. We calculate a sequence of $l=2$ modes for a number
of spherical \poly~\sms. An interesting feature of $w$-modes is that
the lowest order mode of each sequence has a frequency similar to that
of lowest order mode of a \bh.  For higher modes, the spectrum
diverges from the \bh~spectrum, but shows remarkable similarity to
that of the strongly damped modes of the toy problem. As carriers of
\gw~information, $w$-modes may be important and observable in the
burst of \grav~\rad that follows the formation of a \ns. They should
also be essential in solving the problem of the completeness of the
outgoing-wave normal modes of radiating systems.\\[5mm]

\normalsize{\bf Nonradial \oscs~of \nss: A new branch of strongly damped
normal modes}
--- Markus Leins, Hans-Peter Nollert, Michael H. Soffel, \prd{48}{3467}{1993}

\abs We study small, \nr~\oscs of \nss~in a general \rel~\pert~treatment,
considering different values for the central density of the star. We
adapt two techniques used previously for the determination of \qnm~of
\bhs, allowing us for the first time to determine without \app~solutions
which satisfy purely outgoing \bcs~at spatial infinity. We confirm the
existence of strongly damped complex normal modes found by Kokkotas
and Schutz ($w$-modes). In addition, we identify a new branch of
strongly damped modes ($w_{II}$-modes). Our new modes are much more
similar to \qnms~of \bhs~than the $w$-modes known before.\\[5mm]

\normalsize{\bf Axial modes for \rel~stars}
--- Kostas D. Kokkotas, \mnras{268}{1015}{1994}
+ {\bf Erratum}, \mnras{277}{1599}{1995}

\abs For the class of stellar \oscs~named {\it odd}-parity of {\it axial}
modes, Chandrasekhar \& Ferrari have found that, under appropriate
conditions, there exists a spectrum with \freqs~describing
\qn~\oscs~having long damping times. In this work, we use different
(and more appropriate for this kinf od problem) numerical methods to
show the existence of an additional part of the axial mode spectrum
with \qnm~\freqs~having short damping times, for {\it precisely} the
same \sms~that Chandrasekhar \& Ferrari have considered. This part of
the axial modes spectrum is similar to the $w$-modes found earlier by
Kokkotas \& Schutz.\\[5mm]

\normalsize{\bf A new numerical approach to the \osc~modes of \rel~stars}
--- Nils Andersson, Kostas D. Kokkotas, Bernard F. Schutz,
\mnras{274}{1039}{1995}

\abs The \osc~modes of a simple \poly~\sm~are studied. Using a new numerical
approach (based on integration for complex coordinates) to the problem
for the stellar exterior we have computed the eigen\freqs~of the
highly damped $w$-modes. The results obtained agree well with the
recent ones of Leins, Nollert \& Soffel. Specifically, we are able to
explain why several modes in this regime of the complex \freq~plane
could not be identified within the WBK approach of Kokkotas \& Schutz.
Furthermore, we have established that the `kink' that was a prominent
feature of the spectra of Kokkotas \& Schutz, but which did not appear
in the results of Leins et al., was a numerical artefact. Using our
new numerical code we are also able to compute, for the first time,
several of the slowly damped ($p$)-modes for the considered \sms.  For
very compact stars, we find, somewhat surprisingly, that the damping
of these modes does not decrease monotonically as one proceeds to
higher \osc~\freqs. The existence of low-order modes that damp away
much faster that anticipated may have implications for questions
regarding stellar stability and the lifetime of \grav-wave
sources. The present results illustrate the accuracy and reliability
of the complex-coordinate method and indicate that the method could
prove to be of great use also in problems involving rotating
stars. There is no apparent reason why the complex-coordinate approach
should not extend to rotating stars, whereas it is accepted that all
previous methods will fail to do so.\\[5mm]

\normalsize{\bf Two simple models for \gw~modes of compact stars}
--- Nils Andersson \grg{28}{1433}{1996}

\abs Two simple model problems relevant for the \gw~modes of \rel~stars
are discussed. It is shown that the entire mode-spectrum can be
obtained if one considers the modes as arising because of the trapping
of \gws~by the spacetime curvature. The stellar fluid need play no
dynamical role.  Furthermore, it is shown that two distinct families
of \gw~modes exist.  The first corresponds to waves trapped inside the
star, while the second is similar to acoustic waves scattered off a
hard sphere. An infinite number of the first kind of modes exist, but
the latter family will only have a few members.\\[5mm]

\normalsize{\bf On the \osc~spectra of ultracompact stars: An extensive
survey of \grav-wave modes}
--- Nils Andersson, Yasufumi Kojima, Kostas D. Kokkotas  \apj{462}{855}{1996}

\abs An extensive survey of \grav-wave modes for uniform density stars
is presented. The study covers ranging in compactness from $R/M = 100$
to the limit of stability in \gr: $R/M = 9/4$. We establish that polar
and axial \grav-wave modes exist for all these stellar models. Moreover,
there are two distinct families of both axial and polar modes. We
discuss the physics of these modes and argue that one family is
primarily associated with the interior of the star, while the second
family is mainly associated with the stellar surface. We also show
that the problem of axial \perts~has all the essential features of the
polar problem as far as \gws~are concerned. This means that the axial
problem is much more important that has previously been assumed. We also
find some surprising features, auch as avoided crossings between the
polar \grav-wave modes and the Kelvin $f$-mode as the star becomes
very compact. This seems to suggest that the $f$-mode should be considered
on equal footing with the polar $w$-modes for ultracompact stars. All
modes may have the main character of trapped modes inside the curvature
potential barrier for $R/M < 3$.\\[5mm]

\normalsize{\bf Space-time modes of \rel~stars}
--- Nils Andersson, Kostas D. Kokkotas, Bernard F. Schutz,
\mnras{280}{1230}{1996}

\abs The problem of \rel~stellar \puls~is studied in a somewhat ad hoc
\app~that ignores all fluid motions. This Inverse Cowling Approximation
(ICA) is motivated by two observations. (1) For highly damped ($w$-mode)
\oscs~the fluid plays very little role. (2) If the fluid motion is
neglected, the problem for polar \osc~modes becomes similar to that for
axial modes. Using the ICA, we find a polar mode spectrum that has all
features of the $w$-mode spectrum for the full problem. Moreover, in the
limit of superdense stars, we find the ICA spectrum to be qualitatively
similar to that of the axial modes. These results clearly show the
importance of \gr~for the \pul~modes of compact stars, and that there are
modes whose existence does not depend on motions of the fluid at all,
namely pure `space-time' modes.\\[5mm]

\normalsize{\bf Gravitational waves and pulsating stars: What can we learn
from future observations?}
--- Nils Andersson, Kostas D. Kokkotas, \prl{77}{4134}{1996}

\abs We present new results for pulsating stars in \gr. First, we show
that the so-called \grav~modes of a \ns~can be excited when a
\gw~impinges on the star. Numerical simulations suggest that the modes
may be astrophysically relevant, and we discuss whether they will be
observable with future \gw~detectors. We also discuss how such
observations could lead to estimates of both the radius and the mass
of a \ns, and thus put constraints on the nuclear \eos.\\[5mm]

\normalsize{\bf Relativistic stellar \puls~with near-zone \bcs}
--- Lee Lindblom, Gregory Mendell and James R. Ipser, \prd{56}{2118}{1997}

\abs A new method is presented here for evaluating approximately the
\pul~modes of \rel~\sms. This \app~relies on the fact that
\grav~\rad~influences these modes only on time scales that are much longer
that the basic hydrodynamical time scale of the system. This makes it
possible to impose the \bcs~on the \grav~potentials at the surface 
of the star rather than in the asymptotic wace zone of the \grav~field.
This \app~is tested here by predicting the \freqs~of the outgoing
\nr~hydrodynamic modes of nonrotating stars. The real parts of the 
\freqs~are determined with an accuracy that is better than our knowledge
if the exact \freqs~(about 0.01\%) except in the most \rel~models where
it decreases to about 0.1\%. The imaginary parts of the \freqs~ are
determined with an accuracy of approximately $M/R$, where $M$ is the
mass and $R$ is the radius of the star in question.\\[5mm]

\normalsize{\bf Gravitational waves from pulsating stars: Evolving the
\pert~\eqs~for a \rel~star}
--- Gabrielle Allen, Nils Andersson, Kostas D. Kokkotas, Bernard,
F. Schutz
\prd{58}{124012}{1998}

\abs We consider the \perts~of a \rel~star as an initial value problem.
Having discussed the formulation of the problem (the \pert~\eqs~and
the appropriate \bcs~at the center and the surface of the star) in
detail we evolve the \eqs~numerically from several different sets of
initial data.  In all considered cases we find that the resulting
\gws~carry the signature of several of the star's \pul~modes.
Typically, the fluid $f$-mode, the first two $p$-modes and the slowest
damped \grav~$w$-mode are present in the signal. This indicates that the
\pul~modes may be an interesting source for detectable \gws~from colliding
\nss~or supernovae. We also survey the literature and find several 
indications of mode presence in numerical simulations of rotating core
collapse and coalescing \nss. If such mode-signals can be detected by
future \gw~antennae one can hope to infer detailed information about
\nss. Since a perturbation evolution should adequately describe the
late time behavior of a dynamically excited \ns, the present work can
also be used as a bench-mark test for future nonlinear simulations.\\[5mm]

\normalsize{\bf Pulsation modes for increasingly \rel~polytropes}
--- Nils Andersson, Kostas D. Kokkotas, \mnras{297}{492}{1998}

\abs We present the results of a numerical study of the fluid $f$-,
$p$- and the \grav~$w$-modes for increasingly \rel~\nr~polytropes.
The results for $f$- and $w$-modes are in good agreement with previous
data for uniform density stars, which supports an understanding of
the nature of the \gw~modes based on the uniform density data. We show
that the $p$-modes can become extremely long-lived for some \rel~stars.
This effect is attributed to the change in the perturbed density as 
the star becomes more compact.\\[5mm]

\normalsize{\bf Towards \gw~asteroseismology}
--- Nils Andersson, Kostas D. Kokkotas, \mnras{299}{1059}{1998}

\abs We present new results for pulsating \nss. we have calculated
the eigen\freqs~of the modes that one would expect to be the most
important \gw~sources; the fundamental fluid $f$-mode, the first
pressure $p$-mode and the first \gw~$w$-mode, for twelve realistic
\eoss. From these numerical data we have inferred a set of `empirical
relations' between the mode \freqs~and the parameters of the star (the
radius $R$ and the mass $M$). Some of these relations prove to be
surprisingly rebust, and we show how they can be used to extract
the details of the star from observed modes. The results indicate
that, should the various \pul~modes be detected by the new generation
of \gw~detectors that come online in a few years, the mass and the radius
of \nss~can be deduced with errors no larger that a few per cent.\\[5mm]

\normalsize{\bf The imprint of the \eos~on the axial $w$-modes of oscillating
\nss}
--- Omar Benhar, Emanuele Berti, Valeria Ferrari, {\bf gr-qc/9901037}

\abs We discuss the dependence of the pulsation \freqs~of the axial
\qnms~of a \nr~\ns~upon the \eos~describing the star interior.
The continued fraction method has been used to compute the complex
\freqs~for a set of \eoss~based on different physical assumptions
and spanning a wide range of stiffness. The \num~results show that
the detection of axial \gws~would allow to discriminate between models
underlying the different \eoss, thus providing relevant information
on both the structure of \ns~matter and the nature of the hadronic
interactions.\\[5mm]

\normalsize{\bf The inverse problem for pulsating \nss: A ``fingerprint
analysis `` for the supranuclear \eos}
--- Kostas D. Kokkotas, T.A. Apostolatos and Nils Andersson,
{\bf gr-qc/9901072}

\abs We study the problem of detecting, and infering astrophysical
information from, \gws~from a pulsating \ns. We show that the fluid
$f$ and $p$-modes, as well as the \gw~$w$-modes may be detectable from
sources in our own galaxy, and investigate how accurately the
\freqs~and damping rates of these modes can be infered from a noisy
\gw~data stream. Based on the conclusions of this discussion we
propose a strategy for revealing the supranuclear \eos~using the
\ns~fingerprints: the observed \freqs~of an $f$ and a $p$-mode. We
also discuss how well the source can be located in the sky using
observations with several detectors.\\[5mm]

\normalsize{\bf Construction of astrophysical initial data for
\perts~of \rel~stars}
--- Nils Andersson,  Kostas D. Kokkotas, Pablo Laguna, Phillippos
Papadopoulos and Michael S. Sipior, {\bf gr-qc/9904059}

\abs We develop a framework for constructing initial data sets for
\perts~about spherically symmetric matter distributions. This framework
faciliates setting initial data representing astrophysical sources of
\grav~\rad~involving \rel~stars. The procedure is based on
Lichnerowicz-York's conformal approach to solve the constraints in
Einstein's \eqs. The correspondence of these initial data sets in
terms of the standard gauge \pert~variables in the \RW~\pert~variables
is established, and examples of initial data sets of merging \nss
under the close-limit \app~are pesented.\\[5mm]

\newpage
\begin{center}
{\Large \bf Papers on Particles orbiting Black Holes and Neutron Stars}\\[1cm]
\end{center}

\normalsize{\bf Gravitational field of a particle falling in 
\ss geometry analyzed in tensor harmonics}
--- Frank J. Zerilli, \prd{2}{2141}{1970}

\abs We are concerned with the pulse of \grav~\rad~given off when a
star falls into a ``black hole'' near the center of out galaxy. We
look at the problem of a small particle falling in a \ss~background
(``black hole'') and examine its spectrum in the high-\freq~limit.  In
formulating the problem it is essential to pose correct \bcs:
\grav~\rad~not only escaping to infinity but also dissappearing down
the hole.  We have examined the problem in the \app~of linear
\perts~from a \ss~background geometry, utilizing the decomposition
into tensor spherical harmonics given by Regge and Wheeler (1957) and
by Mathews (1962). The falling particle contributes a
$\delta$-function source term (geodesic motion in the background
\ss~geometry) which is also decomposed into tensor harmonics, each of
which ``drives'' the corresponding \pert~harmonic. The power spectrum
radiated in infinity is given in the high \freq~\app~in terms of the
traceless transverse tensor harmonics called ``electric'' and
``magnetic'' by Mathews.\\[5mm]

\normalsize{\bf Stellar resonant \oscs~coupled to \gws}
--- Yasufumi Kojima, \ptp{77}{297}{1987}

\abs Gravitational waves emitted by a particle moving in a circular
orbit around a stellar model are examined. First-order \pert~\eqs~are
solved both inside and outside the star. A resonant \osc~of the star
occurs when the \freq~of the \gws~produced by the particle coincides
with the \qnm~ of the star. The amplitude of the \gw~at the resonance
exhibits a sharp intrinsic peak depending on the \eos~used in the
equilibrium configuration.\\[5mm]

\normalsize{\bf Gravitational \rad~from a particle in circular orbit
around a black hole. I. Analytic results for the nonrotating case}
--- Eric Poisson, \prd{47}{1497}{1993}

\abs [...] This paper is the first in a series that will carry out
detailed calculations relevant to [...] binaries, in the case where
one body is a small-mass \bh~of \ns~and the other is a much more massive
\bh, and the orbit is circular (aside from its gradual inspiral).
These papers will focus primarily on the emitted waveforms and
especially their phasing -- which is crucial for extraction of
information from the detectors' measurements. This first paper is
restricted to the case where the massive \bh~is \nr. The paper begins
by bringing the already well-developed formalisms for computing
waveforms (the ``Regge-Wheeler'' and ``Teukolsky'' formalisms) into a
combined form that is particularly well suited both for high accuracy
numerical calculations (carried out in paper II), and for analytic
calculations (this paper).[...] It is shown that propagation of the
waves through the intermediate zone (which connects the near zone and
the wave zone) distorts the waveforms and changes their power (and
hence phasing), at post$^{3/2}$-Newtonian order, in ways that have not
previously been computed -- except abstractly and nonconcretely as
formal ``tail terms'' in the waves. It is demonstrated that these 
post$^{3/2}$-Newtonian corrections will be of considerable importance
for the extraction of information from the waveforms that LIGO/VIRGO
expects to measure.\\[5mm]

\normalsize{\bf Gravitational \rad~from a particle in circular orbit
around a black hole. I. Numerical results for the nonrotating case}
--- Curt Cutler, Lee Samuel Finn, Eric Poisson, Gerald Jay Sussman, 
\prd{47}{1511}{1993}

\abs [...] We consider the particular case of a compact object
(e.g., either a \ns~or a stellar mass \bh) in a circular orbit about a
much more massive Schwarschild \bh. In this limit, the
\grav~\rad~luminosity can be calculated by integrating the Teukolsky
\eq.  Numerical integration us used to obtain accurate estimates of
the luminosity $dE/dt$ as a function of the orbital radius
$r_0$. These estimated are fitted to a post-Newtonian expansion of the
form $dE/dt = (dE/dt)_N\sum_k a_kx^k$, where $(dE/dt)_N$ is the
standard \quad~formula expression and $x \equiv (M/r_0)^{1/2}$. From
our fits we obtain values for the expansion coefficients $a_k$ up
through order $x^6$. While our results are in excellent agreement with
low-order post-Newtonian calculations, we find that the post-Newtonian
expansion converges slowly. Corrections beyond $x^6$ may be needed to
achieve the desired accuracy for the construction of template
waveforms.\\[5mm]

\normalsize{\bf Gravitational \rad~emitted when a mass falls onto a
compact star}
--- A. Borelli, \ncb{112}{225}{1997}

\abs In this paper we study the energy spectrum related to the axial
\perts~of a compact star when a particle falls spiralling onto it.
We find that both slowly-damped \qnms~and strongly damped $w$-modes
are excited, and that a part of the energy in the process is associated
to these $w$-modes. Our analysis will show substantial difference 
between the energy spectrum of compact stars and \bhs.\\[5mm]

\normalsize{\bf Stellar pulsations excited by a scattered mass}
--- Valeria Ferrari, L. Gualtieri, A. Borelli, gr-qc/9901060

\abs We compute the energy spectra of the \grav~signals emitted when
a mass $m_0$ is scattered by the \grav~field of a star of mass $M \gg m_0$.
We show that, unlike \bhs~in similar processes, the \qnms~of the
star are excited, and that the amount of energy emitted in these modes
depends on how close the exciting mass cna get to the star.\\[5mm]

\normalsize{\bf Gravitational waves from a test particle scattered
by a \ns: Axial mode case}
--- Kazuhiro Tominaga, Motoyuki Saijo, \prd{60}{024004}{1999}

\abs Using a metric \pert~method, we study \gws~from a test particle
scattered by a spherically symmetric \rel~star. we calculate the
energy spectrum and the waveform of \gws~for axial modes. Since metric
\perts~in axial modes do not couple to the matter fluid of the star,
emitted waves for a \ns~show only one peak in the spectrum, which
corresponds to the orbital \freq~at the turning point, where the
\grav~field is strongest. However, for an ultracompact star (the
radius $R \le 3M$), another type of resonant periodic peak appears in
the spectrum. This is just because of an excitation by a scattered
particle of axial \qnms, which were found by Chandrasekhar and
Ferrari. This excitation comes from the existence of the potential
minimum inside the star. We also find for an ultracompact star many
small periodic peaks at the \freq~region beyond the maximum of the
potential, which would be due to a resonance of two waves reflected by
two potential barriers (Regge-Wheeler type and one at the center of
the star). Such resonant peaks appear neither for a normal neutron
star nor for a \ss~\bh. Consequently, even if we analyze the energy
spectrum of \gws~only for axial modes, it would be possible to
distinguish between an ultracompact star and a normal neutron star (or
a \ss~\bh).\\[5mm]

\normalsize{\bf Excitation of the \op~\qnms~of compact objects}
--- Zeferino Andrade and Richard H. Price, gr-qc/9902062

\abs The \grav~\rad~generated by a particle in a close unbounded
orbit around a \ns~is computed as a means to study the importance 
of the $w$ modes of the \ns. For simplicity, attention is restricted
to \op~(``axial'') modes which so not couple to the \ns's fluid modes.
We find that for realistic \ns~models, particles in unbounded orbits
only weakly excite the $w$ modes; we conjecture that this is also the
case for astrophysically interesting sources of \ns~\perts. We also find
that for cases in which there is significant excitation of \quad~$w$
modes, there is comparable excitation of higher multipoles.\\[5mm]